\documentclass[twoside,openright]{report}
\usepackage{suthesis-2e}

\usepackage[utf8]{inputenc}
\usepackage{lipsum}
\usepackage{rotating}
\usepackage{graphicx}
\usepackage{color}
\usepackage{url}
\usepackage{caption}
\usepackage{subcaption}
\usepackage{makecell}
\usepackage{soul} 
\usepackage{diagbox} 
\usepackage{multirow}
\usepackage{hhline}
\usepackage[usenames,dvipsnames,svgnames,table]{xcolor}
\usepackage{textcomp} 
\usepackage{booktabs}
\usepackage[breaklinks=true, pagebackref=True]{hyperref}

\usepackage{breakcites}
\usepackage{adjustbox}

\usepackage{natbib}

\usepackage{mathtools}
\usepackage{float}
\usepackage{xspace}
\usepackage{amsmath}
\usepackage{amssymb}
\usepackage{dsfont}

\DeclareMathOperator*{\argmax}{arg\,max}

\renewenvironment{description}[1][0pt]
 {\list{}{\labelwidth=.25cm \leftmargin=#1
  }}
 {\endlist}

\usepackage[toc, acronym, numberedsection=autolabel]{glossaries}
\usepackage{glossaries-extra}
\GlsXtrEnablePreLocationTag{(page }{(pages }

\usepackage{minitoc} 

\newcommand{\MASK}[1]{}

\usepackage{tikz}
\usetikzlibrary{fit,shapes.geometric}

\newcounter{nodemarkers}
\newcommand\circletext[1]{%
    \tikz[overlay,remember picture]
        \node (marker-\arabic{nodemarkers}-a) at (0,1.5ex) {};%
    #1%
    \tikz[overlay,remember picture]
        \node (marker-\arabic{nodemarkers}-b) at (0,0){};%
    \tikz[overlay,remember picture,inner sep=2pt]
        \node[draw,ellipse,fit=(marker-\arabic{nodemarkers}-a.center) (marker-\arabic{nodemarkers}-b.center)] {};%
    \stepcounter{nodemarkers}%
}

\usepackage{subfiles} 

\makeglossaries

\newglossaryentry{wasabia}{type=\acronymtype, name=\glsadd{wasabi}{WASABI},
  description={Web Audio Semantic Aggregated in the Browser for Indexation},
  see=[Glossary:]{wasabi}}
\newglossaryentry{wasabi}{
    name=WASABI,
    first={Web Audio Semantic Aggregated in the Browser for Indexation (WASABI) \citep{meseguerbrocal_2017} \glsadd{wasabia}},
    description={a semantic database of song metadata collected from various music databases \citep{meseguerbrocal_2017} \url{https://wasabi.i3s.unice.fr/}}
}

\newglossaryentry{dalia}{type=\acronymtype, name=\glsadd{dali}{DALI},
  description={Dataset of Aligned Lyric Information},
  see=[Glossary:]{dali}}
\newglossaryentry{dali}{
    name=DALI,
    first={Dataset of Aligned Lyric Information (DALI) \citep{meseguerbrocal_2018}\glsadd{dalia}},
    description={our proposed dataset with synchronized audio, lyrics, and notes \citep{meseguerbrocal_2018}}
}

\newglossaryentry{mira}{type=\acronymtype, name=\glsadd{mir}{MIR},
  description={Music Information Retrieval},
  see=[Glossary:]{mir}}
\newglossaryentry{mir}{
    name=MIR,
    first={Music Information Retrieval (MIR) \glsadd{mira}},
    description={Music Information Retrieval is an interdisciplinary research field for understanding music audio signals that combines theories, concepts and techniques from music theory, computer science, signal processing perception and cognition}
}

\newglossaryentry{dfta}{type=\acronymtype, name=\glsadd{dft}{DFT},
  description={Discrete Fourier transform},
  see=[Glossary:]{dft}}
\newglossaryentry{dft}{
    name=DFT,
    first={Discrete Fourier transform (DFT) \glsadd{dfta}},
    description={Discrete Fourier transform is a frequency transformation that represents a signal as a sum of $N$ complex-valued Fourier coefficients (magnitude and phase) for sinusoids of varying frequency called “frequency bins”.}
}

\newglossaryentry{cqta}{type=\acronymtype, name=\glsadd{cqt}{CQT},
  description={Constant-Q transform},
  see=[Glossary:]{cqt}}
\newglossaryentry{cqt}{
    name=CQT,
    first={Constant-Q transform (CQT)~\citep{Brown_1991} \glsadd{cqt} },
    description={Constant-Q transform is a frequency transformation where the frequency bins are logarithmically spaced and with equal center frequencies-to-bandwidth ratios}
}

\newglossaryentry{mlsa}{type=\acronymtype, name=\glsadd{mls}{MLS},
  description={Log amplitude Scale},
  see=[Glossary:]{mls}}
\newglossaryentry{mls}{
    name=MLS,
    first={Log-amplitued Mel bands coefficients\glsadd{mls} },
    description={Mel log amplitude Scale is a frequency transformation that mimics the nonlinear critical bands of the human ear by applying a bank of triangular filters to the power of the spectrogram \gls{dft}}
}

\newglossaryentry{ml}{
    name=machine learning,
    description={Machine Learning is a research discipline that designs methods for enabling computers to learn to do particular tasks without being explicitly programmed to do so. Those methods learn from a collection of data where they automatically discover the needed patterns to carry out the desired tasks}
}

\newglossaryentry{ssma}{type=\acronymtype, name=\glsadd{ssm}{SSM},
  description={Self-Similarity Matrices},
  see=[Glossary:]{ssm}}
\newglossaryentry{ssm}{
    name=SSM,
    first={Self-Similarity Matrices (SSM) \glsadd{ssma}},
    description={are a \gls{mir} feature that, given a song described as a set of sequential elements $X = {x1,x2,...,xk}$, it computes a similarity measure between all the possible combinations of the set, creating a matrix that directly highlights similar elements}
}

\newglossaryentry{ncca}{type=\acronymtype, name=\glsadd{ncc}{NCC},
  description={Normalized Cross-Correlation},
  see=[Glossary:]{ncc}}
\newglossaryentry{ncc}{
    name=NCC,
    first={Normalized Cross-Correlation (NCC) \glsadd{ncca}},
    description={Normalized Cross-Correlation is a measure of similarity of two series as a function of the displacement of one relative to the other. It deals with singals that have different energy levels scaling the cross-correlation to a factor that is related to the energy of both signals}
}

\newglossaryentry{ismira}{type=\acronymtype, name=\glsadd{ismir}{ISMIR},
  description={International Society for Music Information Retrieval},
  see=[Glossary:]{ismir}}
\newglossaryentry{ismir}{
    name=ISMIR,
    first={International Society for Music Information Retrieval (ISMIR) \glsadd{ismira}},
    description={The International Society for Music Information Retrieval is the world's leading research forum on processing, searching, organising and accessing music-related data}
}

\newglossaryentry{svpa}{type=\acronymtype, name=\glsadd{svp}{SVP},
  description={Singing Voice Probability vector},
  see=[Glossary:]{svp}}
\newglossaryentry{svp}{
    name=SVP,
    first={Singing Voice Probability vector (SVP)\glsadd{svpa}},
    description={Singing Voice Probability vector is a function over time $\hat{p}(t)$ extracted from a song with $\hat{p}(t) \rightarrow 1$ when there is voice and $\hat{p}(t) \rightarrow 0$ when not}
}

\newglossaryentry{vasa}{type=\acronymtype, name=\glsadd{vas}{VAS},
  description={Voice Annotation Sequence},
  see=[Glossary:]{vas}}
\newglossaryentry{vas}{
    name=VAS,
    first={Voice Annotation Sequence (VAS)\glsadd{vasa}},
    description={The Annotation Voice Sequence vector is a function over time $\mathit{vas}(t)$ extracted from the annotation with $\mathit{vas}(t) = 1$ when the annotations have voice and $\mathit{vas}(t) = 0$  when not}
}

\newglossaryentry{svda}{type=\acronymtype, name=\glsadd{svd}{SVD},
  description={Singing Voice Detector},
  see=[Glossary:]{svd}}
\newglossaryentry{svd}{
    name=SVD,
    first={Singing Voice Detector (SVD)\glsadd{svda}},
    description={A Singing Voice Detector is a \gls{ml} model based on \gls{cnn} that compute the \gls{svp}}
}

\newglossaryentry{dtwa}{type=\acronymtype, name=\glsadd{dtw}{DTW},
  description={Dynamic Time Warping},
  see=[Glossary:]{dtw}}
\newglossaryentry{dtw}{
    name=DTW,
    first={Dynamic Time Warping (DTW)\glsadd{dtwa}},
    description={Dynamic Time Warping is a measure of similarity of two series that warps them onto a common set of instants such that the distance between them is the smallest. It uses dynamic programming to find monotonic alignment such that the sum of a distance-like cost between aligned feature vectors is minimized}
}

\newglossaryentry{dnna}{type=\acronymtype, name=\glsadd{dnn}{DNN},
  description={Deep Neural Networks},
  see=[Glossary:]{dnn}}
\newglossaryentry{dnn}{
    name=DNN,
    first={Deep Neural Networks (DNN)\glsadd{dnna}},
    description={Deep Neural Networks are machine learning algorithms organized in consecutive layers built on the output from the previous layer. They introduce a nonlinearity function between layers for model complex relationships between input and output}
}

\newglossaryentry{cnna}{type=\acronymtype, name=\glsadd{cnn}{CNN},
  description={Convolutional Neural Network},
  see=[Glossary:]{cnn}}
\newglossaryentry{cnn}{
    name=CNN,
    first={Convolutional Neural Network (CNN)\glsadd{cnna}},
    description={Convolutional Neural Network is a neural network architecture in which at least one layer is a convolutional layer i.e. a convolutional filter passes along an input matrix}
}

\newglossaryentry{cuneta}{type=\acronymtype, name=\glsadd{cunet}{C-U-Net},
  description={Conditioned-U-Net},
  see=[Glossary:]{cunet}}
\newglossaryentry{cunet}{
    name=C-U-Net,
    first={Conditioned-U-Net (C-U-Net)\glsadd{cunet}},
    description={our proposed conditioned architecture for \gls{source_separation}. It is based on a \gls{unet} architecture and adds \gls{film} layer to the encoder for controlling its behavior}
}

\newglossaryentry{filma}{type=\acronymtype, name=\glsadd{film}{FiLM},
  description={Feature-wise Linear Modulation},
  see=[Glossary:]{film}}
\newglossaryentry{film}{
    name=FiLM,
    first={Feature-wise Linear Modulation (FiLM)\glsadd{cunet}},
    description={layers conditions the network computation by applying an affine transformation to intermediate features}
}

\newglossaryentry{epf}{
    name=error probability function,
    description={is a function over time that predicts if a given label is an error or not with respect to an audio segment}
}

\newglossaryentry{multimodal}{
    name=multimodal,
    description={is a \gls{ml} discipline that studies how to use data from different domains that observe a common phenomenon toward resolving complex tasks}
}

\newglossaryentry{sl}{
    name=supervised learning,
    description={is a machine learning paradigm where the machine uses labeled data to discover functions that map input-output $f(x) = \hat{y}$}
}

\newglossaryentry{fully}{
    name=fully-connected,
    description={is a deep nueral network architecture where layers contain a set of connected neurons that act in parallel where each neuron is connected to all neurons in the previous layer}
}

\newglossaryentry{hidden}{
    name=hidden layer,
    description={is a deep nueral net intermediate layer that has as input the output of another layer and as output an intermediate features}
}

\newglossaryentry{sv}{
    name=singing voice,
    description={is a \gls{mir} topic that focus on the analysis of everything that is realted with the singing voice}
}

\newglossaryentry{medley}{
    name=MedleyDB,
    description={is a \gls{mir} multitrack dataset with 122 songs, with mix, stems of different instruments. The dataset is annotated in melody F0 (for 108 tracks), instrument activations and genre (for all tracks)}
}

\newglossaryentry{f0}{
    name=$f_0$,
    description={is a \gls{mir} task that estimates the fundamental note frequencies in polyphonic music computing a matrix over time with where each frame stores the note likelihoods}
}

\newglossaryentry{jamendo}{
    name=Jamendo,
    description={is a \gls{mir} dataset with 93 creative-commons licensed music pieces annotated by voice and no-voice}
}

\newglossaryentry{teacher}{
    name=teacher,
    description={is one of the agents of the \gls{teacher-student} and it is in charge of automatically labeling the training data of another system, the \gls{student}}
}

\newglossaryentry{student}{
    name=student,
    description={is one of the agents of the \gls{teacher-student} and it is trained using the labeled data from a \gls{teacher} acquiring its knowledge by mimicking the “teacher behaviour”}
}

\newglossaryentry{unet}{
    name=U-Net,
    description={is a \gls{cnn} architecture with a mirror encoder/decoder based that adds \gls{skip} connections between layers at the same hierarchical level in the encoder and decoder}
}

\newglossaryentry{dropout}{
    name=dropout,
    description={disables a random selection of a fixed number of neurons enabeling the neurons to be useful in conjunction with several random subsets of neurons}
}

\newglossaryentry{batch_norm}{
    name=batch normalization,
    description={is a deep neural network technique that normalizes the output of one layer before applying the activation function}
}

\newglossaryentry{conditioning}{
    name=conditioned learning,
    description={is a \gls{ml} paradigm where we want to process some information in the context of another. For that, we create a generic model that changes its behavior instead of having a dedicated model for each possible context information}
}

\newglossaryentry{data_aug}{
    name=data augmentation,
    description={consists of a series of techniques that artificially enriches or “augments” the training to better approximate the real-world and prevent overfitting}
}

\newglossaryentry{skip}{
    name=residual/skip connection,
    description={is a deep neural network technique that connects the output of one layer with the input of an earlier layer, so that each new layer deviate from the identity function, which still goes through the net. This leaves the outputs of the previous layers unchanged just that we could now do additional transformations}
}

\newglossaryentry{cleansing}{
    name=data cleansing,
    description={is a \gls{ml} strategy for cleaning erroneous labels in datasets}
}

\newglossaryentry{source_separation}{
    name=source separation,
    description={is a \gls{ml} task that aims to separate the different sources that appear in an audio mixture}
}

\newglossaryentry{teacher-student}{
    name=teacher-student paradigm,
    description={is a method for acquiring knowledge between \gls{ml} models where one systems (the \gls{teacher}) transferns its knowledge to another system (the \gls{student}). The \gls{teacher} automatically labels the training data of the \gls{student} acquiring the its knowledge by mimicking the “teacher behaviour”. This is done in the context of model compression or insufficient labeled training}
}

\dept{INFORMATIQUE, TÉLÉCOMMUNICATIONS ET ÉLECTRONIQUE DE PARIS}

\begin{document}

 \title{MULTIMODAL ANALYSIS: \\ Informed content estimation \\ and audio source separation}
 \author{Gabriel MESEGUER BROCAL}
 \principaladviser{Geoffroy PEETERS}

 \beforepreface
 \prefacesection{Abstract}

 This dissertation proposes the study of multimodal learning in the context of musical signals. Throughout, we focus on the interaction between audio signals and text information.
 Among the many text sources related to music that can be used (e.g. reviews, metadata, or social network feedback), we concentrate on lyrics.
 The singing voice directly connects the audio signal and the text information in a unique way, combining melody and lyrics where a linguistic dimension complements the abstraction of musical instruments.
 Our study focuses on the audio and lyrics interaction for targeting source separation and informed content estimation.

 Real-world stimuli are produced by complex phenomena and their constant interaction in various domains. Our understanding learns useful abstractions that fuse different modalities into a joint representation.
 Multimodal learning describes methods that analyse phenomena from different modalities and their interaction in order to tackle complex tasks. This results in better and richer representations that improve the performance of the current machine learning methods.

 To develop our multimodal analysis, we need first to address the lack of data containing singing voice with aligned lyrics. This data is mandatory to develop our ideas. Therefore, we investigate how to create such a dataset automatically leveraging resources from the World Wide Web. Creating this type of dataset is a challenge in itself that raises many research questions.
 We are constantly working with the classic ``chicken or the egg'' problem: acquiring and cleaning this data requires accurate models, but it is difficult to train models without data.
 We propose to use the teacher-student paradigm to develop a method where dataset creation and model learning are not seen as independent tasks but rather as complementary efforts.
 In this process, non-expert karaoke time-aligned lyrics and notes describe the lyrics as a sequence of time-aligned notes with their associated textual information.
 We then link each annotation to the correct audio and globally align the annotations to it.
 For this purpose, we use the normalized cross-correlation between the voice annotation sequence and the singing voice probability vector automatically, which is obtained using a deep convolutional neural network.
 Using the collected data we progressively improve that model. Every time we have an improved version, we can in turn correct and enhance the data.

 Collecting data from the Internet comes with a price and it is error-prone.
 We propose a novel data cleansing (a well-studied topic for cleaning erroneous labels in datasets) to identify automatically any errors which remain, allowing us to estimate the overall accuracy of the dataset, select points that are correct, and improve erroneous data.
 Our model is trained by automatically contrasting likely correct label pairs against local deformations of them.
 We demonstrate that the accuracy of a transcription model improves greatly when trained on filtered data with our proposed strategy compared with the accuracy when trained using the original dataset.
 After developing the dataset, we center our efforts in exploring the interaction between lyrics and audio in two different tasks.

 First, we improve lyric segmentation by combining lyrics and audio using a model-agnostic early fusion approach.
 As a pre-processing step, we create a coordinate representation as self-similarity matrices (SMMs) of the same dimensions for both domains. This allows us to easy adapt an existing deep neural model to capture the structure of both domains.
 Through experiments, we show that each domain captures complementary information that benefit the overall performance.

 Secondly, we explore the problem of music source separation (i.e. to isolate the different instruments that appear in an audio mixture) using conditioned learning. In this paradigm, we aim to effectively control data-driven models by context information.
 We present a novel approach based on the U-Net that implements conditioned learning using Feature-wise Linear Modulation (FiLM).
 We first formalise the problem as a multitask source separation using weak conditioning.
 In this scenario, our method performs several instrument separations with a single model without losing performance, adding just a small number of parameters.
 This shows that we can effectively control a generic neural network with some external information.
 We then hypothesize that knowing the aligned phonetic information is beneficial for the vocal separation task and investigate how we can integrate conditioning mechanisms into informed-source separation using strong conditioning. We adapt the FiLM technique for improving vocal source separation once we know the aligned phonetic sequence.
 We show that our strategy outperforms the standard non-conditioned architecture.

 Finally, we summarise our contributions highlighting the main research questions we approach and our proposed answers.
 We discuss in detail potential future work, addressing each task individually. We propose new use cases of our dataset as well as ways of improving its reliability, and analyze our conditional approach and the different strategies to improve it.

\prefacesection{Acknowledgments}

This thesis is dedicated to Elena for her unconditional love and support every step of our way. 12.

I have a lot of many amazing people to thank. Without them I would not have gotten to this point, their presence have made this possible.
I apologise in advance to those who I am forgetting. Sorry and thank you!

First of all, to my family. My parents, brother, and sister, who have always provided me unwavering support. You have brought me to where I am.

To all my professors who have guided me from the beginning of this journey. Especially to José Manuel Iñesta, Perfecto Herrera, Frédéric Bevilacqua and foremost Geoffroy Peeters for giving me the opportunity of doing a Ph.D and helping me during these years.

To my friends around the world from the different epochs of my life -Jona, Juan, Santi, Daps, Rob, Jakab, Filippo, Felipe, Derek, Matias, Martin, Magda, Tom, Phil, Adriana, Benjamin, Joseph, and Gino- because it does not matter when or where.

To the wonderful ISMIR community and all colleagues I have made there -Moha, Magdalena, Uri, Jordi, Delia, Helena, Zafar, Umut, and Javier-. It is a pleasure to constantly learn from you.

To my Spotify colleagues -Nicola, Simon, Keunwoo, Brian, Juanjo, Marco, Vincent, Ching, Andreas, David and Sebastian- for the stimulating experience of working with them. Especially to Rachel, being in this paragraph does not exclude you from the previous ones.

To IRCAM for being a unique place, to all the people who work there, and to the amazing researchers I had the pleasure to discuss with -Jordan, Dogac, Pierre, Philippe, Leo, Luc, Mathieu, Daniel, and Hugo- and my close friends -Guillaume, Alice, Hadrien and Aurelien- for lunches, dinners, drinks, translations, share hardships, and many other non-tangible things.

Finally, I am grateful to the French National Research Agency for providing the majority of funding for my PhD under the contract ANR-16-CE23-0017-01 (WASABI project).
\afterpreface

 \sloppy 

\graphicspath{{figs/}{introduction/figs/}}

\chapter{Introduction}
\label{sec:c_intro}

\section{Multimodal learning}
\label{sec:c_multimodal}

Real word stimuli are usually produced by various simultaneously occurring complex phenomena each expressed in its own domain that constantly interact.
Our understanding of these stimuli usually involves the fusion of different modalities into a joint representation.
\textbf{\Gls{multimodal} learning} aims at discovering the interaction between domains by developing methods to analyze phenomena from different modalities toward solving complex tasks.
Formally, it is the discipline that studies how to use data from different domains/modalities that observe a common phenomenon toward learning/resolving complex tasks ~\citep{Ramachandram_2017, baltruvsaitis_2018}.
\Gls{multimodal} learning gives the possibility to capture patterns that are not visible when working with individual modalities on their own, consolidating heterogeneous and disconnected data from various domains, and gaining an in-depth understanding of natural phenomena.
This produces more robust and richer representations to improve the performance of the current machine learning methods~\citep{baltruvsaitis_2018}.
For instance, if we want to study a musical artist, we can analyze his music. However, we will not have a complete vision since we are missing other dimensions e.g. lyrics, scores, video clips, album reviews, or interviews, that contribute to our idea of what a musical artist is and complement its purely musical dimension.

However, this comes with a certain cost and complexity~\citep{Atrey_2010}.
It is much harder to discover relationships across modalities than relationships among features in the same modality, each dimension is captured in a different way, resulting in totally different types of data and the modalities may be correlated or independent.
While correlated domains should work in a complementary manner and independent domains have to provide additional cues, the interaction between dimensions is hardly ever linear but has complex relationships and involves different abstraction levels.
Additionally, each modality might have a different relevance for accomplishing a particular task or there might be the absence of modalities at some instants resulting in missing values.
For example, we can have a better understanding of a song if we analyze not only its audio and but also the comments of its creator (e.g. an interview). However, we first will need to identify the part of the interview where the artist talks about that song, extract the relevant information, learn how it aligns with the audio, and complements it. While a comment about the intention of a melody can be the key to understand the song, a simple anecdote about the recording session can be seen as noise.

\begin{figure}[t!]
  \centerline{
    \includegraphics[width=.55\textwidth]{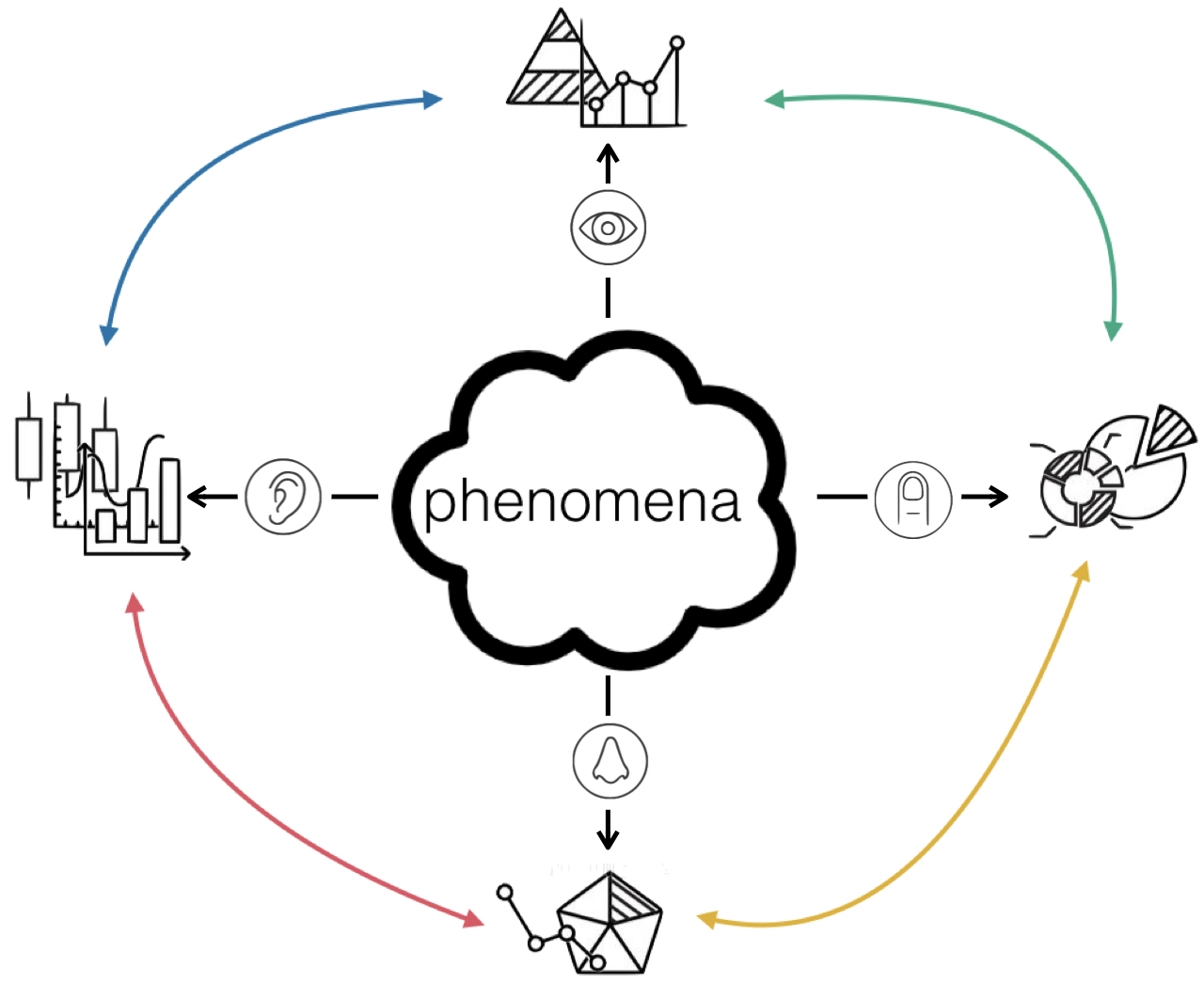}}
  \caption[Multimodal illustration]{Our understanding of complex phenomena extracts information from different dimensions that constantly interact. For instance, we can process the same phenomena through the senses of smell, vision, ear, or touch. We extract different knowledge from each domain and combine them to improve our understanding of reality.}
  \label{fig:multimodal}
\end{figure}

A good \gls{multimodal} learning model must satisfy certain properties that can be summarized in three main questions \textbf{how?}, \textbf{what?} and \textbf{when?} \citep{Bengio_2013, baltruvsaitis_2018}. These questions condition and define each \gls{multimodal} learning approach.

\subsection{How?}

It refers to how the \gls{multimodal} system is constructed i.e. defining the model and techniques used for designing the \gls{multimodal} system. It explores how the knowledge learned from one modality can help another modality. There are two main approaches: model-agnostic and model-based.
We can design models that are agnostic to the fact that the task is \gls{multimodal} or being explicitly dependent on it, addressing the interaction between modalities in their construction.
While in \textbf{model-agnostic} methods the \gls{multimodal} learning is not directly dependent on a specific technique, \textbf{model-based} methods explicitly approaches the multimodality in its architecture design.
Techniques that allow \textbf{model-based} architectures include 1) \textit{multiple kernel learning (MKL)}, a support vector machines (SVM) extension that use kernels for different modalities; 2) \textit{graphical models} with generative models for joint probability (variations of hidden Markov models, dynamic Bayesian networks or Boltzmann Machines) and discriminative models for conditional probability (conditional random fields); and 3) \textit{deep neural networks} for end-to-end training of \gls{multimodal} representation with a wide variety of designs.
Selecting one approach conditions how the following questions are addressed.
Currently, \textit{deep neural networks} are the most popular choice.
Table \ref{table:multimodal comparison} compares both \textbf{model-agnostic} and \textbf{model-based} methods.

Furthermore, we can also distinguish approaches regarding how the different modalities are used i.e. purely \gls{multimodal} or contextual relation. In \textbf{purely \gls{multimodal}} tasks, the different domains that look at the same phenomena work together to solve an external task. In \textbf{contextual} relation, one or more domains are used as `context information' or guide to improve the representation of another. The context domains provide clues of \textit{what} and \textit{where} to `look at'. This context information is an assistant but has a strong influence on the process. It helps to extract better content.

\subsection{What?}

It refers to what information is combined to have a better understanding of the phenomena.
At a high level, it defines the sources of information to be used.
Accessing to proper \gls{multimodal} data is essential to carry on \gls{multimodal} analysis.
Currently, there is a renewed interest in \textbf{\gls{multimodal} learning} thanks for the development of deep learning approaches.
Since they require large training datasets to be successful, researchers have created several large \gls{multimodal} annotated datasets~\citep{bernardi_2016}.
The most active communities (natural language processing and image processing) are those which use explicitly aligned datasets where there is a direct connection between sub-components of each modality.
However, there is still a lack of labeled \textbf{\gls{multimodal}} datasets for many \gls{multimodal} tasks that hinder their growth.
Creating \gls{multimodal} datasets is a challenging task as it requires annotations which often are time-consuming and difficult to acquire.

On deeper levels, this question involves all the aspects related to data representation such as how to exploit correlations (complementary and contradictory elements), deal with different levels of noise, discover independency and redundancy, establish the confidence of each modality, or find intermodality and intramodality relationships. This is challenging due to the heterogeneity of \gls{multimodal} data.
We can compute \gls{multimodal} representations either by considering each modality separately (each modality exists in its own space), but enforcing certain similarity or structure constraints to \textbf{coordinate} them, or by defining a \textbf{joint} representation that projects all the modalities into the same representation space.
While \textbf{model-agnostic} approaches tend to create coordinate representations, \textbf{model-based} methods compute joint representations.

This question also concerns the \textbf{translation} challenges i.e. how to translate one modality to another. It is common to define an example-based dictionary where elements from different modalities are directly linked. This allows retrieving information from one modality given a query from another. This idea is extended to models that can generate a translation between modalities. This goes beyond direct connections between elements and requires the ability to understand modalities to generate a new target sequence.

\begin{table}[H]
  \centering
  \caption[Multimodal approaches comparison]{A comparison between \textbf{model-agnostic} and \textbf{model-based} methods.}
  \resizebox{\textwidth}{!}{%

      \begin{tabular}{ p{.5\textwidth} | p{.5\textwidth} }

        \textbf{Model-based Multimodal Learning} & \textbf{Model-agnostic Multimodal Learning} \\
        \hline
        Features are learned from data and can be shared within dimensions. &
        Features are manually designed and require prior knowledge about the underlying problem and data.  \\ \hline


        Implicit dimensionality reduction within architecture. &
        Feature selection and dimensionality reduction are often explicitly performed.\\ \hline

        Have more flexiblilty for exploring different fusion types. Fusion architecture can be learned during training.  &
        Typically performs early or late fusion. Rigid fusion architecture usually handcrafted. \\ \hline

        Easily scalable in terms of data size and number of modalities. &
        Early fusion can be challenging and not scalable. Late-fusion rules may need to be defined.\\\hline



      \end{tabular}%
      }
      \label{table:multimodal comparison}
\end{table}

\subsection{When?}

 It refers to the problem of when to fuse the data. It can have two meanings:
\begin{enumerate}
  \item  \textbf{Horizontal - alignment:} it refers to which moment in the time should we fuse the different domains. It is related to \textbf{alignment} i.e. identifying the direct relations between sub-components from different modalities. Imagine the audio of a song and its lyrics. Although the lyric gives information about the theme of the song, it does not say anything about the characteristics of the audio signal unless it is aligned in time with it.
  This aspect also deals with missing data or modality problems but also with synchronization issues (at the input and the output level) for dynamic processes.  To tackle this challenge we need to measure similarities between different modalities and deal with possible long-range dependencies and ambiguities. While explicit alignment has a final goal of aligning sub-components between modalities, implicit alignment is used as an intermediate step for another task.

  \item \textbf{Vertical - fusion:} at which depth of the system should we integrate the information from multiple modalities. It is one of the most researched aspects of \gls{multimodal} machine learning. While \textbf{model-agnostic} methods fusion the different modalities independetly of the machine learning method, in \textbf{model-based} approaches, the fusion is dependent on the technique itself, explicitly addressing it in their construction.

  \begin{itemize}
      \item \textbf{Model-agnostic} fusion methods include \textit{early fusion - feature level} where features from the different domains are merged immediately after they are extracted, creating a higher dimensional space; \textit{late fusion - decision level} where each dimension is treated independently with parallel systems and the parallel decisions are merged to obtain the final decision; and \textit{hybrid fusion} that combines early and late fusion techniques. Model-agnostic approaches can be implemented using almost any unimodal domain.

      \item \textbf{Model-based} approaches are designed to have \gls{multimodal} fusion architecture at different depths of the process~\citep{Karpathy_2014}. Each \gls{multimodal} learning problem defines its own architecture.
  \end{itemize}

\end{enumerate}

\subsection{Multimodal tasks}

\Gls{multimodal} machine learning enables a wide range of applications, from human activity recognition and medical applications to autonomous systems and image and video description.
The combination of image (or video) and text is one of the most common \gls{multimodal} approaches.
There is a large number of works that investigate captioning and description for both image~\citep{bernardi_2016, Vinyals_2015, Xu_2015, Kiros_2014} and video~\citep{venugopalan_2014}.
\Gls{multimodal} learning has been also used for retrieving images after providing a text description~\citep{socher_2014}, reading lips into phrases~\citep{chung_2017}, aligning books to movies to provide rich descriptive explanations~\citep{Zhu_2015} or combining audio and video for speech recognition\citep{Ngiam_2011}.
Researchers also investigate the fusion of images and text into a joint representation~\citep{Srivastava_2014, Srivastava_2012, Higgins_2017}.
\citep{kaiser_2017} proposed to train a single model that learns multiple task-specific encoders and decoders that combine images, audio, and text to perform image classification, image captioning, and machine translation.
In audio, the most studied scenario is the fusion of audio and text for automatic speech recognition (ARS)~\citep{Graves_2006} (to transcribe the audio signal into text) or for text-to-speech synthesis~\citep{Oord_2016}.
Researchers have studied the co-occurrence of audio and visual events to train an audio network to correlate with visual~\citep{aytar_2016} or to find audio-visual correspondence task~\citep{Arandjelovic_2017}.
We refer to \citep{Ramachandram_2017}, \citep{Atrey_2010} and \citep{baltruvsaitis_2018} for a detailed survey on the different \gls{multimodal} approaches.

\section{Multimodality in music}

The most natural way to perceive music is through its acoustic rendering.
However, through history music has been described and transmitted in several other different forms~\citep{essid_2012}.
Before being able to store it, music was materialized and exchanged as musical-scores.
Since the growth of communication, music is essential to a wide range of disciplines.
For instance, it conveys emotions in movies and becomes visual art in audiovisual installations or album cover designs.
It is also widely described using text in editorial metadata or other social web content such as user-tags, reviews, or ratings.
We can even capture our mental perception of music~\citep{gulluni_2011}.
Not to mention that musicians have always performed music with motion and precise gestures.
Even if it has always been an encouragement to consider music content beyond audio signals~\citep{Liem_2011}, it has been treated mostly only through its acoustic dimension, not benefiting profoundly from the other perspectives.

The field that studies these music audio signals is \gls{mir}.
\gls{mir} is an interdisciplinary research field dedicated to the understanding of music that combines theories, concepts, and techniques from music theory, computer science, signal processing perception, and cognition.
Nowaday, researchers in \gls{mir} have a growing interest in understanding music through its various facets. 
Most of the studied music \gls{multimodal} tasks fall into one of these four categories: \textbf{classification}, \textbf{similarity}, \textbf{synchronization},  and \textbf{time-dependent} representation~\citep{simonetta_2019}.

\begin{figure}[t!]
  \centerline{
    \includegraphics[width=\textwidth]{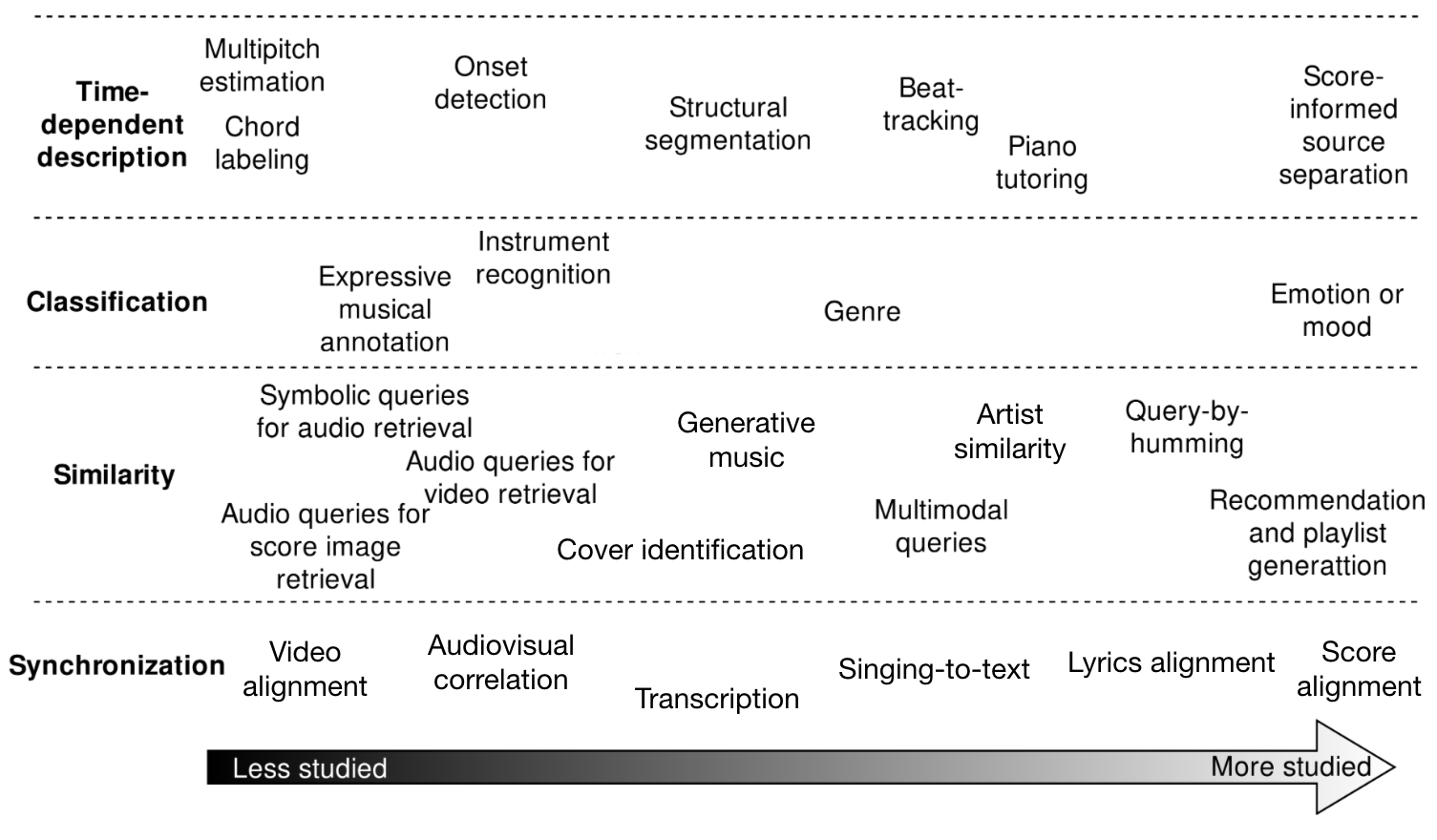}}
  \caption[Music multimodal tasks outline]{Outline of the different music multimodal tasks divided in 4 macro-tasks and plotted along a \textit{less-more} studied axis. Figure reprinted and adapted from~\citep{simonetta_2019}}
  \label{fig:multimodal_categories}
\end{figure}

\textbf{Classification} consists in assigning one or more labels to a song.
Although also commonly studied as a single domain, mood and genre classification are one of the most addressed \gls{multimodal} scenarios.
The multimodel attempts hybridize audio and lyrics to exploit the complementary information between musical features and Latent Semantic Analysis (LSA) text topics~\citep{Laurier_2008, Mayer_2008}.
Currently, mood and genre are awakening a new interest in the community.
While reviews or user-feedbacks are used for genre classificaiton~\citep{Oramas_2017b}, embedding music and lyrics produces advantageous models for mood classificaiton~\citep{Su_2017, Huang_2016, Xue_2015}.

\textbf{Similarity} refers to methods that measure the similarity between the content of different modalities.
It includes mostly retrieving documents through a query.
Multimodality arises from the fact that the query may be from a different domain than the retrieved document, for instance, query-by-lyrics~\citep{Muller_2007}.
Different domains can be combined to create better representations, e.g. artist similarity ranked from acoustic, semantic, and social view data~\citep{mcfee_2011} or more complete descriptions via embedding spaces from biographies, audio signal and available feedback data~\citep{Oramas_2017}.
We can even think in mapping the audio into a representation in the mind of the musicologist for complex electro-acoustic music~\citep{gulluni_2011}.
Another interesting task is generating new information in one domain given a query from another, by means of `generative models'.
For instance when generating images from audio and generate audio from images with a deep generative adversarial network~\citep{Chen_2017}.
The similarity between domains can also be implicitly modeled inside classification methods.

\textbf{Synchronization} tasks focus on the alignment in time or space between elements in different domains.
Lyrics and score alignment are the most popular problems followed by singing-to-text and score transcription.
Annotated chords progression modeled with Hidden Markov Models have been proved useful for improving lyric alignment \citep{Mauch_2010, Mauch_2012} or aligning the audio to them~\citep{McVicar_2011}.
Audiovisual correlation in music videos defines semantic relationships between the stream of audio and video~\citep{gillet_2007}.
Note how synchronization can be a pre-step for defining more complex similarity relationships between domains or for solving classification problems.
Additionally, some tasks such as singing-to-text and score transcription need a direct similarity measure of local elements to be solved.

\textbf{Time-dependent} tasks compute time-dependent descriptions of the music e.g. onset detection. The different domains are combined to enhance the description, for instance by using the video of the musician performing a piece to detect playing activity of the various instrument in multi-pitch estimation~\citep{dinesh_2017}.
Other examples are structure segmentation where we identify the music piece structure using video, lyrics, or scores~\citep{zhu_2005, cheng_2009, gregorio_2016} or audiovisual drum transcription that exploits both modalities~\citep{gillet_2005} or using the video as event detection guide~\citep{mcguinness_2007}.
Nevertheless, score-informed source separation is probably the most studied task~\cite{Ewert_2014}. We guide the separation using a musical score which is often strongly correlated in time and frequency with music.

In the next chapters, we provide an exhaustive literature review for the \gls{multimodal} tasks that are related to our work.
For further details on music \gls{multimodal} tasks and methods, we refer to \citep{essid_2012} and \citep{simonetta_2019}.

\section{Problem formalization}
\label{sec:c_formalization}

Research on multimodality receives a growing interest in \gls{mir}.
Multimodal \gls{mir} is an exciting field that tackles music problems more globally, exploiting the natural multidimensions of music. Nevertheless, it still grows at a much slower rate than other fields in the community.
Existing multimodal music approaches and datasets are not standardized and researchers are still establishing the foundations of it.

In this dissertation, we explore a defined \gls{multimodal} scenario, combining music audio and text information.
Text can refer to many different textual sources: editorial reviews, social web content, user-tags, or ratings.
Among all of them, we focus here on lyrics.
In popular music, lyrics have a direct connection to the audio signal via the singing voice, which is one of the most salient components in a musical piece~\citep{demetriou_2018}.
The singing voice acts as a musical instrument and at the same time conveys semantic meaning through the lyrics~\citep{humphrey_2018}.
It is the central element around which songs are composed, defining the lead melody and creating relationships between sound and meaning, adding a linguistic dimension that complements the abstraction of the musical instruments.
This connection tells stories and conveys emotions, improving our listening experience.
Some musicians even accentuate this connection by composing music that reflects the literal meaning of lyrics e.g. descending scales would accompany lyrics about going down, or happy and energic music would accompany lyrics about joy.
For these reasons, the singing voice is a very motivating and useful multimodal scenario.

\begin{figure}[t!]
  \centerline{
    \includegraphics[width=.9\textwidth]{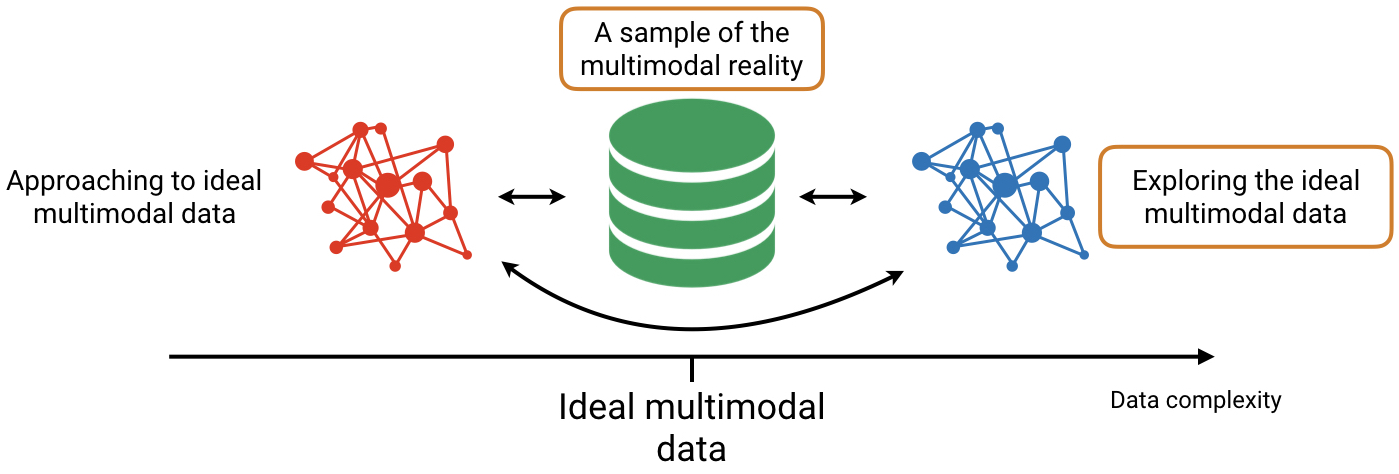}}
  \caption[Problem formalization]{We center our efforts in creating a multimodal dataset that acts as a sample of the multimodal reality we investigate, audio with lyrics aligned in time. We can then use it in two different directions, for either developing models that can produce such type of multimodal data or exploring a \gls{multimodal} formalization showing that this is beneficial to the performance of the models. We opt for the latter.}
  \label{fig:formalization}
\end{figure}

Our goal is to develop methods that use both lyrics and audio information to improve downstream \gls{mir} tasks.
Due to the relatively under-development of \gls{multimodal} analysis in music and in order to tackle the different \gls{mir} tasks, we need first to address generic \gls{ml} aspects.
All \gls{ml} problems have three core elements: the example \textbf{data} from which a system learns generic patterns to solve the task, the \textbf{system} itself (with all its components e.g. optimization or losses), and the \textbf{evaluation} process to check that the system behaves as expected. Since the emergence of neural networks, there have been exponential advances in the representation capabilities of \textbf{systems}. But datasets and \textbf{evaluation} techniques have surprisingly grown at a much slower rate~\citep{Sun_2017}.
Recenetly, fields like \textbf{active}, \textbf{weakly-supervised} or \textbf{semi-supervised} learning have appeared.
Nevertheless, there is still an absent of large and good quality datasets for music \gls{multimodal} analysis, limiting the development of new approaches.
Hence, we first investigate how to automatically create a large and good quality dataset with lyrics and vocal notes aligned in time.
The dataset is a sample of the multimodal reality we aim to investigate, but its automatic creation is often error-prone.
We then tackle questions related to the \textbf{evaluation} and to the problem of both training and evaluating in the presence of label noise, proposing a \textit{self-supervised} method to automatically identify possibly wrong labels.

Once the dataset is defined, it can be used in two different directions (see Figure~\ref{fig:formalization}). On one hand, we can use it to dig into tasks that can automatically transform the current data into the desired multimodal data, e.g. exploring tasks such as automatic lyrics alignment or singing voice transcription systems. On the other hand, we can investigate how to improve downstream \gls{mir} tasks, showing that a \gls{multimodal} formalization that exploits the natural multiple dimensions of music is beneficial for the performance of the models.
Finally, it can be used to train models that tackle both scenarios at the same time.
In this thesis, we focus only on exploring how to improve \gls{mir} tasks once we have access to the aligned data.
The two \gls{mir} tasks we study are: \textit{structure segmentation} and \textit{source separation}.
To use the audio and the lyrics, we study the conditioning of models which allows to guide the resolution of a problem based on external information (see Chapter~\ref{sec:cunet}).
Lastly and following the previous formalization, we define our \gls{multimodal} analysis of lyrics and music as follows:

\begin{itemize}
  \item \textbf{When?} to properly explore the relationship between the audio signal and its `meaning' (lyrics), we need an explicit alignment between lyrics and the audio. We develop our dataset having in mind this goal: to obtain a large amount of songs with their lyrics aligned in time.
  Since \textit{`when?'} can also refer to which moment in the learning process we are combining the lyrics and audio, we use \textbf{model-agnostic} fusion for structure segmentation (see Chapter~\ref{sec:structures}) and \textbf{model-based} approaches to condition the singing voice source separation with respect to the phoneme information (see Chapter~\ref{sec:vunet}).

  \item \textbf{What?} during the course of our work we explore several directions to use lyrics and audio. We first transform the audio signal to highlight vocal areas for the creation of the dataset in an \textbf{agnostic} way (see Chapter \ref{sec:dali_creation}) to adapt the lyrics alignment to the audio. We develop also a \textbf{joint} representation for structure segmentation and use the text information as prior knowledge (\textbf{context}) about the audio signal to condition a singing voice source separation model.

  \item \textbf{How?} The dataset creation belongs to the \textbf{semi-supervised}, \textbf{active} and \textbf{weakly} learning paradigms (see Chapter \ref{sec:dali_creation}). Although having access to this kind of data opens the door to many generative methods (e.g. automatically generating lyrics given a particular melody), the selected \gls{mir} tasks, \textit{structure segmentation} and \textit{source separation}, are moslty studied in a \gls{sl} paradigm using discriminative models.
  Our main learning machines are deep neural networks due to their flexibility and ability to learn a shared representation (see Chapter \ref{sec:tools}).
\end{itemize}

\section{Dissertation Summary and Contributions}

At the start of this work, the question of how to approach a \gls{multimodal} analysis of lyrics and audio remained open, and the current solutions study the use of both in a weakly aligned way.
The recent success of data-driven methods in many \gls{mir} tasks and the renewed interest in \gls{multimodal} analysis promise exciting times.
The goal of building a data-driven approach to explore the rich interaction between lyrics and audio gives rise to the need for large amounts of annotated data.
Despite the importance the singing voice has on how we enjoy music, there is a lack of large datasets of this kind and a relatively small amount of \gls{multimodal} work applied to this task. This opens a challenging area of research.
In this dissertation we address the following questions:

\begin{enumerate}
  \item How can we obtain large amounts of labeled data where lyrics and its melodic representation are aligned in time with the audio to train data-driven methods?
  \item How can we automatically identify and fix errors in these labels?
  \item How can we exploit the inherent relationships between lyrics and audio to improve the performance for lyrics segmentation?
  \item How can we effectively control data-driven models? Can we use prior knowledge about the audio signal defined by the lyrics to improve the isolation of the singing voice from the mixture?
\end{enumerate}

The remainder of this dissertation is organized as follows.
In Chapter~\ref{sec:tools}, we give an overview of core tools used to develop our ideas, including further discussion about \gls{sl} and relevant concepts from deep neural networks.
Chapter~\ref{sec:dali_description} provides a detailed description of our multimodal dataset with lyrics and vocal notes aligned in time at different levels of granularity. It also outlines the different versions and the characteristics of the data.
Chapter~\ref{sec:dali_creation} describes how we create the dataset where we explore Active learning and Weakly-supervised Learning techniques, creating an interaction between the dataset creation and model learning that benefits each other.
In Chapter~\ref{sec:dali_errors}, we deepened into the labeling errors and propose automatic solutions to several types of issues. However, we cannot measure if the new labels are better or worse.
In Chapter~\ref{sec:dali_correction}, we propose a novel data cleansing method for automatically knowing the current status of the dataset. Our method exploits the local structure of the labels to find possible errors in vocal note event annotations. This chapter is the last chapter concerning our multimodal dataset.
Chapter~\ref{sec:structures} explores lyrics segmentation as a first scenario to use text and audio, showing that they capture complementary structure.
Chapter~\ref{sec:cunet} provides a first approximation to conditioning learning for music source separation. We present there a novel approach for performing multitask source separation effectively controlling a generic neural network to perform several instruments isolations.
Chapter~\ref{sec:vunet} extends this approach to singing vocal source separation, using prior knowledge about the phonetic characteristics of the signal using strong conditioning to improve vocal separation.
Finally, we conclude and give directions for future work in Chapter~\ref{sec:con_fut}.

The \gls{multimodal} analysis of lyrics and audio is without a doubt still an open question.
We provide a new grain of sand here to help research grow this field.
We develop dataset-focused strategies and contribute a new dataset.
We also explore conditioning techniques that show that \gls{multimodal} formalizations that exploit the natural multidimensionality of music help to solve problems satisfactorily.
We are optimistic that this work will help future researchers to tackle this challenging topic with more resources and ideas.

\graphicspath{{figs/}{tools/figs/}}

\chapter{Tools}
\label{sec:tools}

In this thesis we use a common collection of tools to develop our ideas.
Broadly speaking, we employ techniques from the field of machine learning.
In this chapter, we give a high-level overview of this field and a more specific definition of the various tools we will use.

\section{Machine learning}
\label{sec:machine_learning_tools}


\textbf{\Gls{ml}} is a research discipline that designs methods for enabling computers to learn to do particular tasks without being explicitly programmed to do so.
Instead of defining any custom algorithm with specific logic, \gls{ml} methods learn its own logic from data, where they automatically discover the needed patterns to carry out the desired tasks~\citep{Goodfellow_2016}.
Within \gls{ml}, and based on the kind of data available, there are several ways of solving tasks.
They differ in their availability of accessing prior knowledge of what the output of the model should be.

\textbf{\Gls{sl}} uses a collection of labeled data where we specify the desired output for a given input. Using the labeled data, we can directly evaluate the accuracy of a model e.i. measuring how correct the answers are~\citep{Goodfellow_2016}.
However, the labels are not always available.
\textbf{Unsupervised learning} stands as a solution to these cases. There, models use unlabelled data (that is, just the input) to infer the natural structure within the set of data points.
Since we do not know the labels, most unsupervised learning methods have no specific way to measure model performances.
\textbf{Semi-supervised learning} takes a middle ground between the two previous approaches and combines both, labeled and unlabelled data.
It uses labeled datasets (usually smaller than unlabeled datasets) to extract knowledge, allowing to infer the labels of the larger unlabeled set.
Finally, \textbf{reinforcement learning} trains models focusing on the optimal way of making decisions.
In this paradigm, we provide feedback (rewards or penalties) for guiding models when they perform actions.
In this thesis, we mainly use \gls{sl} along with a specific \textbf{semi-supervised learning} method, called the \gls{teacher-student} (see Chapter~\ref{sec:teacher-student}).

\subsection{Supervised Learning}
\label{sec:supervised}

\Gls{sl} uses labeled data to discover functions that map input-output pairs.
When we talked about labeled data we refer to the scenario in which each input value is tagged with the answer the model needs to find on its own.
Thus, the learning process consists of identifying patterns in the input data that correlate with the desired target output.
Given a set $\mathcal{S} = (x_1, y_1), (x_2, y_2), ..., (x_{|\mathcal{S}|}, y_{|\mathcal{S}|})$ of input-output pairs where $x_n \in \mathcal{X}$ is the input space and $y_n \in \mathcal{Y}$ the target space, we aim to find a function $f_S$ with controllable parameters $\theta$ that capture the relationship between $x$ and $y$ so that:

\begin{equation}
  f_S(x_i, \theta) = y_i' \approx y_i
\end{equation}

\noindent where $y_i'$ is the output of the model and $y_i$ the real answer we aim to obtain.
We assume that the training pairs $(x_i, y_i)$ are drawn from an unknown joint probability distribution $p(x,y)$, of which we only know the training set $\mathcal{S}$.
We approximate the probability distribution $p(x,y)$ with $f$ by adjusting the parameters $\theta$ based solely on $\mathcal{S}$.
The ultimate goal of a trained model / learned function  $f_S(x_i, \theta)$ is to take new unseen data $x_i$ (not in $\mathcal{S}$) and correctly determine $f_S(x_i, \theta) = y'_i \approx y_i$ based only on the prior knowledge acquired. We call this process inference or prediction.
Note that the ``correct'' output is determined entirely during the training phase using only the data points of $\mathcal{S}$.
It is frequent to assume that the pairs $(x_i, y_i) \in \mathcal{S}$ are true, meaning that the label $y_i$ is the correct answer to $x_i$.
Nonetheless, this often does not hold.
Noisy and/or incorrect labels will certainly reduce the effectiveness of the model, and sometimes there is no clear-cut way to assign univocal labels (e.g., some chord or mood labels).
We study this matter in detail on Chapter~\ref{sec:dali_correction}.

Because of we have access to the labeled data, we can directly evaluate the accuracy of a trained model.
However, this does not necessarily reflect real-world performances since the data to which we have access may not contain all the cases we face in the real world.
The error of a model can be broken down into three distinct parts.
The first part is the irreducible error due to the noise in $p(x,y)$.
This error is intrinsic to the phenomenon being modeled and cannot be eliminated through good modeling practices.
The other two types of errors are related to the training dataset $\mathcal{S}$ and our model definition $f(\theta)$.

When creating $\mathcal{S}$, we want it to be a representative and well balanced (each class label is equally represented) description of the unknown joint probability distribution $p(x,y)$.
Our datasets are ``samples'' taken from an unfathomable reality.
Ideally, they would capture that reality in its essential aspects and guarantee good models.
But our sampling techniques and limitations lead us to unrepresentative samples.
As a result, $\mathcal{S}$ is often far from $p(x,y)$ and does not contain all the possibles $x$ nor the outputs $y$.
This produces a variation in performance between the direct evaluation compute on $\mathcal{S}$ and the real-word performance.
This difference is known as the \textit{variance} error and measures the amount by which the performance may vary for the various sets we can draw from $p(x,y)$.
It decreases augmenting the size and representativity of the training \textbf{data $\mathcal{S}$}.
If we define the expected peformance of our selected function $f(\theta)$ across all possible draws from $p(x,y)$ as $E[f(\theta)]$ and $f_S(\theta)$ as the actual performance on $\mathcal{S}$, we can formalize the \textit{variance} as:

\begin{equation}
  \mathit{Var}_\theta = E[(f_S(\theta)-E[f(\theta)])^2]
\end{equation}

Finally, there is always also the \textit{bias} error related to a specific task independently of the training data $\mathcal{S}$.
It refers to the constant inherent error to our particular formalization of the problem $f(\theta)$.
The \textit{bias} describes how far our selected $f(\theta)$ is from the ideal unknown function $f^*$ that describes perfectly the joint probability distribution $p(x,y)$.
It decreases augmenting the complexity of the \textbf{model} defined by the parameters $\theta$.
We can formalize \textit{bias} as:

\begin{equation}
  \mathit{Bias}_\theta = E[f(\theta)]-f^*
\end{equation}

\begin{figure*}[ht!]
  \centerline{
    \includegraphics[width=.6\textwidth]{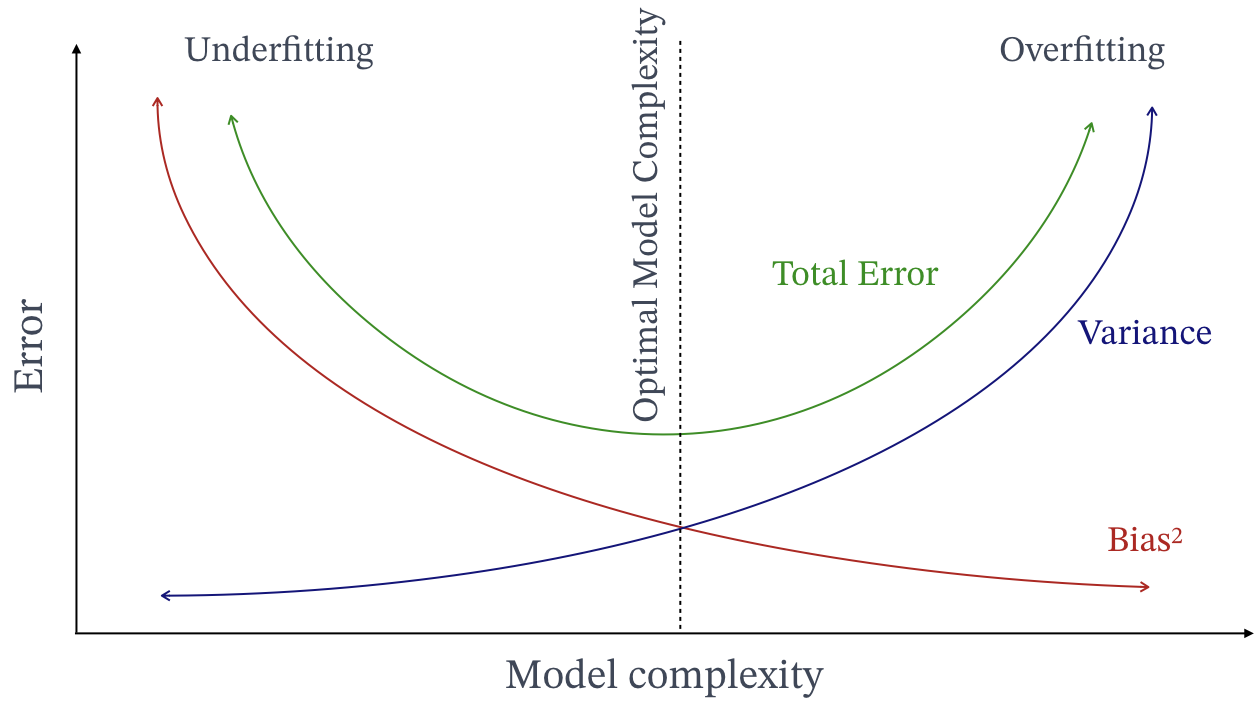}}
  \caption[Bias-variance trade-off]{Bias-variance trade-off. Complex models with high \textit{variance} learns a function that fits only the training set $\mathcal{S}$. Simple models with high \textit{bias} oversimplify the relationship between $x$ and $y$. Neither of these scenarios captures the actual structure in $p(x,y)$.}
  \label{fig:trade-off}
\end{figure*}

The goal of any \gls{sl} model is to achieve low \textit{bias} and low \textit{variance}.
The proper level of model complexity $\theta$ is generally determined by the nature of $\mathcal{S}$.
In an ideal scenario, we would be able to develop the perfect model using infinite training data, thereby eliminating all errors due to \textit{bias} and \textit{variance}.
That is it, the training process will result in a well-approximate minimum over the unknown $p(x,y)$.
Adjusting $\theta$ typically involves using a task-specific ``loss function'' $\mathcal{L}$ which measures the agreement between the output of the model $y'$ and the target $y$, i.e. the error that measures the model performance whose output is $y'$ with respect to the target output $y$.
Instead of computing the loss function for a single example, we average it for many training examples (ideally the whole training set $\mathcal{S}$).
This function is called ``cost function''.
It is also usual to use the term ``objective function'' to refer to any function optimized during training.
The choice of an objective function $\mathcal{L}$ depends on the characteristics of the target space $\mathcal{Y}$.
We minimize $\mathcal{L}$ over the training set $\mathcal{S}$ by adjusting $\theta$ being the total error due to both, \textit{bias} and \textit{variance}.
Unfortunately, we cannot directly calculate the contribution of each term (\textit{bias} and \textit{variance}) because we do not know the actual target joint probability distribution $p(x,y)$.
Moreover, \textit{bias} and \textit{variance} typically move in opposite directions of each other in balance known as \textit{bias-variance trade-off} (see Figure~\ref{fig:trade-off}).
In practice, we move between simple models (or with rigid underlying structure) that oversimplify the relationship between $x$ and $y$, reducing \textit{variance}, but potentially introducing \textit{bias} known as ``underfit''; to more complex models with reducing \textit{bias}, but potentially introducing \textit{variance} known as ``overfit''.
Since current models tend to be complex (with a larger number of $\theta$), the success of a model depends on $\mathcal{S}$.
If it is small, or not uniformly spread throughout different possible scenarios, complex models will ``overfit'', i.e. learning a function that fits only $\mathcal{S}$ very well capturing randomness in the data and going beyond the true signal into the noise, without learning the actual trend or structure in $p(x,y)$.
This results in unnecessarily complicate the relationship between $x$ and $y$ and therefore tends to generalize poorly.

We measure the evolution of the \textit{bias-variance trade-off} by dividing the training set $\mathcal{S}$ into three sets: training, validation, and test.
The training set is the actual set used for training our model.
The test set measures the expected performance of our model in the real-world. Ideally, we want our test set to a be different draw to $\mathcal{S}$ of $p(x,y)$, which will reflect a more precise performance. However, this is not always possible.
The validation set is used for finding the optimal model complexity with the minimum error.
We do so by comparing during training how the error evolves in the training set against the validation set.
The training set increases the model complexity, i.e. minimizing the $bias^2$.
Testing each version of our model in the validation set, we can have an approximation of the $variance$.
Additionally, the validation set is also used for tuning the internal control parameters of our model.

\subsection{Neural Networks}
\label{sec:neural}

\begin{figure*}[ht!]
  \centerline{
    \includegraphics[height=.2\textheight]{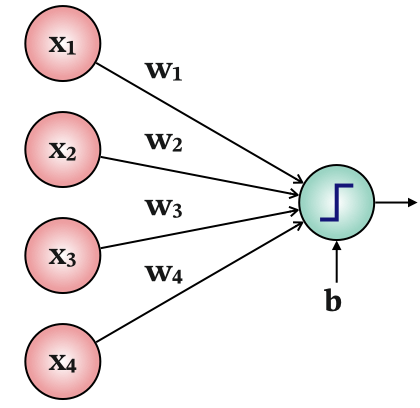}}
  \caption[A neuron]{A neuron computes a vector to scalar operation and apply a non-linearity operation to the result.}
  \label{fig:neuron}
\end{figure*}

Most of the machine learning algorithms used in this thesis come from the \gls{dnn} class of models.
\gls{dnn} models break down and distribute tasks onto \gls{ml} algorithms that are organized in consecutive layers built on the output from the previous layer.
Loosely inspired by the brain (where the name `neural network` arises) where neurons are associated one to another passing information, \gls{dnn} algorithms consist of a sequence of non-linear processing stages passing information to each other.
The basic unit is a ``neurons'' (see Figure~\ref{fig:neuron}):

\begin{equation}
  h(x) = \sigma (Wx + b)
\end{equation}

where $\sigma$ is a non-linear activation function, $x \in \mathbb{R}^D$ is the input of the layer, $W \in \mathbb{R}^{1xD}$ the weight matrix, $b$ the \textit{bias} vector.
The bias term $b$ helps models to represent patterns that do not necessarily pass through the origin.
The weights $W$ perform a linear transformation of the input data $x$. Indeed, without a non-linear activation function, a \gls{dnn} architecture is just a group of linear transformations.
Both $W$ and $b$ are parameters that the model has to learn during the training process.
The activation function $\sigma$ is the key element and introduces non-linear properties to the network for learning complex relationships between input and output.

\begin{figure*}[ht!]
  \centerline{
    \includegraphics[width=0.6\textwidth]{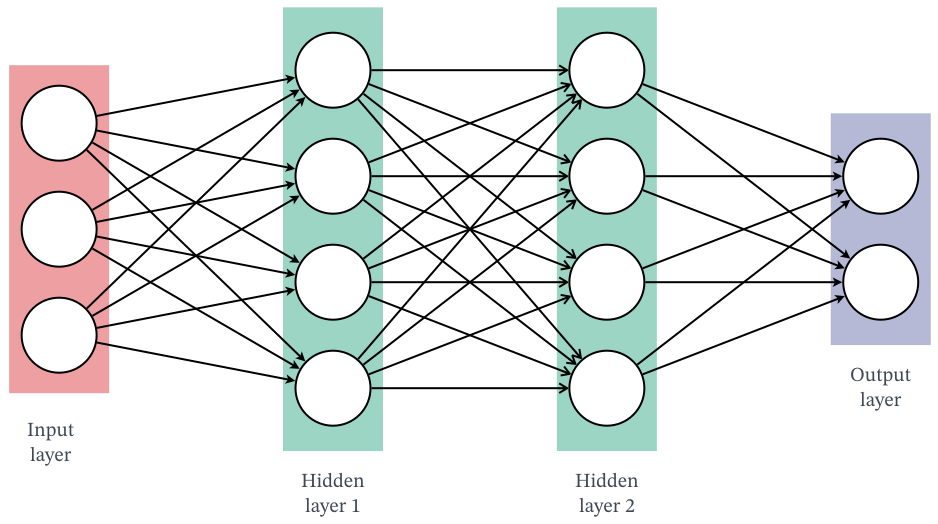}}
  \caption[Hidden layers]{Deep neural networks consist of a set of layers that progressively transform the input information into the output data, discovering the necessary abstractions for solving a task.}
  \label{fig:hidden}
\end{figure*}

\gls{dnn} models are composed of layers.
There are three main types of layers: the \textit{input} layer that receives $x$, the \textit{output} layer that generates $y'$ and at least one \textit{`\gls{hidden}'}.
\textit{\Glspl{hidden}} have as input the output of another layer (not the original one $x$) and output also intermediate features (not the final output $y'$).
\textit{\Glspl{hidden}} are in charge of capturing complex relationships by progressively computing a more `abstract' representation of the input $x$, i.e. the first layer detects a first abstraction of $x$, the second layer an abstraction of the first abstraction, the third layer abstraction of those abstractions, and so progressively.
Each layer consists of a set of simply connected neurons that act in parallel (see Figure~\ref{fig:hidden}).
Each neuron in a layer is connected to all neurons in the previous layer.
It receives the inputs and computes its own activation value (a vector-to-scalar function), capturing a different input combination~\citep{Goodfellow_2016}.
This makes each neuron independent to the rest of neurons of the same layer (they do not share any connections).
This architecture is called \textit{``\gls{fully}''}.

\gls{dnn} models make predictions by ``forward propagating'' the input data through the network layer by layer to the final layer which outputs a prediction $y'$.
The final prediction can be viewed as a long series of equations of the input.
This process is known as \textbf{forward propagation}.

The variable $\theta$ defines the total number of parameters to learn.
We adjust them by minimizing an objective function $\mathcal{L}$ using \textbf{Backpropagation}~\citep{Rumelhart_1986}.
Backpropagation is the method that computes the gradient of $\mathcal{L}$, measuring the deviation between the network's output $y'$ and the target output $y$ with respect to $\theta$.
At the heart of backpropagation is an expression for the partial derivative $\frac{\partial \mathcal{L}}{\partial w}$ of the cost function $\mathcal{L}$ with respect to any parameter $\theta$ in the network.
In other words, backpropagation tells how much the objective function $\mathcal{L}$ changes when a parameter changes, i.e. how the overall behavior of the net is affected by each parameter.
Similar to forward propagation, the model error (differences between $y'$ and $y$) is ``back propagated'' layer by layer through the output to the input layer updating each parameter progressively.
Using the partial derivatives $\frac{\partial \mathcal{L}}{\partial w}$, we update $\theta$ using gradient descent methods.
These methods minimize functions by iteratively moving in the error surface in the direction of steepest descent, as defined by the negative of the gradient down toward a minimum error value.
The most utilized gradient descent method is stochastic gradient descent (SGD). At each training iteration, SGD updates each parameter by subtracting the gradient of the loss with respect to $\mathcal{L}(\theta_t)$, scaled by the \textbf{``learning rate''} $\eta$. The resulting product is called the gradient step in the error surface:

\begin{equation}
  \theta_{t+1} = \theta_{t} - \eta \nabla \mathcal{L}(\theta_t)
\end{equation}

Current $\mathcal{S}$ used for training \gls{dnn} cannot be employed all at once. Instead, we train over a tiny subset called a \textbf{``minibatch''}. Minibatches are sampled randomly at each iteration of the gradient descent.
The size of the \textbf{minibatch} (also named just as ``batch size'') conditions the \textbf{learning rate} value.
The optimal value also depends on the morphology of $\mathcal{S}$.
Minibatches cause the objective function to change stochastically at each iteration of optimization. If it is small and the learning rate large, we can move far from the desired minimum error value. On the other hand, it is is too small, we may never reach it.

\begin{figure*}[ht!]
  \centerline{
    \includegraphics[width=.8\textwidth]{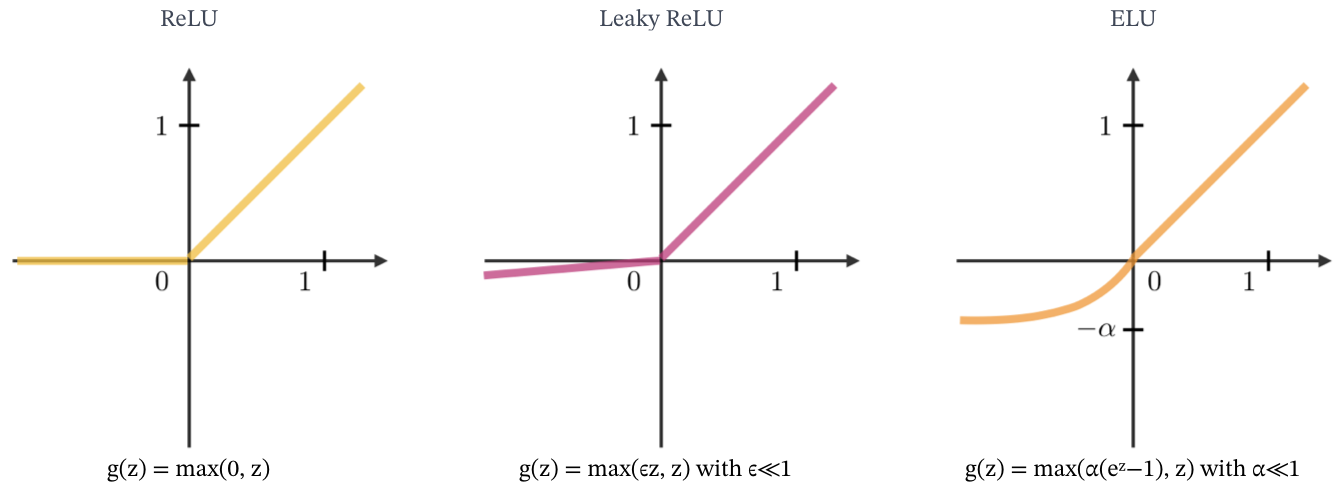}}
  \caption[non-linear activations]{Rectifier non-linearities activation function. The rectified linear unit layer (\textbf{ReLU}) represents a nearly linear function with a threshold operation where values below zero are set to zero. \textbf{LeakyReLU} and \textbf{ELU} are alternatives with a non-zero derivate for negative values. Hence, we can backpropagate the error also in these values. Figure copyright \href{https://stanford.edu/~shervine/teaching/cs-230/cheatsheet-convolutional-neural-networks}{CS 230 Deep learning}.}
  \label{fig:non-linear}
\end{figure*}

When designing a \gls{dnn} there are many choices to be made such as the number of layers (depth of the net) and the number of neurons in each one.
The number of neurons defines how many different input combinations.
The number of layers is connected with the capacity of the model to find hierarchical transformation with more and more abstract representations.
Both are related to the complexity of the model and the \textit{bias}-\textit{variance} trade-off.
Increasing them increases modeling power but also exacerbates overfitting.
To overcome the overfitting, it is usual to use regularization techniques that simplify the model.
These factors penalize on the model's complexity to ensure that the optimized neural network’s \textit{variance} is not too high.
Another important ingredient is the non-linearity $\sigma$ applied at each neuron.
This is essential for the model not to reproduce a linear combination of the inputs and discover complex relationships.
There are many different functions such as logistic or hyperbolic tangent.
However, the most usual choices are among rectifier non-linear functions (see Figure~\ref{fig:non-linear}) due to its computational efficiency, i.e. its tendency to produce sparse representation and to reduce the vanishing gradient problem when the gradients of the loss function approaches zero, making the network hard to train~\citep{Nair_2010, Glorot_2011}.

Many of the concepts presented in this section uses the \textit{\gls{fully}} architecture as an illustrative example are common to other \gls{dnn} architectures.

\subsection{Convolutional neural networks}
\label{sec:convolutional}


The main limitation of \textit{\gls{fully}} architectures is that they do not scale well.
For instance, if we want to process an image of 64x64x3 (64 wide, 64 high and 3 color channels), we will need 12288 learnable weights for a single neuron.
Furthermore, we are almost certain that we want to have many of such neurons to compute different combinations and several \glspl{hidden} to obtain complex relationships.
As a result, we are increasing the complexity and the number of parameters $\theta$ of our model which would quickly lead to overfitting.
\gls{cnn} architectures are ordinary \gls{dnn} that make the explicit assumption that the inputs are images~\citep{LeCun_1995}.
This allows efficient implementations, reducing vastly the number of parameters.
\gls{cnn} are also made up of neurons with learnable weights and biases, that perform a vector-to-scalar operation followed by a non-linearity activation.
We also train them with an objective function $\mathcal{L}$ that expresses a single differentiable score.

To constrain the architecture noticeably, \gls{cnn} take advantage of the image characteristics.
A digital image is a three dimensional function $i(w, h, d)$, where $w$, $h$ and $d$ are spatial coordinates.
Any coordinate is usually called a `pixel` and its amplitude $i(w, h, d)$ `intensity`.
Connectivity between pixels and spatial correlation (a pixel depends on both itself and its surrounding) are fundamental concepts that define the basic image components.
These components make possible more complex attributes to finally create concepts such as a dog or a nose.
In summary, pixel position and neighborhood have semantic meanings and elements of interest can appear anywhere in the image.

\begin{figure*}[ht!]
  \centerline{
    \includegraphics[width=.9\textwidth]{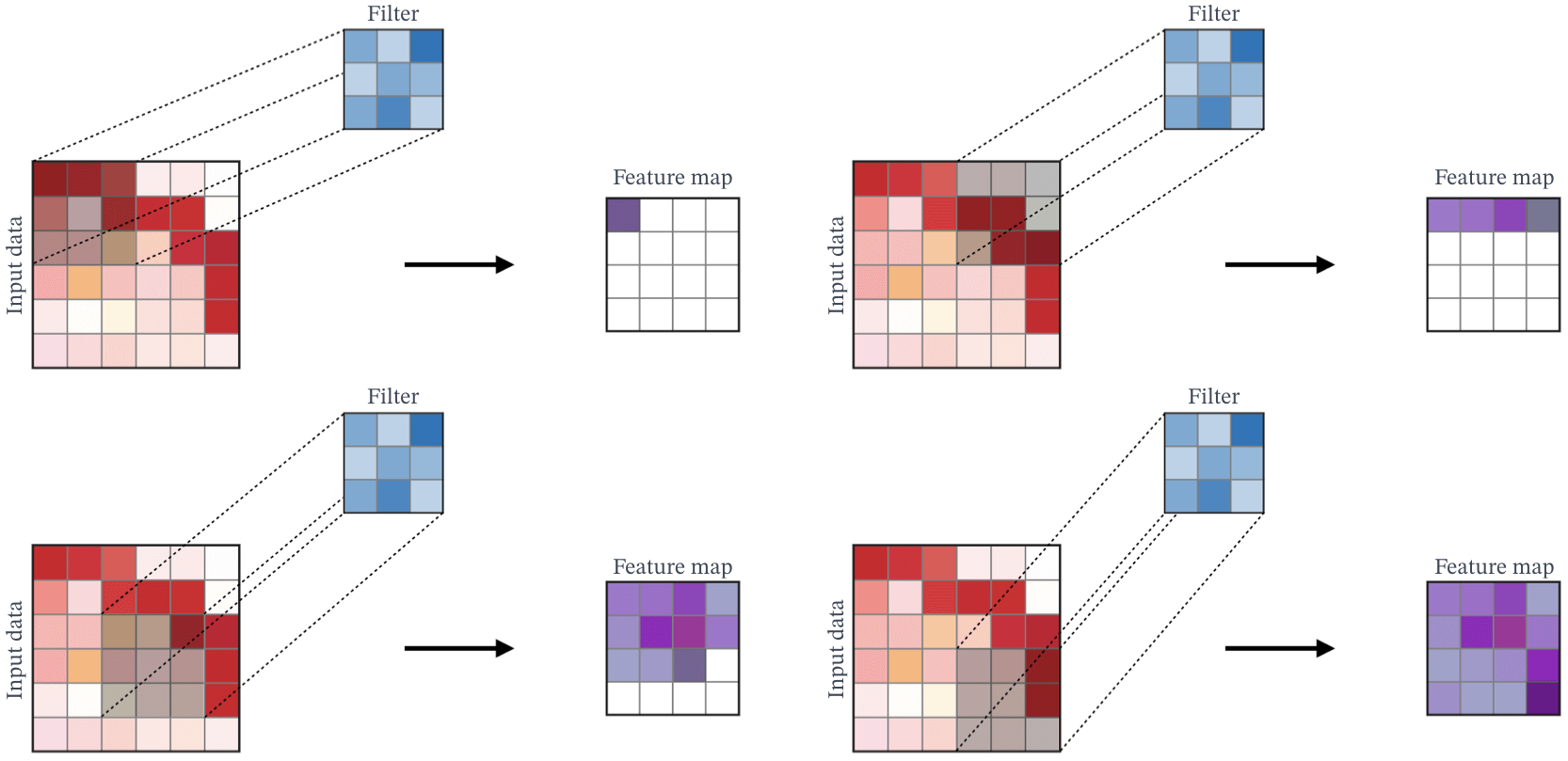}}
  \caption[Convolutional operation]{A convolution operation consists of aplying a filter that scans the input image with respect to its dimensions. Figure copyright \href{https://stanford.edu/~shervine/teaching/cs-230/cheatsheet-convolutional-neural-networks}{CS 230 Deep learning}.}
  \label{fig:conv}
\end{figure*}

The core concept of \gls{cnn} is filtering. Filters (also called kernels) have been used since the foundations of the image processing domain.
They are in charge of detecting image attributes, defining in which locations they occur and how strongly they seem to appear.
We apply a filter over the whole image using a convolution operation (see Figure~\ref{fig:conv}).
When convolving a filter, we slide it over the whole image.
At each location, we compute an element-wise multiplication between each filter element and the input elements it overlaps, summing up the result to obtain the output in the current filter location.
As a result, we obtain a matrix that captures the activations of the filter over the whole image, i.e. whether a certain feature is present at a given location in the image.
If something moves in the input image, its activation will also move by the same amount in the output.

\begin{figure*}[ht!]
  \centerline{
    \includegraphics[width=.9\textwidth]{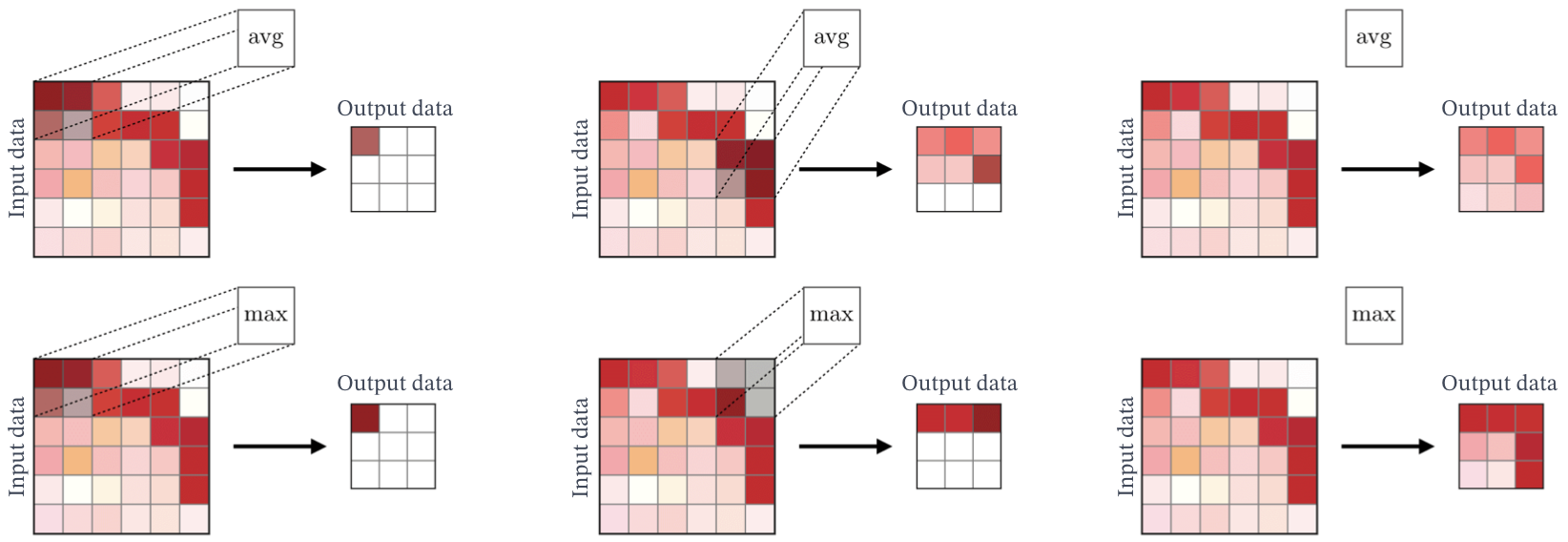}}
  \caption[Pooling operation]{Two pooling strategies. \textbf{Max pooling}: we select the maximum value of the filter. \textbf{Mean pooling}: we average the values of the filter. Figure copyright \href{https://stanford.edu/~shervine/teaching/cs-230/cheatsheet-convolutional-neural-networks}{CS 230 Deep learning}.}
  \label{fig:pool}
\end{figure*}

\begin{figure*}[ht!]
  \centerline{
    \includegraphics[width=.5\textwidth]{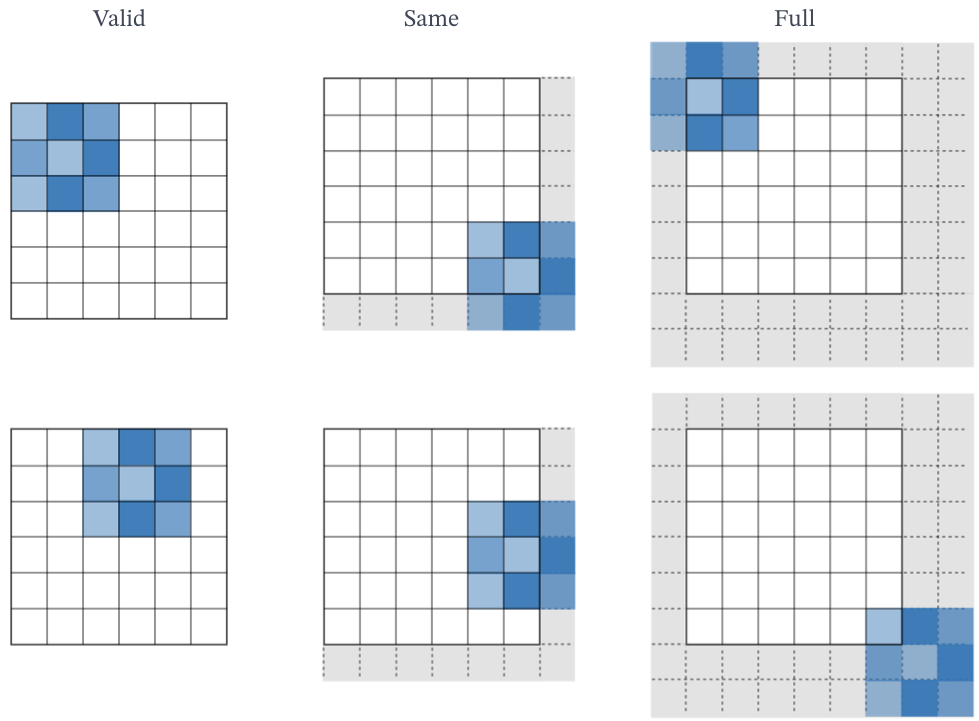}}
  \caption[Padding operation]{Three common padding strategies. The images show two positions of the same filters are in blue. The grey area indicates the zero-padding added for each case. \textbf{Valid}: No padding, we drop the last convolution if after applying the stride a part of the filter is `outside` the input image. \textbf{Same}: padding such that output feature map has the same dimension that the input image.  \textbf{Full}: maximum padding such that the last convolutions on each axis are applied on the limits of the input. Figure copyright \href{https://stanford.edu/~shervine/teaching/cs-230/cheatsheet-convolutional-neural-networks}{CS 230 Deep learning}.}
  \label{fig:padding}
\end{figure*}

When performing a convolution, there are several aspects to define~\citep{dumoulin_2016}.
First of all, we have to define the \textbf{dimensions of a filter}. It usually has three dimensions \textit{width}, \textit{height} and \textit{depth}.
While the \textit{depth} dimension matches the \textit{depth} of the input image, the \textit{width} and \textit{height} are consideribly smaller than the input \textit{width} and \textit{height}.
We frequently use square filters for these dimensions.
Other shapes are also possible when we want to emphasize a particular dimension.
\textbf{Stride} denotes the number of pixels on each axis\footnote{Since the \textit{depth} dimension matches the \textit{depth} of the input we slide only in the \textit{width} and \textit{height} axes} by which the filter moves after each operation. Strides bigger than one have less overlapped  information and downsample the output.
\textbf{Zero-padding} concatenates zeros to each side of input boundaries. This is done to obtain outputs with the same or higher dimension of the input (see Figure~\ref{fig:padding}).
Zero-padding is essential to perform a \textit{`transposed convolution`} operation~\citep{Zeiler_2010}.
Transposed convolutions are used when we want to change the order of the dimensions, e.i. having a bigger output than input.
The output shape of a convolutional layer is defined by these parameters.
It is usual to apply a special filter called \textbf{pooling} that does a final downsampling operation after a convolution operation. Pooling reduces the size of our array while keeping the most important features. It also produces spatial invariance and makes the features robust against noise and distortion.
Pooling downsamples each \textit{depth} independently, reducing only the height and width.
The most used pooling strategies are \textit{max} and \textit{average} where the maximum and average value is, respectively, taken (see Figure~\ref{fig:pool}).
The digital image processing domain has developed many handcrafted filters such as the Laplacian filter for highlighting regions of rapid intensity change (edge detection).

\begin{figure*}[ht!]
  \centerline{
    \includegraphics[width=.8\textwidth]{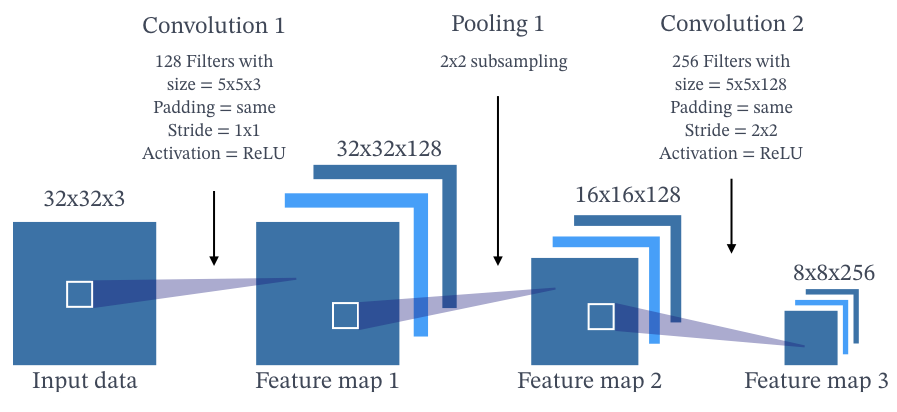}}
  \caption[Convolutional neural network example]{\glspl{cnn} are composed of a set of convolutional layers. Each layer computes a convolution operation, a non-linearity transformation and most of the time also a downsampling phase. Each filter at a given convolution is slid over the input producing a feature map that stores filter activations. Feature maps are arranged along the depth dimension.}
  \label{fig:cnn}
\end{figure*}

\begin{figure*}[ht!]
  \centerline{
    \includegraphics[width=\textwidth]{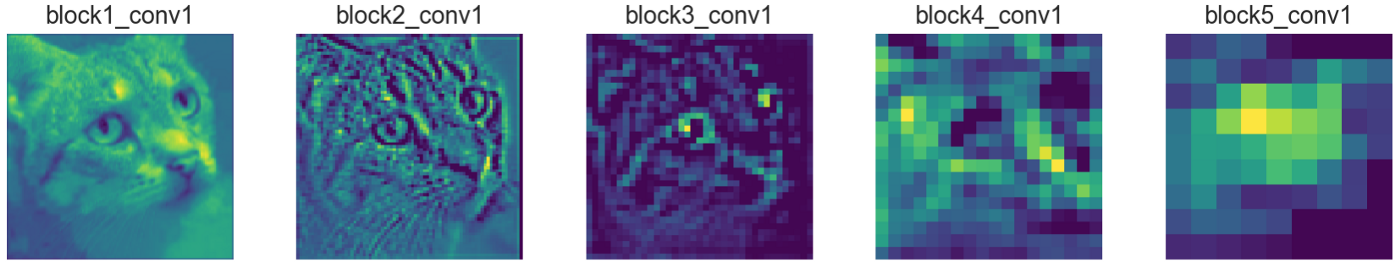}}
  \caption[Feature maps]{Visualization of one feature map per layer obtained using VGG16~\citep{Simonyan_2014}. The first convolution layers (block1\_conv1  and block2\_conv1)  compute feature maps that retain most of the original information, detecting low-level visual features like edges. As we go deeper in the network, the feature maps are more abstract but the original image is still visable (block3\_conv1). It is interesting to see how important aspects such as the eyes and noise are more active. These layers focus on the classes in the image and less in the image itself. The deepest convolutional layers (block5\_conv1) produce sparser feature maps, meaning the filters detect complex elements that may be not presented in every image. Figure copyright \href{Applied-Deep-Learning-with-Keras/notebooks/Part 4 (GPU) - Convolutional Neural Networks.ipynb}{Applied Deep Learning}}
  \label{fig:feature_maps}
\end{figure*}

Summing up, filters are sparse (with only a few elements we can transform the whole input), robust to spatial transformations and they reuse parameters (the same filter are applied to multiple locations).
This is ideal for dealing with images and efficiently reduce the number of parameters of \gls{dnn} architectures.
A \gls{cnn} is, in essence, a set of convolutional layers, each one composed by a convolution, a non-linearity operation and a downsampling phase (see Figure~\ref{fig:cnn}).
The non-linearity operation is applied after we slide the filter over the input.
We can downsample either by using a stride (the downsampling is computed directly in the main convolution itself) or applying pooling operation after the non-linearity operation.
At a given layer, we apply many different convolutions in parallel, each one with a different filter.
The main characteristic of \gls{cnn} architectures is that filters are not hand-designed but learned as part of the training process using the backpropagation algorithm. Hence, the values of a filter are learnable weights that are trained for detecting the important features without any human supervision, playing the role of a feature extractor.
Each convolution transforms the input into a tensor of filter activations arranged along the depth dimension called ``feature maps''.
We then apply the non-linearity.
In a \gls{cnn}, each convolutional layer learns filters of increasing complexity.
Adding many layers increase the abstraction capacity of the net (see Figure~\ref{fig:feature_maps}).
The first convolution layer extracts low-level visual features like oriented edges, lines, end-points, and corners.
The middle layers learn filters that detect parts of objects.
The last layers have higher representations: they learn to recognize full objects, in different shapes and positions.
\gls{cnn}s learn such complex features by building on top of each convolutional layer.

When defining a \gls{cnn} architecture, there are many design choices such as the number of layers, filter sizes, the number of filters, stride, padding or non-linearity.
This is task-depending and conventions are constantly updated.
After the convolution layers, it is common to add several \gls{fully} layers to find patterns in the obtained high-level features. For that, we flatten the tensor into a 1D vector.
This becomes quite standard for classification problems where the last \gls{fully} layer represents all the possible classes.
\gls{cnn} architectures are trained also with backpropagation and gradient descent.

\gls{cnn} models are the most popular deep learning architecture.
Complex architectures that stack multiple and different convolutional layers have revolutionized the digital image processing domain.
They are also widely used in other domains such as recommender systems, speech recognition, natural language processing, or \gls{mir}.
\gls{cnn} architectures are the main tool we employ to develop our ideas in the next chapters.

\subsection{Autoencoders}
\label{sec:autoencoders}

\begin{figure*}[ht!]
  \centerline{
    \includegraphics[width=.7\textwidth]{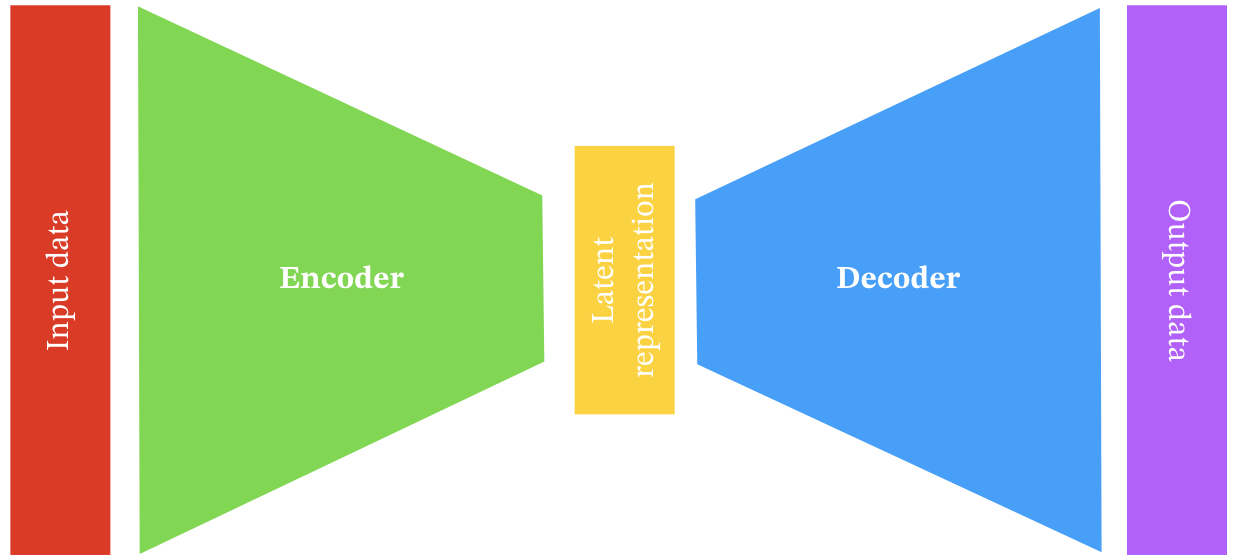}}
  \caption[Autoencoders example]{First the input passes through the encoder to project the data into the latent-space. Then the decoder produces the output only using  the latent-space data. We train the system minimizing the differences between the input and output.}
  \label{fig:autoencoders}
\end{figure*}

Autoencoders are \gls{dnn} architectures that have as target value $y$ the input $x$~\citep{Ballard_1987}.
They compress the input into a lower-dimensional representation and reconstruct the output from it.
The lower-dimensional representation (also called latent-space representation) serves as a compact “summary/compression” of the input.
In practice, it is an internal \gls{hidden} that describes the input only by a few variables.
Autoencoders have two components: the encoder and the decoder (see Figure~\ref{fig:autoencoders}).
The encoder compresses the input into the latent-space $e(x) = l$.
Alternatively, the decoder reconstructs the original input using the latent-space information $d(l) = x'$ only.
In fact, autoencoders aim to learn an approximation to the identity function $d(e(x)) = x' \approx x$.
Rather than a direct identity function, they add constraints for learning useful properties of the data.
Commonly, the decoder architecture is the mirror image of the encoder but this is not mandatory.
The only requirement is that the input and output must have the same dimensions so that the ``loss function'' $\mathcal{L}$ can directly compare point-wise them.

Autoencoders are one of the most popular unsupervised learning architectures.
They belong to the self-supervised family because they generate their own labels from the training data.
Autoencoders were originally used as compression techniques with losses (the final output is a close but degraded representation of the original). Soon, they were applied as robust denoising methods~\citep{Lu_2013} by simply adding noise to the inputs and using as target the original noise-free data.
They are also used as feature learning by removing the decoder and adding new layers for performing a particular task.
This is usually combined with transfer learning, i.e. transferring the learned variables of an architecture to another architecture.
In this case, the encoder must have the same architecture as the target dedicated net (the final net that performs the task).
Once the autoencoder is trained, we use the weights of the encoder to initialize the weights of the target dedicated net~\citep{Masci_2011, Zhuang_2015}.
This helps overcoming the problem of insufficient label data in a \gls{sl} task.
Modern autoencoders use stochastic mappings $p_{encoder}(h|x)$ and $p_{decoder}(x|h)$ for generative modeling, i.e. being able to generate new samples from the learned distribution.
The most well-known example is Variational Autoencoder~\citep{Diederik_2014}

\subsection{U-Net}
\label{sec:unet}
\begin{figure*}[ht!]
  \centerline{
    \includegraphics[width=.55\textwidth]{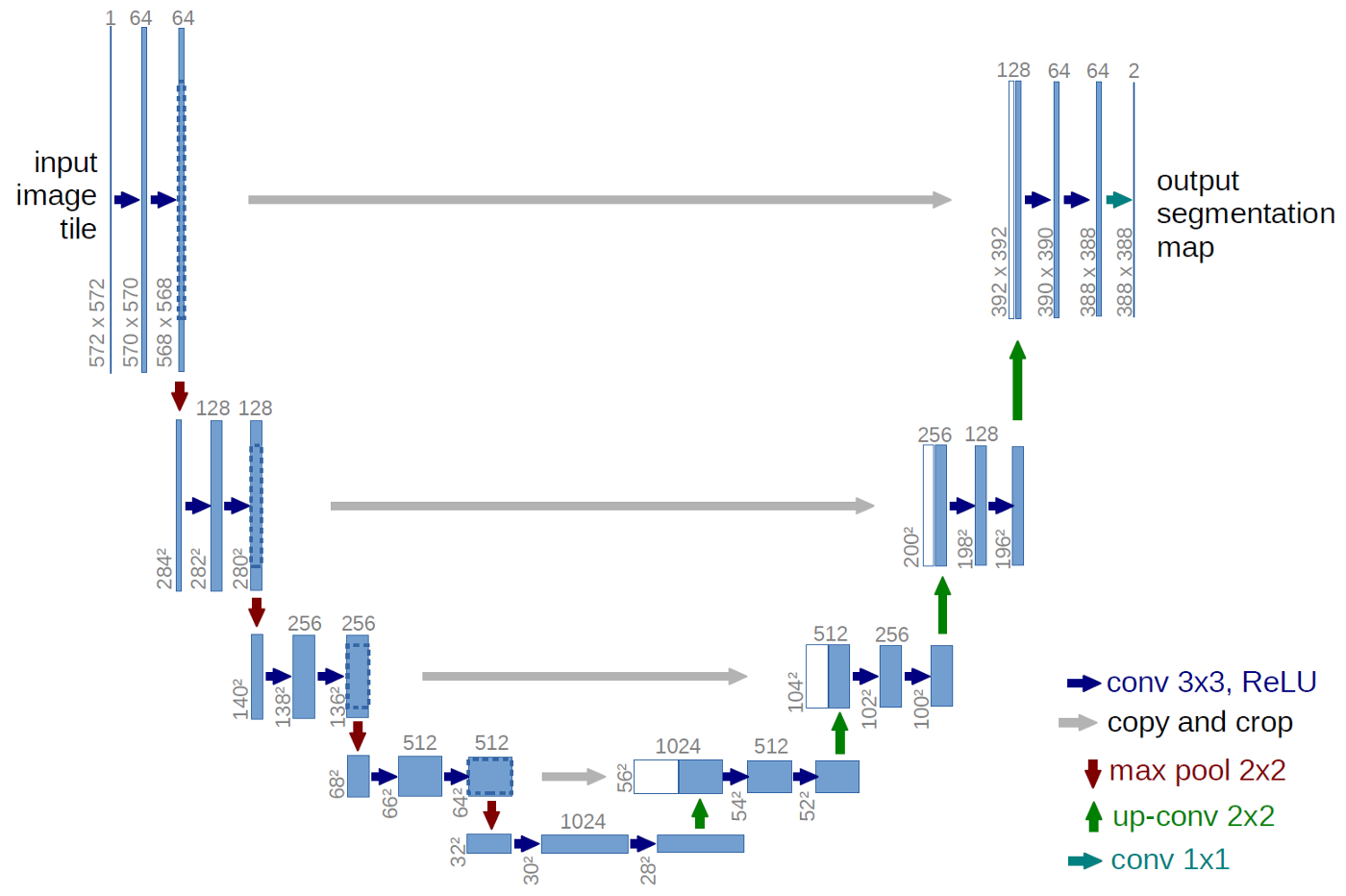}}
  \caption[U-Net architecture]{Original \gls{unet} architecture. It follows a encoder decoder schema with \glspl{skip} between layers at the same hierarchical level. The output is not the original input but rather a modification that highlights or isolated a particular aspect of the input. Figure copyright~\citep{Ronneberger_2015}}
  \label{fig:unet}
\end{figure*}

Inspired by autoencoders, the \gls{unet} architecture~\citep{Ronneberger_2015} has also an encoder/decoder mirror architecture based on \gls{cnn} (see Figure~\ref{fig:unet}).
Each convolutional block in the encoder halves the size of the input and doubles the number of channels.
The decoder obtains the original size of the input by a stack of transposed convolutional operation.
Both encoder and decoder have the same number of blocks.
This architecture adds \glspl{skip} (see Section~\ref{sec:skip}) between layers at the same hierarchical level in the encoder and decoder, i.e. the input of each decoder block is both the output of the previous block and the output of the corresponding encoder layer.
This ensures that the encoded features are directly used in the reconstruction.
The output of the \gls{unet} is not the input but a modification that highlights or isolates a particular aspect or a specific target location.
Notice how this formalization is no longer unsupervised learning but rather \gls{sl} because it requires labeled data.
The \gls{unet} architecture is one of the main tools we employ in this thesis.

\subsection{Additional deep neural network components}
\label{sec:additional_dnn}

In this section, we detail a set of common techniques used to speed-up the training, avoid overfitting and create more robust neural networks.
These techniques are employed in our models.

\subsubsection{Dropout}
\label{sec:dropout}
\Gls{dropout} prevents over-fitting during training time.
At each training iteration, we ``drop'' a random selection of a fixed number of the units~\citep{Srivastava_2014}.
Dropped neurons are disabled, not participating in the training process. Dropped neurons at one step are usually active at the next step.
Using \gls{dropout} prevents neurons co-adaptation (i.e. to be dependent on a small number of previous neurons).
We explicitly force every neuron to be able to operate independently by learning robust features useful in conjunction with several random subsets of neurons.
We avoid \gls{dropout} in the output layer (it is where we want to specialize each neuron to do something concrete).
\Gls{dropout} is not applied during inference.

\subsubsection{Batch Normalization}
\label{sec:batch_norm}
\Gls{batch_norm} improves the performance and stability of neural networks~\citep{Ioffe_2015}.
We usually standardize (zero mean and standard deviation of one) the input data so that each feature has the same contribution, reducing the sensitivity to small changes.
\Gls{batch_norm} extends this idea by normalizing/standardizing activations in intermediate layers.
At each iteration of the training process and given a minibatch, we normalize the output of one layer before applying the activation function.
The normalization is done using the mean and standard deviation of the values in the current batch.
We then feed it into the following layer.
\Gls{batch_norm} also adds two learnable parameters: a shift factor $\gamma$ and scale factor $\beta$.
These parameters restore the representation power of the network to take advantage of the non-linearity function in the case it cannot learn with that zero-mean and unit-variance constraint.
They also control the needed mean and the variance of the layer which helps our optimization algorithm.
In inference, we usually use an average of the accumulated mean and variance during training.

\Gls{batch_norm} reduces the amount by what the hidden unit values shift around, giving the same importance at each input feature.
It optimizes the training because networks learn faster (converge quickly), allows higher learning rates, reduces the sensitivity to the initial starting weights and keeps a controllable range of values avoiding saturations for some non-linearity activations~\citep{Goodfellow_2016}.

\subsubsection{Data Augmentation}
\label{sec:data_aug}

Overfitting happens because we have too few examples to train on. As a result, our model finds an overcomplex function that does not generalize.
In the hypothetical case of having access to all the instances of the unknown joint probability distribution $p(x,y)$, we would not overfit because we would see every possible instance.
Nevertheless, we only have access to $\mathcal{S}$. \Gls{data_aug} artificially enriches or ``augments'' the training set $\mathcal{S}$ by generating new instances.
We aim to generate realistic $\mathcal{S}$ instances.
The transformations should be learnable by the model, and not being simple noise.
\Gls{data_aug} can rapidly increase the size of our training set, reducing overfitting.
It is only performed on the training data, we do not modify the validation or test set.

There are many \gls{data_aug} techniques.
Traditional methods apply random transformations to the existing instances in $\mathcal{S}$.
This technique is very effective for image classification task.
Dedicated strategies designed for particular tasks also enhance the accuracy and generalization ability~\citep{Mauch_2013}.
New augmentation techniques explore how to learn augmentations that best improve the ability of the net to correctly perform a task.
These methods achieve state-of-the-art results~\citep{Perez_2017, Cubuk_2019}.

\subsubsection{Residual/skip connections}
\label{sec:skip}

\begin{figure*}[ht!]
  \centerline{
    \includegraphics[width=.4\textwidth]{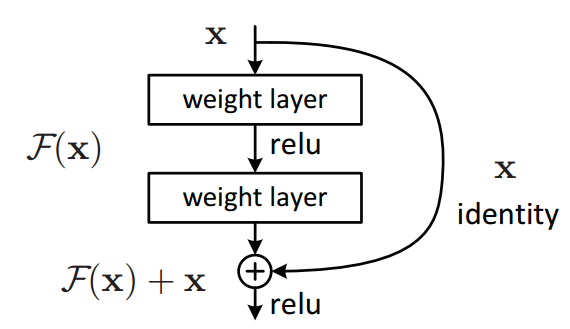}}
  \caption[Residual/skip connections]{\glspl{skip} connects the output of one layer with the input of another. The final output corresponds to the identity function in its simplest form. Figure copyright~\citep{He_2016}}
  \label{fig:skip}
\end{figure*}

A \gls{skip} ``connects'' the output of one layer with the input of an earlier layer~\citep{He_2016} (see Figure~\ref{fig:skip}). These connections can skip multiple layers.
Adding more layers increases the complexity and expressiveness of the network but also makes them much more expressive, difficult to train and adds more unpredictability.
New layers define new independent functions.
A new independent function does not guarantee increasing the expressive power of the network.
This is only guaranteed when larger function classes contain the smaller ones (nested functions).
This is the core idea behind \gls{skip}.
Each additional layer should contain the identity function as one of its elements.
Thereby, rather than parameterizing around a function $f(x)$ that outputs zero in its simplest form (weights are zeros), we parameterize around a function that outputs $x$ (the identity) in its simplest form.
With that each new layer deviate from the identity function, which still goes through the net.
This leaves the outputs of the previous layers unchanged just that we could now do additional transformations.
It also helps in a better gradient propagation.
The \gls{skip} makes the gradient to pass unchanged to a previous connected layer, and also to the intermediate block to update its weights.

\Glspl{skip} are implemented as summation or concatenation.
Using element-wise summation can be seen as feature refinement through the various layers of the network.
It is a compact solution that keeps the number of features fixed across blocks.
On the other side, concatenating allows the subsequent layers to re-use middle representations, maintaining the original information.
It has a better gradient propagation for deep architectures but it can lead to an exponential growth of the parameters.

\section{Conclusion}
In this chapter, we reviewed the collection of tools we use to develop our ideas.
In the following chapters, we delve into the topics covered here explaining some aspects furthermore, adding some more specific and advanced tools and showing how they have effectively applied for our tackled problems.
We use them during the creation of our dataset and for exploring two \gls{mir} tasks: lyrics segmentation and source separation.
When working with the audio signal, we employ spectral representations. Although they differ from traditional images in many aspects (e.g. the spatial correlation is much more complex and `pixels' at a particular frequency do not only depend on their closer ones but also on `pixels' at far frequencies, the harmonics), these spectral representations can be seen as images. Hence, we mostly employ \gls{cnn} architectures.

\graphicspath{{figs/}{dali_1_description/figs/}}
\chapter{DALI: A Dataset of Audio with Lyric Information Aligned In Time}
\label{sec:dali_description}

The central topic of this thesis is the \gls{multimodal} analysis of \gls{sv} investigating music and lyrics.
Our research focuses on the vocals of a song.
We are interested in the direct interaction between the audio signal and the lyrics.
Thus, we need a specific kind of data: audio signals and their matched lyrics aligned in time.
Nevertheless, there is a lack of large and good quality datasets of this kind.
Lyrics aligned in time can be found for commercial purposes (\href{http://www.lyricfind.com}{LyricFind}, \href{https://www.musixmatch.com}{Musixmatch} or \href{http://www.music-story.com}{Music-story}).
Yet, they are private, not accessible outside host applications, come without audio, have only aligned text lines and do not contain vocal melody symbolic notation (notes).
To carry out our research we need to have access to this kind of data.
Thereby, the first contribution of this thesis is the \gls{dali}: a large dataset with time-aligned vocal notes and lyrics at four levels of granularity:  notes (with their correspondent underlying phonemes), words, lines, and paragraphs.
\gls{dali} has 5358 songs for the first version and 7756 for the second one.

In this chapter, we first define the dataset itself and discuss why \gls{dali} is needed. Then we explain the developed tools that come with it and analyze the information presented.

\section{Motivation}
\label{sec:dali_description_motivation}

  Many \gls{mir} tasks are complex predictions estimated at a particular time instant such as note estimation or instrument recognition.
  To solve these tasks, researchers usually formulate their solutions in a \gls{sl} setting.
  In this paradigm, models use labeled data to discover functions that map input-output pairs (see Chapter~\ref{sec:tools}).
  Having large, good quality and reality representative of the real world datasets is essential to success in any supervised learning problem.
  The generalization of a model (i.e. to correctly determine the output of unseen input data) depends critically on the number of labeled examples~\citep{Sun_2017}.

  The image processing domain has constantly improved thier models thanks to benchmark reference datasets such as MNIST~\citep{lecun_1998}, CIFAR~\citep{Krizhevsky_2009}, YouTube-8M~{\citep{Abu_El_Haija_2016}} or ImageNet~\citep{Deng_2009}. These datasets are used, standardized and accessible by everybody enabling model comparisons.

  Nevertheless, in \gls{mir} there is a lack of benchmark reference datasets.
  There are various reasons for this absence: legal problems, label complexity (each audio segment has fine time and/or frequency resolution), task diversity (the same audio excerpt can have many different labels related to the task at hand) and the need for expert knowledge (with possible disagreements).
  There are two main paths for creating datasets: either doing so manually or reusing/adapting existing resources.
  Although the former produces precise labels, it is time-consuming, and the resulting datasets are often rather small.
  Since existing resources with large data do not meet the \gls{mir} requirements, datasets that reuse/adapt them are usually noisy and biased.
  \begin{figure*}[ht]
   \centerline{
     \includegraphics[width=\textwidth]{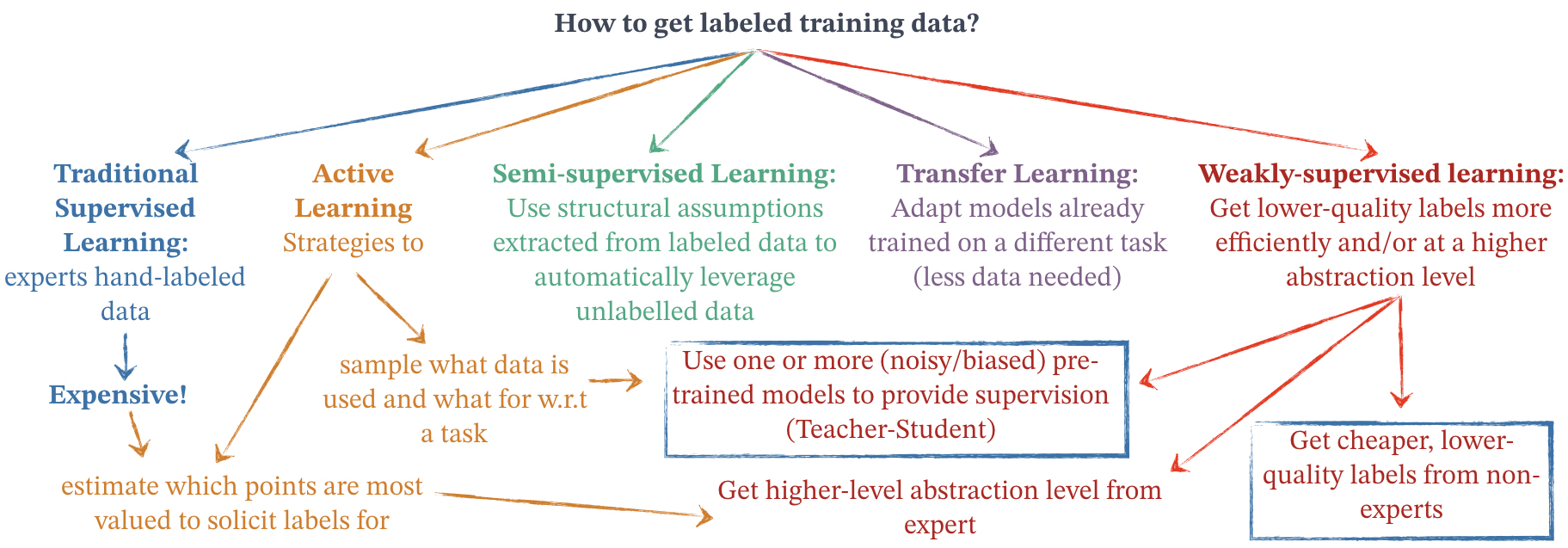}
   }
   \caption[Insufficient data solutions]{Schema of the different paradigms to deal with the problem of insufficient labeled data. \gls{dali} is framed inside Weakly supervised learning approches where cheap labels are obtained from non-experts. In Chapter~\ref{sec:dali_creation} we will explore how to use pre-trained models to provide supervision for creating the \gls{dali} dataset.}
   \label{fig:training_data}
  \end{figure*}

  Currently, many new areas have appeared to face the issue of insufficient labeled data (see Figure~\ref{fig:training_data}).
  \textbf{Semi-supervised learning} uses labeled data together with a large amount of unlabeled data~\citep{Zhu_2005s}. Hereabouts, systems automatically leverage unlabeled data through deriving insights from the labeled one.
  \textbf{Active learning} estimates the most valuable points for which to solicit manual expert labels and explore strategies to automatically select what data is the most valuable to use and what for given a particular task~\citep{Settles_2008, Krause_2016}.
  \textbf{Weakly supervised learning}~\citep{Mintz_2009, Mnih_2012, Xiao_2015} deals with low-quality labels (or at a higher abstraction level that needed, for instance having labels only at the audio excerpt and not the frame level) to infer the desired target information.

  Many \gls{mir} datasets include \emph{musical note events} labels, where a note event consists of a start time, end time and pitch.
  They are useful for a number of applications that bridge between the audio and symbolic domain, including symbolic music generation and melodic similarity.
  Instruments such as the piano produce relatively well-defined note events, where each key press defines the start of a note.
  Other instruments, such as the singing voice, produce more undefined note events, where the time boundaries are often related with changes in lyrics or simply as a function of our perception~\citep{furniss2016michele}, and are therefore harder to annotate correctly.
  Reference datasets such as \gls{medley}~\citep{Bittner_2014} or MusicNet~\citep{thickstun_2017} with full audio tracks and musical note events labels and/or fundamental frequency have a great impact on the \gls{mir} research community.

  Datasets providing note event annotations are created in a variety of ways.
  The most predominant \gls{mir} datasets are manually created, where notes are manually labeled by music experts, requiring the annotator to specify the start time, end time and pitch of every note event manually, aided by software such as Tony~\citep{mauch2015computer}.
  However, there are not many of such type and the final dataset is rather small as it consumes a large number of resources.
  Recently, large datasets created reusing and/or leveraging MIDI files from the Internet have been proposed~\citep{meseguerbrocal_2018, Donahue_2018, Raffel_2016, Benzi_2016, Fonseca_2017, Donahue_2018, Nieto_2019, Maia_2019, Yesiler_2019}.
  Yet, these sources are not designed for \gls{mir} needs, producing in noisy labels emerging the question of how unambiguous and accurate they are.
  Note data has also been collected automatically using instruments which ``record'' notes while being played, such as a Disklavier piano in the MAPS~\citep{emiya2009multipitch} and MAESTRO~\citep{hawthorne2018enabling} datasets, or a hexaphonic guitar in the GuitarSet dataset~\citep{xi2018guitarset}.
  Data collected in this way is typically quite accurate, but may suffer from global alignment issues~\citep{hawthorne2018enabling} and can only be achieved for these special types of instruments.
  Another approach is to play a midi keyboard in time with a musical recording, and use the played midi events as the note annotations~\citep{su2015escaping} but this requires a highly skilled player in order to create accurate annotations.

  In this thesis we are intesested in tasks related to \gls{sv} which, despite being one of the most important elements in popular music, it is a lesser-studied topic in \gls{mir} community.
  Although many \gls{sv} problems (e.g. singing voice detection or lyrics alignment) have been widely studied in the \gls{mir} community, \gls{sv} was introduced as a standalone topic  only a few years ago when it~\citep{Goto_2014, Mesaros_2013}.
  This topic specially suffers from lack of benchmark dataset.
  Currently, researchers working in \gls{sv} use small datasets.
  Each one is designed following different methodologies~\citep{Fujihara_2012}.
  Large datasets remain private~\citep{Humphrey_2017, Stoller_2019} or are vocal-only captures of amateur singers recorded on mobile phones that involve complex pre-prossessing~\citep{Smith_2015, Kruspe_2016, gupta2018semi}.
  This absence of reference datasets is a critical point that has been always neglected preventing the \gls{sv} community from training state-of-the-art \gls{ml} algorithms and comparing their results.

  \begin{table*}[ht]
      \caption{Comparison of the different datasets with lyrics aligned in time.}
      \label{table:datasets}
      \footnotesize
      \centering
      \begin{tabular}{|l|l|l|l|l|}
          \hline
          \textbf{Dataset} & \textbf{Number of songs}   & \textbf{Language} & \textbf{Audio type} & \textbf{Granularity} \\ \hline
          \citep{Iskandar_2006} & No training. 3 tests songs & English & Polyphonic & Syllables \\ \hline

          \citep{Wong_2007} & \begin{tabular}[c]{@{}l@{}}14 songs divided into \\ 70 segments with 20s long \end{tabular} & Cantonese   & Polyphonic  & Words\\ \hline

          \citep{Muller_2007} & 100 songs & English & Polyphonic & Words\\ \hline

          \citep{Kan_2008} & 20 songs & English & Polyphonic & \begin{tabular}[c]{@{}l@{}}Section\\ Lines\end{tabular} \\ \hline

         \citep{Mesaros_2010} & \begin{tabular}[c]{@{}l@{}} Training: 49 fragments \\ $\sim$25 seconds  \\ for adapting a phonetic model  \\ Testing: 17 songs \end{tabular} & English&  \begin{tabular}[c]{@{}l@{}} Training: A Capella \\ Testing: Vocals \\ after source separation \end{tabular} & Lines \\ \hline

          \citep{Hansen_2012} & 9 pop music songs  & English  & \begin{tabular}[c]{@{}l@{}} Both, Polyphonic \\ A Capella \end{tabular}  & \begin{tabular}[c]{@{}l@{}}Words \\ Lines \end{tabular} \\ \hline

         \citep{Mauch_2012} & 20 pop music songs  & English  &  Polyphonic  & Words \\ \hline

          \textbf{DAMP} dataset, \citep{Smith_2015}  & \begin{tabular}[c]{@{}l@{}} 34k \textbf{amateur} versions \\ of 301 songs \end{tabular}& English& Amateurs A Capella  & \begin{tabular}[c]{@{}l@{}}Not time-aligned \\ lyrics \\ only textual \\ lyrics\end{tabular} \\ \hline

          \textbf{DAMPB} dataset, \citep{Kruspe_2016}  &\begin{tabular}[c]{@{}l@{}} A \textbf{DAMP} subset with \\ 20 performances of 301 songs \end{tabular}  & English & Amateurs A Capella  & \begin{tabular}[c]{@{}l@{}} Words \\ Phonemes \end{tabular} \\ \hline

          \citep{Dzhambazov_2017}  & \begin{tabular}[c]{@{}l@{}}70 fragments \\ of 20 seconds\end{tabular}& \begin{tabular}[c]{@{}l@{}} Chinese\\ Turkish\end{tabular} & Polyphonic & Phonemes\\ \hline

          \citep{Lee_2017} & 20 pop music songs  & English  &  Polyphonic  & Words \\ \hline

          \citep{gupta2018semi}   & \begin{tabular}[c]{@{}l@{}}A \textbf{DAMP} subset with  \\ 35662  segments of 10s long \end{tabular} & English & Amateurs A Capella  & Lines  \\ \hline

          \begin{tabular}[c]{@{}l@{}}$\text{Jamendo}_{\text{aligned}}$, \\ \citep{Ramona_2008} \\ \citep{Stoller_2019} \end{tabular} & 20 Creative commons songs & English & Polyphonic & Words \\ \hline

          \begin{tabular}[c]{@{}l@{}}\textbf{DALI v1} \\ \citep{meseguerbrocal_2018} \end{tabular}& 5358 songs in full duration  & Many  &  Polyphonic  & \begin{tabular}[c]{@{}l@{}} Notes, words,\\ lines and \\ paragraphs\end{tabular}  \\ \hline

          \textbf{DALI v2} & 7756 songs in  full duration  & Many  &  Polyphonic   & \begin{tabular}[c]{@{}l@{}} Notes, words, \\ phonemes,\\ lines and \\ paragraphs\end{tabular}  \\ \hline

      \end{tabular}
  \end{table*}

\newpage
Table~\ref{table:datasets} contains an overview of public datasets with lyrics aligned in time.
Most of these datasets are created in the context of lyrics alignment task.
In this task, researchers try to assign start and end times to every fragment of textual information.
Lyrics are inevitably language-dependent.
Researchers have created several datasets for different languages: English~\citep{Kan_2008, Iskandar_2006, gupta2018semi}, Chinese~\citep{Wong_2007, Dzhambazov_2017}, Turkish~\citep{Dzhambazov_2017}, German~\citep{Muller_2007} and Japanese~\citep{Fujihara_2011}.
Most datasets contain polyphonic popular music.
There are many datasets with A Capella music~\citep{Kruspe_2016}. However, it is always difficult to migrate the methods to the polyphonic case~\citep{Mesaros_2010}.
Datasets do not always contain the full duration audio track but often a shorter version~\citep{gupta2018semi, Dzhambazov_2017, Mesaros_2010, Wong_2007}. If the tracks are complete, respective datasets are typically small.
One of the goals of this thesis is to build a large and public dataset with audio, lyrics, and notes aligned in time, the \gls{dali} dataset.

\begin{figure*}[ht]
    \centerline{
        \includegraphics[width=0.95\textwidth]{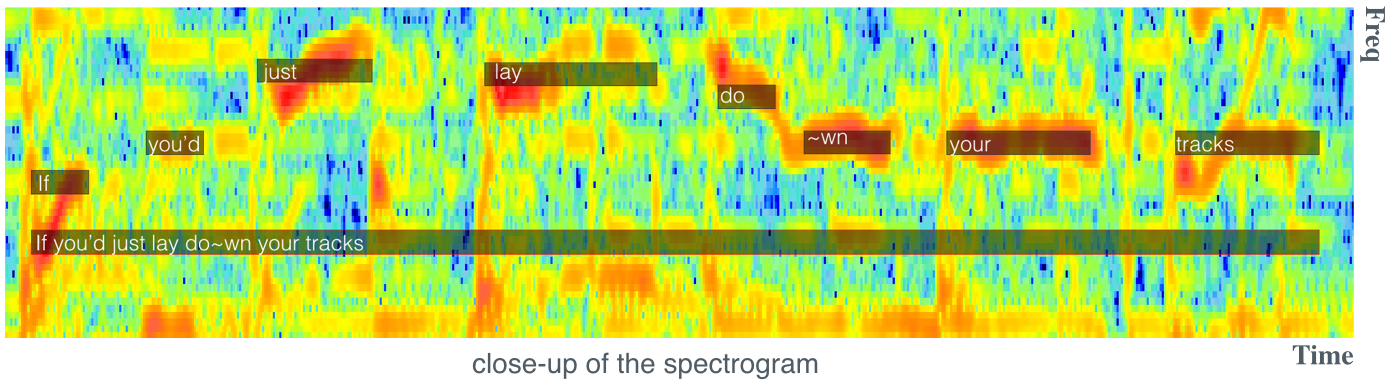}
    }
    \caption[An example a manual annotation superimposed to the spectrogram of the track.]{
    An example of the manual annotation overlap with its spectrogram. The close-up of the spectrogram illustrates the alignment for a small excerpt at two levels of granularity: notes and lines.}
    \label{fig:ex_annotation}
\end{figure*}

\subsection{Our proposal}
\begin{figure*}[ht]
        \centerline{
            \includegraphics[width=0.7\textwidth]{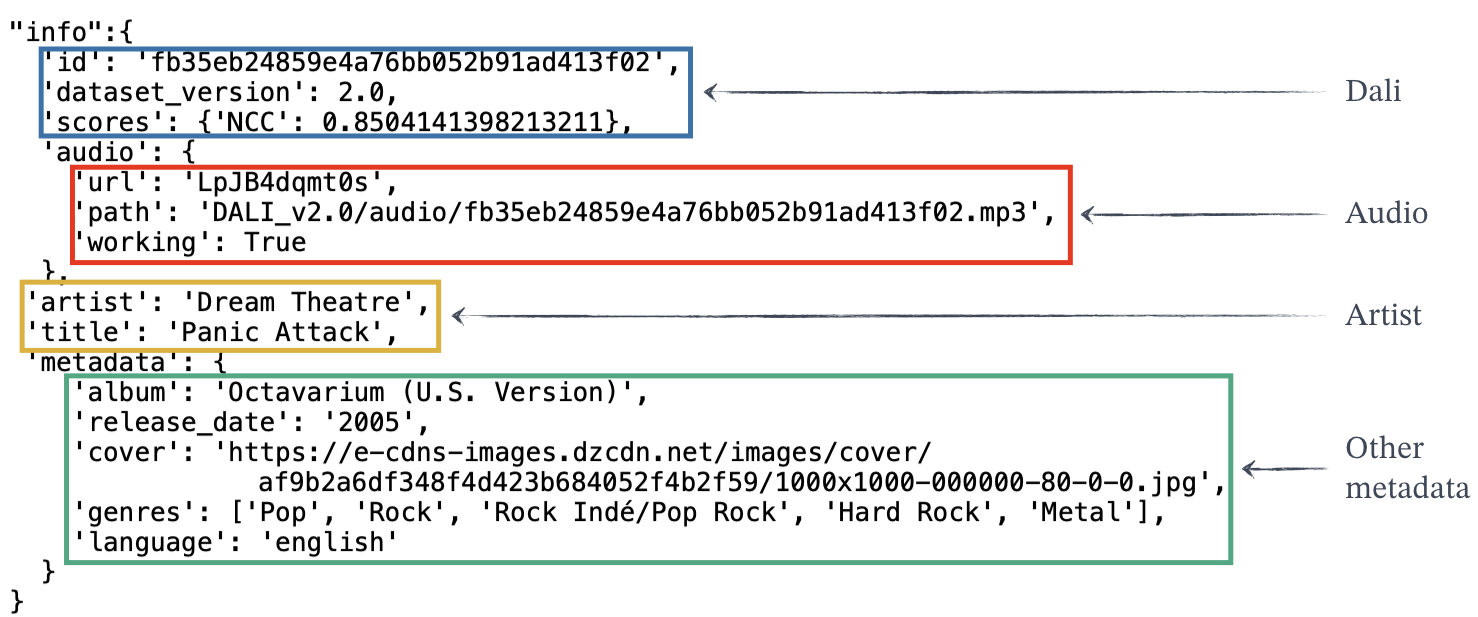}
        }
        \caption[DALI metadata example]{Example of metadata annotations.}
        \label{fig:metadata}
    \end{figure*}

    The \gls{dali} dataset: a large \textbf{D}ataset of synchronised \textbf{A}udio, \textbf{L}yrics and p\textbf{I}tch aims to serve as a reference dataset for the singing voice community.
    We presented it in 2018 at the \gls{ismir} conference~\citep{meseguerbrocal_2018}.
    It contains the audio of full songs each with -- their audio in full-duration, -- their time-aligned vocal melody and -- their time-aligned lyrics.
    Thus, it contains \emph{musical note events} and lyrics aligned information.
    Lyrics are described according to four levels of granularity: \textbf{notes} (the phonemes underlying a given note), \textbf{words}, \textbf{lines}, and \textbf{paragraphs}.
    It also provides additional metadata such as genre, language and musician and some \gls{multimodal} information like album covers or links to video clips (see Figure~\ref{fig:metadata}).

    \begin{table}[t]
      \centering
      \caption[DALI dataset general overview]{\gls{dali} dataset general overview}
      \label{table:dali_overview}
      \small
      \begin{tabular}{ c | c | c | c | c | c }
        \hline
        V & Songs & Artists & Genres & Languages & Decades \\
        \hline
        \hline
        1.0 & 5358 & 2274 & 61 & 30 & 10 \\
        \hline
        2.0 & 7756 & 2866 & 63 & 32 & 10 \\
        \hline
        Multitracks & 512 & 247 & 32 & 1 & 7 \\
        \hline
      \end{tabular}
    \end{table}

    The \gls{dali} dataset has not been created manually.
    Rather, we leveraged existing open-source karaoke resources where non-expert users manually annotated the lyrics and melody of a song.
    From these resources, we developed a system that finds the corresponding audio tracks and aligns the annotations to it.
    Our approach consists of constant interaction between dataset creation and learning models where they benefit from each other (this is described in detail in Chapter \ref{sec:dali_creation}).
    \gls{dali} has 5358 songs for the first version and 7756 for the second one.
    There are also 512 songs in multitrack version (\textit{M}) with two stems (\textbf{vocals} and \textbf{accompaniment}) and the final \textbf{mix} (see Table~\ref{table:dali_overview}).

    \section{Definition}
    \label{sec:dali_description_definition}

    The \gls{dali} dataset is a collection of songs described as a sequence of time-aligned lyrics, each one linked to its audio in full-duration.
    Annotations define a direct relationship between the audio and the lyrics represented as text information at different hierarchical levels.
    This is very useful for a wide variety of \gls{mir} problems such as lyrics alignment and transcription, melody extraction, structure analysis, hierarchical interaction or vocal \gls{source_separation}.

    Time-aligned lyrics are described at four levels of granularity: \textbf{notes}, \textbf{words}, \textbf{lines} and \textbf{paragraphs}, from the deepest to the highest.
    The lyrics are described as a sequence of characters for all levels.
    For the multitrack and the second version, the \textbf{word} level also contains the lyrics as sequence of \textbf{phonemes}.
    Lyrics at the \textbf{note} level correspond to the syllable (or group of syllables) sung, and the frequency defines the musical notes for the vocal melody.
    The different granularity levels are vertically connected, i.e. one level is associated with its upper and lower levels.
    For instance, we know the words of a particular line, or which paragraph a line belongs to.
    In Figure~\ref{fig:ex_annotation}, we illustrate an example with two levels of granularity: a line and its corresponding notes.

    \subsection{Formal definition}

    In \gls{dali}, songs are defined as:
    \begin{equation}
        S = \{A_\mathit{notes}, A_\mathit{words}, A_\mathit{lines}, A_\mathit{paragraphs}\}
    \end{equation}

    \noindent where each granularity level $g$ with $K$ elements is a sequence of aligned segments:
    \begin{equation}\label{eq:dali-formal}
      A_{g} = (a_{k, g})_{k=1}^{K_g} \textrm{ where } a_{k,g} = (t_k^0, t_k^1, f_k, l_k, i_k)_g
    \end{equation}

    \noindent with $t_k^0$ and $t_k^1$ being a text segment's start and end times (in seconds) with $t_k^0 < t_k^1$, $f_k$ a tuple $(f_\mathit{min}, f_\mathit{max})$ with the frequency range (in Hz) covered by all the notes in the segment (at the note level $f_\mathit{min} = f_\mathit{max}$, a vocal note), $l_k$ the actual lyric's information and $i_k = j$ the index that links an annotation $a_{k, g}$ with its corresponding upper granularity level annotation $a_{j, g+1}$.
    The text segment's events for a song are ordered and non-overlapping - that is, $t_k^1 \le t_{k+1}^0 \forall k$.
    Note how the annotations define a unique connection in time between the \textbf{musical} and \textbf{textual} domains.

    \section{Dataset analysis}
    \label{sec:dali_description_analysis}

    \gls{dali} has $5358$ songs for its first version~\citep{meseguerbrocal_2018}, $7756$ for the second one and $512$ for the multitrack subset.
    This means a total of $344.9$, $488.1$ and $35.4$ hours of music respectively with $176.9$, $247.2$ and $14.1$ hours with vocals.
    In terms of annotations, there are more than $3.6$ and $8.7$ million $a_{k, g}$ for version $1$ and $2$ and $486$k for the multitrack.
    The average $a_{k, g}$ per song is $679$, $710$ and $950$.
    There are, on average, $2.36$ ($2.71$) songs per artist and $119$s ($115$s) with vocal durations per song, in v1 (v2) (see Table~\ref{table:dali_details}).

    As seen in Table~\ref{table:dali_overview}, \gls{dali} has a great range of artists, genres, languages and decades.
    Most of the songs are from popular genres like Pop or Rock and the 2000s.
    The most predominant language is English but there are also many songs in German and French.

    \begin{table*}[t]
      \centering
      \caption[DALI dataset statistics]{Statistics for the different \gls{dali} dataset versions. One song can have several genres.}
      \label{table:dali_details}
      \footnotesize
      \begin{tabular}{ c | c | c | c | c | c | c }
        \hline
        V  &  \makecell{Average \\ songs per \\ artist} & \makecell{Average duration \\ per song} & Full duration  & \makecell{Top 3 \\ genres} & \makecell{Top 3 \\ languages} & \makecell{Top 3 \\ decades}  \\
        \hline
        \hline
        1.0 &  2.36 & \makecell{Audio: 231.95s \\ With vocals: 118.87s} & \makecell{Audio: 344.9hrs  \\ With vocals: 176.9hrs}  &  \makecell{Pop: 2662 \\ Rock: 2079 \\ Alternative: 869} & \makecell{ENG: 4018 \\ GER: 434 \\ FRA: 213}   & \makecell{00s: 2318 \\ 90s: 1020 \\ 10s: 668}		\\
        \hline
        2.0 &   2.71 & \makecell{Audio: 226.78s \\ With vocals: 114.73s} & \makecell{Audio: 488.1hrs  \\ With vocals: 247.2hrs}  &  \makecell{Pop: 3726 \\ Rock: 2794 \\ Alternative: 1241} & \makecell{ENG: 5913 \\ GER: 615 \\ FRA: 304}   & \makecell{2000s: 3248 \\ 1990s: 1409 \\ 2010s: 1153} \\
        \hline
        M &   2.07 & \makecell{Audio: 220.83s \\ With vocals: 98.97s} & \makecell{Audio: 35.4hrs  \\ With vocals: 14.1hrs}  &  \makecell{Rock: 312 \\ Pop: 258 \\ Alternative: 162} & ENG: 512   & \makecell{2000s: 188 \\ 1990s: 103 \\ 1980s: 93} \\
        \hline
      \end{tabular}
    \end{table*}

    Finally, using the correlation scores described at Chapter~\ref{sec:NCC}, we propose to split \gls{dali} into 3 sets: train, validation, and test (see Table~\ref{table:dali_split}). However, depending on the task at hand (e.g. analyzing only English songs or lyrics alignment) other splits are possible.

    \begin{table}[t]
      \centering
      \caption[DALI proposed split]{Proposed split with respect to the time correlation values described in Chapter \ref{sec:NCC}}
      \label{table:dali_split}
      \small
      \begin{tabular}{ c | c | c }
        \hline
         & Correlations & Tracks   \\
        \hline
        \hline
        Test & $\mathit{NCC}_t >= .94$  &  \makecell{1.0: 167  \\ 2.0: 402}  \\
        \hline
        Validation & $.94 > \mathit{NCC}_t >= .925$ &  \makecell{1.0: 423 \\ 2.0: 439} \\
        \hline
        Train &  $.925 > \mathit{NCC}_t >= .8$ &  \makecell{1.0: 4768  \\ 2.0: 6915}  \\
          \hline
      \end{tabular}
    \end{table}

    \begin{figure*}[ht]
        \centering
        \begin{subfigure}[c]{\textwidth}
            \includegraphics[width=.9\textwidth]{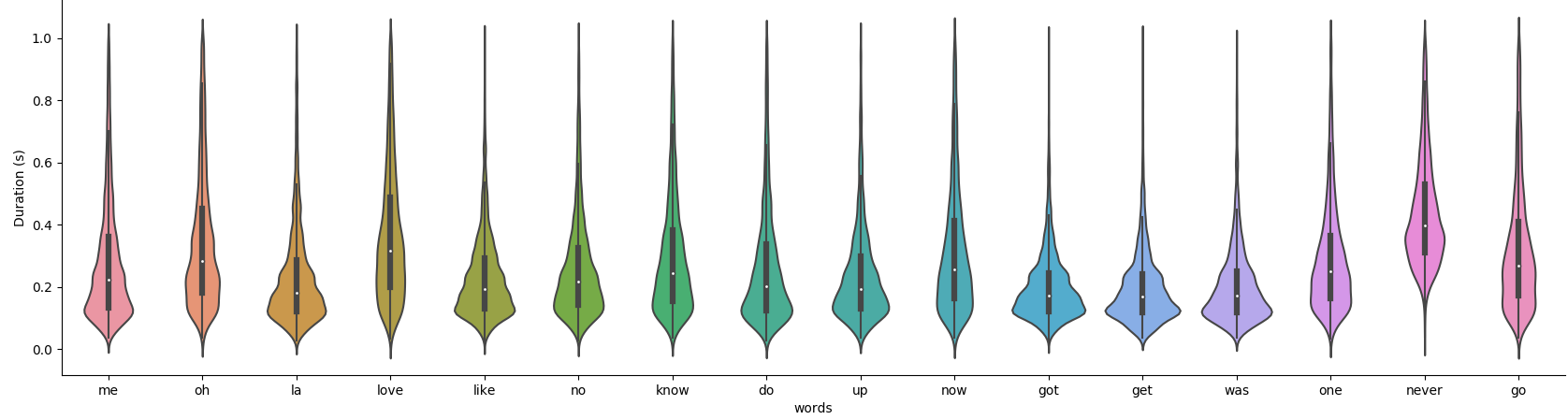}
        \end{subfigure}
        \begin{subfigure}[c]{\textwidth}
            \includegraphics[width=.9\textwidth]{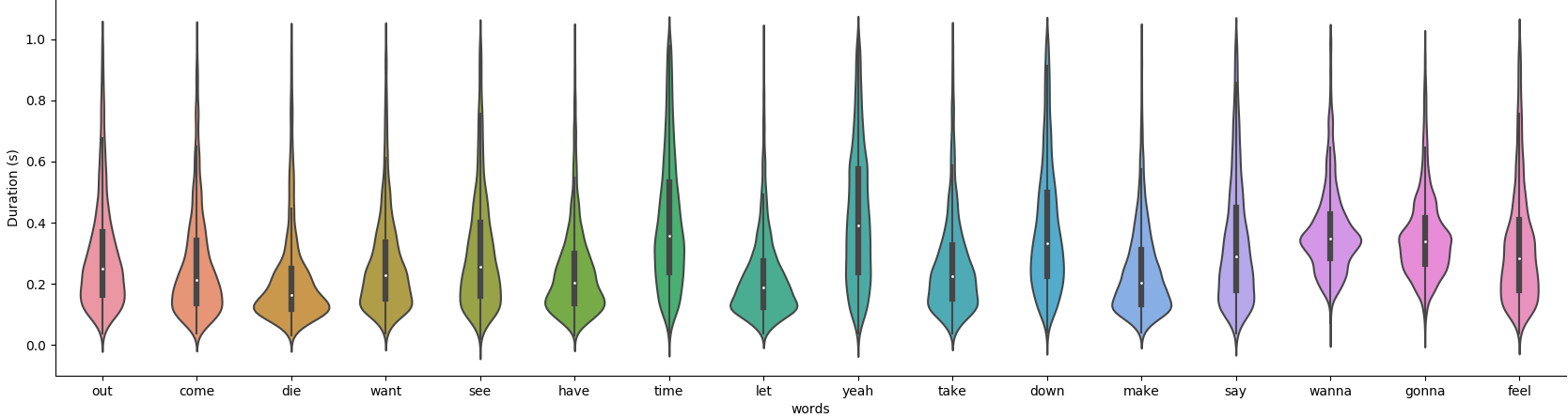}
        \end{subfigure}
        \caption[Duration of the most common words]{Distribution of the duration in seconds of the most common 32 words (after removing pronouns and articles) for the second version of \gls{dali}.}\label{fig:words}
    \end{figure*}

    \section{Working with DALI}

    \begin{figure*}[ht]
        \centerline{
            \includegraphics[width=\textwidth]{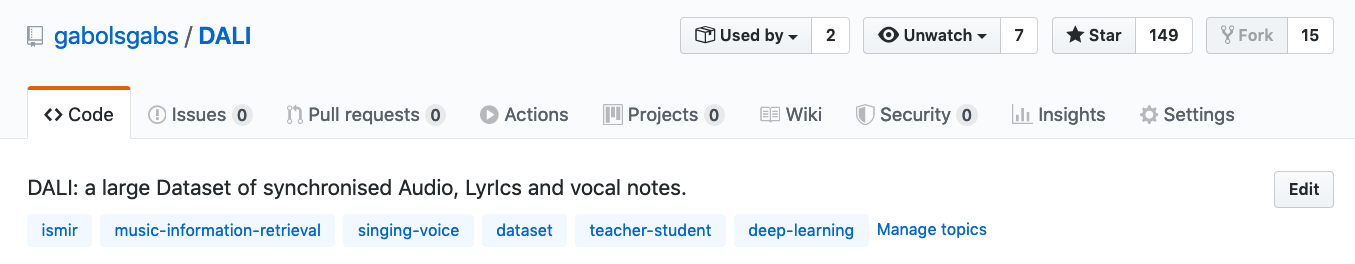}
        }
        \caption[DALI tools github]{The \gls{dali} package is available at \url{https://github.com/gabolsgabs/DALI} and contains all the necessary tools for working with the annotations.}
        \label{fig:github}
    \end{figure*}

    \subsection{Tools}

    The richness of \gls{dali} renders the data complex. Therefore, it would be difficult to use in a raw format such as {\tt JSON} or {\tt XML}.
    To overcome this limitation, we have developed a specific {\tt Python} package that has all the necessary tools to access the dataset.
    It can be found at \url{https://github.com/gabolsgabs/DALI} (see Figure~\ref{fig:github}) and easily be installed using pip\footnote{\url{https://pypi.org/project/DALI-dataset/}}.

\begin{figure*}[ht]
        \centerline{
            \includegraphics[width=\textwidth]{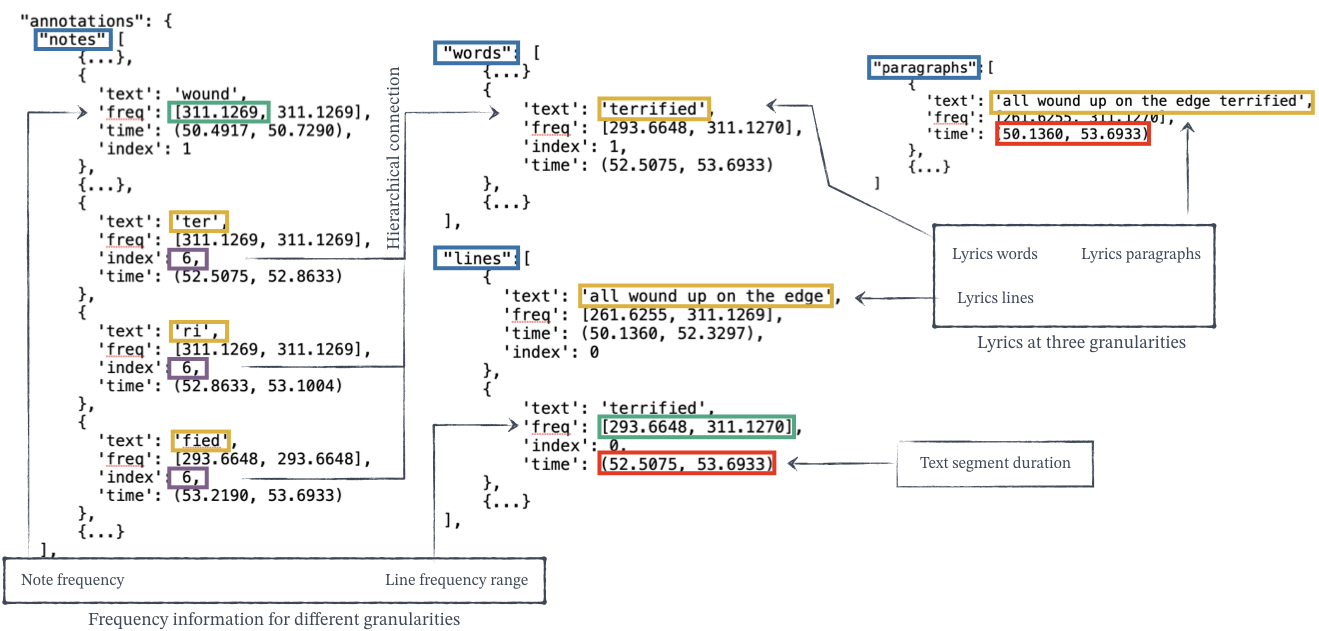}
        }
        \caption[Annotations class data example]{Annotations class data example with four levels of granularity. Note how the word 'terrified' is sung in three different notes. Thanks to the index each level of granularity is connected with its upper one.}
        \label{fig:class_example}
    \end{figure*}

    A song is represented as the Python class,  {\tt \textbf{Annotations}} (see Figure~\ref{fig:class_example}).
    {\tt \textbf{Annotations}} instances have two attributes {\tt \textbf{info}} and {\tt \textbf{annotations}}.
    The attribute {\tt \textbf{info}} contains the metadata, the scores that guide the quality of the annotations (see Chapter~\ref{sec:NCC} and Chapter~\ref{sec:global_in_freq}) and links to the audio.

    The attribute {\tt\textbf{annotations}} contains the aligned segments $a_{k,g}$.
    We can work in two modes {\tt \textit{horizontal}} and {\tt \textit{vertical}}, and easily change from one to the other.
    The {\tt \textit{horizontal}} mode stores the granularity levels in isolation, providing access to all its segments.
    The {\tt \textit{vertical}} mode connects levels vertically across the hierarchy.
    A segment at a given granularity contains all its deeper segments e.g. a line has links to all its words and notes, allowing the study of hierarchical relationships.

    \begin{figure*}[ht]
        \centerline{
            \includegraphics[width=0.95\textwidth]{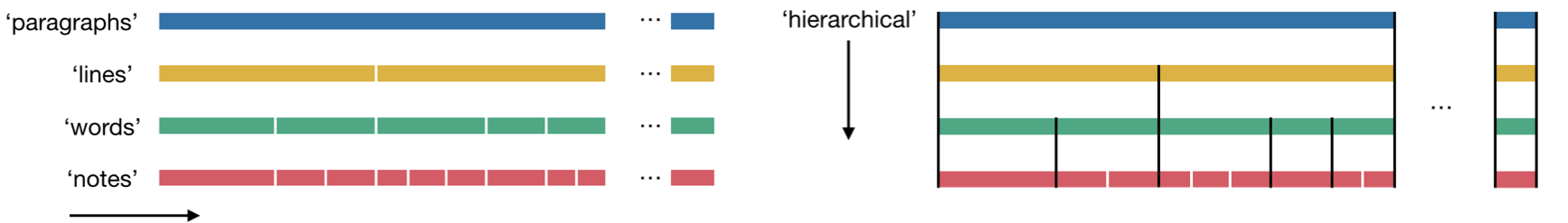}
        }
        \caption[Two annotation modes]{Annotations are presented in two different modes. On the left, the {\tt \textit{horizontal}} mode that stores the granularity levels in isolation. On the right, the {\tt \textit{vertical}} mode that connects all levels hierarchically.}
        \label{fig:annot_modes}
    \end{figure*}

    The package also includes a group of additional tools.
    There are general tools for reading the whole dataset and automatically retrieving the audio from the internet.
    We also provide tools for working with individual granularity levels i.e. transforming the data into vectors or matrices with a given time resolution, to manually correct the global parameters or to re-compute the alignment for different audio than the original one (see Chapter~\ref{sec:NCC}).
    For a detail explanation of how to use \gls{dali}, we refer to the tutorial at \url{https://github.com/gabolsgabs/DALI}.

    \subsection{Distribution}

    \begin{figure*}[ht]
        \centerline{
            \includegraphics[width=0.95\textwidth]{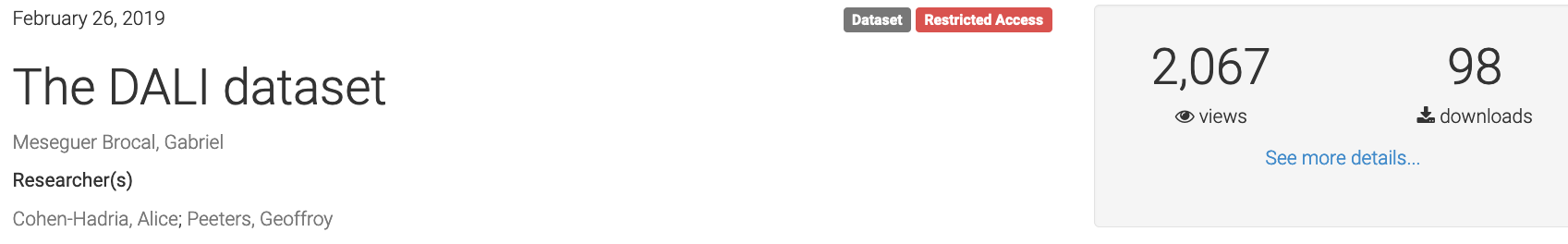}
        }
        \caption[DALI data zenodo]{The actual \gls{dali} annotations are available at \url{https://zenodo.org/record/2577915} and can be downloaded after agreeing to use them only for research.}
        \label{fig:zenodo}
    \end{figure*}

    Each \gls{dali} dataset version is presented as a set of gzip files.
    Each file encloses an instance of the class {\tt \textbf{Annotations}}.
    The different versions can be found at \url{https://zenodo.org/record/2577915} (see Figure~\ref{fig:zenodo}). They are distributed as open-source under an {\tt Academic Free License, AFL} \footnote{https://opensource.org/licenses/AFL-3.0}.
    Each version is described following the MIR corpora description~\citep{Peeters_2012} and has a fingerprint (a {\tt MD5} checksum file~\citep{MD5}) that verifies the integrity of it.

    Finally, the \gls{dali} dataset is also part of mirdata~\citep{Bittner_2019}, which provides a standard framework for MIR datasets as well as a fingerprint (also a {\tt MD5} file) per {\tt \textbf{Annotations}} instance and audio track that verifies their integrity.

    \subsection{Reproducibility}
    One of the main problems in \gls{dali} is the restriction on sharing the audio of each song. This complicates the comparison of the results or may end up in misaligned annotations if different audio is used. We suggest three ways to overcome this issue:

    \begin{enumerate}
        \item to use the tools provided to retrieve the audio we use directly from YouTube. Unfortunately, some of the links may be broken and not all the audio might be available.
        \item to use a different audio version and reproduce the alignment techniques as in Chapter~\ref{sec:NCC}. We provide all the tools for the task and grant a model (second generation (see Chapter~\ref{sec:second-generation})) for computing the singing voice activation vector needed.
        \item to send us the computation needed to be run on our audio. The user has to agree to distribute the new feature to other users (at {\tt zenodo}) as the main melody representation (\gls{f0}) computed in Chapter~\ref{sec:global_in_freq}.
    \end{enumerate}

    Finally, the multitrack version is not distributed.


\section{Conclusions}

In this chapter we detailed our first contribution, the \gls{dali} dataset. We formally defined it and performed an statistical analysis of its content to get to know better its peculiarities. We have also introduced the developed tools that help us to work with this complex data.
We use \gls{dali} for tackling our research problems in the following chapters.


\graphicspath{{figs/}{dali_2_creation/figs/}}

\chapter[Creating DALI: a real case scenario]{Creating DALI: a real case scenario\raisebox{.3\baselineskip}{\normalsize\footnotemark}}\footnotetext{Some of the work reported here was done in collaboration with Alice Cohen-Hadria who implemented the Singing Voice detection model and trained the base line versions.}

\label{sec:dali_creation}

Creating the \gls{dali} dataset represents a challenging research question.
We start with songs manually annotated by non-expert users into notes and lyrics of the vocal melody.
These annotations come without audio and they are only described by artist name and song title.
Also, the annotations are not always accurate enough to be used as a \gls{mir} dataset.
To create a clean dataset, we need to 1) find the corresponding audio used and 2) improve the quality of the annotations’ time information.

For each annotated songs, we retrieve a set of audio candidates from YouTube.
Each one is turned into a \gls{svp} over time using a \gls{svd}, based on a deep \gls{cnn} architecture.
We find the best candidate and correct the annotations' time information by comparing this \gls{svp} to the annotated \gls{vas}, derived from the time-aligned lyrics.
The quality of this matching is restricted by the performances of the \gls{svd} system.
Whereas our original model retrieves good annotations, it does not align properly the annotations to it.
To improve the \gls{svp}, we adopt a \gls{teacher-student}~\citep{Hinton_2014, Li_2014}.
In this paradigm, a first model named the \gls{teacher} is trained using a clean and well-annotated controlled dataset.
The \gls{teacher} is then used to select the best-aligned tracks defining a new training set (annotation/audio matches).
This set is used to train a new \gls{svd} system, the \gls{student}.
Using the “knowledge" learned by the \gls{teacher} on clean data, the \gls{student} has proved to perform better than the \gls{teacher} on the \gls{svd} task.

Our method is well motivated by Active learning and Weakly-supervised Learning (see Figure~\ref{fig:training_data}).
It establishes a loop whereby dataset creation and model learning interact, benefiting each other by progressively improving our model using the collected data.
At the same time, we correct and enhance the data every time we update the model.
This process creates an improved \gls{dali} after each iteration.

\section{From karaoke annotations to structured MIR data}
\label{sec:raw_annot}

\gls{dali} stands as a solution to the absence of large reference dataset with lyrics and vocal notes time-aligned.
These types of annotations are hard to obtain and very time-consuming to create.
We opt for reusing/adapting existing resources.
Concretely, our solution is to look outside the field of \gls{mir}.
We turn our attention to karaoke video games where users sing along with the music.
They win points singing accurately which is measured by comparing the sung melody with time and frequency-aligned references.
Therefore, large datasets of time-aligned melodic data and lyrics exist.
Apart from commercial karaoke games, there are several active and large karaoke open-data communities.
In those, non-expert users exchange text files that contain the reference annotations without any professional revision.
We retrieved 13,339 of these karaoke annotation files.
However, they need to be adapted to the requirements of \gls{mir} applications\footnote{Standardized format with a time in seconds, frequency in hertz and the normalized annotations with characters in the utf8 format}.

\begin{figure}[ht!]
    \centering
    \includegraphics[width=.5\linewidth]{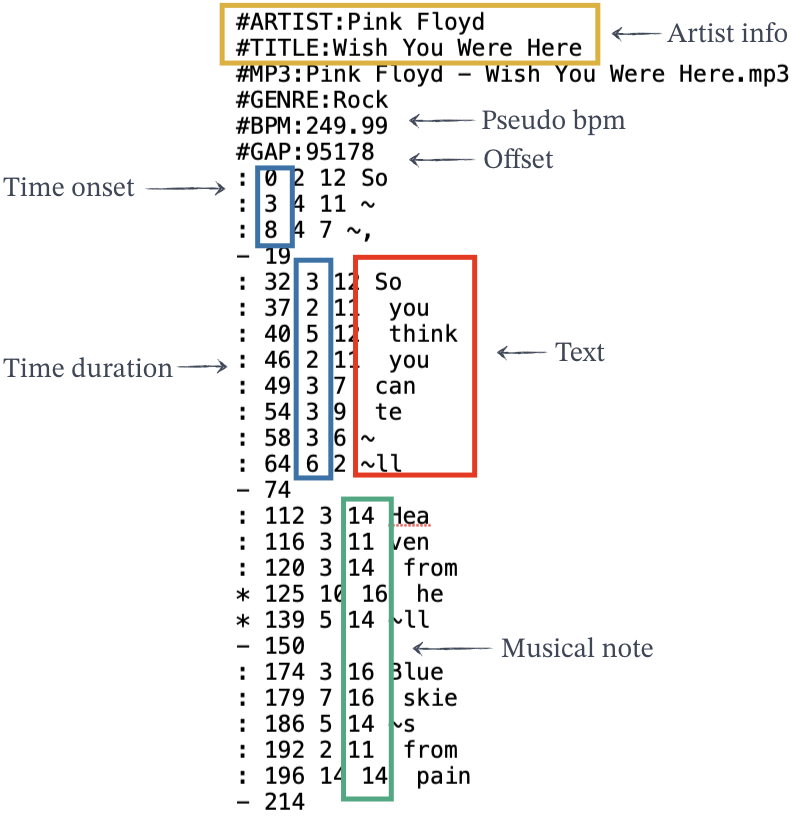}
    \caption[Raw karaoke data example]{Example of the raw data contained in a karaoke file}
    \label{fig:raw_annot}
\end{figure}

\noindent As illustrated in Figure~\ref{fig:raw_annot}, karaoke users create and exchange annotations in text files. Each file contains::

\begin{itemize}
  \setlength\itemsep{0em}
   \item the {\tt song\_title} and {\tt artist\_name}.
  \item a sequence of triplets {\tt \{time, musical-note, text\}} with annotations,
  \item the {\tt offset\_time} (start of the sequence) and the {\tt frame rate} (annotation time-grid),
\end{itemize}

We first transform the raw information into useful data obtaining the time in seconds and the frequency note.
Then, we create the different levels of granularity: \textbf{notes} (and textual information underlying a given note), \textbf{words}, \textbf{lines} and \textbf{paragraphs}.
The note, word, and line levels are encoded in the retrieved files. We deduct the paragraph level as follows.

\begin{figure}[ht!]
    \centering
    \includegraphics[width=.5\linewidth]{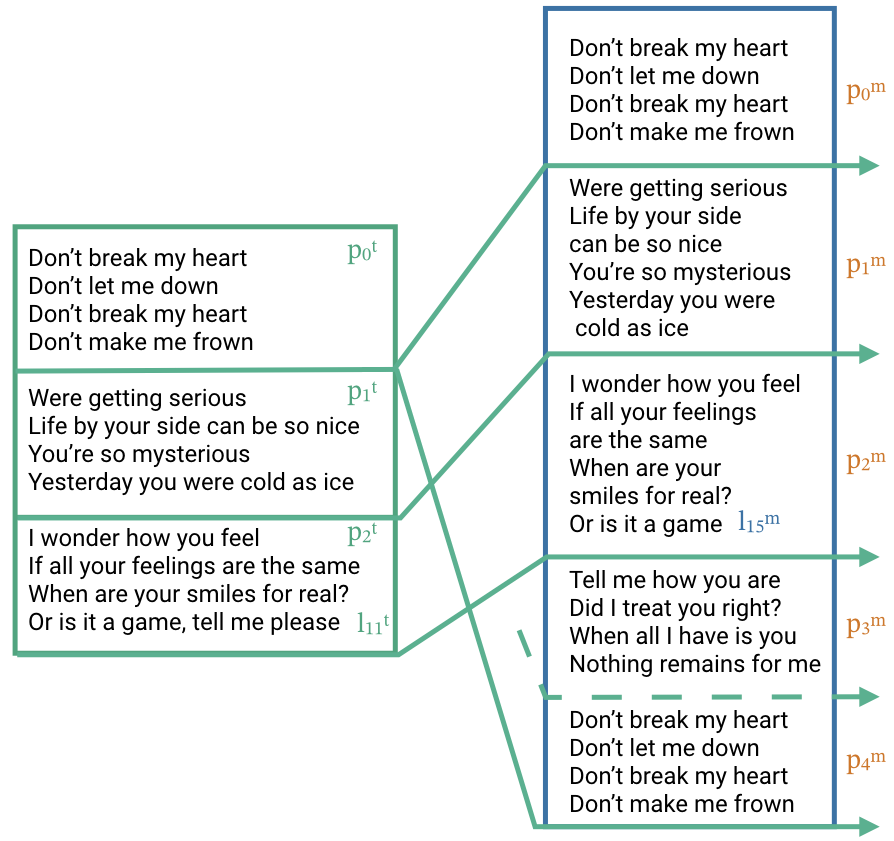}
    \caption[Creating the paragraph level]{[Left] Target lyrics lines and paragraphs as provided in \gls{wasabi} [Right] The melody paragraphs $p^{m}$ are created by merging the melody lines $l^{m}$ into an existing target paragraph $p^{t}$.
    Note how line $l_{11}^{t}$ in $p_{2}^{t}$ has no counterpart in $l_{*}^{m}$ and chorus $p_{3}^{m}$  does not appear in any $p_{*}^{t}$.}
    \label{fig:lyrics}
\end{figure}

\textbf{The paragraph level.} Using the {\tt  song\_title} and {\tt  artist\_name}, we connect each raw annotation file to the \gls{wasabi} dataset, a semantic database of song metadata collected from various music databases.
\gls{wasabi} provides lyrics grouped in lines and paragraphs, in a text-only form.
We created the paragraph-level by merging the two representations (melodic note-based annotations from karaoke annotations and text-only annotations from \gls{wasabi}) in a text to text alignment.
Let $l^{m}$ be our existing raw \textit{lines} and $p^{m}$ the \textit{paragraph} we want to obtain.
Similarly, $p^{t}$ represents the \textit{target} paragraph in \gls{wasabi} and $l^{t}$ its \textit{lines}.
Our task is to progressively merge a set of $l^{m}$ such that the new $p^{m}$ is maximally similar to an existing  $p^{t}$.
This is not trivial. $l^{m}$ and $l^{t}$ differ in some regards: $l^{m}$ tends to be shorter, some lines might be missing in one domain, and $p^{t}$ can be rearranged/scrambled.
An example of the merging system is shown in Figure \ref{fig:lyrics}.

\textbf{The phoneme information.} The phonetic information is computed only for the word level.
We use the Grapheme-to-Phoneme (G2P) \footnote{https://github.com/cmusphinx/g2p-seq2seq} system by CMU Sphinx\footnote{https://cmusphinx.github.io/wiki/} at Carnegie Mellon University.
This model uses a {\tt tensor2tensor} transformer architecture that relies on global dependencies between input and output.
The final phoneme level has the text information transcribed into a vocabulary of 39 different phoneme symbols defined in the Carnegie Mellon Pronouncing Dictionary (CMUdict)\footnote{https://github.com/cmusphinx/cmudict}.
The phoneme information is only available for multitrack and version two.

\textbf{The metadata.} Additionally, \gls{wasabi} provides extra multi-modal information such as cover images, links to video clips, metadata, biography, and expert comments.

\begin{table*}[t]
  \centering
  \caption[Term used for creating DALI]{Overview of terms: definition of each term used in this chapter. $\mathit{NCC}_t$ is defined at Section~\ref{sec:NCC}.}
  \label{table:terms}
  \footnotesize
  \begin{tabular}{ r  l }

    Term & Definition  \\
    \toprule
    \footnotesize{\tt Notes} & time-aligned symbolic vocal melody annotations. \\
    \footnotesize{\tt Annotation} & basic alignment unit as a tuple of:  \\
     & time (start and duration in frames), musical note (with 0 = C3) and text. \\
    \footnotesize{\tt  A file with annotations}  & group of annotations that define the alignment of a particular song. \\
    \footnotesize{\tt Offset\_time} ($o$)  & the start of the annotations. \\
    \footnotesize{\tt Frame rate} ($\mathit{fr}$) & the reciprocal of the annotation grid size.\\
    \footnotesize{\tt Voice annotation sequence} ($\mathit{vas}(t) \in \{0,1\}$) & a vector that defines when the singing voice (SV) is active \\
    &  according to the karaoke-users' annotations. \\
    \footnotesize{\tt Predictions} ($\hat{p}(t) \in [0,1]$) & probability sequence whether or not singing voice is active at any frame \\
    &   provided by our singing voice detection. \\
    \footnotesize{\tt Labels}  & label sequence of well-known ground truth datasets checked by the MIR community. \\
    \footnotesize{\tt Teacher}  & SV detection (SVD) system used for selecting audio candidates \\
    & and aligning the annotations to them. \\
    \footnotesize{\tt Student}  & new SVD system trained on $\mathit{vas}(t)$ of the subset selected \\
    & by the Teacher after $\mathit{NCC}(\hat{o}, \hat{\mathit{fr}}) \geq T_{\mathit{corr}}$. \\
  \end{tabular}
\end{table*}

\section{Finding the correct audio}
\label{sec:linking_auido}

The annotations are now ready to be used. Nevertheless, they come without audio.
Linking each annotation file with its proper audio track is a mandatory step to perform any audio-related task.
The same {\tt  song\_title} and {\tt  artist\_name} may have many different versions (studio, radio, edit, live or remix) and each one can have a different lyrics alignment.
Hence, we need to find the correct version used by the karaoke-users to do the annotations.
The \gls{wasabi} metadata provides exactly all the versions for a pair  {\tt  song\_title} and {\tt  artist\_name}.
With this information, we query YouTube to recover a collection of audio candidates.
This is similar to other works where authors used chroma features and diagonal matching to align jazz solos and audio candidates from YouTube~\citep{Balke2018}.

\noindent To find the correct audio used by the karaoke-users we need to answer to three questions:

\begin{enumerate}
  \item \textit{Is the correct audio among the candidates?}
  \item \textit{If there are more than one, which is the best?}
  \item \textit{Do annotations need to be adapted to the final audio to perfectly match it?}
\end{enumerate}

Moreover, the fact that users are amateurs may lead to errors and some annotations have to be discarded. This introduces a fourth question: \textbf{\textit{are annotations good enough to be used?}}.

To answer these questions, we need to measure the accuracy of an audio candidate to an aligned annotation.
Therefore, we need to find a common representation for both audio and text.

\subsection{Working with Audio and Lyrics part 1: Annotations as audio}
\label{sec:annot-as-audio}

In this section, we review our first attempts to find the correct audio candidates. These attempts were guided by the idea that the annotations can be seen as ``audio features'' i.e. we focus on how to transform them to fit audio representation spaces.

\subsubsection{Lyrics-alignment.}
We first explored lyrics alignment techniques~\citep{Fujihara_2012, Kruspe_2016, Dzhambazov_2017} which aims at automatically synchronizing sung lyrics with their written versions.
In other words, these techniques aim at determinating \textbf{where} lyrics appeared in the audio.
From a more technical point of view, it is the problem of finding the correct temporal location in the audio of limited textual units.

The problem starts with a given audio signal and its corresponding lyrics.
The goal is to assign start and end times to every fragment of textual information.
These fragments can have different levels of granularity: phonemes, syllables, words, phrases and paragraph and the difficulty of the tasks increases with granular levels.

Most of the approaches work with polyphonic popular music.
Methods developed for A Capella music~\citep{Kruspe_2016} are always difficult to migrate to the polyphonic music~\citep{Mesaros_2010}.
Lyrics are inevitably language-dependent.
They have been several studies for English~\citep{Kan_2008, Iskandar_2006}, Chinese~\citep{Wong_2007, Dzhambazov_2017}, Turkish~\citep{Dzhambazov_2017}, German~\citep{Muller_2007} and Japanese~\citep{Fujihara_2011}.
In theory, these systems can be adapted to any language but no experiments have been conducted to prove this.

There are several dimensions over which methods can be classified, being \textbf{features} employed one of them.
Most of the approaches use phonetic features following a speech recognition paradigm.
That is, for each phoneme an acoustics model is created which aims to capture the traits of a specific phoneme.
Due to the lack of annotated and isolated data, there has been a lot of effort in adapting speech models to the particularities of singing voice~\citep{Mesaros_2010} or particular singers~\citep{Fujihara_2011}.
There are few approaches that do not use phonetic features: methods that align the fundamental frequency (F0) of the singing voice with the tone of each word (only for tonal languages such as Cantonese)~\citep{Wong_2007} or use the phoneme duration with prior structures (detected by chord and rhythm)~\citep{Kan_2008}.

Studies can be also grouped along the \textbf{alignment} technique used.
There are two approaches: forced or non-forced.
\textbf{Forced} alignment \textbf{aligns all} the textual elements with audio segments.
It is a method inherited from the speech community and it is the most widely used.
Usually, the whole lyrics is expanded to a network of phonetic models (including a silence element).
Each model yields a likelihood according to an input features vector.
Thus, the audio track is synchronized with the lyric net forcing all elements in the net to have a connection with an audio segment.
Most of the forced methods use techniques such as Hidden Markov Models (HMMs) or Viterbi algorithm.
Recently, more sophisticated methods such as Dynamic Bayesian Networks (DBN) have been proposed for dealing not only with the transitions between phonemes but also with the complementary context~\citep{Dzhambazov_2017}.
In contrast, \textbf{non-forced} systems have the freedom to \textbf{not align all} the textual unit with audio extracts.

Finally, there is a set of studies that use the complementary context around lyrics for improving the performance.
Elements such as structure~\citep{Iskandar_2006, Lee_2008}, melodic phrases and metric circles~\citep{Dzhambazov_2017}, chord progressions~\citep{Mauch_2011}, MIDI files~\citep{Muller_2007} or manually-annotated segmentation labels (such as Chorus and Verse)~\citep{Lee_2008} are used as cue for alignment.

When we first tackled this problem the recent techniques based on \gls{dnn} did not exist.
Some authors proposed to use Automatic Speech Recognition (ASR) imperfect transcription to align lyrics and A Capella singing signals, in a semi-supervised way~\citep{gupta2018semi}.
The authors transfered a method conceived for A Capella audio to the polyphonic case. They adopted a neural network acoustic model trained on a large number of solo singing vocals by using its weights as initialization for the training with polyphonic music~\citep{gupta2019acoustic}.
New methods also use connectionist temporal classification (CTC) loss to train a model that extracts a character probability matrix from the audio signal.
This matrix is used to align the target lyrics to it~\citep{Stoller_2019}.

The problem with these techniques is that they are complex and have in the phonetic model its main limitation (there is no phonetic model trained on \gls{sv} rather adaptations from speech). Having a good dataset such as \gls{dali} is a key element for developing good lyrics alignment algorithms. They also assume that the pair audio-lyrics are correct which is not our case. Finally, with these techniques it is difficult to know if the resulting lyrics are well aligned. Hence, we discarded this direction.

\subsubsection{Score-alignment.}
Annotations can be transformed into a sequence of musical notes.
This allows us to formalize the problem as a score alignment approaches~\citep{Cont_2007, Soulez_2003, Raffel_2016}.
Score alignment techniques are similar to the previous ones but they focus on the harmonic distribution of the spectrogram instead of the phonetic traits.

%

The problem with these techniques is that they assume that every event in the audio has a representation in the musical score.
This is not our case because there is a lack of information for the non-vocal areas (we only have notes for the singing voice).
Silence in the score does not match with silence in the audio signal.
Furthermore, the musical background is not represented whatsoever in our score sequences.
They also presume that the musical score is correct which again is not our case.
Thus, these systems do not solve any of our issues and result in misaligned sequences.

\subsubsection{Dominant melody estimation.}
A natural solution to this issue is to assume that the dominant melody in the audio corresponds only to the singing voice, and then align both signals.
We investigate this idea using Melodia~\citep{Salamon_2012}, since it was one of the state-of-the-art dominant melody detection algorithms\footnote{At the moment these experiments were carried out, new \gls{dnn} architectures such as~\citep{Bittner_2017} did not exist.}.
Nevertheless, after several experiments, we found that the estimated melody is not sufficiently precise and does not always correspond to the vocal melody.
Hence, both sequences can not be properly aligned.

Accordingly, we did not persist in this direction. However, we will explore some of these ideas in future works, especially in the scenario in which we already have the annotations globally well alligned with the audio.
In this scenario, annotated notes are a key element to solve local alignment problems.

\subsection{Working with Audio and Lyrics part 2: Audio as annotations}
\label{sec:audio-as-annot}

Instead of focusing on how to transform the annotations to fit the audio representation spaces, our solution is to focus on how to transform the audio into the annotation space.
We do that by converting the audio into an \gls{svp} over time $\hat{p}(t)$:

\begin{equation*}
  \hat{p}(t) = \begin{cases}
      1, & \text{ if there is singing voice at time frame $t$}  \\
      0, & \text{ if there is no singing voice at time frame $t$}
  \end{cases}
\end{equation*}

We denote by $\hat{p}(t)$ the \textbf{predictions}. The model in charge of doing so is an \gls{svd} system and it is described in next section.
Likewise, we can transform the lyrics annotations into $\mathit{vas}(t)$ into a \gls{vas} with value $1$ when there is vocal annotations and $0$ otherwise:
\begin{equation*}
  vas(t) = \begin{cases}
      1,& \text{if there is singing voice} \\
      0,              & \text{otherwise}
  \end{cases}
\end{equation*}

We only focus on the vocal segments contained in $A_{g}$ (see Chapter~\ref{sec:dali_description_definition}), discarding the frequency and text information:
\begin{equation}\label{eq:dali-align}
  A_{g} = (a_{k, g})_{k=1}^{K_g} \textrm{ where } a_{k,g} = (t_k^0, t_k^1)_g
\end{equation}
This type of problem is typically not estimated at the segment observation level, but rather at the \emph{frame} level.
The annotations of a given granularity level $A_{g}$ are divided into an evenly spaced time grid $(r_i = H \cdot i)_{i=0}^m$ where $H$ is a constant defining the spacing between time stamps ($H = 14$ ms to have the same time resolution as $\hat{p}(t)$).
Although we can construct $\mathit{vas}(t)$ at any level of granularity, we only work at the note level because it is the finest.

\begin{figure*}[ht]
  \centerline{
    \includegraphics[width=.9\textwidth]{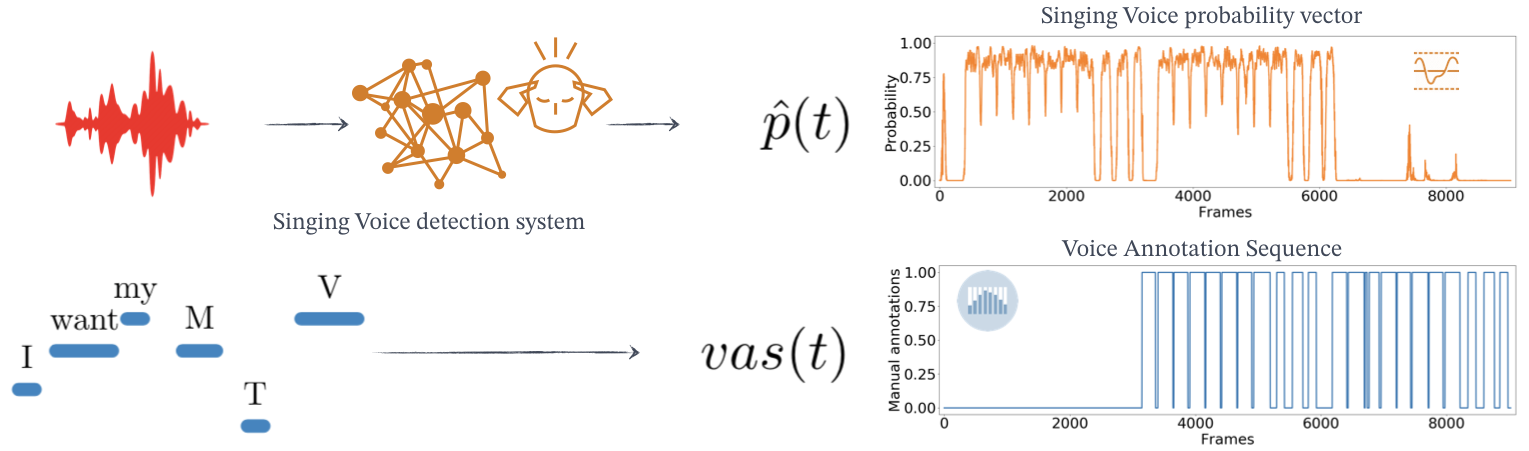}}
  \caption[Computing the $\hat{p}(t)$ and $\mathit{vas}(t)$]{Illustration of the computation of $\hat{p}(t)$ and $\mathit{vas}(t)$}
  \label{fig:two_domains}
\end{figure*}

These two vectors can be directly compared (see Figure~\ref{fig:two_domains}).
Our hypothesis is that with a highly accurate \gls{svd} system and exact annotations, both vectors should be identical. Consequently, it should be reasonably easy to use them to find the correct audio.

\section{The Singing Voice detection problem}
\label{sec:svd}

The problem of transforming the audio into \gls{svp} over time is called the \textit{singing voice detection} task.
In contrast with the \textit{alignment} tasks, in the \textit{singing voice detection} we aim to know, from the audio signal analysis, the probability of having a singing voice or not.

\subsection{Previous approaches}
Most approaches for the \textit{singing voice detection} task share a common architecture.
Short-time observations are used to train a classifier that discriminates observations (per frame) in vocal or non-vocal classes.
The final stream of predictions is then post-processed to reduce artifacts.

\textbf{Classification approach.}

Early works explore classification techniques such as Support Vector Machines (SVMs) or K-Nearest Neighbors (KNN) based on different audio descriptors~\citep{rocamora2007comparing}. In~\citep{Ramona_2008}, the authors use as audio features the centroid, width, asymmetry, slope, decreasing, flux and similar temporal statistical moments, along with their first and second derivatives and 13-order Mel Frequency Cepstral Coefficient (MFCC) resulting in a raw feature vector of 116 components, at each time frame.
After this extraction process, an SVM is used to classify each frame into a singing or non-singing class.
The same work also presents the \gls{jamendo} dataset, used in future experiments.
On this dataset, the authors report a 71.8\% accuracy without post-processing and 82.2\% accuracy with a Hidden Markov Model (HMM) post-processing.
Similarly, MFCC features extracted from a predominant melody extraction, like the pitch fluctuation feature and MFCC of Re-Synthesized Predominant Voice are also used~\citep{Mauch_2011}.
As in the previous method, classification is performed using SVMs applied to the frame level vector of features.
They test their model on 100 songs from the RWC Music Dataset~\citep{Goto_2002} reporting accuracy of 87.2\%.

\textbf{Using specific voice properties.}
Other approaches also try to use specific vocal traits~\citep{Regnier_2009}.
Here, the particularities of vocal vibrato and tremolo (average rate, average extent, and presence of both modulations) are exploited to discriminate instrumental and singing voice segments.
This approach achieves a recall of the singing voice class of 83.57\% on the \gls{jamendo} dataset.
Some authors adopt speech recognition systems for the particularities of singing voice~\citep{Berenzweig_2001}.
Given a time frame, they create a feature vector, composed of 13 perceptual linear predictions (PLP) coefficients.
The idea is to exploit the resemblance between singing voice and spoken voice when compared with non-vocal music segments.
From the PLP coefficients, the authors extract different features.
They use a dataset composed of 246 15-second fragments recorded at random from FM radio in 1996. The authors report a 73,9\% accuracy on the singing / non-singing classification task.

\textbf{Singing voice detection as a preprocessing step.}
In~\citep{Berenzweig_2002}, Multi-Layer Perceptron (MLP) on each feature vector (13 perceptual linear prediction (PLP) coefficients) was employed to distinguish between singing and non-singing segments. To provide context, 5 consecutive frames are presented to the network. This segmentation in vocal and non-vocal segments is then used to perform an artist classification, the target task of this study.
Similarly, to perform audio lyrics alignment, authors proposed a vocal activity detection method based on an HMM on top of a Gaussian mixture model (GMM) which models the power of harmonic components~\citep{Fujihara_2011}. They then compare the given harmonic structure with only those in the dataset that have similar F0 values. This first step helps at isolating the vocal segment, to reconstruct them as a separate audio track and then perform the lyrics alignment.
Lyrics transcription~\citep{Mesaros_2013} or \gls{source_separation}~\citep{Simpson_2015} trained then to obtain ideal binary masks also use the \textit{singing voice detection} as a pre-processing-step.

\textbf{Deep learning.}
Over the past few years, works have focused on the use of \gls{dnn} techniques.
Some researchers propose the use of Recurrent Neural Networks (RNN)~\citep{Lehner_2015} on a 60-dimensional feature vector (30-dimensional MFCC and their first order derivative) along with the cross-correlation of each filter-bank spectrum of a time frame (called a fluctogram), spectral contraction and spectral flatness. They reported 89.42\% accuracy on the \gls{jamendo} dataset.
Bidirectional Long Short-Term Memory (BLSTM), a class of recurrent neural network, were also used in~\citep{Leglaive_2015}. Authors chose to represent the audio signal as a combination of 1) a harmonic part and 2) a percussive part, using a double stage HPSS as proposed in~\citep{Tachibana_2010}. Each part is then transformed into a Mel-spectrogram and passes to the BLSTM network. They report a 91,5\% accuracy on the \gls{jamendo} dataset.
A comprehensive discussion of these approaches can be found at~\citep{Lee_2018}.

Finally, the use of \gls{cnn} is also popular. They were introduced in combination with data augmentation to increase the size of the training set~\citep{Schluter_2015} or trained on weakly labeled data (each 30 seconds excerpt is labeled as containing singing voice or not, but not at the frame of the excerpt) along with a three steps training strategy~\citep{Schluter_2016}.
A Constant-Q input model trained on a very large private dataset (mined from Spotify resources) obtains an accuracy of 87,8\% on the \gls{jamendo} dataset~\citep{Humphrey_2017}.

\subsection{Our model}
\begin{figure*}[ht]
  \centerline{
    \includegraphics[width=\textwidth]{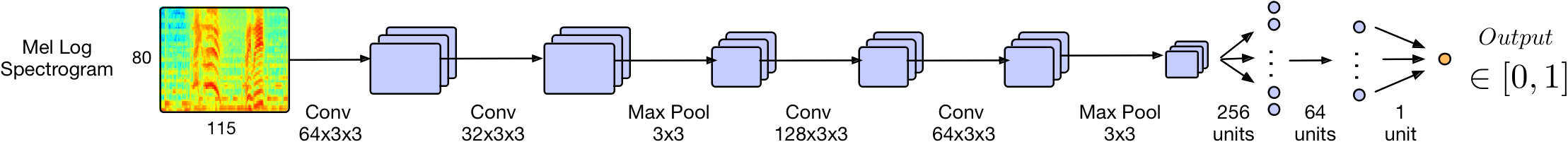}}
  \caption[SVD architecture]{Architecture of our Singing Voice Detection (SVD) system using \gls{cnn} proposed by~\citep{Schluter_2015}.}
  \label{fig:convNet_voice_detection}
\end{figure*}

The model we use is based on the \gls{cnn} proposed by~\citep{Schluter_2015}.
It follows a standard \gls{cnn} architecture with two \gls{fully} layers and a final neuron that provides $\hat{p}(t)$ for the center time-frame of the patch.
The input is a sequence of patches of 80 \gls{mls} over 115 consecutive time frames (0.014 seconds per frame).
\gls{mls} applies a bank of triangular filters to the power of the \gls{dft} spectral coefficients (see Figure~\ref{fig:freq_domains}). These filters are designed based on the Mel-scale to be more discriminative at lower frequencies and less at higher frequencies. This mimics the nonlinear critical bands of the human ear. We then compute the logarithm of the magnitude in each new bin.
Figure \ref{fig:convNet_voice_detection} shows the architecture of the network.
The model is trained on binary target using a binary cross-entropy loss-function, ADAMAX optimizer, mini-batch of 128, and 10 epochs.
We chose this architecture because it has proved to be easy to implement, fast and more important, robust for the singing voice detection task.

\section{Normalized cross-correlation}
\label{sec:NCC}

At this stage, audio and annotation are described as vectors over time $\hat{p}(t) \in [0,1]$ and $\mathit{vas}_{o,\mathit{fr}}(t) \in \{0,1\}$.(see Figure~\ref{fig:two_domains}).
We measure their similarity using the \gls{ncc}. This similarity is not only important in recovering the right audio and finding the best alignment, but also in filtering imprecise annotations. (see Section~\ref{sec:linking_auido}).

We measure the similarity between $\hat{p}(t) \in [0,1]$ and $\mathit{vas}_{o,fr}(t) \in \{0,1\}$ using the \gls{ncc}, which is the normalized version (between 0 and 1) of the cross-correlation.
The cross-correlation measures the similarity between two digital sequences by sliding one -$y$- over the other -$x$-.
We use it when we search a shorter sequence in a longer sequence or when there is a relative displacement between $x$ and $y$ (our case).
At each value of the lag $l$, we calculate the correlation between them i.e. their degree of similarity.
This results in a new function that describes where $y$ best matches with $x$.
The highest correlation coefficient represents the best fits position $l$ between the two sequences.
This is the position that interests us.


\begin{equation}
  R_{x,y}(l) =  \sum_{l=-\infty}^{\infty} x(t) y(t-l)
\end{equation}

Since we are interested in global alignment we found this technique more precise than others such as Dynamic Time Warping (DTW). Indeed, DTW finds the minimal cost path for the alignment of two complete sequences. To do so, it can locally warps the annotations which usually deforms them rather than correcting them. It is also costly to compute and its score is not directly normalized, which prevents us from selecting the right candidate.

The $\mathit{vas}(t)$ depends on the parameters {\tt  offset\_time} ($o$) and the {\tt  frame rate} ($\mathit{fr}$),  $\mathit{vas}_{o,\mathit{fr}}(t)$. Hence, the alignment between $\hat{p}$ and $\mathit{vas}$ depends on their correctness.
While $o$ defines the beginning of the annotations, $\mathit{fr}$ controls the time grid size.
When changing $\mathit{fr}$, the grid size is modified by a constant value that compresses or stretches the annotations as a whole respecting the global structure. Our $\mathit{NCC}$ formula is as follows:

\begin{equation}
    \mathit{NCC}(o, \mathit{fr}) = \frac{\sum_t \mathit{vas}_{o,\mathit{fr}}(t-o) \hat{p}(t)}{\sqrt{\sum_t \mathit{vas}_{o,\mathit{fr}}(t-o)^2}  \sqrt{\sum_t \hat{p}(t)^2}}.
\end{equation}

\noindent For a particular $\mathit{fr}$ value, $\mathit{NCC}(o, \mathit{fr})$ can be used to estimate the best $\hat{o}$ to align both sequences.
We obtain the optimal $\hat{\mathit{fr}}$ using a brute force search in an interval $\alpha$\footnote{we use $\alpha = \mathit{fr}*0.05$} of values around $\mathit{fr}$:

\begin{equation}
(\hat{\mathit{fr}}, \hat{o}) = \argmax_{\mathit{fr} \in [\mathit{fr}-\alpha, \mathit{fr}+\alpha] ,~o } \mathit{NCC}(o, \mathit{fr}).
\end{equation}

This automatically obtains the $\hat{o}$ and $\hat{\mathit{fr}}$ values that best align $\hat{p}(t)$ and $\mathit{vas}_{o,\mathit{fr}}(t)$ and yields a similarity value between 0 and 1.

\begin{figure*}[ht]
  \centerline{
    \includegraphics[width=.8\textwidth]{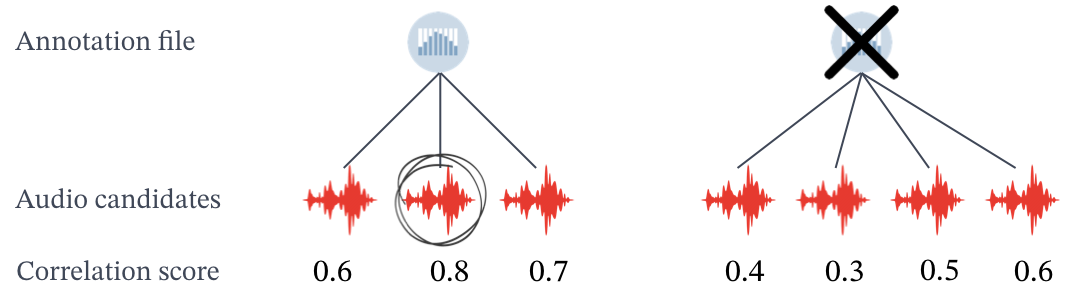}}
  \caption[Finding the correct audio example]{In order to discover its best alignment for each candidate, annotations are modified by changing its {\tt offset\_time} and {\tt frame rate}.
    The candidate with the highest \gls{ncc}  value is our final audio track.
    This similarity is not only important in recovering the correct audio and finding the best alignment but also in determining if annotations are accurate enough. Fixing a threshold we can filter imprecise annotations.}
  \label{fig:filtering}
\end{figure*}

This similarity automatically obtains the $\hat{o}$ and $\hat{fr}$ values that best align $\hat{p}(t)$ and $\mathit{vas}_{o,fr}(t)$. $\mathit{NCC}(o, fr)$ scores between 0 and 1.
We compute the $\mathit{NCC}(\hat{o}, \hat{fr})$ for all the audio candidates of a given $\mathit{vas}_{o,fr}$. Using this score we can both find the best candidate (highest score) and establish if the best candidate is good enough to be kept.
To this end, we fix a threshold $\mathit{NCC}(\hat{o}, \hat{fr}) \geq T_{\mathit{corr}}$ to keep only the accurate matches and discarding those for which the correct audio could not be found, and those with insufficiently accurate annotations (see Figure~\ref{fig:filtering}).

\begin{figure*}[ht!]
     \centerline{
     \includegraphics[width=\textwidth]{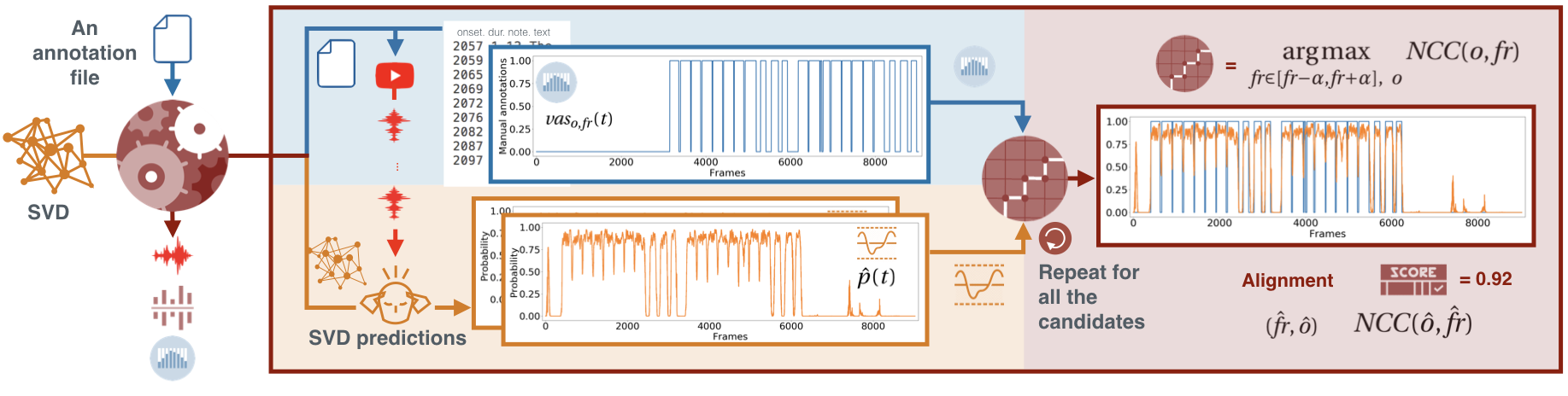}}
     \caption[Finding the correct audio summarization]{The input is an $\mathit{vas}_{o,fr}(t)$ (blue part) and a set of audio candidates retrieved from Youtube. The similarity estimation method uses a SVD model (orange part) to convert each candidate in a $\hat{p}(t)$ (orange part). We measure the similarity between the $\mathit{vas}_{o,fr}(t)$ and each $\hat{p}(t)$ using the  cross-correlation method $\argmax_{fr, o } NCC(o, fr)$ (garnet part). The output is the audio file with the highest $\mathit{NCC}(\hat{o}, \hat{fr})$ and the annotations aligned to it using $\hat{fr}$ and $\hat{o}$.}
     \label{fig:ncc}
\end{figure*}

The value of $T_{\mathit{corr}}$ has been empirically found to be $T_{\mathit{corr}}=0.8$.
We have set up a high threshold to ensure that a good proportion of our chosen audio and labels are quite well annotated.
This strategy is similar to the ones used in active learning, where, instead of labeling and using all possible data, we find ways of selecting the accurate data. The whole process is summarized in Figure~\ref{fig:ncc}.

At this point and after manually examining the obtained alignments, we noticed that the quality of the process strongly depends on the quality of $\hat{p}(t)$.
The $\hat{p}(t)$ obtained with the baseline SVD systems is sufficient to correctly identify the audio (although false negatives still exist), but not to align the annotations.
Thus, we need to improve $\hat{p}(t)$.
With increasing $\hat{p}(t)$, we will find more suitable matches and align the annotations more precisely (more accurate $\hat{o}$ and $\hat{fr}$).

There are two possibilities to do so: to develop a novel \gls{svd} system or to train\footnote{We train each new model from scratch, not using transfer learning.} the existing architecture with better data.
Since \gls{dali} is considerably larger (around 2000) than similar datasets (around 100), we choose the latter.
This idea re-uses all of the labeled data created in the previous step to train a better \gls{svd} system.

\section{Improving DALI: The teacher student paradigm}
\label{sec:teacher-student}

In this section we show how we can take advantage of the data we just retrieved and aligned using a \gls{teacher-student}.

\subsection{Previous work}
The two main agents of this paradigm are: the \textit{'\gls{teacher}'} and the \textit{‘\gls{student}’}.
The \textit{\gls{teacher}} is trained with labels of well-known ground truth datasets (often manually annotated) and used to label some unlabeled data.
The new labels given by the teacher are used for training the \textit{\gls{student}(s)}.
\textit{\Gls{student}} indirectly acquires the desired knowledge by mimicking the ``\textit{\gls{teacher}'s} behavior".
This paradigm was originally introduced as a model compression technique to transfer knowledge from larger architectures to smaller ones~(\citep{Bucilua_2006}).
Small models (the students) are trained on a larger dataset labeled by large models (the teachers).
A more general formalization of this knowledge distillation trains the student in both the teachers' labels and the training data~\citep{Hinton_2014}.
In the context of Deep Learning, the teacher can also automatically remove layers in the architecture of the student, automating the compression process~\citep{Ashok_2017}.

The \gls{teacher-student} is also used as a solution to overcome the problem of insufficient labeled training data, for instance for speech recognition~\citep{Watanabe_2017, Wu_2017} or and multilingual models~\citep{Cui_2017}.
Since manual labeling is a time-consuming task, the \gls{teacher-student} explores the use of unlabeled data for supervised problems.
The teachers (trained on labeled datasets automatically) label unlabeled data on a (usually) larger dataset used for training the students.
This way of applying the \gls{teacher-student} is one most popular methods employed in \textbf{semi-supervised} learning.
This paradigm is still relatively undeveloped in \gls{mir}. One of the few examples that applies it to automatic drum transcription is~\citep{Wu_2017}, in which the teacher labels the student dataset of drum recordings.

All of these works report that the students improve over the performance of the teachers. Therefore, this learning paradigm meets our requirements.

\subsection{Our Teacher student}

\begin{figure*}[ht!]
  \centerline{
    \includegraphics[width=0.8\textwidth]{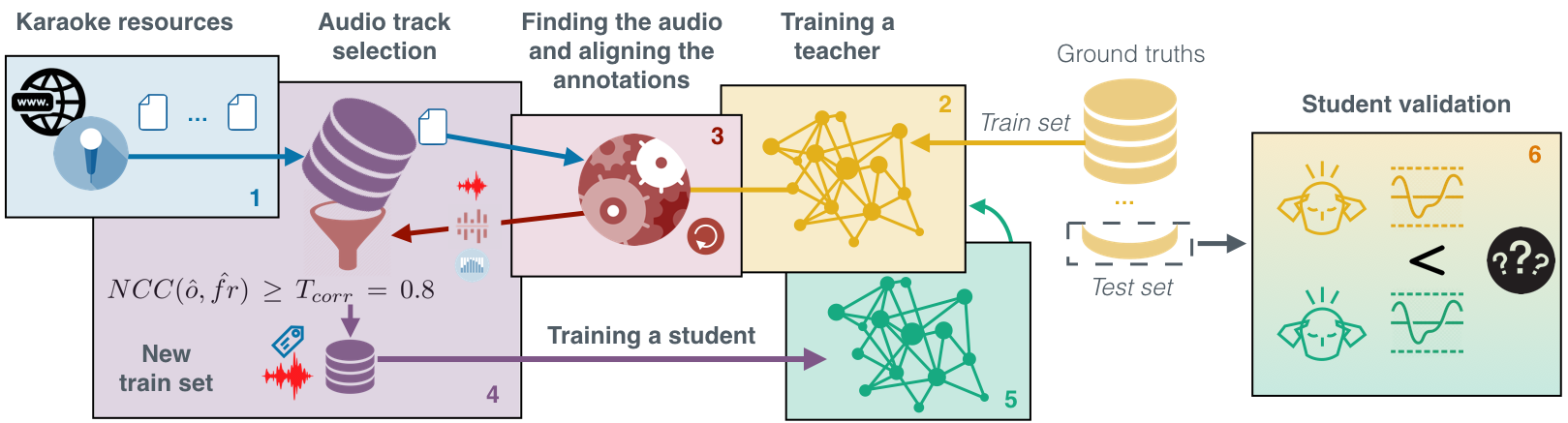}}
  \caption[Creating DALI with the teacher-student paradigm]{The \gls{dali} dataset creation using the \gls{teacher-student}.}
  \label{fig:teacher-student}
\end{figure*}

Our goal is to improve our \gls{svd} system. We use the teacher to select the retrieved audio and align the annotation to it. This new data is used for training a new \gls{svd} system.
Our hypothesis is that if $\hat{p}(t)$ becomes better, the $\argmax \text{ }NCC(o, fr)$ will find better matches and align more precisely the annotations to the audio (more accurate $\hat{o}$ and $\hat{fr}$).
As a result, we obtain a better \gls{dali} dataset.
This larger dataset can then be used to train a new \gls{svd} system which again, can be used to find more and better matches improving and increasing the \gls{dali} dataset.
This can be repeated iteratively.
After our first iteration and using our best \gls{svd} system, we reach 5358 songs.
We then perform a second iteration that defines the current 7756 songs.

Our teachers do not label directly the input training data of the students but rather select the audio and align the annotations to it (see Section \ref{sec:svd}).
Similar to noisy label strategies such as active learning or weakly learning, we use a threshold to separate good and bad data points.
But, instead of doing this dynamically during training, we filter it statically once an \gls{svd} model is trained.
This process is summarized in Figure~\ref{fig:teacher-student} and detailed in \citep{meseguerbrocal_2018}:

\begin{description}[.15cm]
  \item[1- Blue.]
  The retrieved karaoke annotation files are converted to an annotation voice sequence $\mathit{vas}(t)$.

  \item[2- Yellow.]
  We train a SVD, the \textbf{teacher}, either using clean ground-truth datasets or after the first iteration using \gls{dali} annotations (green arrow).

  \item[3- Garnet.]
  With the teacher $\hat{p}(t)$ and the $\mathit{vas}(t)$ we compute the $\mathit{NCC}$ to find the best audio candidate and alignment parameters $\hat{fr}, \hat{o}$ (see Section \ref{sec:NCC}).

  \item[4- Purple.]
  We select the pair audio-annotation with $\mathit{NCC}(\hat{o}, \hat{fr}) \geq T_{\mathit{corr}} =  0.8$.
  This set defines a new training set (and a \gls{dali} version).

  \item[5- Green.]
  Using the new data we train a new \gls{svd}, called the \textbf{student}\footnote{We retrained from scratch, \textbf{not} adapting the previous models.}.

  \item[6- Yellow-Green.]
  The two systems, teacher and student, are compared on the clean ground-truth test set to check that the new system performs indeed better than the previous one.

\end{description}

To train a new a \textbf{student} (step 5) we need to define the true value we want to model (target values $p$ to be minimized in the loss $\mathcal{L}(p,\hat{p})$ of the \gls{svd} model).
There are three choices. We can either use:

\begin{itemize}
  \setlength\itemsep{0em}
  \item[a)] the predicted value $\hat{p}$ as provided by the \textbf{teacher} (this is the usual \gls{teacher-student}).
  \item[b)] the value of the $\mathit{vas}$ corresponding to the annotations after aligning them using $\hat{fr}$ and $\hat{o}$.
  \item[c)] a combination of both: keeping only the frames for which $\hat{p}(t)=vas(t)$.
\end{itemize}

Up to now, and since $\mathit{vas}$ has been found more precise than $\hat{p}$, we only investigated option \textbf{b)}. This differs from other works where they use as target the output of the teacher (option \textbf{a)}). In our approach, the teacher `filters' and `corrects' the source of knowledge from which the student has to learn. Metaphorically speaking, instead of telling him exactly which `sentences', our teacher tells the student which `books' to read.

\begin{figure*}[ht]
  \centerline{
    \includegraphics[width=0.9\textwidth]{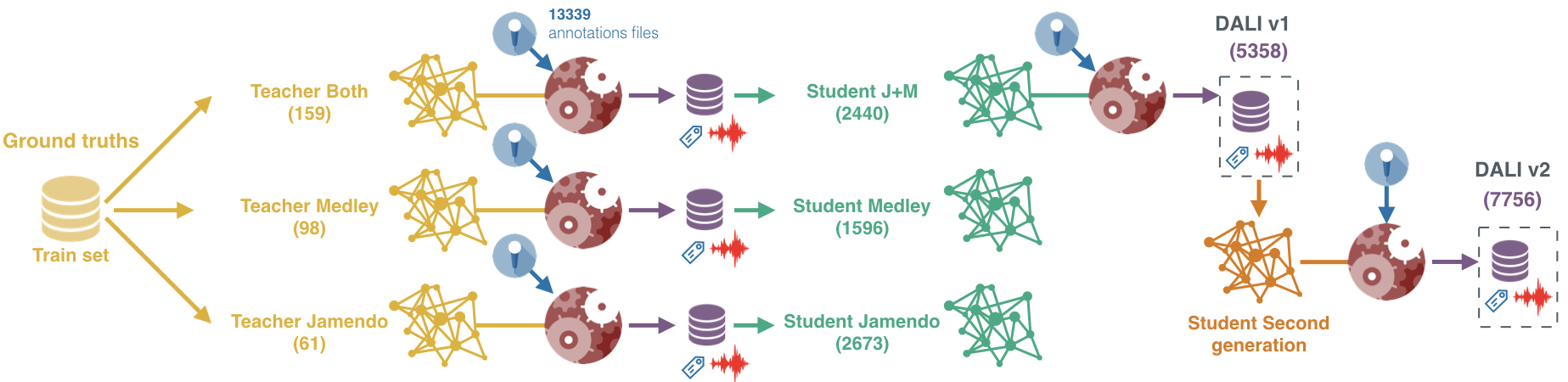}}
  \caption[The different DALI versions]{We create three SVD systems (\textit{teachers}) using the ground truth dataset (\gls{jamendo}, \gls{medley} and Both).
  The three systems generate three new datasets (\gls{dali} v0) used to train three new SVD systems (the \textit{students} first generation).
  Now, we use the best student, \textit{J+M}, to define \gls{dali} v1 released in~\citep{meseguerbrocal_2018}.
  \gls{dali} v1.0 is used to train a second-generation student that creates \gls{dali} v2 in~\citep{meseguerbrocal_2020}\\}
  \label{fig:second_generation}
\end{figure*}

This process incrementally adds more good audio-annotations pairs. We perform this three times as summarized in Figure~\ref{fig:second_generation}.
With this process, we are simultaneously improving the \gls{svd} model and the dataset.
Besides, this is also an indirect way of examining the quality of annotations: a well-performing system trained only with this data show us that the time alignment of the annotation is correct enough for this task.

\newcommand{\M}[0]{\textit{MedleyDB}}
\newcommand{\Ja}[0]{\textit{Jamendo}\xspace}
\newcommand{\Me}[0]{\textit{MedleyDB}\xspace}
\newcommand{\both}[0]{\textit{J+M}\xspace}
\newcommand{\SeGe}[0]{\textit{Second Generation}\xspace}
\newcommand{\JTr}[0]{\textit{J\_Train}\xspace}
\newcommand{\MTr}[0]{\textit{M\_Train}\xspace}
\newcommand{\bothTr}[0]{\textit{J+M\_Train}\xspace}
\newcommand{\JTrN}[0]{\textit{J\_Train}}
\newcommand{\MTrN}[0]{\textit{M\_Train}}
\newcommand{\bothTrN}[0]{\textit{J+M\_Train}}
\newcommand{\JTe}[0]{\textit{J\_Test}\xspace}
\newcommand{\MTe}[0]{\textit{M\_Test}\xspace}
\newcommand{\bothTe}[0]{\textit{J+M\_Test}\xspace}
\newcommand{\JTT}[0]{\textit{J\_Test+Train}\xspace}
\newcommand{\MTT}[0]{\textit{M\_Test+Train}\xspace}

\subsection{First generation}
\subsubsection{Ground-Truth datasets}
\begin{figure*}[ht]
  \centerline{
    \includegraphics[width=.4\textwidth]{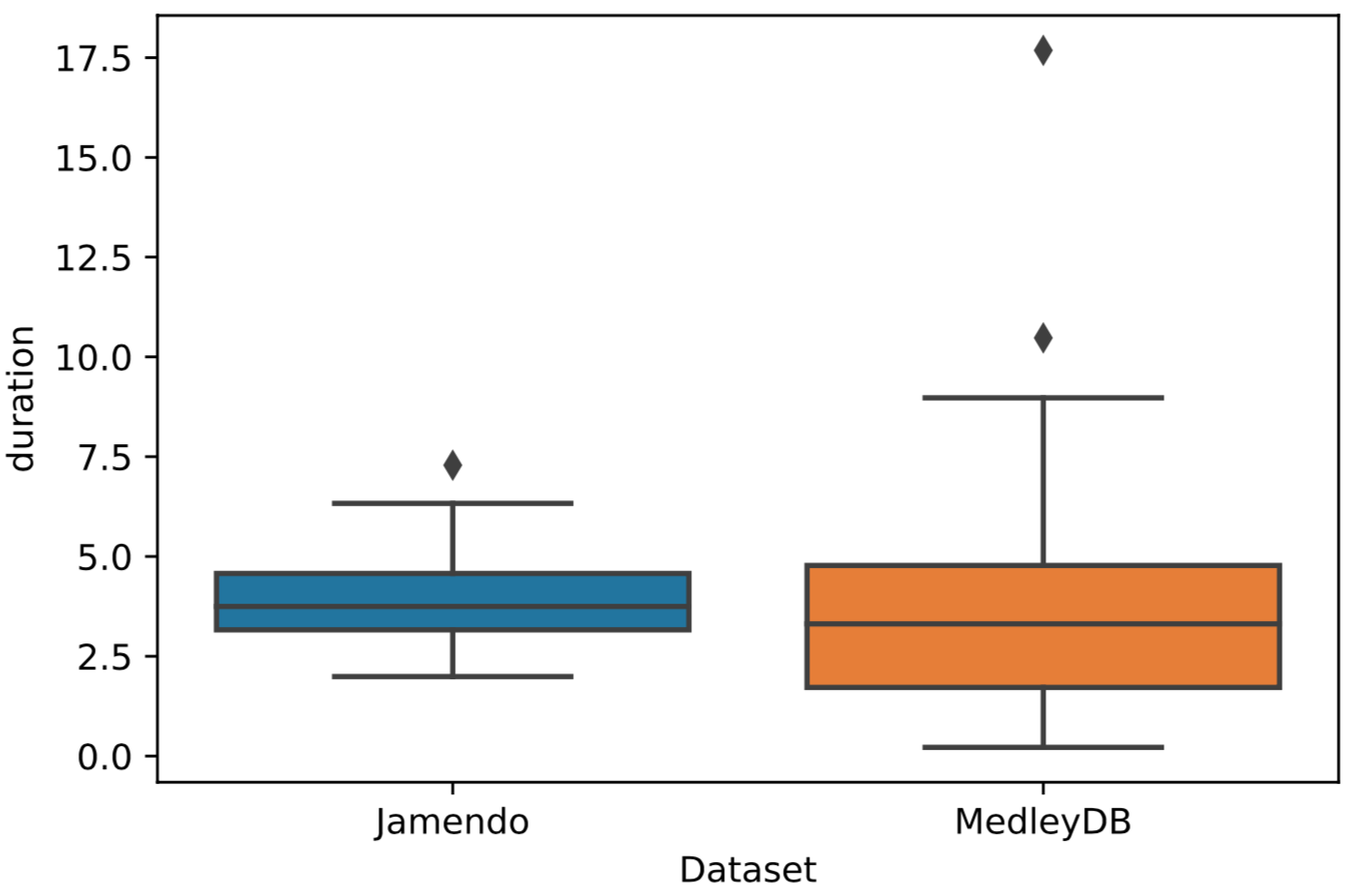}}
  \caption[Box plot of duration of Jamendo and MedleyDB datasets]{Box plot of duration of \gls{jamendo} and \gls{medley} datasets datasets}
  \label{fig:datasets}
\end{figure*}

We use three ground-truth datasets to train the first teachers: \Ja~\citep{Ramona_2008}, \Me~\citep{Bittner_2014} and a third one that merges both \both.
They are accurately labeled but small.
In Figure~\ref{fig:datasets} we present the mean duration of \gls{jamendo} and \gls{medley}. We can see that \gls{medley} contains track with many durations. 

\gls{jamendo} is a music website that offers free music streaming and download.
Crawling this website, the authors downloaded 93 tracks with Creative Commons license~\citep{Ramona_2008}. Each track has been manually annotated by the same person.
The dataset is split in a training set of 61 songs, 16 songs are kept for the validation set and 16 songs for the testing set.

\gls{medley}~\citep{Bittner_2014} contains 122 songs in multi-tracks, with mix, stems of different instruments.
The dataset is annotated in melody F0 (for 108 tracks), instrument activations and genre (for all tracks).
Among the 122 songs, 52 are instrumental only, 70 contains vocals.
The melody annotations were generated semi-automatically.
A monophonic pitch tracking algorithm was used on the separated stems, the pYin algorithm.
This gave the authors a good initial estimate of the F0 curve. For the automation of the instrument activations, a standard "envelope follower" technique on each stem was used\footnote{It consists of half-wave rectification, compression, smoothing and down-sampling.}. After this automatic step, the annotations were refined by five human musician annotators.

Each dataset is split into a train, validation and test part using an artist filter\footnote{No artist who appears in the training set can appear in the test set.}. We keep \Ja~and \Me~for testing the different SVD systems.

\subsubsection{The Teachers}
 We train three teachers using the training part of each ground-truth set. 
The teachers select the audio and align the annotations as described in Section \ref{sec:NCC} creating three new datasets (DALI v0) with 2440, 2673 and 1596 items for the teacher \both, \Ja~and \Me~ respectively.

\begin{table*}[th!]
    \centering
    \caption[Audio candidates intersection]{Audio candidates intersection in percentage for filtering threshold = 80.}
     \begin{tabular}{l | c  | c  | c |  c }
        {} &  \both &  \Ja &  \Me &  Three\\
        \midrule
        \both (2440)      & 100      & 91.4    & 61.6     & 58.8   \\ \hline
        \Ja    (2673)     & 83.4     & 100     & 55.2     & 53.6   \\ \hline
        \Me    (1596)     & 94.2     & 92.4    & 100      & 89.8   \\ \hline
    \end{tabular}
    \label{table:intersection}
\end{table*}

Table \ref{table:intersection} presents the intersection of the three sets.
It indicates how many tracks of each new set are also in the other two sets.
For example, the 89.8\% (bottom right) of the selected tracks using the \Me~teacher are also present within the ones selected by the \both~teacher or the \Ja~teacher.
Also, the 91.4\% (top second left) of the selected tracks using the \Ja~teacher are within the ones selected by the \both~teacher.
This table shows that the three teachers agree most of the time on selecting the correct audio for a given annotation.

\subsubsection{The Students}
We train three different students. Among the possible target values: $\hat{p}$ given by the \textbf{teacher} -as common in the \gls{teacher-student}-, $\mathit{vas}$ after being aligned using $\mathit{NCC}$ or a combination of both; we use the $\mathit{vas}$. We have found this vector to be more accurate than $\hat{p}$. In our approach, the teacher `filters' and `corrects' the source of knowledge from which the student learns.
Each student is trained with different data since each teacher may find different audio-annotations pairs and different alignments (each one gets a different $\hat{p}$ which leads to different $\hat{fr}, \hat{o}$ values).

We hypothesize that if we have a more accurate $\hat{p}$, we can create a better \gls{dali}. In Table \ref{table:teacher-student} and \ref{table:align-teacher-student}, we observe how the students outperform the teachers in both the singing voice detection task and the alignment experiment. Thus, they produce better $\hat{p}$. Furthermore, we assume that if we use these \gls{svd} systems, we will retrieve better audio and have a more accurate alignment.
For this reason we use the \textbf{student} based on \textbf{\both} that obtains the best results to create \gls{dali} version one with 5358 songs\citep{meseguerbrocal_2018}.

\subsection{Second generation}
\label{sec:second-generation}

We now use as a teacher the best student of the first iteration, the student \textbf{\both}.
We again repeat the process described in Section \ref{sec:NCC}. In this case, the teacher is not trained on any ground-truth but with \gls{dali} v1 and using the well aligned $\mathit{vas}$ as a target.
We split \gls{dali} v1 (5358 tracks) into three sets: 5253 for training, 100 for validation (the ones with higher NCC) and 105 for testing.
The test set has been manually annotated (right $\hat{fr}$ and $\hat{o}$) and constitutes our ground-truth for future experiments\footnote{Note that this split is different from the proposed in Section \ref{sec:dali_description_analysis} because of the nature of the experiments carried out}.

The new \gls{svd} (student of the second generation) obtains even better results in both the singing voice detection task and the alignment experiment (see Section~\ref{sec:analysis}). Hence, we assume that another repetition of the process will output a better \gls{dali}. Indeed, this is the \gls{dali} v2 with 7756 audio-annotations pairs.

\section{Experiments for validating the new Singing Voice Detection systems}
\label{sec:analysis}

We validate the performance of each model on two different tasks: the singing voice detection and the $\mathit{vas}$ alignment result of the $ \argmax_{\mathit{fr} \in [\mathit{fr}-\alpha, \mathit{fr}+\alpha] ,~o } \mathit{NCC}(o, \mathit{fr})$.
We demonstrate that generally, students perform much better than their teacher for both tasks.

\subsection{Singing voice detection}
This task validates the accuracy of the predictions $\hat{p}(t)$ in the singing voice classification task. 
Results are indicated in Table~\ref{table:teacher-student} where e.g. `\textit{S(T(\bothTr))} (2673)' refers to the student trained on the 2673 audio-annotation pairs and the $\mathit{vas}$ values obtained with the Teacher \textit{T(\bothTr)} trained on \bothTr set.

We evaluate the performances of the various \gls{svd} models using the test and test+train (when possible) set for the \Ja~and \Me~ground-truths.
We measure the frame accuracy for each model as followed.
Firstly, we compute the threshold that defines which $\hat{p}(t)$ values are considered 1 (there is singing voice) or 0 (no singing voice) using the validation set of each ground-truth.
The threshold might be different for each model and/or dataset.
Then, we compute the binary accuracy score per track and finally the mean and standard deviation.

\begin{table*}[th!]
  \centering
  \caption[Results for the singing voice detection task]{Performances of \textit{\textbf{singing voice}} detection, measured as mean accuracy and standard deviation, for the teachers and students. Number of tracks in brackets. Nomenclature: T = Teacher, S = Student, J = \Ja, M = \Me, \both = \Ja+ \Me, 2G = second generation, the name of the teacher used for training a student in bracket.}
  \label{table:teacher-student}
  \footnotesize
  \begin{tabular}{c | c  | c || c | c }
    \hline
    \backslashbox{SVD system}{Test\_sets}  & \JTe (16)    & \MTe (36)   & \textbf{\JTT} (77)  & \textbf{\MTT} (98) \\
    \hline
    \hline
    \textit{T(\JTr)} (61)         		       & 88.95\% $\pm$ 5.71     & 83.27\% $\pm$ 16.6     &  -                     & 81.83\% $\pm$ 16.8 \\
    \textit{S(T(\JTrN))} (2673)     	     & 87.08\% $\pm$ 6.75     & 82.05\% $\pm$ 15.3     & 87.87\% $\pm$ 6.34     & 84.00\% $\pm$ 13.9 \\
    \hline
    \hline
    \textit{T(\MTr)} (98)        		       & 76.61\% $\pm$ 12.5     & 84.14\% $\pm$ 17.4     & 76.32\% $\pm$ 11.2     & - \\
    \textit{S(T(\MTrN))} (1596)    	     & 82.73\% $\pm$ 10.6     & 79.89\% $\pm$ 17.8     & 84.12\% $\pm$ 9.00     & 82.03\% $\pm$ 16.4 \\
    \hline
    \hline
    \textit{T(\bothTr)}  (159)   				       & 83.63\% $\pm$ 7.13   	& 83.24\% $\pm$ 13.9     &   -     & - \\
    \textit{S(T(\bothTrN))} (2440) 		     & 87.79\% $\pm$ 8.82 		& 85.87\% $\pm$ 13.6     & 89.09\% $\pm$ 6.21     & 86.78\% $\pm$ 12.3 \\
    \hline
    \textit{2G(S(T(\bothTrN)))} (5253) 	   & \textbf{93.37\% $\pm$ 3.61} 		& \textbf{88.64\% $\pm$ 13.0}     & \textbf{92.70\% $\pm$ 3.85}    & \textbf{88.90\% $\pm$ 11.7} \\
    \hline
  \end{tabular}
\end{table*}

\begin{description}[.15cm]\setlength{\parindent}{1pc}
\setlength{\parskip}{.005cm plus0mm minus0mm}
    \item[Baseline SVD.]
    We first test the teachers.
    \textit{T(\JTr)}~obtains the best results on \JTe (89\%).
    \textit{T(\MTr)}~obtains the best results on \MTe (84\%)\footnote{There is a constant effect observed in all the experiments: a great result variability while testing on \M.
    We hypothesize that this is due to having a great number of instrumental songs in \Me~(62 out of 122).}
    In both cases, since training and testing are performed on the two parts of the same dataset, they share similar audio characteristics.
    Therefore, these results are artificially high.

    To best demonstrate the generalization of the trained SVD systems, we need to test them in a cross-dataset scenario, namely training and testing on different datasets.
    Comparing the performances on different test sets gives a sense of how good is the real generalization of the model.
    Indeed, in this scenario, the results are quite different.
    Applying \textit{T(\JTr)}~on \MTT the results decrease to 82\% (a 7\% drop).
    Moreover, when applying \textit{T(\MTr)}~on \JTT the results decrease to 76\% (an 8\% drop).
    Consequently, we can say that the teachers do not generalize very well.

    Lastly, the \textit{T(\bothTrN))}~trained on J+M\_train performs worse on both \JTe (84\%) and \MTe (83\%) than their non-joined teacher (89\% and 84\%).
    This result is surprising and remain unexplained.
    Unfortunately, we cannot prove its generalization since this system cannot be tested on \MTT nor \JTT because it has been trained with the training tracks.

    \item[First students.]
    We now test the students.
    We hypothesize that students improve the results from the teachers due to the fact they have been trained using more data.
    Especially, we assume that their generalization to unseen data will be better.
    It is important to note that students are always evaluated in a cross-dataset scenario since their training set does not contain any track from \Ja~or \M.
    Hence, there are not artificially high results.

    The student \textit{S(T(\JTr))} based on \textit{T(\JTr)} outperforms its teacher.
    When applied to \MTT, it reaches 84\% which is slightly higher than the performances of the \textit{T(\JTr)}~directly (82\%).
    It also reaches similar results in the \JTT 88\% than its teacher in its own test set (89\%).

    Likewise, the student \textit{S(T(\MTr))} based on \textit{T(\MTr)}~is also better than its teacher in the cross dataset scenario \JTT with 84\% while T(\MTr)~(76\%).
    This case is even better than the previous one since there is an 8\% improvement.
    It also gets 82\% on the \MTT that is close to the 84\% from its teacher in its own test set.

    Finally, it is also true for the performances computed with the student \textit{S(T(\bothTrN))} based on \textit{T(\bothTrN)}.
    This system can be only compared with its teacher in the test dataset (not the test+train).
    Here, when applied either to \JTe or \MTe, it reaches 87\% (88\% on \Ja~and 86\% on \M) which is above the \textit{T(\bothTrN)}~(83.5\%).
    Also, this student performs as good (or above) as the other two previous teachers, \textit{T(\JTr)}~on \JTe (89\%) and \textit{T(\MTr)}~on \MTe (84\%).
    Besides, if we focus on the test+train section, this student outperforms any system on any dataset with 89\% on \JTT and 87\% on \MTT.
    These are very interesting results because this is the best student but it has not been trained with the best teacher (which is \Ja).
    This SVD system is the one used for defining the first DALI dataset.

    \item[Second student.]
    In this scenario, we hypothesize that the new student improves the results from previous students or teachers not only because it sees more data (5253) but also because this data is better than any previous one.
    To that end, we rely on the alignment results shown in Section \ref{sec:alignment}, which demonstrates that the students produce also a better annotation alignment, therefore a more accurate target value while training.

    In this second iteration, there is only one system: the \SeGe \textit{2G(S(T(\bothTrN)))} trained on the \gls{dali} dataset version 1 defined and aligned by the student \textit{S(T(\bothTrN))}.
    The new \gls{svd} system confirms our hypothesis and outperforms notably any existing system.
    It gets 93\% on \JTe and 89\% on \MTe which are the best results on these test sets.
    These results are even higher than the artificially high ones obtained by the system trained directly on them where \textit{T(\JTr)}~gets 89\% on \JTe and \textit{T(\MTr)}~83\% on \M.
    The \SeGe~is also the one that generalizes the best.
    It reaches 93\% on \Ja\_test+train and 89\% on \Me\_test+train, which is more precise than our best previous system (the student based on \bothTrN) that has 89\% and 87\% respectively.

\end{description}

This experiment proves that students work much better in the cross-dataset scenario (real generalization) when the train-set and test-set are from different datasets.
It is important to note that the accuracy of the student networks is higher than that of the teachers, even if they have been trained on imperfect data.

\subsection{On alignment}
\label{sec:alignment}

We hypothesize that a more accurate $\hat{p}$ leads to a better \gls{dali}. To prove this, we measure the precision of the $o$ and $fr$ computed by each SVD system.
We have manually annotated 105 songs of \gls{dali} v1.0, i.e. finding the $\hat{\mathit{fr}}$ and $\hat{o}$ values that give the best global alignment.
These songs are our ground-truth data.
We measure how far the estimated $\hat{fr}$ and $\hat{o}$ diverge from the manually annotated ones.
We name these deviation $\mathit{offset}_d$ and $\mathit{fr}_d$.

\begin{table*}[th!]
  \footnotesize
  \caption[Results for the alignment task]{\textit{\textbf{Alignment}} performances for the teachers and students. Mean offset deviation in seconds, mean frame deviation in frames and pos = position in the classification.}
  \label{table:align-teacher-student}
  \footnotesize
  \begin{tabular}{l | c | c | c || c | c | c  }
    \hline
    {} &  mean offset rank  & pos & mean $\mathit{offset}_{d}$ &  mean fr rank & pos & mean $\mathit{fr}_{d}$ \\
    \hline
    \hline
    \textit{T(\JTr)} (61)         		             & 2.79 $\pm$ .48 & 4 & 0.082 $\pm$ 0.17 & 1.18 $\pm$ .41 & 4 & 0.51 $\pm$ 1.24\\
    \textit{S(T(\JTrN))} (2673)    	             & 2.37 $\pm$ .19 & 3 & 0.046 $\pm$ 0.05 & 1.06 $\pm$ .23 & 3 & 0.25 $\pm$ 0.88\\
    \hline
    \hline
    \textit{T(\MTr)} (98)        		             & 4.85 $\pm$ .50 & 7 & 0.716 $\pm$ 2.74 & 1.89 $\pm$ .72 & 7 & 2.65 $\pm$ 2.96\\
    \textit{S(T(\MTrN))} (1596)    	             & 4.29 $\pm$ .37 & 6 & 0.164 $\pm$ 0.10 & 1.30 $\pm$ .48 & 5 & 0.88 $\pm$ 1.85\\
    \hline
    \hline
    \textit{T(\bothTr)}  (159)   				           & 3.42 $\pm$ .58 & 5 & 0.370 $\pm$ 1.55 & 1.47 $\pm$ .68 & 6 & 1.29 $\pm$ 2.29\\
    \textit{S(T(\bothTrN))} (2440) 		           & 2.23 $\pm$ .07 & 2 & 0.043 $\pm$ 0.05 & 1.04 $\pm$ .19 & 2 & 0.25 $\pm$ 0.85\\
    \hline
    \textit{2G(S(T(\bothTrN)))} (5253) 		      & \textbf{1.82 $\pm$ .07} & 1 & \textbf{0.036 $\pm$ 0.06} & \textbf{1.01 $\pm$ .10} & 1 & \textbf{0.21 $\pm$ 0.83}\\
    \hline
  \end{tabular}
\end{table*}

Results are indicated in Table~\ref{table:align-teacher-student}.
As in the previous table, `\textit{S(T\_J\_train)} (2673)' refers to the student trained on the 2673 audio-annotation pairs obtained with \textit{T(\JTr)}.
We estimate the average $\mathit{offset}_d$, $\mathit{fr}_d$ and the mean rank of each SVD.
The mean rank averages over songs the indiviual rank of each system for each song i.e. how far they are from the ground truth value.
For instance, four systems \textit{a}, \textit{b}, \textit{c} and \textit{d} with offset deviations 0.057, 0.049, 0.057 and 0.063 seconds are ranked as: \textit{b} = 1st, \textit{a} = 2nd, \textit{c} = 2nd and \textit{d} = 3rd, respectively.
The mean rank per model is the average of all individual ranks per song.
Finally, the \textit{position} value is the result of classifying the systems according to their mean rank value.

For this experiment we observe the same tendency than before: students outperform their teacher and the \SeGe~is the one that achieves the best results.

\begin{description}[.15cm]\setlength{\parindent}{1pc}
\setlength{\parskip}{.005cm plus0mm minus0mm}
    \item[Baseline SVD.]
    The main motivation to improve $\hat{p}$ is that the alignment we observed with the baseline SVD systems was not good enough.
    This experiment quantifies this judgment.
    The T(\MTr) and T(\bothTr) are ranked last in finding both the right offset and frame rate.
    Their values are considerably different from the ground-truth which produces an unnacceptable alignment.
    Remarkably, the T(\JTr) is much better and its results are comparable to the student networks.
    %
    %

    \item[First students.]

    Each student exceeds its teacher with consistently higher rank and lower deviations.
    S(T(\bothTrN)) is the best student.
    This is surprising because we use not particularly well-aligned data to train it (its teacher T(\bothTrN) is placed 5th and 6th for offset and fr).
    Yet it scores almost as well as the S(T(\JTrN)), which was trained with better-aligned data (its teacher is the best one).
    We presume an error tolerance in the singing voice detection task. This tolerance is not critical when below an unknown value, but crucial above it: S(T(\_\MTrN)) (trained with the most misaligned data) is far worse than the other students.

    \item[Second student.]
    As before, the \SeGe \textit{2G(S(T(\bothTrN)))} is the best system.
    It is placed first for both rankings and has the lowest deviation.
    However, the increase is moderate. We presume that we are reaching the limit of the alignment precision that we can achieve with the $\mathit{NCC}$, Section \ref{sec:NCC}.

    These results, together with the ones at Table \ref{table:teacher-student}, prove that \gls{dali} is improving at each iteration.

\end{description}


\section{The multitracks}
\label{sec:multitracks}

The multitrack version of \gls{dali} was built differently.
In the \gls{wasabi} project, there are 2k multi-tracks for popular music.
The intersection between this set and the retrieved annotations is 863 multi-tracks.
Nevertheless, these multi-tracks do not have any kind of standardization meaning that there are cases in which the sources are grouped by RAW files, other by STEMS files\footnote{While RAW files define individual tracks in the mixture, for instance, each part of a drum kit is stored in a different file, STEMS files merge all the RAW files of a given instrument into a single file.} or even several STEMS together in a track.
Additionally, they come without any metadata or label, just the name in the form track\_1, track\_2, ..., track\_n making difficult to work with them.

\begin{figure*}[ht]
  \centerline{
    \includegraphics[width=0.65\textwidth]{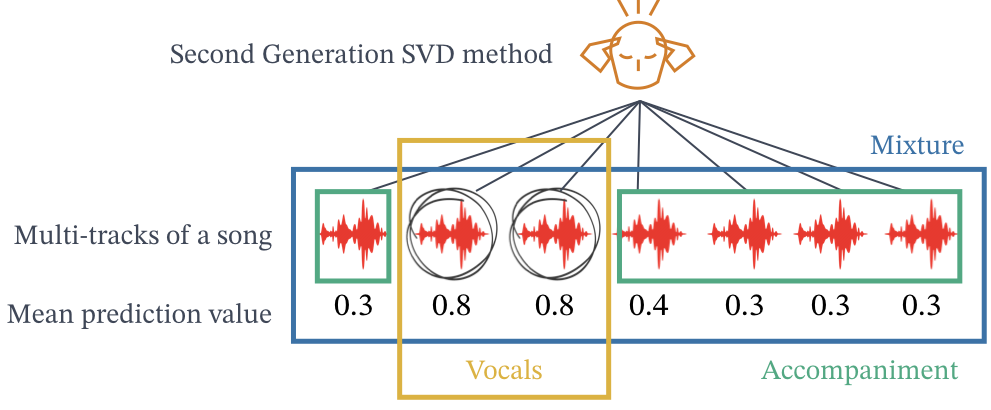}}
  \caption[Defining the multitracks version]{Method used for creating the \textit{vocals}, \textit{accompaniment} and \textit{mixture} version}
  \label{fig:multitracks}
\end{figure*}

Since in Chapter~\ref{sec:vunet} we focus on the vocal source sepration, we need to have three sources: \textit{vocals}, \textit{accompaniment} and the \textit{mixture}.
To do so, we use our best \gls{svd}, the Second Generation (see Section~\ref{sec:second-generation}) as follow.
Given all the tracks of a song, we apply this model to each one computing a different $\hat{p}$.
Assuming that there must be at least one track with vocals (this subset of the multi-tracks is overlapped with \gls{dali}) we compute the mean of each $\hat{p}$.
Finding the maximum of those mean and using a tolerance around this value (2\%) we can find the tracks that include vocals.
We then sum those tracks to produce the \textit{vocal} one. The rest of the tracks are joined to define the \textit{accompaniment} and all the tracks are merged to produce the final mixture (see Figure~\ref{fig:multitracks}). Lastly, we align the annotations using the procedure describe in Section~\ref{sec:NCC}.

We manually verify the resulting sources. We have found that only 512 multi-tracks were correct. Out of the 351 wrong ones, there were 30 where the original tracks were cut in the middle of the song and 321 where the original tracks did not contain the vocals isolated but rather mixed with other instruments (usually guitar or keyboards).
In the latter group, 69 songs are mostly good but vocals are combined with something else at the end of the track (when the chorus is repeated adding new instruments), 128 are mixed with an instrument in the background during the whole song and 154 are mixed with other instruments with a loudness similar to the vocals.

For future versions, we plan to prepare the rest of the instruments so that we can work with all the instruments that form the \textit{accompaniment} section.
This will allow us to deepen more into the work done in Chapter~\ref{sec:cunet}.

\section{Conclusion}
In this chapter, we explained our methodology to create the \gls{dali} dataset.
We defined a loop where dataset creation and model learning interact in a way that benefits each other.
Our approach is motivated by the Teacher-Student paradigm.
The time-aligned lyrics and notes come from Karaoke resources where non-expert users manually annotated the lyrics as a sequence of time-aligned notes with their associated textual information.
From the textual information, we derived the different levels of granularity.
We then linked each annotation to the correct audio and globally align the annotations to it using the \gls{ncc} on the \gls{vas} and \gls{svp}.
To improve the alignment, we iteratively updated the \gls{svd} using the teacher-student paradigm.
Through experiments, we showed that the students outperform the teachers notably in two tasks: alignment and singing voice detection.
In our case, we showed that, in the context of deep learning, it is better to have imperfect but large datasets rather than small and perfect ones.

\gls{dali} is a great challenge.
It has a large number of imperfect annotations that have the potential to make our field move forward.
But we need to solve the issues still presented in the dataset.
This requires ways to automatically identify and quantify them (see Chapter~\ref{sec:dali_correction}), which is costly and time-consuming.
On the other hand, we have deep learning models with imperfections that are difficult to quantify.
Therefore in \gls{dali}, we deal with two sources of imperfect information.
This puzzle is also common to other machine learning domains.
Our solution creates a loop where machine learning models are employed to filter and enhance imperfect data, used then to improve the models.
We prove that this loop benefits both: the model creation and the data curation.
We believe that \gls{dali} can be an inspiration to our community to not regard model learning and dataset creation as independent tasks but rather as complementary processes.

However, there is room for improvement.
In this first iteration of the \gls{dali} dataset, our goal was to find the correct audio candidates and globally align the annotations to it.
But, we still have false positives.
Moreover, once we have a good global alignment, we can try to solve local issues due to errors in the annotation process (see Chapter~\ref{sec:dali_errors}).

\graphicspath{{figs/}{dali_3_errors/figs/}}

\chapter{Errors in DALI}
\label{sec:dali_errors}

So far, we can only guarantee that each new \gls{dali} version has better audio-annotations pairs and a more accurate global alignment than the previous one.
We can also indirectly confirm that the current annotation timing $(t_k^0, t_k^1)$ is good enough to create a state-of-the-art \gls{svd} system.
But the dataset is still quite noisy and we know very little about the type of errors that are there.
After a manual analysis in detail, we classify those errors in two groups: \textbf{global} errors that affect the song as a whole and \textbf{local} errors to non-professional users.
In this chapter, we analyze these errors, propose solutions, suggesting some ways of indirectly measure the quality of the annotations based on a set of proxies.

\section{Global errors}

These errors affect the song as a whole and are usually due to problems in the transformation of the raw data (see Section~\ref{sec:raw_annot}) and the global alignment technique (see Section~\ref{sec:NCC}).
They are the least frequent issues.
Additionally, during our inquiry we have found almost no false positives meaning that the final audio we kept is the right one. On the other hand, there are still many true negatives that are not chosen due to the restrictive threshold we are using $T_{corr}=0.8$. There is the possibility of using a less restrictive threshold. However, we do not know how this affects the number of false positives.

\subsection{Global errors in time}
\label{sec:global_in_time}


The most common global time errors are those which have misaligned sections despite the audio-annotation pair having a high \gls{ncc} and good $\hat o$ and $\hat{\mathit{fr}}$ values.
In these cases, each section has a different offset. This can be solved by simply using the same \gls{ncc} technique but instead of applying it to the whole audio track, doing it so locally for each paragraph.
Thus, we isolate each $\mathit{vas}(t)$ of $A_\mathit{paragraphs}$ and the corresponding $\hat{p}(t)$ segment around the corresponding time extract.
We perform the \gls{ncc} described in Section~\ref{sec:NCC} on the two new vectors.

Moreover, there are songs with one or more missed sections. This is likely to occur when several vocalists sing at the same time or when a chorus is repeated at the end of the song but this is not indicated in the lyrics. The solution to these errors are not tackled and remain for future work. We just mark the lines that have possible errors as described in Section~\ref{sec:multitrack_errors}.

\subsection{Global errors in frequency}
\label{sec:global_in_freq}
\begin{figure*}[ht]
 \centerline{
   \includegraphics[width=0.9\textwidth]{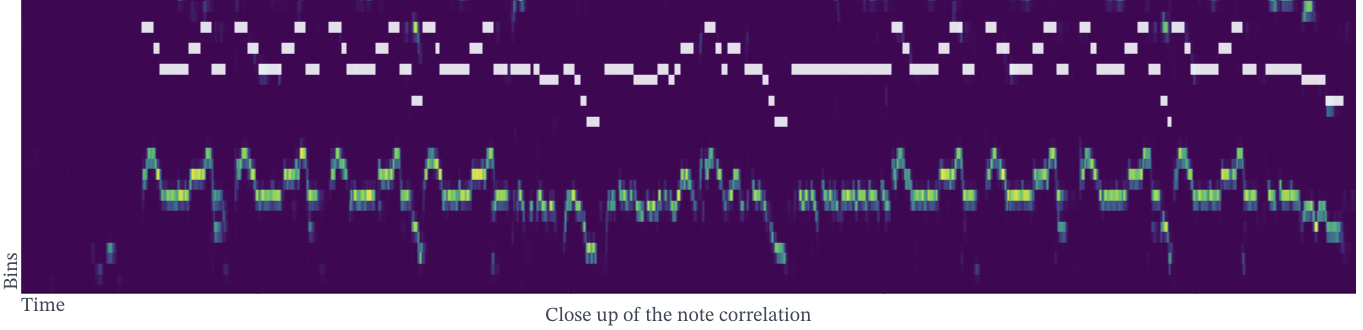}
 }
 \caption[In white the annotations as matrix overlapped with the $f_0$]{Annotations as matrix overlapped with the \gls{f0}. Note how the annotations are shifted in frequency by a constant factor.}
 \label{fig:annot_matrix}
\end{figure*}

The raw annotations store the notes as interval differences with respect to an unknown reference frequency (see Figure~\ref{fig:raw_annot}). Most of the songs use C4, which is the reference  used for transforming these differences into frequencies (see Section~\ref{sec:raw_annot}). But, this is not always the case and some songs use a different reference as shown in Figure~\ref{fig:annot_matrix}.

To solve the global frequency errors, we perform a new correlation in frequency between the note level $(a_{k, \mathit{notes}})_{k=1}^{K_\mathit{notes}}  = (t_k^0, t_k^1, f_k, l_k, i_k)_{\mathit{notes}} \rightarrow (t_k^0, t_k^1, f_k)_{\mathit{notes}}$ and a \gls{f0} extraction~\citep{Doras_2019}, that extracts the main pitch for each instant.
Pitch defines how we perceive the periodicity of music signals and it is highly related to the notion of ``musical notes''.
It permits us to classify musical events as high or low.
The extracted \gls{f0} is a matrix over time where each frame stores the pitch likelihoods obtained directly from the original audio.
We compress the original \gls{f0} representation to 6 octaves, 1 bin per semitone and a time resolution of $0.058$s. Similar to the process done in Section~\ref{sec:NCC}, we then transform the annotations $(a_{k, n})$ into a matrix. Unlike the previous correlation, we are measuring correlation along the frequency axis and not in the time one. We then simply transpose all the $f_k$ in the $a_{k, \mathit{note}}$ by the same value. The transposition factor covers all the frequency range in the \gls{f0} matrix. We find the frequency factor that maximizes the energy between the annotated frequencies $a_{k, \mathit{note}}$ and the estimated \gls{f0}. This defines the correct global position of the annotations of the whole set $S = \{A_\mathit{notes}, A_\mathit{words}, A_\mathit{lines}, A_\mathit{paragraphs}\}$.

\begin{figure*}[ht]
 \centerline{
   \includegraphics[width=0.9\textwidth]{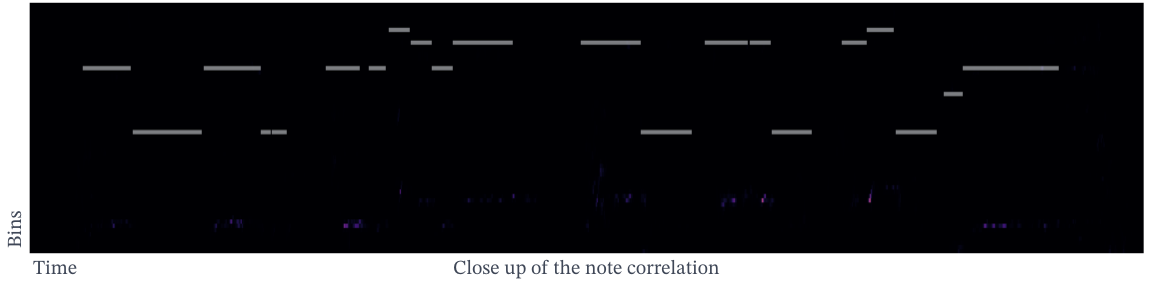}
 }
 \caption[Low $f_0$ correlation example]{Overlap between \gls{dali} annotations (white lines) and the \gls{f0} matrix with pitch likelihood distributions. Although the annotations are correct, the system that computes the \gls{f0} does not find any $f_0$ in this region (the black background means low likelihood for all the possible $f_0$). This results in a low-frequency correlation.}
 \label{fig:f0_corr}
\end{figure*}

Besides, we calculate the new “flexible” versions of the melody metrics \textit{Overall Accuracy, Raw Pitch Accuracy, Raw Chroma Accuracy, Voicing Recall, Voicing False Alarm}~\citep{Bosch_2019}. These metrics add new extra knowledge for understanding the quality of the annotations. Together with the \gls{ncc}, these can guide us to know which annotations are good and which ones are of lesser quality.
We can assume that a high \textit{Raw Pitch Accuracy} suggests a good alignment. However, low metrics do not necessarily indicate a bad one since this can be due to errors in the \gls{f0} extraction (see Figure~\ref{fig:f0_corr}).

\subsection{Multitracks}
\label{sec:multitrack_errors}

The most usual mistakes are some vocal segments without any annotations or some annotation lines in silent segments. To find vocals segments without annotations, we simply look for audio segments without annotations that have high energy in the isolated \textbf{vocals}. We perform the inverse process to find the annotations lines in silence segments. That is, to look for audio segments with annotation $a_{k, \mathit{lines}}$ that have low energy. We add those areas to the original annotations to be taken into account in future chapters. There are on average $0.26$ $a_{k, \mathit{lines}}$ on possible silence segments and $2.169$ audio segments with vocals (usually ohs) without annotations.

\section{Local errors}
\label{sec:local_errors}

\begin{figure*}[ht]
 \centerline{
   \includegraphics[width=.9\textwidth]{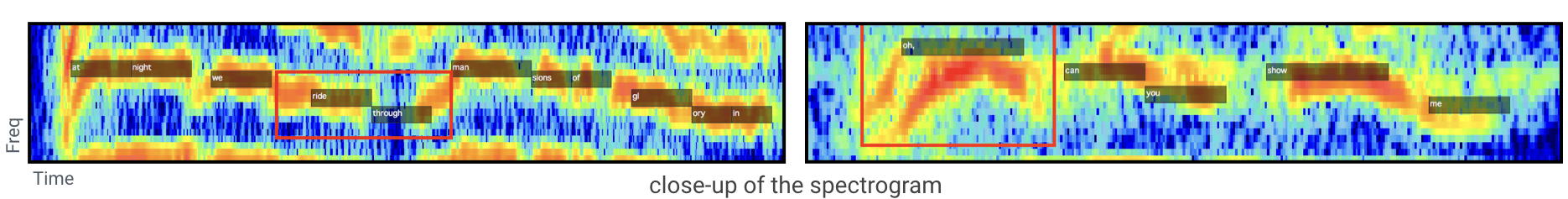}
 }
 \caption[Local erros in DALI]{Type of errors still presented in DALI. [Left] Mis-alignments in time. [Right] Mis-alignments in frequency. These problems are difficult to detect.}
 \label{fig:dali_issues}
\end{figure*}

Local errors occur because the karaoke users that did the anotations are non-professionals.
They cover local segment alignments, text misspellings and note frequency errors (see Figure~\ref{fig:dali_issues}):

\begin{itemize}
  \item \textbf{Time}: errors in the positions of the start ($t_k^0$) or end ($t_k^1$). Notes are placed in the wrong position in time or have the wrong duration.
  \item \textbf{Frequency}: errors in $f_k$. These errors are quite arbitrary, including octave, semitones and other harmonic intervals such as major third, perfect fourth or perfect fifth.
  \item \textbf{Text}: misspellings in $l_k$. Additionally, there are also errors in the phoneme level due to the automatic process employed (Chapter~\ref{sec:raw_annot}).
  \item \textbf{Missing notes or melodies}: notes can be annotated where there are no actual notes in the audio, and conversely, notes in the audio can be missed all together (during humming, ohs or similar parts).
\end{itemize}

Local errors are difficult to detect, and they can cause every note event to be wrong in some way.
At the individual time-frame level, notes with incorrect start/end times will have errors at the beginning/ending frames, but can still be correct in the central frames.
To quantify such local issues, it would be necessary to manually review all the annotations one by one. This is demanding and time-consuming. Indeed, this is what we aim to avoid. Beside, correcting them requires expert knowledge and doing it manually is unfeasible (for one song there are on average $372$ notes). 

We address local errors concentrating on the note level $a_{k, \mathit{notes}}$. The solutions found for this level will be then propagated to the rest of levels.
For the following experiments, we discard their text dimension $l_k$, $(a_{k, \mathit{notes}})_{k=1}^{K_\mathit{notes}}  = (t_k^0, t_k^1, f_k, l_k, i_k)_{\mathit{notes}} \rightarrow (t_k^0, t_k^1, f_k)_{\mathit{notes}}$. Nevertheless, this dimension can give many useful insights for future work.
Given the current set of notes $a$ and the audio signal $x$, we aim to find a function $g()$:

\begin{equation}
  g(a, x) = a' \approx a*
\end{equation}

\noindent where $a'$ is a new annotation expected to be close to the hidden correct one $a*$. The $a'$ has to have the same number of elements as $a$ and $g()$ can only modify the values $(t_k^0, t_k^1)$ of each note without overlap between notes.

\subsection{Local alignment}

\begin{figure*}[ht]
 \centerline{
   \includegraphics[width=.9\textwidth]{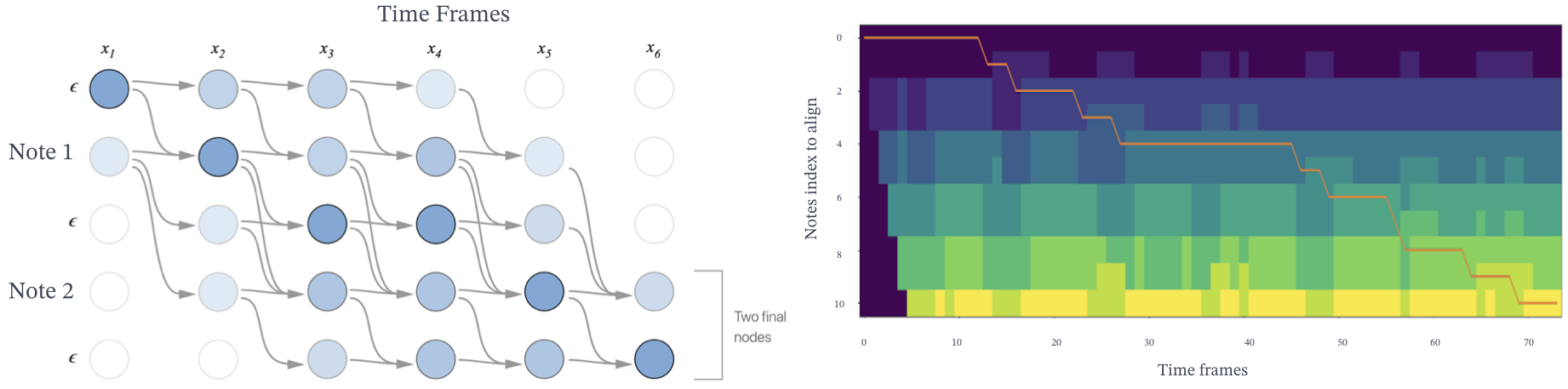}
 }
 \caption[Alginment example]{[Left] Graph alignment example with all the valid alignment paths. In the y-axis, the target notes to align (a sequence of two notes). In the x-axis the time evolution of the observation probability $x$ with six-time steps. The intensity of each node represents the value in $x$ for the target state at a time $t$. The blank state $\epsilon$ can go either to itself or the next state. On the other hand, no-blank states can go either to themselves, the next blank state or the next no-blank state. Reprinted from \url{https://distill.pub/2017/ctc/}. [Right] A real accumulation matrix obtained with Viterbi and the final path that minimizes the cost obtained after backtracking.}
 \label{fig:path}
\end{figure*}

Our task is to compute local modifications on the time series $(a_{k, \mathit{notes}})_{k=1}^{K_\mathit{notes}}$ to find a monotonic\footnote{Advances in time in the signal correspond to the same position or advance in the target sequence} alignment where the distance between $a$ and $x$ is the smallest. We target it using alignment techniques.
These aim to order sequences so that we can identify regions of similarity.

After a first exploration of the different techniques, we decide to deepen in the Connectionist Temporal Classification (CTC) loss~\citep{Graves_2006}.
This loss computes internally a conditional probability $p(y|x) = \sum_{a \in A_{x,y}} \prod_{t=1}^{T} - log(p_t(a_t|x))$\footnote{The loss then propagates the error to update $x$ to maximize the alignment.}, where $y$ is the \textbf{sequence of states} to align, $x$ the \textbf{observation probability} matrix of each state at a given time $t$ and
$\sum_{a \in A_{x,y}}$ all the valid alignments of $y$ given $x$.
We find these alignments with the Viterbi algorithm~\citep{Ryan_1993}, a dynamic programming technique that evaluates the best way to arrive at each node at a given time.
This results in an accumulation matrix that can be backtracked to find the best alignment.
The \textbf{observation probability} $x$ describes the audio as state probabilities over time and is obtained in such a way that we can direct evaluation $p(y_t|x_t)$ at a particular node. The \textbf{sequence of states} is an ordered set of states (e.g. a phrase like 'This is an example'), drawn from a finite and limited alphabet (e.g. the letters usually with a white-space character). CTC adds the blank state to the alphabets for codifying repetitions, allowing to distinguish between consecutive states with the same value.
This formalization has interesting advantages.
It describes the signal to be warped ($y$) as a graph with defined states and possible transitions.
It requires a probability formalization to directly evaluate $p_t(a_t|x)$ where one signal is described a conditional probability of the other.
Finally, it scales well to long sequences.

Figure~\ref{fig:path} illustrates the process.
At $x_1$, we evaluate all the possible initial states in $y$ (either the first blank character or the note 1).
We advance in time to $x_2$. Starting at the positions defined by the previous points, we evaluate the possible advances in $y$ (if we are in a blank state we can either repeat the state or move to the next state, if we are in note 1 state we can repeat the state, move to the next blank state or the next note).
We reproduce this process iteratively until covering $x$.
At a given time-step, a state can be linked to states with several previous steps. To optimize this process, we compute an accumulation matrix using Viterbi where we keep only the link that defines the most probable path to arrive at one state at a particular time.
In the end, we obtain the most promising alignment of $y$ in $x$ backtracking this matrix.

\begin{figure*}[th]
    \centering
    \begin{subfigure}[c]{\textwidth}
        \centerline{\includegraphics[width=.9\textwidth]{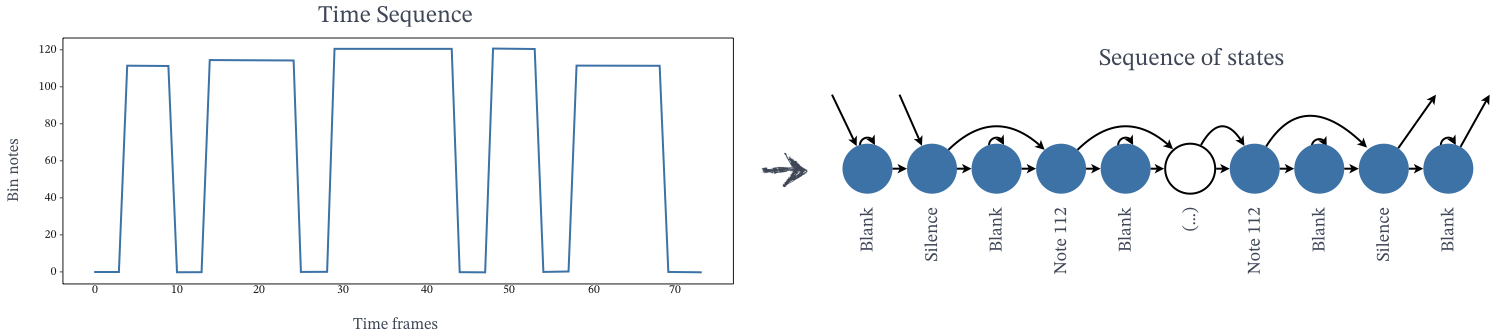}}
        \caption[Notes annotations to sequence of states]{We transform the note annotations into a sequence of states, losing the duration. In between these states, we add the blank state. This state encodes the repetition of the previous step, allowing distinctions between two consecutive notes with the same value. In this particular example, there is always a silence between notes but it is not always the case.}\label{fig:two}
    \end{subfigure}

    \begin{subfigure}[c]{\textwidth}
      \includegraphics[width=.9\textwidth]{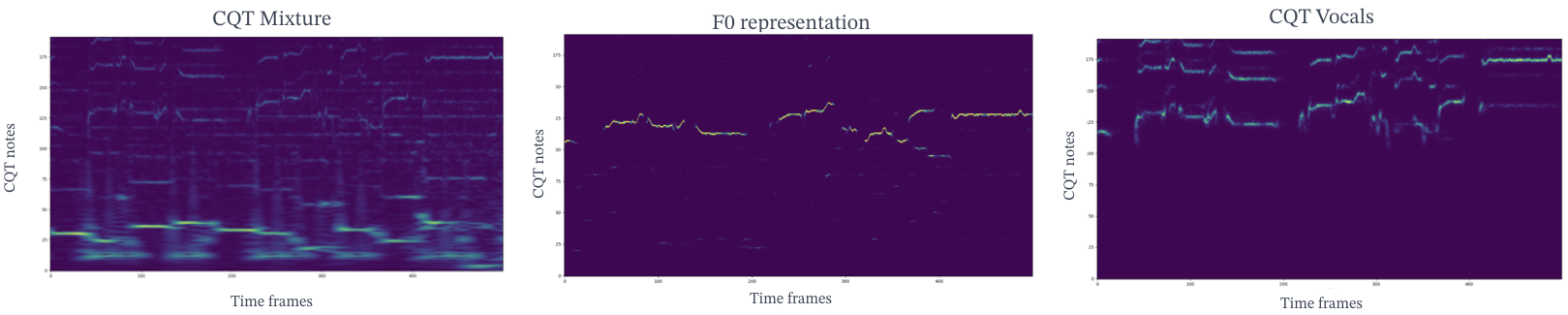}
      \caption[Observation probability matrix]{There are many possible observation probability matrices for performing the alignment.}\label{fig:one}
    \end{subfigure}

    \caption[Elements of the local errors alignment]{The main elements for alignment used are: [Top] the observation probability matrix of each state at a given time. [Bottom] A sequence of states to align.}\label{fig:two_elements}
\end{figure*}

Thereby for our problem, we need first to transform $a_{k, \mathit{notes}}$ into a graph, where each state is the frequency information $f_k$ (quantized to bins notes) and the silence state (see Figure~\ref{fig:two}). We add the blank state between states and limited the transition between note states so that a note can only go to a blank state or the next note but not repeat itself. This does not affect the results and simplifies the computation. Note how the graph removes the original time information ($t_k^0, t_k^1$) contained in the annotations. This reduces the length of the sequence which is faster and easier to control.
We need then to define the \textbf{observation probability}. A natural choice is to obtain the \gls{f0}, where the audio is described as likelihoods over time of the fundamental frequencies.
To do so, we first artificially create the silence likelihood at each time step by computing $1 - \max_{bins} f_0(t)$. We then apply a $softmax$ so that the sum of all the bins (plus silence) at each time step is $1$.
Using this formalization we can easily compute new alignments using $g(a,x) = a'$.

\begin{figure*}[ht]
 \centerline{
   \includegraphics[width=.9\textwidth]{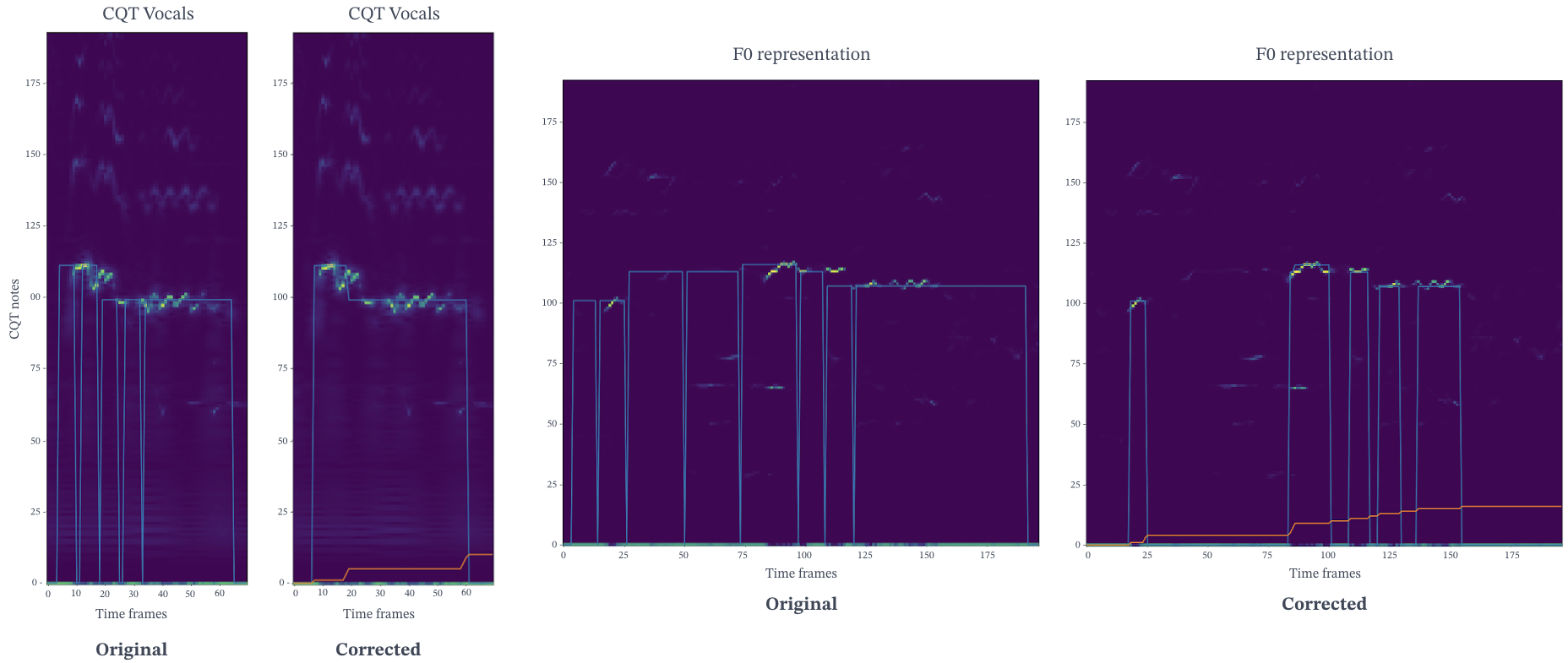}
 }
 \caption[Alginment examples]{[Left] Example of how errors in the observation probability, $x$ can induce dangerous missalignments. [Right] First alignment example where the new $a'$ seems to be better than the original. Nonetheless, some new notes are shorter that they should be. }
 \label{fig:decoding}
\end{figure*}

Figure~\ref{fig:decoding} illustrated two new $a'$.
In the first example, we can observe how the new annotations seem to be more precise than the original ones. However, analyzing the path (orange small line) we realize that the first note is compressed radically and the second one covers the original duration of both notes. This effect is even more severe for the 3rd, 4th and 5th states. Here, the first note covers all the duration and the other two are compressed to a minimum value. In contrast, the same experiment was carried out with \gls{dtw}. Here the effect is more dangerous because it merges notes with the same $f_k$ making it impossible to distinguish consecutive ones.
The second example proves how errors in the observation probability, $x$ can induce dangerous missalignments. The \gls{f0} estimation misses some essential notes which distorts the whole alignment.

\subsubsection{Alignment configurations.}

\begin{figure*}[ht!]
  \centerline{
    \includegraphics[width=\textwidth]{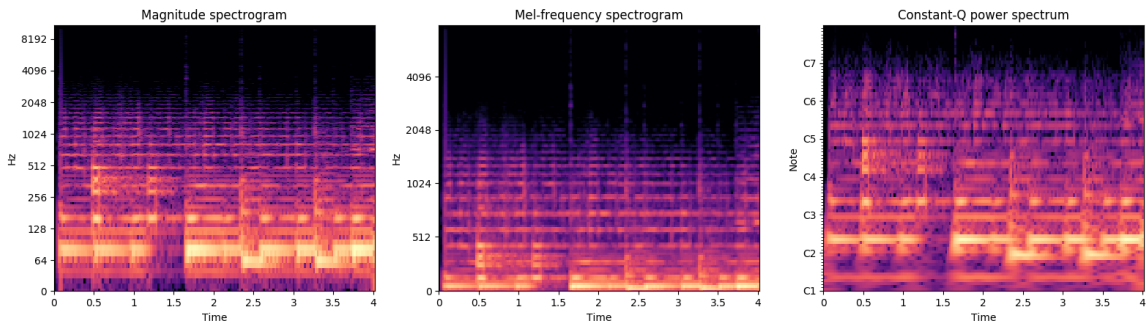}}
  \caption[Frequency domain examples]{The three frequency domain transformations used in this thesis for an illustrated example extracted from a 4-seconds audio segment.}
  \label{fig:freq_domains}
\end{figure*}

As shown in Figure~\ref{fig:decoding}, we have tested many configurations.
We can use as \textbf{observation probability} many different audio representations $x$ (see Figure~\ref{fig:one}).
Appart from the \gls{f0}, we can directly use the normalized log magnitude \gls{cqt} as a likelihood function.
\gls{cqt} provides suitable representation of a musical signal without introducing possible errors. The frequency bins of the \gls{cqt} are logarithmically spaced and with equal center frequencies-to-bandwidth ratios~\citep{Brown_1991} (see Figure~\ref{fig:freq_domains}). Formally, the center of each bin $f_b = f_{min} 2^{b/p}$, where $f_{min}$ is the lowest frequency we aim to capture, $b$ the number of bins per octave. For music signals, it is common to set $p$ to a multiple of 12 because there are 12 semitones per octave in the Western music scales. Hence, each $f_b$ corresponds to a semitone frequency. The bandwidth of each filter $p^{th}$ is defined as $\triangle f_p = f_p (2^{1/b} - 1)$, producing constant center bin-to-bandwidth ratio $Q = \triangle f_p/f_p$. The \gls{cqt} produces complex-valued coefficients.
We work the mixture audio or the isolated vocals\footnote{\Gls{source_separation} techniques are treated in detail in Chapter~\ref{sec:vunet}.}~\citep{Jansson_2017}.
All the values of normalized \gls{cqt} are between $0$ and $1$. This can be seen as a pseudo-likelihood function, that we transform into probabilities as before (artificial silence + softmax). The only difference is that the silence is computed on the isolated vocals for both \glspl{cqt} (mixture and vocals).

We can also combine the \gls{f0} and the \gls{cqt} into a single observation probability matrix. Since both have the same dimensions, we do an element-wise mean.
We usually use more than $12$ bins per octave to have a fine frequency resolution.
We need to define tolerance in the annotation to evaluate observations $x_t$ to define the final contribution. Since real notes do not stay constant in frequency but rather evolve with variations around the central frequency and annotations define single bins, to properly evaluate $p(y_t|x_t)$.
A small tolerance results in short notes with long silences. On the other hand, a large tolerance artificially creates long notes.
Additionally, to be robust to common octave errors in individual notes, we can add the contribution of each octave bin below and above.
This can be extended to reduce the states and \textbf{observation probability} into a single octave, focusing only in the correctly class note without its pitch.

Finally, we have formalized the problem focusing only on the note information. But, we can also use the sequence the text information $l_k$ in $(a_{k, \mathit{notes}})_{k=1}^{K_\mathit{notes}}  = (t_k^0, t_k^1, f_k, l_k, i_k)_{\mathit{notes}} \rightarrow (t_k^0, t_k^1, l_k)_{\mathit{notes}}$ as target states. The observation probability matrix will be then the character information obtained directly from the audio~\citep{Stoller_2019}. Another different configuration is to combine both dimensions, where each state is defined by $(l_k, f_k)$. We will need also to consider matrices with two different representation of the audio signal $x$, one for audio and one for text information.

Instead of obtaining a single alignment using all the notes in $a_{k, n}$, we can do individual alginments of smaller segments, e.i. computing $p(y|x) = \sum_{A \in A_{x,y}} \prod_{t=1}^{T} p_t(a_t|x)$ using as target $y$ the sequences define by each element in the different levels of hierarchy.
Thereby, we compute as many alignmets as lines $a_{k, l}$ has, using for each one the notes contained in a line.
We have done this also using also the paragraph level $a_{k, paragraph}$.
The simplicity of the alignment allows many further modifications in its computation:
\begin{itemize}
  \item We can weight the observation matrix $x$ by the annotation $a$ i.e. introducing penalties to observations that a given time $t$ differs from the original annotations, $a_k$. This is done by transforming $a_{k, n}$ into a matrix (with one at the note positions and the desired weight elsewhere) that is multiplied with the observation matrix $x$.
  \item We can also try to solve the errors in the states $y$. This is done using a beam search decoding\footnote{A tree algorithm that searches many (a fixed number of beams) potentially optimal paths in parallel} that creates different versions of the sequence of states $y$ (with variations such as major third, perfect fourth or perfect fifth) at each time $t$, keeping only a fixe number of possible paths for future examination.
  \item Since this alignment is a Markovian process, we can introduce weights in the transition between states according to the duration of each note. That means that every time we are evaluating the best way to arrive at a node state for creating the accumulation matrix, we can multiply the contribution of each linked states by a factor. These factors are defined by the relative duration of each note in the sequence $y$. This integrates the duration of the notes.
  \item Another way to integrate note durations is by adding another penalty to favor paths that produce notes with similar durations as the ones in the $a_k$. When computing the backtracking we evaluate several alignments (not only the most promising one) favoring the ones with similar notes durations. This helps in avoiding conflict transitions that result in notes with a very distinct duration from the annotated one.
\end{itemize}

Although these variations are quite powerful and produce interesting results.
Nevertheless, it is quite challenging to establish which one works the best, i.e. which $a'$ is really the closest to the hidden correct $a*$. We have analyzed manually many configurations and we cannot conclude with certitude which one is the best. For some examples, a particular configuration seems to help but it introduces errors for others.
Quantifying how the accurateness of the new $a'$ is challenging and still requires many manually reviewing (annotations have to be analyzed one by one every time we have a new configuration). This is extremely demanding and time-consuming and not scalable. We need then to address a new research question: \textbf{Is the new $a'$ better than before? How much? Which $a_{k, n}$ are getting better and which worse?}. This is the central topic of the Chapter~\ref{sec:dali_correction}.

\section{Conclusion}
In this chapter, we analyzed and improved over the errors presented in \gls{dali}.
The types of issues are quite diverse and difficult to quantify.
We introduced solutions to globals time errors that still persist as well as a correlation based on the note annotations and the \gls{f0} for solving the global frequency shifts.
This correlation can be used as an extra insight into the quality of the annotations.
Local errors are a great challenge. They do not have constant types of corruption nor class dependencies. We explored the internal CTC alignment technique as a solution to solve them.
The performance of this technique is heavily affected by factors such as the \textbf{observation probability} used and its adjustable configurations.
However, we cannot measure if the new annotations are better or worse.
In the next Chapter, we explore an automatic way to measure the quality of the annotations that can be used to evaluate each new $a'$.

\graphicspath{{figs/}{dali_4_noisy/figs/}}

\chapter[Training with noisy data]{Training with noisy data\raisebox{.3\baselineskip}{\normalsize\footnotemark}}\footnotetext{The work reported here was done in collaboration with Rachel Bittner who helped in framing and formalizing the problem and trained the different transcription models used for the validation.}
\label{sec:dali_correction}

Labeled data is necessary for training and evaluating \gls{sl} models, but the process of creating labeled data is error prone.
Labels may be created by human experts, by multiple human non-experts (e.g. via crowd sourcing), semi-automatically, or fully automatically.
Even in the best case scenario where data is labeled manually by human experts, labels will inevitably have inconsistencies and errors.
The presence of label noise is problematic both for training and for evaluation. During training, it causes models to converge slower (requiring much more data) or overfit the noise, resulting in poor generalization.
Label noise in evaluation sets results in unreliable metrics with artificially low scores for models with good generalization and artificially high scores for models which overfit data having the same kind of noise.
This is also the case of the \gls{dali} dataset.
Due to its nature (non-expert user annotations) and the way it was created (see Chapter~\ref{sec:dali_creation}), the assigned labels are considerably noisy.
Having control over the errors is essential to be able to work with databases like \gls{dali}. This allows us to know which data points are likely wrong or correct.
We can then correct them or select only the correct points.
Nevertheless, manual quantification
is demanding and time-consuming.

In this chapter, we present a \textbf{self-supervised data cleansing} model which exploits the structured nature of music annotation data in order to predict incorrectly labeled time-frames.
This directly addresses the question of \textbf{\textit{how good are the annotations in \gls{dali}?}} and aim to establish a similarity relationship between labels and the audio signal.
The proposed strategy is designed for multiclass label noise that is time and/or position-dependent.
The model requires examples of correct and incorrect labels. We generate training data for this model by selecting a subset of data with labels we believe to be correct as the ``correct'' examples, and creating artifical deformations of these labels as the ``incorrect'' examples.
This set is then used to train a self-supervised binary classifier that predicts the probability of having a wrong label.
We show that the model can be defined as a simple binary classifier (is the label correct or incorrect?), which can be learned with simpler architectures than those needed to solve the target task (e.g. a perfect note transcription model).
We demonstrate the usefulness of this data cleansing approach by training a transcription model on the original and cleaned versions of the \gls{dali}, improving the performance by almost 10 percentage points.
Finally, we outline several other potential applications of the proposed approach.

\section{Working with noisy data in supervised learning}

\Gls{sl} approaches commonly assume that once a dataset is created the assigned labels are unambiguous and accurate.
Nevertheless, labels may be noisy, inconsistent, subjective or incomplete~\citep{Xiao_2015, Reed_2014, Laine_2016}.
Many studies have shown that label noise has consequences for learning~\citep{Nettleton_2010, Arpit_2017, Zhang_2016, Sukhbaatar_2015, Frenay_2014, Mishkin_2017}, including requiring increased complexity and deteriorating classification performance because models eventually also memorize these wrongly given labels.
Thus, it is important to consider that the labels assigned to data points in a dataset may not be equivalent to the underlying true labels.
Label noise is especially prevalent when using data sources from the Web~\citep{Mahajan_2018}, such as \gls{dali}. 

The process that pollutes labels is referred to as \textbf{label noise}\footnote{Note that this is distinct from feature noise, which affects the value of the observation/features, i.e. the model input in contrast with label noise that concerns the model outputs (target to obtain).}~\citep{Frenay_2014}.
Manual curation strategies for removing label noise from datasets are notably time-consuming (especially when domain expertise is required, like in our case).
Moreover, they are not scalable.
There are a number of automated methods that have been proposed for dealing with label noise, falling into two general categories~\citep{Frenay_2014}.

The first category is \textbf{Learning with Noisy Labels}, where methods are built to retain their classification performance in the presence of label noise.
They explore strategies for filtering the noise or implicitly modeling it as part of the training process\footnote{Note that this is distinct from approaches that, as a result of the label noise, learn inaccurate patterns damaging the performance. Approaches that implicitly model label noise want to have a defining part in the architecture that captures the noise contribution and remove it from the final output.}. 
This approach is model-centric, where the overall goal is to maintain or improve the performance of the specific model.
The second category is \textbf{\gls{cleansing}}, which
is typically a pre-processing step where incorrect labels are identified and removed from the dataset or corrected.

The vast majority of \gls{cleansing} techniques involve training a model (or a subpart of a model) to predict which labels are most likely to be incorrect. This information is used then to remove or correct these data points.
These methods suffer from the classic ``chicken or the egg'' problem; accurate filtering of noisy data requires an accurate model, but it is difficult to train an accurate model on noisy data.
As a result, they typically suffer from either removing too many data points, throwing away good data (particularly points that are difficult -but important- to model) or letting too many noisy datapoints through. Both cases result in reduced model performances~\citep{Nettleton_2010, Arpit_2017, Zhang_2016, Sukhbaatar_2015, Frenay_2014, Mishkin_2017}.

Despite the clear detriment to downstream performance from training on noisy data, \gls{cleansing} techniques
are not universally applied. 
We hypothesize that the lack of usage during training stems from (1) the (typically false) assumption that the labels for datasets are already correct and (2)
the costly process of applying most data cleansing techniques since it is not just the output of one model but rather the output of several models or different configurations of the same model~\citep{Laine_2016}.
Finally, the results of data cleansing are rarely fed back to the dataset itself, and the process needs to be repeated each time the data is reused.

Rather than training a model to fit the labels themselves, we propose a simple \gls{cleansing} approach that trains a model to predict if the label of a datapoint is correct or not.
This can be performed with much simpler models (particularly for complex tasks); determining ``is this the correct answer'' is almost universally much easier than ``what is the correct answer''.
This approach has the advantage that it is independent of downstream inference task.
Additionally, it is easy to create data for which the labels are wrong, by simply distorting correct labels.

\section{Label noise formalization}

\begin{table}
\centering
\small
\caption[Data cleansing formalization variables]{Variables used for the \gls{cleansing} formalization.}
\label{table:formalization_variables}
\begin{tabular}{r | l}
  Varaiable & Definition  \\
  \hline
  $\mathcal{X}$ &  generic input space           \\
  $\mathcal{Y}$ &  generic output space           \\
  $\pi_{\mathcal{X},\mathcal{Y}}$ & distribution that indicates if $\mathcal{Y}$ is an incorrect label for $\mathcal{X}$ \\
  $g(x, y)$ & the probability that $y$ is an incorrect label for $x$\\
  $h: \mathcal{X} \rightarrow  \mathcal{Y}$ &  generic classifier  \\
  $S$ & a generic dataset with pairs ($x_i$, $\hat y_i$)  \\
  $Z$ & a generic dataset with triples ($x_i$, $\hat y_i$, $z_i$)  \\
  $I$ & set of all ($\mathcal{X}, \mathcal{Y}$) in a generic dataset  \\
  $I^+$ & set of all ($\mathcal{X}$, $\mathcal{Y}$, $\mathcal{Z}$) in a generic dataset  \\
  $F$ & a subset of $I$ that contains all of the correctly labeled data points  \\
  $i$ & index over instances in a dataset \\
  $x$ & a generic input instance \\
  $y$ & a generic true output instance \\
  $\hat y$ & a generic label instance \\
  $z$ & a generic label that indicates if $\hat y = y$ \\

\end{tabular}
\end{table}

The problem of label noise can be formalized as follows (see Table~\ref{table:formalization_variables} for a summary of the variables used in this section).
Consider an input/output (data/label) space $\mathcal{X} \times \mathcal{Y}$ endowed with a probability measure $\pi_{\mathcal{X},\mathcal{Y}} : \mathcal{X} \times \mathcal{Y} \rightarrow [0,1]$, and a sample made of i.i.d.\footnote{Independent and identically distributed} points ${(x_i, y_i)}_{i=1}^m$ drawn from the distribution $\pi_{\mathcal{X},\mathcal{Y}}$~\citep{friedman2001elements}.
$\pi_{\mathcal{X},\mathcal{Y}}$ is equal to zero everywhere that $\mathcal{Y}$ is a correct label for $\mathcal{X}$, and one elsewhere i.e. it defines if a label $y_i$ is a truth label for a data point $x_i$.
In practice, we do not have access to the points $(x_i, y_i)$ themselves (nor to $\pi_{\mathcal{X},\mathcal{Y}}$), but instead to a dataset $\mathcal{S}_I = \{ (x_i, \hat{y}_i)\}_{i \in I}$, where $\hat{y}_i$ are possibly wrong labels and $I = \{0, 1, 2, \dots m \}$ is an index.
\\~\\
Thereby, any \gls{sl} task can be formalized as a classifier trained using $\mathcal{S}_I$:

\begin{equation}\label{eq:clf}
h_I: \mathcal{X} \rightarrow \mathcal{Y}
\end{equation}

\subsection{Learning with Noisy Labels}
This set of approaches for dealing with noisy labels considers the scenario where a classifier $h_I$ is trained, using the full dataset $\mathcal{S}_I$ despite the noise.
The methods are built to mitigate the effects of label noise, remaining accurate despite its presence.

Most of the approaches directly model the noisy distribution in the loss function.
For example, by creating noise-robust loss with an additional softmax layer (removed during inference) to predict correct label during training~\citep{Goldberger_2017}, or with a generalized cross-entropy that discards predictions that are not confident enough while training, looking at convergence time and test accuracy~\citep{Zhang_2018}, or inferring the probability of each class being corrupted into another~\citep{Patrini_2017}.
However, this restricts models to specific types of loss functions, noisy types or distributions, assuming that the noise definition is class dependent and conditionally independent of inputs given the true labels.
Finally, models trained in this way usually need larger datasets to capture enough signal from the noise.

\textbf{Semi-supervised learning} uses reliable labeled data together with a large amount of unlabeled data~\citep{Zhu_2005s}.
Hereabouts, systems automatically leverage unlabeled/noise data through deriving insights from the labeled data.
If you have a subset of data you can trust, even several noisy labels can be dealt effectively~\citep{Hendrycks_2018}.
Recent ideas such as \textit{consistency}, the ability of a model to “disagree” with target labels and re-labeling them during training~\citep{Reed_2014} have been successfully implemented.
\textit{Consistency} can be also combined with pseudo-labeling on weakly-augmented unlabeled images, where the pseudo-labeled are only retained if the model produces a high-confidence prediction.
The model is then trained to predict the pseudo-label when fed a strongly-augmented version of the same image~\citep{Kihyuk_2020}.
Authors also use undirected graph to model the relationship between noisy and clean labels~\citep{Vahdat_2017}.
The inference over latent clean labels is intractable and regularized during training.


\subsection{Data Cleansing}
The set of approaches aims to find a subset $F \subseteq I$, such that $F$ contains all of the correctly labeled data points.
\\~\\
Let
\begin{equation}\label{eq:p}
g(x, \hat y) = \pi_{Y | X}(\hat y_i | x_i)
\end{equation}
be the probability that $\hat y$ is the wrong label $y$ for $x$.
Then, the ideal goal of \gls{cleansing} is to find
\begin{equation}\label{eq:filter}
    F = \{i: i \in I, g(x_i, \hat y_i) = 0 \}
\end{equation}

\noindent That is to say, the subset $F$ where all $\hat y_i = y_i$ (a true label for $x_i$).
Given such a subset $F$, we can train classifier $h_F$ without needing to account for label noise.

This approach has several advantages over learning directly with noisy labels.
First, the filtering does not depend on the downstream inference task, thus one \gls{cleansing} method can be applied to filter data used to train many different models.
Second, this allows downstream models to be trained with less complex models, as they do not need to account for label noise.

\subsubsection{Outlier Detection}
Outlier detection-based \gls{cleansing} methods use techniques for detecting outliers in data distributions to remove noisy labels.
In general, they apply some form of anomaly measure $a_I(x, \hat y)$ and filter out points where the anomaly measure falls above a threshold $\tau$:
$$ F \approx \{i: i \in I, a_I(x_i, \hat y_i) < \tau \} $$

For instance, measuring data complexity using its distribution density in feature space to alleviates the negative impact of the noisy labels~\citep{Guo_2018}.

\subsubsection{Model Prediction}
\label{sec:model_prediction}
The vast majority of \gls{cleansing} methods are model prediction-based.
In its simplest form, this category consists of training a model (or ensemble of models) $h_I(x_i): \mathcal{X} \rightarrow \mathcal{Y}$ on the original dataset $\mathcal{S}_I$, and removing points where the label predicted by the model does not agree with the dataset's label:
$$ F \approx \{i: i \in I, h_I(x_i) = \hat y_i \} $$

Note that in many cases, the choice of model $h$ for \gls{cleansing} is often the same as the choice of model used after \gls{cleansing}.
One of the new concepts successfully implemented is \textit{self-ensembling}, which is the consensus on simulated predictions using the outputs of the network-in-training on different conditions~\citep{Laine_2016}.

\textbf{Weakly supervised learning}~\citep{Mintz_2009, Mnih_2012, Xiao_2015} deals with low-quality labels (or at a higher abstraction level than needed) to infer the desired target information.
The \gls{teacher-student} paradigm~\citep{Hinton_2014} is the most employed method\footnote{Note how this formalization is different from the previous usage of this paradigm, a \textbf{semi-supervised} learning method where the \gls{teacher} uses a different dataset than the original $\mathcal{S}_I$ to filter the data of student}.
In this case, the \gls{teacher} is an extra network trained for selecting clean instances to guide the training of an extra network (a \gls{student}) in a mentoring way~\citep{Jiang_2018}.
It can be decoupled into two trained predictors and only update when they disagree~\citep{Malach_2017}.
This mitigates the problem of updating in wrong direction when label is wrong.
Finally, co-teaching uses two nets that complement to each other~\citep{Han_2018}. At each mini-batch data, each network samples its small-loss instances as the useful knowledge, and teaches such useful instances to its peer network.

\textbf{Active learning} estimates the most valuable points for which to solicit labels~\citep{Settles_2008, Krause_2016}.
It has been applied for weighting important data points in loss during training to balance both noise and class imbalance problem.
Weights are learned online during training to maximize performance on a small clean validation set~\citep{Mengye_2018}.

All these ideas have assisted to achieve state of the art results when using noisy but large data.
However, they rarely fed back the learnt knowledge to the dataset itself, and the process needs to be repeated each time the data is reused.
Model are considerable more complex than the original downstream model and they usually learn a trade off between how much to trust the data and how much to refuse~\citep{Reed_2014}, throwing away good data or letting too many noisy datapoints through.

\subsection{Position-Dependence}
The most common and effective \gls{cleansing} approach is to build a model to identify and discard data points with incorrect labels.
Most methods taking this approach do not assume any structure or correlation between different labels.
Thus, the input/output space $\mathcal{X} \times \mathcal{Y}$ considers $\mathcal{X}$ to be multidimensional features and $\mathcal{Y}$ to be non-structured (e.g. as in multi-class classification).
This is appropriate for many common tasks, such as image recognition, $\mathcal{X}$ are images (or features of images) and $\mathcal{Y}$ is a finite set of class labels.

However, in music, labels $\mathcal{Y}$ are often highly structured and time-varying, and the label noise is not random.
For example, musical note-annotations, which we focus on in this work, are locally stable in time and follow certain common patterns.
Typical noise for note events include incorrect pitch values, shifted start times, and incorrect durations, among others.
In \gls{dali}, we face the case where $\mathcal{Y}$ is \emph{position-dependent} - that is, where a label $y \in \mathcal{Y}$ is multidimensional, and varies as a function of the position relative to the corresponding $x \in \mathcal{X}$.
This is also common in other tasks.
For example, in object detection, $\mathcal{X}$ is the set of $(m \times n)$ images, and $\mathcal{Y}$ is an $(m \times n)$ matrix which is 1 inside the bounding box of an object and 0 otherwise. The position of the bounding box may not be well aligned with an object in the corresponding image.
As another example, in speech recognition, $\mathcal{X}$ is an audio recording, which is a function of time, and its labels $\mathcal{Y}$ vary over time. The positions where these labels change in time define the boundaries of \emph{segments}, which may be incorrectly aligned with the audio.
Finally for the note annotations presented in \gls{dali}, \emph{the time-frequency position} happens to be also the label information i.e. position in the y-axis corresponds to the note label.

\subsection{Our Proposed Approach}
\label{sec:our_approach}

We propose a model prediction based approach~\citep{meseguer2020data}, but rather than training a classifier $d: \mathcal{X} \rightarrow \mathcal{Y}$, we directly train a model $g: (\mathcal{X} \times \mathcal{Y}) \rightarrow [0, 1]$ which approximates $h(x, \hat y)$ from Equation~\ref{eq:p}.
Specifically, we model $F$ as:
\begin{equation}\label{eq:Fquality}
F \approx \{i: i \in I, g(x_i, \hat y_i) = 0 \}
\end{equation}

Note that this is mathematically equivalent in the ideal case of the previous model prediction based approaches: given a perfect estimator $h$ which always predicts a correct label $y$, $g(x,y)=\mathds{1}_{h(x)\ne y}$ where $\mathds{1}$ is the indicator function.
However, for complex classification tasks with high numbers of classes and structured labels, modeling $h$ can be much more complex than modeling $g$.
For instance, consider the complexity of a system for automatic speech recognition, versus the complexity needed to estimate if a predicted word-speech pair is incorrect.
Intuitively, you don't need to know the right answer to know if something is right or wrong.
As an anecdotal example, consider the difference in difficulty of answering ``Did they say `apple'?'' versus ``What did they say?''.

This idea is similar to the ``look listen and learn''~\cite{Arandjelovic_2017} concept of predicting the ``corespondance'' between video frames and short audio clips -- two types of structured data.
It is also similar to CleanNet~\cite{Lee_2017cleannet}, where a dedicated model predicts if the label of an image is right or wrong by comparing its features with a class embedding vector. However, this approach operates on global, rather than position-dependent labels.

We generate training data for $g$ in a self-supervised way by directly taking pairs $(x,y)$ from the original dataset as positive examples and creating artificial distortions of $y$ to generate negative examples.
In this section, we study the use of an estimator $g(x, \hat{y})$ for detecting local errors in noisy note-event annotations.
See Figure~\ref{fig:sys_diagram} for an overview of the system.



\section{Data Cleansing for the DALI dataset}
\label{sec:quality_data}

\begin{table}
\centering
\small
\caption{Variables used for data cleansing note events.}
\label{table:dali_variables}
\begin{tabular}{r | l}

  Variable & Definition  \\
  \hline
  $L$ & the set of tracks in DALI \\
  $\ell$ & index over tracks \\
  $I^\ell$ & the set of data points for track $\ell$ \\
  $i$ & index over time frames \\
  $j$ & index over frequency bins \\
  $r_i$ &  the time (s) at time frame $i$ of the CQT \\
  $q_j$ & the frequency (Hz) at frequency bin $j$ of the CQT \\
  $X^{\ell}$ & the CQT of track $l$ \\
  $\hat Y^{\ell}$ & binary note annotations matrix for track $\ell$  (Eq. \ref{eq:dali_labels})\\
  $X_i$ & time index $i$ of $X$ with all its frequency bins \\
  $X_{a:b}$ & a slice of $X$ between time index $a$ and $b$ \\
  $\hat Y_i$ & time index $i$ of $\hat Y$ with all its frequency bins \\
  $\hat Y_{a:b}$ & a slice of $\hat Y$ between time index $a$ and $b$\\
  $D$  & the set of all time indexes in DALI (Eq. \ref{eq:dali_set}) \\
  $g$ &  data cleansing model (Eq. \ref{eq:cleansing}) \\
  $F \subseteq D$ & subset of $D$ based on $g$ \\
  $s(x_i)$ & rough transcription model  \\
  $\kappa(\hat Y_i, s(X_i))$ & agreement function (Eq. \ref{eq:agreement}) \\
  $\mu(\hat y_i)$ & artificial modification function  (Fig. \ref{fig:modifications})\\
\end{tabular}
\end{table}

\begin{figure*}
    \centering
    \includegraphics[width=\textwidth]{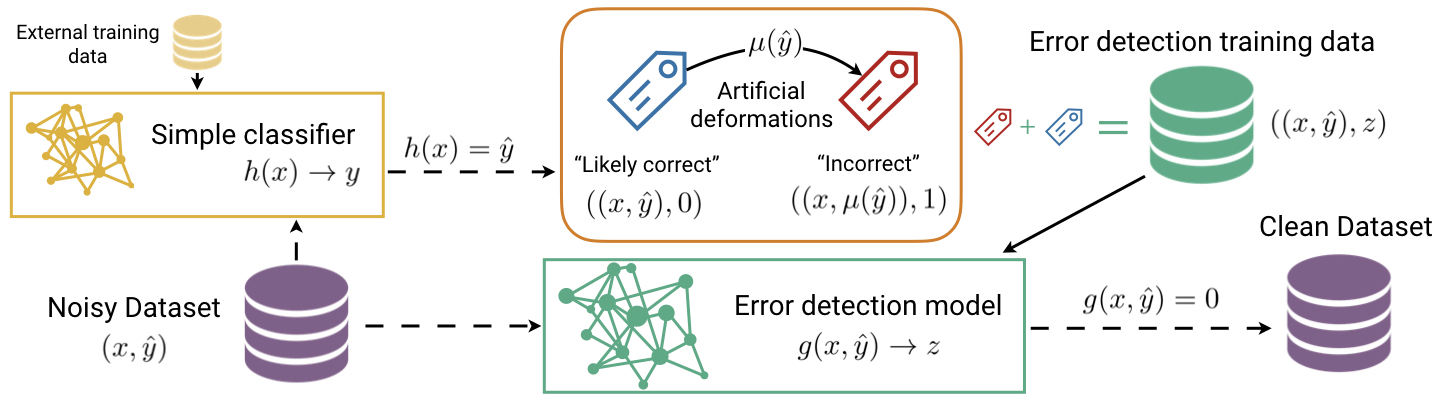}
    \caption[Top level overview of the proposed data cleansing approach]{Top level overview of the proposed \gls{cleansing} approach.}
    \label{fig:sys_diagram}
\end{figure*}

In this section, we study the use of an estimator $g(x, \hat{y})$ for detecting local errors in noisy note-event annotations.
See Table~\ref{table:dali_variables} for a summary of the variables used in this section, and Figure~\ref{fig:sys_diagram} for a top level view of the system.


Due to the nature of the data (non-expert user annotations) and the way \gls{dali} was created, Chapter~\ref{sec:dali_creation}; the note-event labels are considerably noisy.
As detailed in Chapter~\ref{sec:dali_errors}, there are local errors in the positions of the start or end times of note events (notes are placed in the wrong position in time and/or have the wrong duration), as well as having errors in the labeled frequency value (see Figure~\ref{fig:dali_issues}).
There are also extra labeled note events that do not correspond to any real event in the audio signal (e.g. during silence) and conversely, there are some real note events that are not labeled.
The errors that we encounter in \gls{dali} typically do \emph{not} include systematic substitutions of certain labels, and the positional noise is not label-dependent.
Finally, these errors are time and position-dependent with respect to the audio signal.
These type of noisy labels are common to other MIR datasets such as~\cite{Raffel_2016}.
In \gls{dali} version 2, there are more than $2,8$ million note events with an average of $362$ events per track. Correcting them requires expert knowledge and is very time consuming, making manual correction infeasible.
Hence, this dataset serves as a perfect case study for our proposed \gls{cleansing} strategy.

As our input representation, instead of using the raw audio signal itself, we compute the \gls{cqt}~\cite{Brown_1991} as a matrix $X$, where $X_{ij}$ is a time-frequency bin.
The time index $i$ corresponds to the time stamp $r_i = \upsilon \cdot i$ where $\upsilon$ is a constant defining the spacing between time stamps ($\upsilon = 0.0116$s for this particular task), and the frequency index $j$ corresponds to a frequency $q_j$ in Hz.
The \gls{cqt} is a bank of filters transformation centered at geometrically spaced frequencies. This results in a constant ratio of frequency bins ideal for musical purposes.
We use a frequency bin resolution with 6 octaves, 1 bin per semitone, a sample rate of $22050$ Hz and a hope size of $256$, resulting in a time resolution of $11.6ms$.
We compute the \gls{cqt} from the original mixture and from the isolated vocal version derived from the mixture using \gls{source_separation} techniques~\cite{Jansson_2017} (see Chapter~\ref{sec:souce_literature}).
We include the \gls{cqt} of the isolated vocals to boost the information in the signal related to the singing voice, and couple it with the \gls{cqt} of the mixture to include information we may have lost in the separation process.

We define the label target $\hat Y$ as a binary matrix created from the original note-event annotations. For a given track, let $K$ be the set of note annotations, let $t_k^0$ and $t_k^1$ be the start and end time in seconds, and $f_k$ be the frequency in Hz of note $k$. Then $\hat Y$ is defined as:

\begin{equation} \label{eq:dali_labels}
  \hat{Y}_{ij} =
   \begin{cases}
      1, & \text{ if } t_k^0 \le r_i \le t_k^1, \, q_{j-1} < f_k \le q_{j}, \, k \in K \\
      0, & \text{ otherwise}
  \end{cases}
\end{equation}

Hence, $\hat Y$ has the same time and frequency resolution as $X$.
We estimate the label noise at the \textbf{\emph{time frame}} level, rather that at the pixel or note event level.
Let $\ell \in L$ be an index over the set of tracks $L$, $I^{\ell}$ be the index of all time frames for track $\ell$, and $X^{\ell}$ and $\hat Y^{\ell}$ be its \gls{cqt} and label target matrices respectively.
Let $X^{\ell}_i$ and $Y^{\ell}_i$ indicate a time frame of $X^{\ell}$ and $\hat Y^{\ell}$; we use all frequency bins $j$ in the following discussion, so we index the data points according to their time index only.
Finally, let $X^{\ell}_{a:b}$ and $\hat Y^{\ell}_{a:b}$ denote the sequence of time frames of $X$ and $\hat Y$ between time indices $a$ and $b$.


Our goal is to identify the subset of time frames $i \in I^{\ell}$ which have errors in their annotation for each track in a dataset by training a binary \gls{cleansing} model.
Our \gls{cleansing} model is a simple estimator that can be seen as a binary supervised classification problem that produces an \gls{epf}.
Given a data centered at time index $i$, $g$ predicts the probability that the label $\hat Y_i$ is wrong.
Critically, we take advantage of the structured labels (i.e. the temporal context); as input to $g$ we use $X_{a:b}$ and $Y_{a:b}$ to predict if the center frame $\hat Y_{(a+b)/2}$ of $\hat Y_{a:b}$ is incorrect.
That is, we aim to learn $g$ such that:

\begin{equation} \label{eq:cleansing}
  g(X_{a:b}, \hat Y_{a:b}) = \begin{cases}
      0, & \text{ if } \hat{Y}_{(a+b)/2} \text{ is correct}  \\
      1, & \text{ if } \hat{Y}_{(a+b)/2} \text{ is incorrect}
  \end{cases}
\end{equation}
Thus, in order to evaluate if a label $\hat Y_i$ is correct using $n$ frames of context, we can compute $g(X_{i-n:i+n}, \hat Y_{i-n:i+n})$. In the remainder of this work, we will define
\begin{equation}
    g_n(X_i, Y_i) := g(X_{i-n:i+n}, \hat Y_{i-n:i+n})
\end{equation}
as a shorthand.

Let
\begin{equation}\label{eq:dali_set}
D = \bigcup_{\ell \in L} I^{\ell}
\end{equation}
be the set of all time indices of all tracks in $L$. Our aim is to use $g$ to create a filtered index $F$, where:

\begin{equation}\label{eq:filtered_index}
F = \{i \in D : g_n(X_i, \hat Y_i) = 0 \}
\end{equation}

\subsection{Training data}
Let $z_i$ be a binary label indicating whether the center frame $\hat Y_i$ for an input/output pair $(X_{i-n:i+n}, \hat Y_{i-n:i+n})$ is incorrect.
To train $g$, we need to generate examples of correct and incorrect data-label pairs $\left((X_{i-n:i+n}, \hat Y_{i-n:i+n}), z_i\right)$.
We will again introduce a shorthand $((X_i, Y_i), z_i)$ to refer to data points of the form  $\left((X_{i-n:i+n}, \hat Y_{i-n:i+n}), z_i\right)$.

\begin{figure}[th]
 \centerline{
   \includegraphics[width=.7\textwidth]{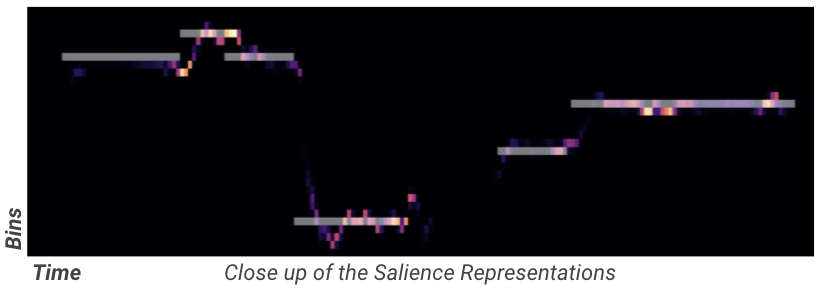}
 }
 \caption[Overlap between the annotations the $f_0$]{Overlap between \gls{dali} annotations (white lines) and the \gls{f0} with pitch probability distributions.}
 \label{fig:salience}
\end{figure}

Since we do not have direct access to the true label $Y_i$ (indeed this is what we aim to discover), we first use a proxy for selecting likely correct data points.
We first compute the output of a pre-trained \gls{f0} estimation model $s(X_i)$ that given $X_i$ outputs a matrix with the likelihood that each frequency bin contains a note~\cite{Bittner_2017}.
$s(X_i)$ is trained on a different dataset and has been proven to achieve state-of-the-art results for this task~\cite{Bittner_2017}.
$s(X_i)$ produces \gls{f0} sequences, rather than note events, which vary much more in time than note events (see Figure~\ref{fig:salience}), so we define an agreement function in order to determine when the labels agree.
$s(X_i)$ is not a perfect classifier, and while its predictions are not always correct, we have observed that when the agreement is high, $\hat Y_i$ is usually correct.  However, we cannot use low agreement to find incorrect examples, because there are many cases with low agreement even though $\hat{Y_i}$ is correct.
Therefore, we only use $\kappa$ to select a subset of ``likely correct'' data points.

\begin{figure}[th]
 \centerline{
   \includegraphics[width=.7\textwidth]{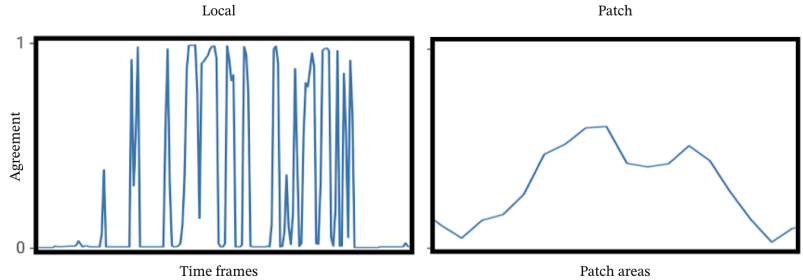}
 }
 \caption[Creating the $((X_i, \hat Y_i), z_i = 0)$ subset]{The two types of agreements $\kappa_l(\hat Y_i, s(X_i))$ used for selecting the $((X_i, \hat Y_i), z_i = 0)$ subset.}
 \label{fig:proxies}
\end{figure}

We compute both ``local'' (single-frame) and ``patch-level'' (multi-frame) agreement, and use thresholds on both to select time frames which are likely correct.
The local agreement, $\kappa_l$ is computed as:
\begin{equation} \label{eq:agreement}
    \kappa_l(\hat Y_i, s(X_i)) = \max_j \left( \hat Y_{ij} \cdot s(X_i)_{j} \right)
\end{equation}
and the patch-level agreement $\kappa_p$ is a $k$ point moving average over time of $\kappa_l$ (see Figure~\ref{fig:proxies}).
For the test set of $g$, we use very strict thresholds and select $(X_i, Y_i)$ pair to be a likely correct if $\kappa_{l} > .999$ and $\kappa_{p} > .85$.
For the training set, we use more relaxed thresholds, and select points with $0.9 < \kappa_l \le 0.999$ and $0.7 < \kappa_p \le 0.85$.
This procedure gives us a set of positive examples in non-silent regions, but does not take into account the silent areas.
In order to select correctly labeled points from silent regions, we take additional  $(X_i, \hat Y_i)$ points from regions with low energy in the isolated vocals and no annotations in a window of length $v$. In this work we use $V = 200$ ($\approx 2,32$ s).

\begin{figure}[th]
 \centerline{
   \includegraphics[width=.7\textwidth]{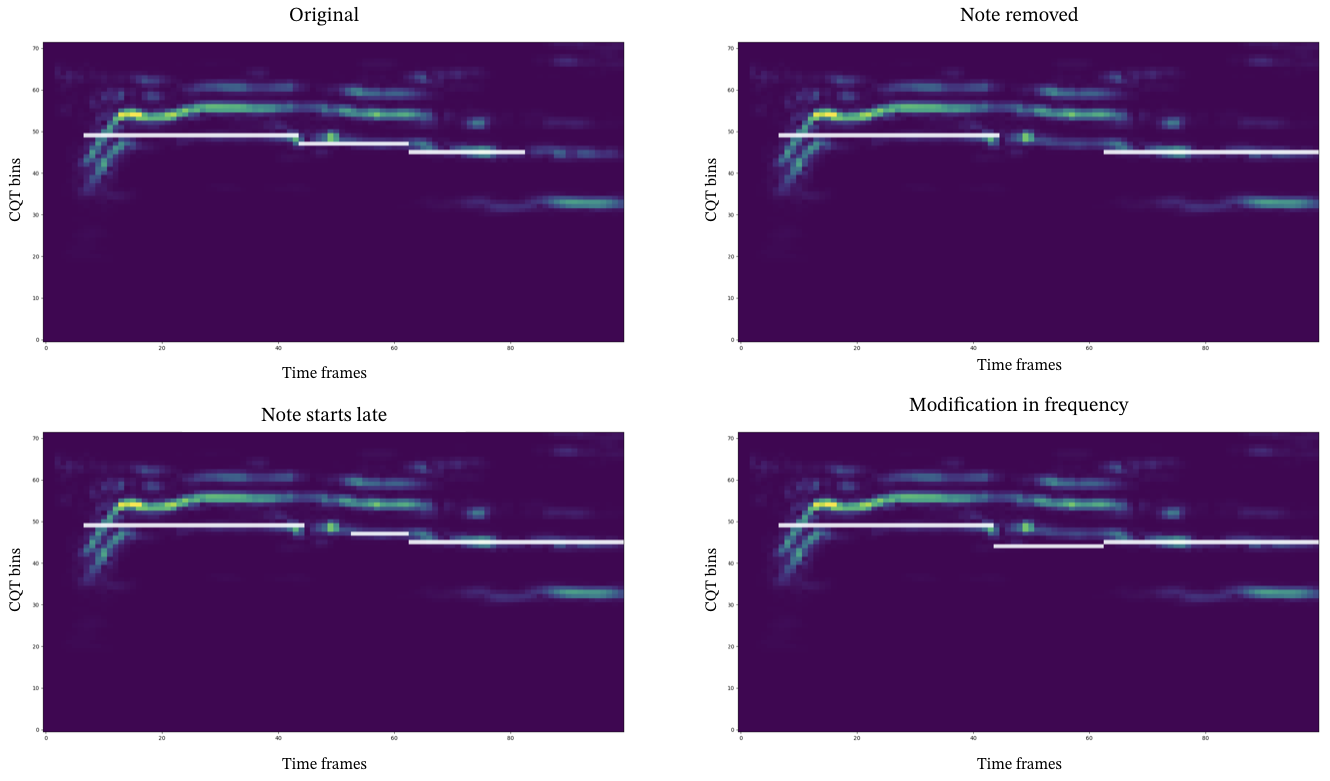}
 }
 \caption[$\hat{Y_i}$  modifications for generating the fake wrong examples]{$\hat{Y_i}$  modifications for generating the fake wrong examples. These modifications are specific to match the characteristics of label noise.}
 \label{fig:modifications}
\end{figure}

Once the ``good'' subset $((X_i, \hat Y_i), 0)$ is defined, we generate the ``error'' subset $((X_i, \hat Y_i), 1)$ where the $Y_i$ is incorrect by artificially modifying  $\hat Y_i$ in pairs from the ``good'' subset.
These modifications $\mu(\hat Y_i)$ are not random but rather specific to match the characteristics of label noise presented in \gls{dali} (issues in the positions of the start or end times, incorrect frequencies, or the incorrect absence/presence of a note).
These $\hat y_i$ should be contextually realistic, meaning that notes should have a realistic duration and should not overlap with the previous or next note (see Figure \ref{fig:modifications}).
The $((X_i, \hat Y_i), 1)$ data points for silent regions are generated by artificially adding random notes to data points where there is low energy in the isolated vocals.
Finally, this gives us a dataset of $\{((X_i, Y_i), z_i) \}$ with which we can train $g$, created in a self-supervised way.
The proposed process is summarized in Figure~\ref{fig:sys_diagram}.






\subsection{Data cleansing model}

\begin{figure}[ht]
 \centerline{
   \includegraphics[width=\textwidth]{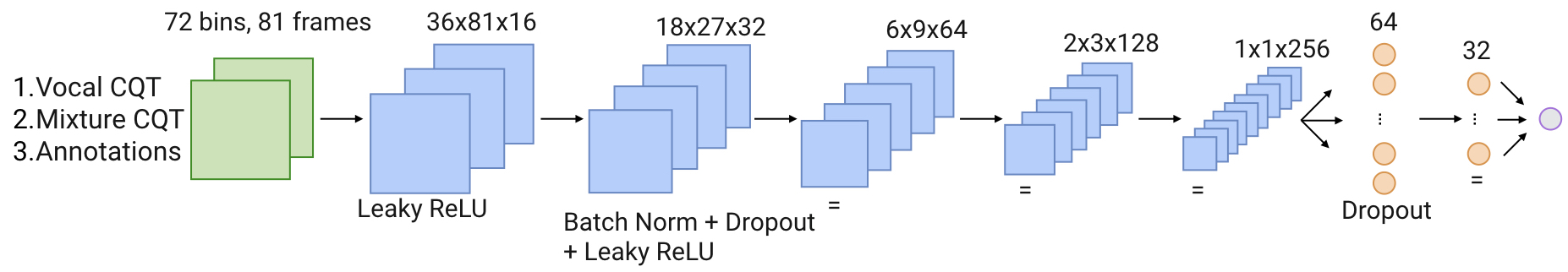}
 }
 \caption[Error detection model for DALI.]{Error detection model for \gls{dali}.}
 \label{fig:quality_model}
\end{figure}

We want to model the \gls{epf} $g_n(X_i, \hat Y_i) = z_i$.
In this work, we use $n=40$.
Hence, the input of the model is matrix with $72$ frequency bins, $81$ time frames ($0.94$ seconds)\footnote{For inference, the network would progress frame-by-frame.} and three channels: the two \glspl{cqt} (mixture and vocals) $\{X_{i-n:i+n}\}$ and the label matrix $\{\hat Y_{i-n:i+n}\}$. 

The proposed model is a standard \gls{cnn} described at Figure~\ref{fig:quality_model}.
It has five convolutional blocks with $3 \times 3$ kernels, `same` mode convolutions, and leaky ReLU activations for the first block and \gls{batch_norm}, \gls{dropout} and Leaky ReLU for the rest.
The strides are [$(2, 1)$, $(2, 3)$, $(3, 3)$, $(3, 3)$, $(2, 3)$] and the number of filters [$16$, $32$, $64$, $128$, $256$] which generates features maps of dimension  $36 \times 81 \times 16$, $18 \times 27 \times 32$, $6 \times 9 \times 64$, $2 \times 3x \times 28$, $1 \times 1 \times 256$.
Then, we have two \gls{fully} layers with $64$, $32$ neurons, ReLu activation and \gls{dropout} and a last \gls{fully} layer with only one neuron and sigmoid activation.

We test our approach using the second version of \gls{dali}~\cite{meseguerbrocal_2020}, which contains 7756 songs.
The error detection model $g$ is trained in a self-supervised way using the data generation method described in Section~\ref{sec:quality_data}.
The trained model $g$ had a frame-level accuracy of 72.1\% on the holdout set, and 76.8\% on the training set.

\section{Validation}
 \label{sec:cleansing_validation}

 \begin{figure}[ht]
  \centerline{
    \includegraphics[width=\textwidth]{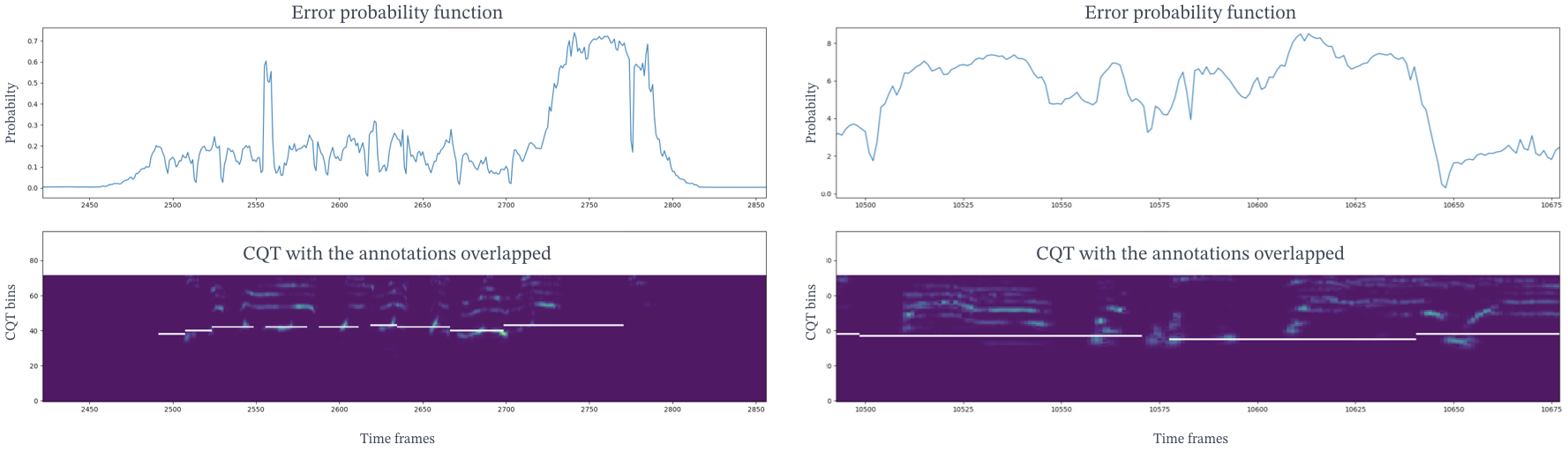}
  }
  \caption[Error probability function examples]{\Gls{epf} examples. Each example contains the output of the \gls{epf} and the annotations overlapped with the CQT of the vocals. [Left side] It also detects when the annotated note is not right. [Right side] The \gls{epf} can spot issues in the duration of the last note and the beginning of the 4th note.}
  \label{fig:quality_example}
 \end{figure}


 Validating the performance of $g$ is challenging, as we only have likely correct and artificially created wrong examples, but we do not have any ``real'' ground truth good and bad examples.
 We first manually verified many random examples of the output of the \gls{epf}, and found that appeared to be strongly correlated with errors in $y_i$.
 However, a manual perceptual evaluation of the \gls{epf} is both infeasible and defeats the purpose of automating the process of correcting errors. Instead, we validate the usefulness of this approach by applying it to model training.
 In this section, we address the question: \textbf{Is \gls{epf} useful? How much?}

 The ultimate goal of a \gls{cleansing} technique is to identify incorrect labels and remove them from the dataset in order to better train a classifier.
 Thus, one way to demonstrate the effectiveness of the \gls{cleansing} method is to see if training a model using the filtered dataset results in better generalization than training on the full dataset.

 \begin{table*}
   \centering
   \resizebox{\textwidth}{!}{%
   \begin{tabular}{c|c|c|c||c|c||c|c}
   \hline
   \multirow{2}{*}{Dataset}  & \multirow{2}{*}{Scores} & \multicolumn{6}{c}{Training data used}         \\
   \cline{3-8}
                             &                         & \multicolumn{2}{c||}{All data}     & \multicolumn{2}{c||}{Filtered}      & \multicolumn{2}{c}{Weighted}       \\
   \hline\hline
   \multirow{2}{*}{MedleyDB} & Raw Pitch Accuracy      & 0.403 $\pm$ 0.148 & 0.391 & 0.453 $\pm$ 0.143 & 0.464 & 0.495 $\pm$ 0.141 & 0.510  \\
   \cline{2-8}
                             & Raw Chroma Accuracy     & 0.456 $\pm$ 0.138 & 0.443 & 0.502 $\pm$ 0.134 & 0.505 & 0.540 $\pm$ 0.131 & 0.532  \\
   \hline\hline
   \multirow{2}{*}{iKala}    & Raw Pitch Accuracy      & 0.413 $\pm$ 0.101 & 0.416 & 0.484 $\pm$ 0.090 & 0.483 & 0.535 $\pm$ 0.092 & 0.545  \\
   \cline{2-8}
                             & Raw Chroma Accuracy     & 0.441 $\pm$ 0.094 & 0.438 & 0.515 $\pm$ 0.088 & 0.520 & 0.546 $\pm$ 0.088 & 0.551  \\
   \hline\hline
   \multirow{2}{*}{Both}     & Raw Pitch Accuracy      & 0.411 $\pm$ 0.112 & 0.414 & 0.478 $\pm$ 0.103 & 0.482 & 0.527 $\pm$ 0.105 & 0.542  \\
   \cline{2-8}
                             & Raw Chroma Accuracy     & 0.444 $\pm$ 0.104 & 0.439 & 0.513 $\pm$ 0.099 & 0.515 & 0.545 $\pm$ 0.098 & 0.551  \\
   \hline
   \end{tabular}%
   }
   \caption[Results for the three models trained using the error function prediction]{Detailed results with the mean, standard deviation and median for the Raw Pitch Accuracy and Raw Chroma Accuracy for the various models trained using the error function prediction.}
   \label{fig:quality_exp_table}
 \end{table*}

 We test this hypothesis in the context of the \gls{f0} estimation.
 This task is a well-established problem in the \gls{mir} community with ground-truth datasets and proven architectures.
 It is also a task that can benefit from the type of errors the \gls{epf} locates (errors in $t_k^0, t_k^1$ and $f_k$).
 The $a_{k, n}$ annotations can be easily transformed into pseudo-fundamental frequency annotations, defining a frequency bin over a set of frames that belong to the same note.
 Note how they differ from the traditional \gls{f0} annotations, that follow the variations around the central frequency.
 Besides, the $a_{k, n}$ are only for vocals.
 For this reason, the final model aims to solve a slightly different and more complex task, vocal note transcription.
 We assume that this is constant for all our experiments.

 To validate the usefulness of $g_n(X_i, \hat Y_i)$ for improving training, we train the Deep Salience vocal pitch model~\cite{Bittner_2017} three times\footnote{We train each new model from scratch, not using transfer learning} using three different training sets.
 The training sets are subsets of DALI, and contain the $\hat{Y_i}$ of all the songs that have a Normalized Cross-Correlation $> .9$~\cite{meseguerbrocal_2018}.
 This results in a training set of 1837 songs. The three sets are defined as follows:

 \begin{enumerate}
     \item \textbf{All} data. Trained using all time frames, $D$ (Eq.~\ref{eq:dali_set}).
     \item \textbf{Filtered} data. Trained using the filtered, ``non-error'' time frames, $F$ (Eq.~\ref{eq:filtered_index}), where the output of $g$ has been binarized with a threshold of 0.5. With this data we tell the model to skip all the noisy labels.
     \item \textbf{Weighted} data.  Trained using all time frames, $D$, but during training, the loss for each sample is weighted by $1 - g_n(X_i, \hat Y_i)$. This scales the contribution of each data point in the loss function according to how likely it is to be correct.
 \end{enumerate}

\begin{figure}[ht]
  \centerline{
    \includegraphics[width=0.65\textwidth]{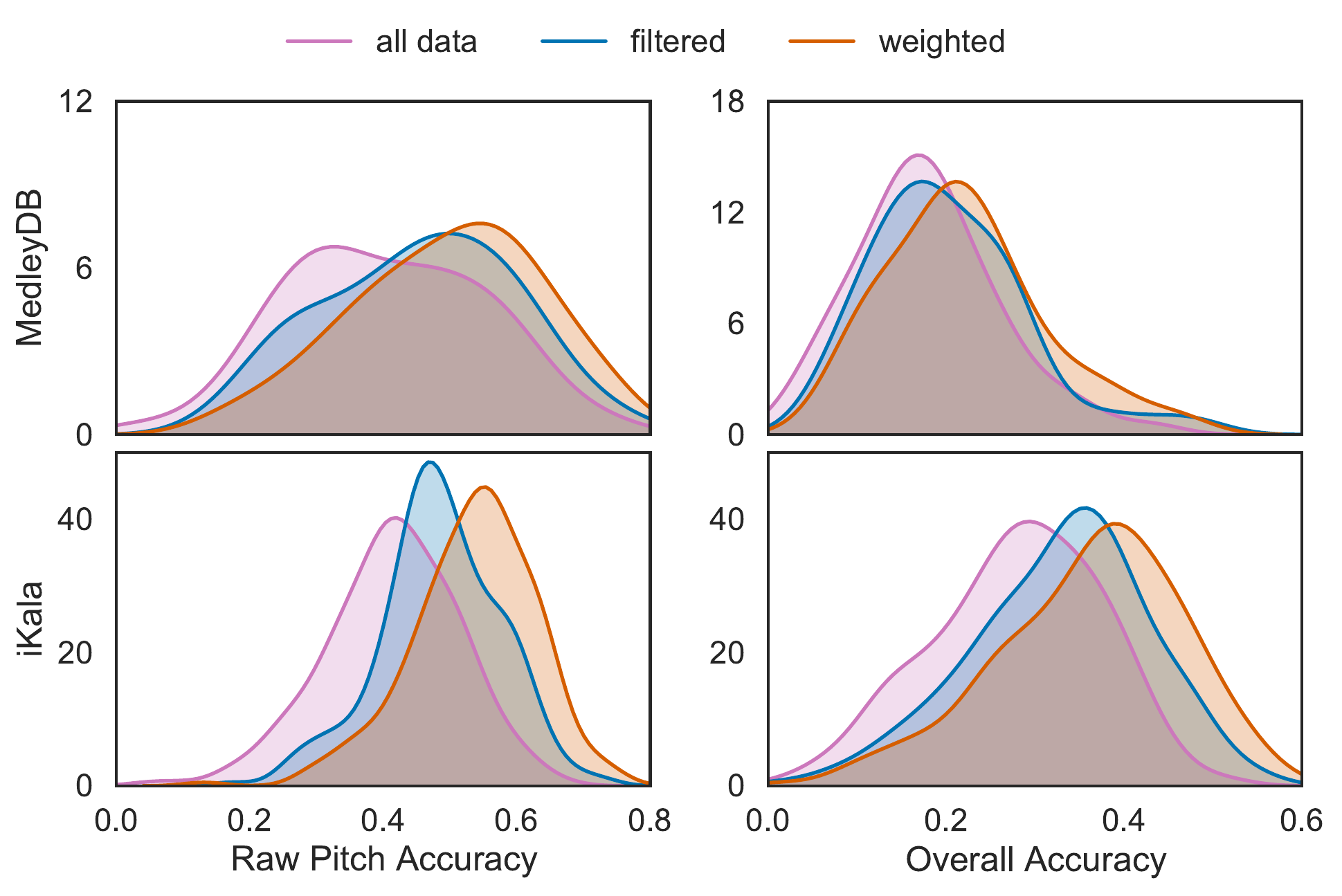}
  }
  \caption[Distribution of scores when training with the error probability function]{Distribution of scores for the three training conditions. Each condition is plotted in a different color. Scores for MedleyDB are shown in the top row and scores for iKala are in the bottom row. Raw pitch accuracy is shown in column 1 and overall accuracy is shown in column 2. The y-axis in all plots indicates the number of tracks.}
  \label{fig:quality_exp_hist}
 \end{figure}

 We test the performance of each model on two polyphonic music datasets that contain vocal fundamental frequency annotations: 61 full tracks with vocal annotations from \gls{medley}~\cite{Bittner_2014} and 252 30-second excerpts from iKala~\cite{Chan_2015}.
 We compute the the generalized\footnote{using continuous voicing from the model output and binary voicing from the annotations} Overall Accuracy (OA) which measures the percentage of correctly estimated frames, and the Raw Pitch Accuracy (RPA) which measures the percentage of correctly estimated frames where a pitch is present, which are standard metrics for this task~\cite{Bosch_2019, salamon2014melody}.
 The distribution of scores for each dataset and metric are shown in Figure~\ref{fig:quality_exp_hist}.

 While the scale of the results are below the current state of the art~\cite{Bittner_2017}.
 This result is likely expected due to the noisiness of \gls{dali} and the fact that vocal note transcription is a different task from fundamental frequency estimation.
 Despite this result, we see a clear positive impact on performance when data cleansing is applied (see Table~\ref{fig:quality_exp_table}).
 The overall trend we see is that training using filtered data outperforms the baseline of training using all the data with statistical significance ($p < 0.001$ in a paired t-test) for all cases, indicating that our error detection model is successfully removing time frames which are detrimental to the model.
 We also see that, overall, training using all the data but using the error detection model to weigh samples according to their likelihood of being correct is even more beneficial than simply filtering.
 This suggests that the likelihoods produced by our error detection model are well-correlated with the occurrence of real errors in the data.
 These trends are more prominent for the iKala dataset than for the MedleyDB dataset -- in particular, the difference between training on filtered vs. weighted data is statistically insignificant for MedleyDB while it is statistically significant ($p < 0.001$ in a paired t-test) for the iKala dataset.
 The iKala dataset has much higher proportion of \emph{voiced} frames (frames with a pitch annotation) than MedleyDB. This suggests that the weighted data is beneficial for improving pitch accuracy, but does not bring any improvement over filtering for detecting whether a frame should have or do not have a pitch (voicing).
 Nevertheless, both conditions which used the error detection model to aid the training process see consistently improved results compared with the baseline.

\begin{figure}[ht]
  \centerline{
    \includegraphics[width=\textwidth]{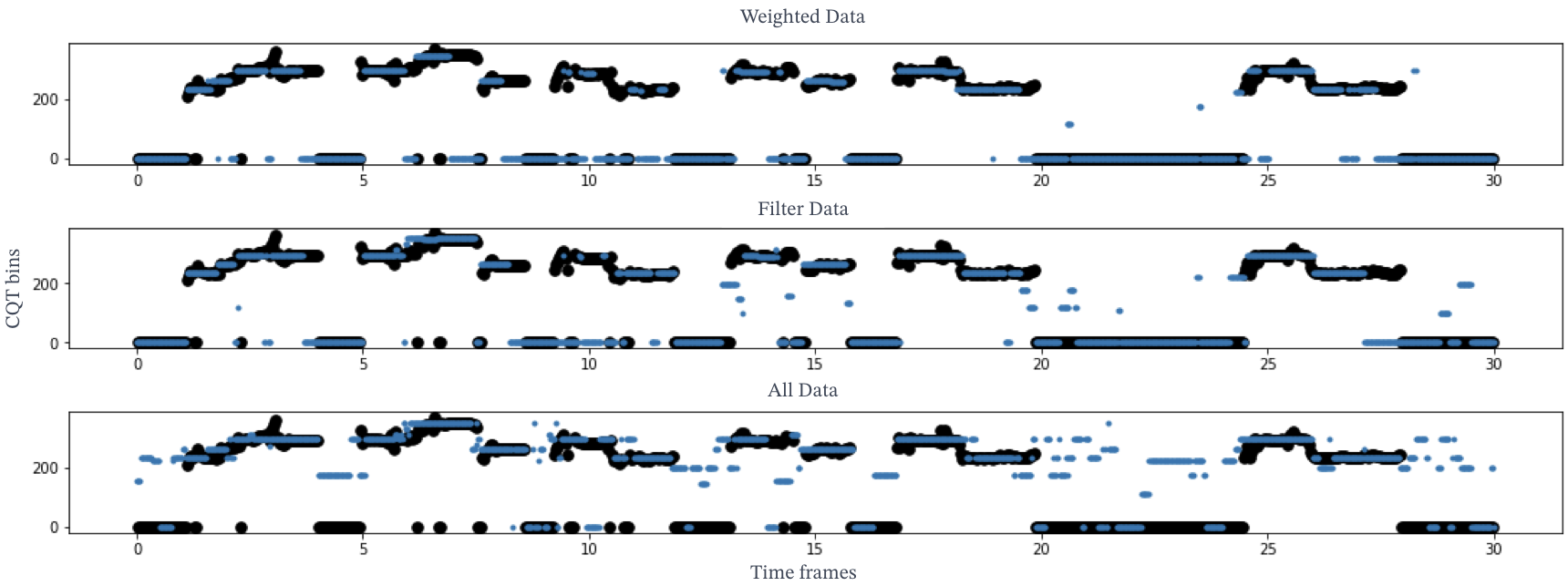}
  }
  \caption[Ouput example for the three models trained using the error function prediction]{Example for each of the three models for an  example  track  from iKala dataset. We can see how both the filtered and weighted versions output a transcription close to the expected one with less sparse output.}
  \label{fig:example_melody}
 \end{figure}

 Figure~\ref{fig:example_melody} illustrates the output of each of the three models for a track from the iKala dataset.
 We can see how the two versions that used the \gls{epf} transcribe a melody closer to the expected output.
 Both models have considerably less dispersion with more coherent notes in time.

 \subsection{Estimated Quality of The DALI Dataset}

\begin{figure}
     \centering
     \includegraphics[width=.6\textwidth]{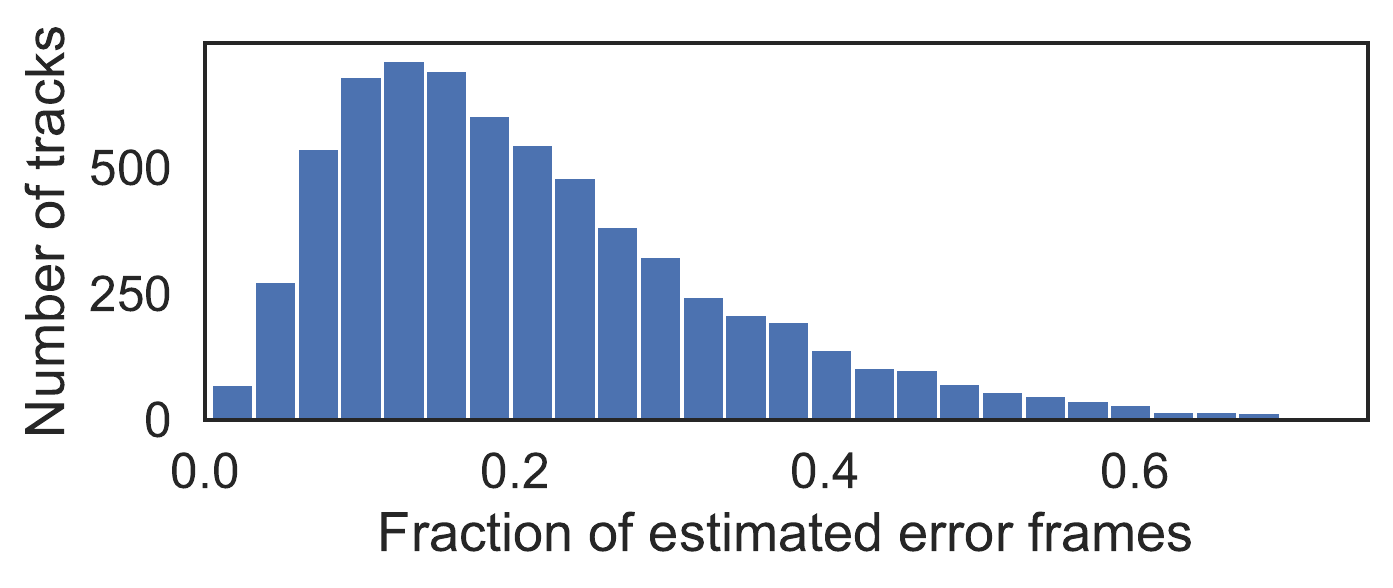}
     \caption{Histogram of the estimated error rate per track.}
     \label{fig:quality_hist}
 \end{figure}

 As a final experiment, we ran the error detection model on the full \gls{dali} dataset (version 2) in order to estimate the prevalence of errors.
 We compute the percentage of frames per-track where the likelihood of being an error is $\ge 0.5$.
 A histogram of the results is shown in Figure~\ref{fig:quality_hist}.
 We estimated that on average, 21.3\% of the frames of a track in \gls{dali} will have an error in the note annotation, with a standard deviation of 12.7\%.
 31.1\% of tracks in \gls{dali} have more than 25\% errors, while 18.2\% of tracks have less than 10\% errors.
 We also measured the relationship between the percentage of estimated errors per track and the normalized cross correlation from the original \gls{dali} dataset~\cite{meseguerbrocal_2018}, and found no clear correlation.
 This indicates that while the normalized cross correlation is a useful indication of the global alignment, it does not reliably capture the prevalence of local errors.

\begin{figure}
     \centering
     \includegraphics[width=.8\textwidth]{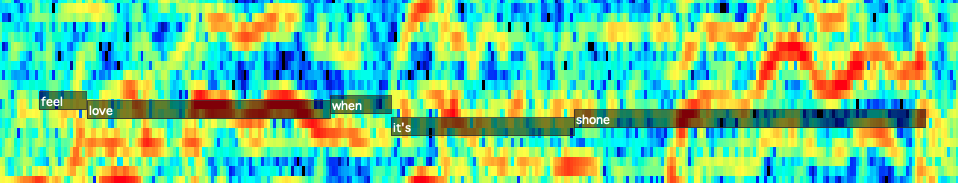}
     \caption[Lowest percentage error]{The CQT of an excerpt of the track in \gls{dali} with the lowest percentage error ($<1\%$ error), with its annotations overlaid.
     }
     \label{fig:qual_ex}
 \end{figure}

 We manually inspected the tracks with a very high percentage of estimated errors ($> 70\%$) and found that all of them were the result of the annotation file being matched to the incorrect audio file (see~\cite{meseguerbrocal_2020} for details on the matching process).
 On the other hand, we found that the tracks with very a low percentage of estimated errors ($< 1\%$) had qualitatively very high quality annotations.
 For example, Figure~\ref{fig:qual_ex} shows an excerpt of the track with the lowest error rate along with a link to listen to the corresponding audio.
 While this is only qualitative evidence, it is an additional indicator that the scores produced by the error detection model are meaningful.
 The outputs of our model on \gls{dali} are made publicly available.

 An error detection model that estimates the quality of a dataset can be used in several ways to improve the quality of a dataset.
 For one, we can use it to direct manual annotation efforts both to the most problematic tracks in a dataset, but also to specific incorrect instances within a track.
 Additionally, one of the major challenges regarding automatic annotation is knowing how well the automatic annotation is working.
 For example, the creation of the \gls{dali} dataset used an automatic method for aligning the annotations with the audio, and it was very difficult for the creators to evaluate the quality of the annotations for different variations of the method.
 This issue can now be overcome by using an error detection model to estimate the overall quality of the annotations for different variations of an automatic annotation method.

 \section{Discussion}
 \label{cleansing_discussion}

 We have introduced our \gls{cleansing} strategy for dealing with noisy labels.
 We showed that training directly a model that detects errors in the labels leads to a big improvement in downstream performance over training on the noisy data.
 In particular, the sample weights filter confidence outperforms when training with a filter version.
 Our approach is particularly interesting for complex prediction tasks with many multiple label classes since the binary problem is theoretically less complex than a model than the multi-class problem.

 Our approach has some advantages over previous ones.
 For instance, it is simple and has a reduced architecture and training procedure.
 The output of the model can be used alongside existing datasets in a self-supervised way.
 Less effort is needed since we are using the dataset to mitigate the effects of noisy labels.
 It is especially useful for tasks that are complex and slow to train.
 Binary models can be simpler and trained much faster than models for the task itself.
 Another advantage is that our approach has a direct reflection on the data regardless of its final purpose, being independent of it.
 This opens a wide range of possible usages.
 For instance, the \gls{epf} gives a general idea of the current state of a dataset moving forward from the idea of datasets as ground truth, instead  treating them as noisy estimates.
 We can also filter the dataset (globally e.g an audio track and locally e.g. a particular segment) depending on the sensitivity of the task, measuring things like \textit{how much noisy points affect this particular task?}.
 It can be also seen as a way of evaluating the accuracy of a model in unlabeled data given a direct sense of the performance in the real world scenario.
 It can be used in a \gls{teacher-student} schema gathering automatically training data from the Web and, instead of using all label provided by the \gls{teacher} as the truth, they can be filtered with our error detection function, removing the noisy labels.
 Finally, in the context of \textbf{active learning}, the \gls{epf} can select the noisiest data points for manually annotations or measure a human annotator's reliability.

 \subsection{Future work}

 Many things remain for future works.
 We plan to use our error detection models as a guide for testing the effectiveness of automatic label correction techniques described in Chapter~\ref{sec:local_errors}.
 We can then automatically decide if the new labels are better or worse than the previous one.

 Formulating the problem as a standalone binary supervised classification problem requires labeled data.
 Given a noisy dataset, the first step is to define the dataset $\mathcal{S}_{I^{+}} = \{((x_i, \hat y_i), z_i)\}_{i \in I}$ to train our error detection model.
 We have explored an initial hypothesis where the true labels are found comparing $\hat{y_i}$ with the estimated \gls{f0} and the noisy labels are artificially created.
 Nonetheless, we can use now our error detection model to identify the true and noisy labels and define a new dataset $\mathcal{S}_{I^{+}} = \{((x_i, \hat y_i), z_i)\}_{i \in I}$.
 This is can be formulated again in a \gls{teacher-student} where the first error detection model is the \gls{teacher} that labels the dataset used to train a new error detection model.

 We also plan to extend our approach to multiple domains with similar label noise such as objects positioned in an image or speech aligned with the text.
 

\section{Conclusions}
In this chapter, we have introduced our novel \gls{cleansing} technique that automatically locates errors in the noisy labels of \gls{dali}.
We have shown how this technique can be effectively used to train a downstream task, improving the performance of same architecture by almost 10\%.

Since the goal of this thesis is not to create a perfect dataset but rather to explore the interaction between audio and lyrics, we did not pursue further in this direction.

With this chapter, we close the work done concerning \gls{dali}, a big \gls{multimodal} dataset with lyrics aligned in time.
We list our main contributions as:

\begin{enumerate}
  \item we created a \gls{multimodal} dataset with lyrics aligned in time at different levels of granularity with different types of audio (original and multi-tracks) leveraging data from the Web, using novel techniques such as the \gls{teacher-student} and a methodology where model creation and data curation help each other.
  \item we showed that the \gls{teacher-student} is a good solution while dealing with unlabeled to improve the performance of models such as our \gls{svd}.
  \item we provided different proxies such as the \gls{ncc} and the \gls{f0} correlation to get to know the quality of the labels.
  \item we explored the CTC internal alignment and several configurations to overcome the local noisy labels still present in the dataset.
  \item we have directly addressed the noisy label issue by developing a novel \gls{cleansing} method for automatically evaluating the quality of the labels. This method is especially interesting for position-dependent multiclass labels with non-defined errors nor class dependencies.
\end{enumerate}

In the following chapters, we use \gls{dali} for exploring direct interaction between lyrics and audio for \gls{mir} tasks.

\graphicspath{{figs/}{structures/figs/}}

\chapter{Improving lyrics segmentation combining text and audio}
\label{sec:structures}

The first \gls{multimodal} case scenario that we addressed with \gls{dali} is lyrics segmentation.
The goal of this task is to detect the boundaries between sections (e.g. chorus, verse) of a song.
That is, given a song composed by $k$ lyrics lines $X = \{x_1, x_2, ..., x_k\}$ we want to find the lines $x_i \in X$ that define the section borders.
The task can be easily formulated in a \gls{sl} way as a binary classifier $p(y_i|x_i)$ with $y_i = 1$ when $x_i$ defines a boundary block and $y_i = 0$ otherwise.
In \gls{dali}, each text line $a_i \in X$ is associated with a concrete audio section.
We hypothesize that the text and audio modalities should capture complementary structure.
In this chapter, we show how a \textbf{\gls{multimodal}} representation composed by audio and text performs significantly better than using text information only.

\section{Segmenting lyrics using text information}
\label{sec:structures_intro}

\begin{figure}
	\centering
  \includegraphics[width=.95\linewidth]{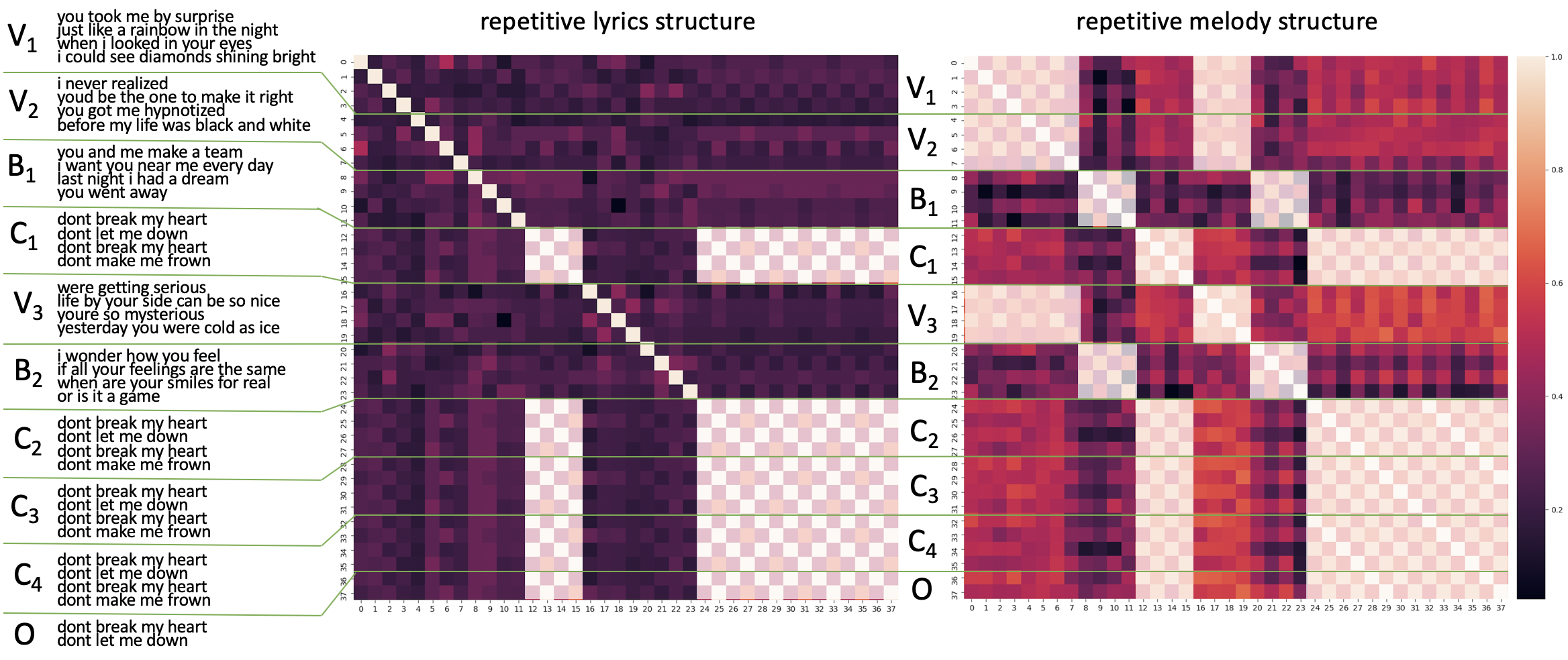}
  \caption[Lyrics segmentation example]{Lyrics (left) of ``Don`t Break My Heart'' by Den Harrow. We illustrated two different SSM based on lyrics (middle), and on audio (right). Lyrics section boundaries (green lines) coincide with highlighted rectangles in lyrics and melody patterns in the SSM. Reprinted from~\citep{fell2018lyrics}}
\label{fig:motivational_example}
\end{figure}

Understanding the structure of the lyrics is an important task for music content analysis \citep{Cheng2009,watanabe2016modeling}.
It permitts splitting a song into semantically meaningful sections, enabling a description per section (rather than a global description of the whole song).
The detection of the lyric structures can be made in two steps.
We first segment in sections, using the repeated patterns presented in the lyrics and then, we label each section (e.g. intro,  verse, chorus).
To better understand the problem, consider the example illustrated in Figure~\ref{fig:motivational_example} where we show the lyrics and the corresponding segmentation into sections.
Horizontal green lines indicate the section boundaries we aim to find.
Even though it is frequently assumed that boundaries correspond to line breaks, this does not always hold.
Additionally, line breaks are usually annotated by users and they are not necessarily identical to the intended segmentation defined by the songwriter.
This motivated researchers to develop methods to automatically segment lyrics.
This task is similar to music structure detection, where we automatically estimate the temporal structure of a music track by analyzing the repetitive audio patterns.

It is habitual to use textual features (e.g., n-grams and characters count), which have been proven effective~\citep{fell2018lyrics, Mahedero2005,watanabe2016modeling, Barate2013}.
The first attempt focused on lyrics with a recognizable structure, identified using relevant heuristics such as the length of each line and the total lenght of the lyrics or the relative position of a section border in the song~\citep{Mahedero2005}.
They tested their algorithm on a small dataset of 30 lyrics, 6 for each language (English, French, German, Spanish and Italian), which had previously been manually segmented.
Authors have also looked for recurrent and non-recurrent groups of lines with a rule-based method to label the different sections~\citep{Barate2013}.
Lyrics segmentation as a binary classification task was introduced as a solution to model repeated patterns~\citep{watanabe2016modeling}.
Given a song composed by $k$ lyrics lines $X=\{x_1, x_2, ..., x_k\}$ this approach looks for $p(y_i|x_i)$ with $y_i = 1$ when $x_i$ defines a boundary between sections and $y_i = 0$ otherwise.
They use as features a \gls{ssm}~\citep{foote2000automatic}, a feature quite popular in \gls{mir} that highlights distinct patterns, revealing the underlying structure.
Given a song described as a set of sequential elements $X=\{x_1, x_2, ..., x_k\}$, \gls{ssm} is the result of computing a similarity measure between the two corresponding elements $SSM_{i,j} = similarity(x_i, x_j)$ for all the possible combinations of the set $X$.
As a result, each element is compared with all the elements in the set, creating a matrix that directly highlights similar elements.
Diagonals (and sub-diagonals) and blocks (or the absence of them) are the two main patterns in a \gls{ssm}.
While diagonals indicate sequences that are repeated, blocks indicate sequences in which all the elements are highly similar to each another.
Both patterns are indicators of sections.
Figure~\ref{fig:motivational_example} highlights two examples of \gls{ssm} with the repetitive structure of the lyrics.
Despite being a local element, each row/column of a \gls{ssm} captures the similarity with all the other elements in the song.

\begin{figure*}
	\centering
	\includegraphics[width=.9\linewidth]{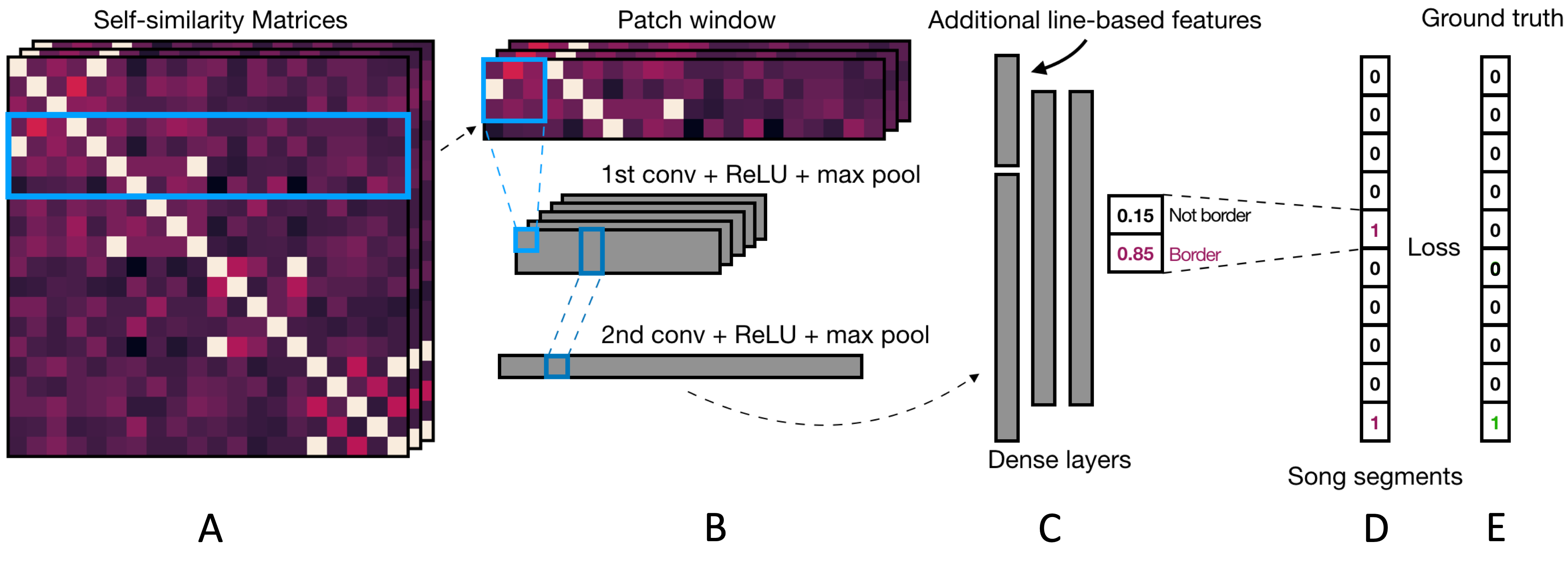}
	\caption[CNN model for lyrics segmentation]{\gls{cnn} model for inferring lyrics segmentation. Reprinted from~\citep{fell2018lyrics}}
	\label{fig:approach}
\end{figure*}

This chapter extends previous work done in~\cite{fell2018lyrics}.
Authors use the binary formalization~\citep{watanabe2016modeling} with a \gls{cnn} to detect section boundaries by leveraging the repeated text patterns conveyed in a \gls{ssm} (see Figure~\ref{fig:approach}).
The model predicts if a lyrics line $y_i$ is a border $p(y_i|x_i) = 1$ or not $p(y_i|x_i) = 0$ evaluating the corresponding \gls{ssm} row plus some additional context rows around $x_i$.
It has two convolutional layers each one with max-pooling after to downsample each feature to a scalar.
The resulting feature vector is concatenated with the line-based features and is used as input to a series of \gls{fully} layers.
The last layer has a single neuron with a $sigmoid$ activation.
The model produces the final probability using a binary cross-entropy loss between the prediction and the ground truth label.
Authors use three different \gls{ssm} respectively based on different line-based text similarity measures on characters, phonetic information or syntax characteristics.
The best results are obtained with the \textbf{string similarity ({\tt str}):} a normalized Levenshtein string edit similarity between the characters of two text lines~\citep{levenshtein1966binary} used also in \citep{watanabe2016modeling}.


However, this approach fails where there is no clear structure in the lyrics, for instance when sentences are never repeated or in the opposite case when they are always repeated.
For instance as we can see in Figure~\ref{fig:motivational_example}, the verses ($V_i$) and bridges ($B_i$) have not repetive pattern in the \gls{ssm} extracted from the lyrics (middle).
The reason is that these verses and bridges have different lyrics.
However and since these lines share the same melody, we can easily see how these patterns arise in the \gls{ssm} extracted from the audio (right) using \textit{IrcamDescriptors}~\citep{Peeters2004}.
These are the cases we target in this chapter.
We hypothesize that since melodies are often repeated, the part of the structure which is not captured in the text may arise from the audio.

\section{Formalization}\label{sec:structures_formalization}
Following the formalization described at Chapter~\ref{sec:c_multimodal}, we use the multimodal information as complementary to improve lyrics segmentation:

\begin{description}[.15cm]
  \item[How is the multimodal model constructed?] We opt for a \textbf{model-agnostic} multimodal learning method that re-uses the architecture of a model proven to perform well for this task when used in one domain.
	\item[What multimodal information?] We aim to construct a joint representation between audio and text where each domain captures complementary structure information.
    In the pre-processing step, both domains are treated independently to obtain a coordinate representation, the \glspl{ssm}. This input is fed to the model that processes it and finds a joint space to infer the boundaries.
  \item[When is the context information used?] This is a traditional early-fusion approach where the two domains are merged as input to the model. Since the baseline model is a \gls{cnn}, we just simply concatenate both representations as extra channel inputs. The model is then in charge of processing it and finds the needed relationships between domains.

\end{description}

\section{Multimodal version using text and audio information}
\label{sec:structures_multimodal}

\subsection{Dataset}

\begin{figure}
	\begin{subfigure}{0.5\textwidth}
		\centering
	  \includegraphics[width=.95\linewidth]{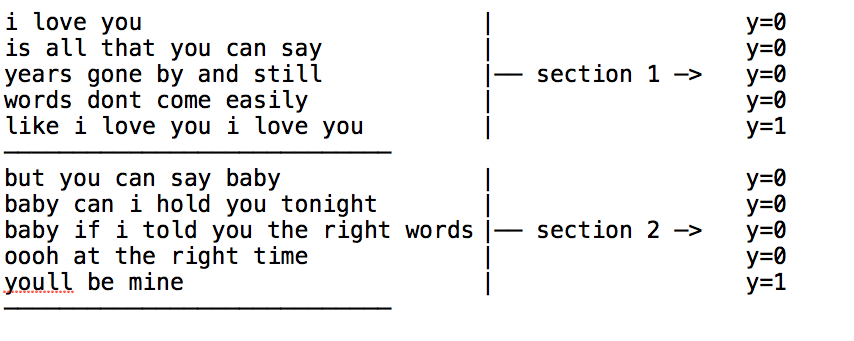}
	  \caption[Target values $y_i$ for lyrics segmentation]{Example of how the target values $y_i$ are assigned to each line.}
		\label{fig:selecting_targets}
	\end{subfigure}
	\begin{subfigure}{0.5\textwidth}
		\centering
		\begin{tabular}{c|c|c} \hline
		Version & Merge value & Number of songs\\
		\hline
		\hline
		full dataset & - & 4784\\
		\hline
		$M^+$ & high (90-100\%) & 1048\\
		\hline
		$M^0$ & med (52-90\%) & 1868\\
		\hline
		$M^-$ & low (0-52\%) & 1868\\
		\hline
		\end{tabular}
		\caption[DALI subsets for multimodal lyrics segmentation]{The \gls{dali} dataset partitioned by $merge(D, W)$}
		\label{tab:datasets}
	\end{subfigure}
\end{figure}

We use the first version of \gls{dali} with 5358 songs.
We focus only on the time and text information of lines $A_{\mathit{lines}} = (a_{k, \mathit{lines}})_{k=1}^{K_\mathit{lines}} \textrm{ where } a_{k, \mathit{lines}} = (t_k^0, t_k^1, l_k)_{\mathit{lines}}$ (see Chapter~\ref{sec:dali_description_definition}).
Since we know which lines are in each paragraph we can easily define the section boundary lines, $(a_i, y_i) = 1$ when the line is the last of a paragraph (see Figure~\ref{fig:selecting_targets}).
To be consistent with previous approaches~\citep{watanabe2016modeling,fell2018lyrics}, we only select songs that have at least 5 sections reducing the number of songs to 4784.

As described earlier (see chapter~\ref{sec:raw_annot}), the raw lines extracted from the karaoke resources have been merged to create the paragraph level using as reference the paragraphs in \gls{wasabi}.
This process may induce errors.
For this reason, we compute a new proxy that indicates the accuracy of the merging process.
Let $D={D_0, D_1, ..., D_u}$ be the paragraphs in \gls{dali} and $W={W_0, W_1, ..., W_v}$ the paragraphs in \gls{wasabi}, we can compute the similarity between them using \textbf{String similarity} {\tt str}:

\begin{equation}
     merge(D, W) = \min_{0 \leq i \leq u}\{ \max_{~ 0 \leq j \leq v}\{\;\text{str}(D_i, W_j)\;\}\;\}
\end{equation}

This metric finds the paragraph in \gls{dali} with the lowest string similarity to a paragraph in \gls{wasabi}.
Note how this is quite restrictive because it only focuses on the worst single paragraph in \gls{dali} and does not take into account the rest.
We partition \gls{dali} into three different subsets based on the different values of $merge$ (see Table~\ref{tab:datasets}).
The $M^+$ subset consists of 50842 lines and 7985 section boundaries.
It has 72\% of the lines in English, 11\% in German, 4\% in French, 3\% in Spanish, 3\% in Dutch and 7\% in other languages.

\subsection{Self Similarities}

Each $a_{k, l} = (t_k^0, t_k^1, l_k, i_k)_{lines}$ connects a text line $l_k$ with a particular audio segment $(t_k^0, t_k^1)$.
We can then add to the text \glspl{ssm} the new \glspl{ssm} that compare audio segments.
The audio \glspl{ssm} captures complementary information such as melodic or harmonic structures that complement the text structures.
Since the audio \glspl{ssm} have also the same dimensions, they can be simply stacked as an extra channel to the original input.
The architecture of the model stays the same.
Instead of comparing the raw audio of each $x_i$ itself, we extract two sets of well-known audio features: the \textbf{Mel-frequency cepstral coefficients} ($\textit{mfcc} \in \mathbb{R}^{14}$)~\citep{mfcc80}, to emphasize the part of the signal that is related (with our understanding) to the musical timbre and the \textbf{Chroma} ($\textit{chroma} \in \mathbb{R}^{12}$)~\citep{chromaFujishima99} to describe the harmonic information of each frame, computing the ``presence'' of the twelve different pitch-classes.

\begin{figure*}[ht!]
 \centerline{
   \includegraphics[width=.7\textwidth]{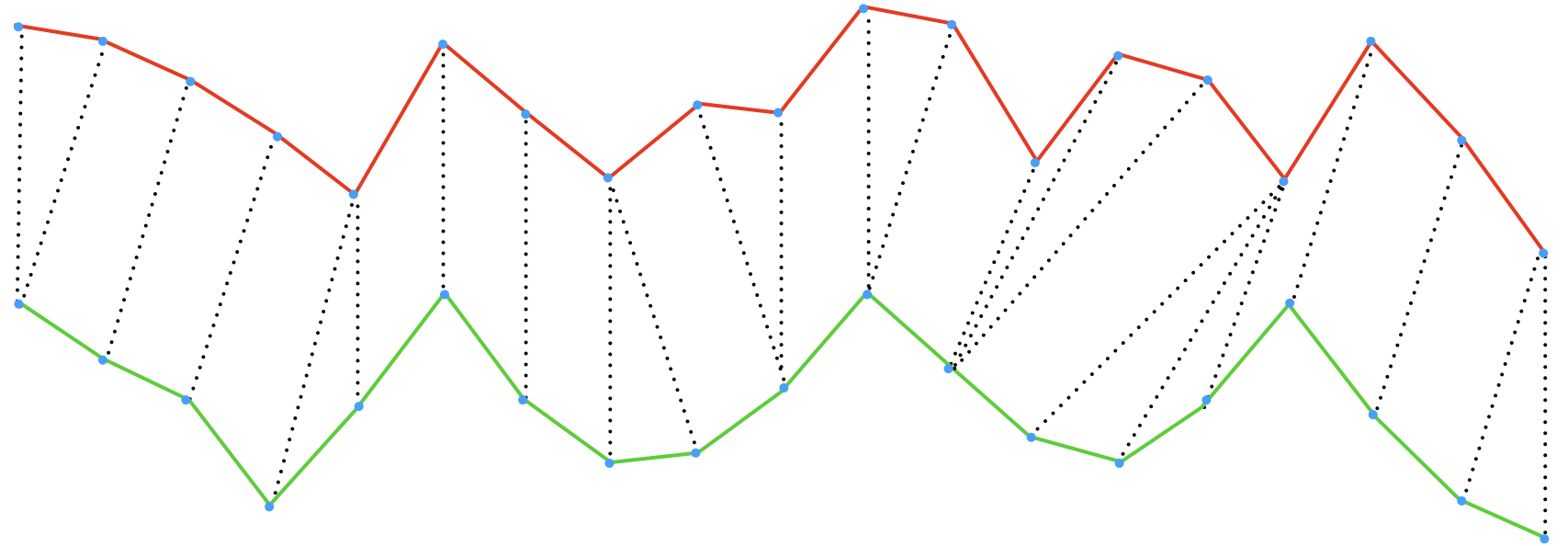}}
 \caption[DTW alignment example]{Illustrative example of the \gls{dtw} alignment between two sequences.}
 \label{fig:dtw}
\end{figure*}

We then compute the audio \gls{ssm} by measuring the similarities between all the audio segments $X=\{x_1, x_2, ..., x_k\}$ defined by the $a_{k, l}$.
Considering that each audio segment may have a different length, we use as similarity the last element of the \gls{dtw} cost matrix.
Given two lines $i$ and $j$ of length $p$ and $q$ respectively described as a sequence of features $x_{i,p} = \{x_{i,0}, x_{i,1}, ..., x_{i,p}\}$ and $x_{j, q} = \{x_{j,0}, x_{j,1}, ..., x_{j,q}\}$, \gls{dtw}~\citep{Sakoe_1978} applies a nonlinear warping to best align their similar areas (even if they are out of phase in the time axis) with a minimal distance between them (see Figure~\ref{fig:dtw}).
\gls{dtw} is formulated as an optimization problem:

\begin{equation}
  DTW(x_{i,p}, x_{j, q}) = \min_\pi \sqrt{ \sum_{(i, j) \in \pi} d(x_{i,p}, x_{j, q}) }
\end{equation}

\noindent where $d(x_{i,p}, x_{j, q})$ is the distance between the feature frames $i$ and $j$ of $x$ and $y$ respectively, and $\pi = [\pi_0, \dots , \pi_K]$ the alignment path between $x_{i,p}$ and $x_{j, q}$ where each element is a pairs $\pi_k = (i_k, j_k)$ with $0 \leq i_k < n$ and $0 \leq j_k < m$.
The path has to be \textbf{monotonic} (it does not go backwards in time), \textbf{contiguous} and covers full sequences, $\pi_0 = (0, 0)$ and $\pi_K = (n - 1, m - 1)$.

The algorithmic solution to this problem can be found efficiently using \textit{dynamic programming}.
The algorithm creates a cumulative matrix where each point cumulative the minimum distances of the adjacent elements.

\begin{equation}
  DTW_\mathit{cost}(x_{i,a}, x_{j,b}) = d(x_{i,a}, x_{j, b}) + min \left\{\begin{matrix} DTW_\mathit{cost}(x_{i,a-1}, x_{j,b})\\DTW_\mathit{cost}(x_{i,a-1}, x_{j,b-1})\\ DTW_\mathit{cost}(x_{i,a}, x_{j,b-1}) \end{matrix}\right\} \quad \footnote{Researchers have explored many other constrains to build this matrix such as \textit{slope weighting} by a factor $w(a-1, b), (a-1,b-1), w(a, b-1)$ or proposing different \textit{step patterns} like $(a-1, b-1), (a-1,b-2), (a-1,b-1)$. These variations can favor different alignment behaviors.}
\end{equation}

Once $DTW_\mathit{cost}$ is computed, we obtain the alignment path by backtracking from the last position $(p-1, q-1)$ to the origin $(0, 0)$ with the minimum cost.
The most significant advantage of \gls{dtw} is its invariance against shifting and scaling in the time axis and its capacity to align both signals where their corresponding similar areas are linked.

\begin{figure}
	\centering
  \includegraphics[width=.9\linewidth]{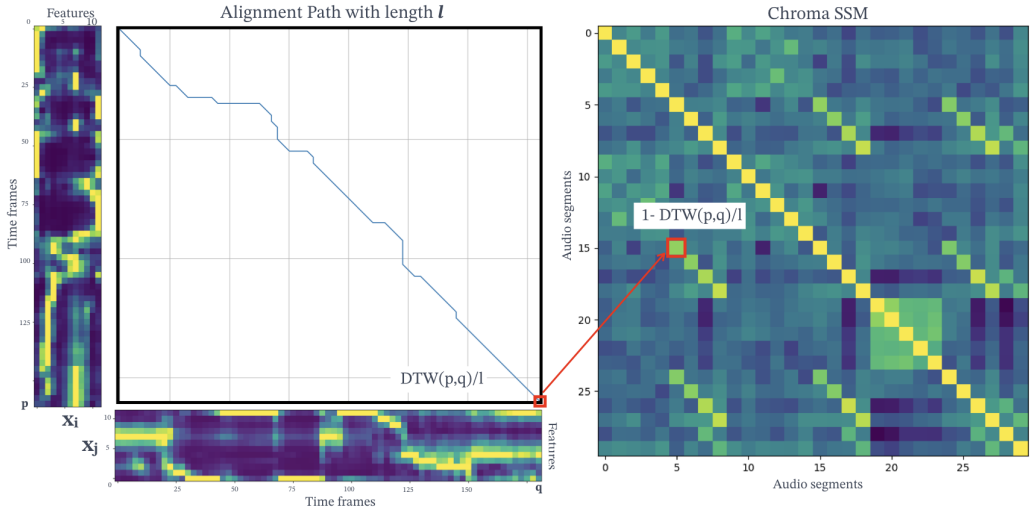}
  \caption[Audio Self-Similarity Matrix computation]{Audio \gls{ssm} computation. [Left]  The \gls{dtw} computes the alignment between two audio features sequences $x_i$ and $x_j$ of lengths $p$ and $q$. We use the inverse of the global alignment cost as a similarity measurement between audio sequences of different lengths. [Right] Repeating this procedure and comparing all the audio segments $X=\{x_1, x_2, ..., x_k\}$ in a song defined by $a_{k, l}$ we create the final \gls{ssm}.}
\label{fig:dtw_example}
\end{figure}

We use as distance between features $d(x_{i,a}, x_{j,b}) = 1 - |cosine(x_{i,a}, x_{j, b})|$\footnote{The $cosine$ of $0^{\circ}$ is $1$ and less than $1$ for any angle in the interval $(0, \pi]$ } which is a normalized value between 0 (no distance between features) and 1 (very different features) and does not depend on the dimensionality of the features.
Because of this normalization, the cost matrix stores an accumulation where each feature frame contributes equally (a normalized value between $0$ and $1$).
Thus, by simply dividing by the length of the path to get to that point we compute a normalized cost.
This allows us to use as similarity to use $s(x_{i,p}, x_{j,q}) = 1 - (DTW_\mathit{cost}(p,q) / l)$ where $l$ is the lenght of the best alignment path between both signals (the minimal path of the cost matrix) instead of $s(x_{i,p}, x_{j,q}) = 1 / DTW_\mathit{cost}(p,q)$ which depends on signal lenghts.
The final $s(x_{i,p}, x_{j,q})$ has always a value between 0 and 1, which corresponds to the mean cost contribution to the final path per feature frame.
Thereby, the different $DTWs_\mathit{cost}$ that compare the audio segments $X=\{x_1, x_2, ..., x_k\}$ has the same range of values allowing us to direclty compute the audio \gls{ssm} (see Figure~\ref{fig:dtw_example}).
We employ two \gls{ssm} matrices $ssm_\mathit{mfcc}$ and $ssm_\mathit{chroma}$.

\section{Evaluation of the multimodal lyrics segment detector}
\label{sec:exp_multimodal}

We compare different versions of the original model~\citep{fell2018lyrics} (see Figure~\ref{fig:approach}).
Each one is trained with different input data: \textbf{$text$}-based , \textbf{$audio$}-based, and \textbf{$multi$}modal-based with both text and audio features.
We use as dataset the $M^+$ subset.
We split the data randomly into training and test sets using a 5-fold cross-validations.
The cross validation is performed twice every time with different random initialization.
The results are depicted in Table~\ref{tab:results_multimodal}.

\begin{table}
    \centering
    \begin{tabular}{l|l|c|c|c} \hline
    \textit{Model} & \textit{Features} & \textit{P} & \textit{R} & $F_1$\\ \hline
    $text$  & \{str\}    & 78.7 &	64.2 &	70.8 \\
        \hline
    \multirow{3}{*}{$audio$}
        & \{mfcc\} & 79.3 &	55.9 &	65.3 \\
        & \{chroma\}     & 76.8 &	54.7 &	63.9  \\
        & \{mfcc, chroma\}        & 79.2 &	63.8 &	70.4 \\
        \hline
    \multirow{3}{*}{$multi$}
        & \{str, mfcc\}    & 80.6 &	\textbf{69.0} &	73.8\\
        & \{str, chroma\}    & 82.5 &	\textbf{69.0} &	74.5 \\
        & \{str, mfcc, chroma\}    & \textbf{82.7} &	\textbf{70.3} &	\textbf{75.3} \\
        \hline
    \end{tabular}
    \caption[Lyrics segmentation results]{Results with multimodal lyrics lines on the $M^+$ dataset in terms of Precision (\textit{P}), Recall (\textit{R}) and F-measure$(F_1)$ in \%. Note that the $text_\mathit{str}$ model is the same configuration as in~\cite{fell2018lyrics}, but trained on different dataset.}
    \label{tab:results_multimodal}
\end{table}

If we focus on the $F_1$, the model $text_\mathit{str}$ trained on $M^+$ performs similary (70.8\%) to the $audio$ one (70.4\% for $audio_\mathit{mfcc, chroma}$).
However, each indivual features perform worse with 65.3\% for $audio_\mathit{mfcc}$ and 63.9\% for $audio_\mathit{chroma}$.
As the $mfcc$ feature models timbre and instrumentation, whilst the chroma feature models melody and harmony.
They provide complementary information that benefits the model with both features, $audio_\mathit{mfcc, chroma}$.

Most importantly, the overall best performing model is a combination of the $text$ and $audio$ features~\citep{fell_nechaev_2021}.
It achieves the best results in all three evaluation metrics: $precision$, $recall$, and a $F_1$ of 75.3\%.
This shows that text and audio modalities capture complementary structures that are beneficial for the lyrics segmentation.
Additionally, each $multi$ version also outpeforms the $text$ and $audio$ models. $multi_\mathit{str, mfcc}$ and $multi_\mathit{str, chroma}$ achieve a performance of 73.8\% and 74.5\%.

\begin{table}
    \centering
    \begin{tabular}{cllccc} \hline
    Dataset & Model & Features & \textit{P} & \textit{R} & $F_1$\\ \hline
    \multirow{3}{*}{$M^+$}
        & $text$  & \{str\}    							& 78.7 &	64.2 &	70.8 \\
        & $audio$ & \{mfcc, chroma\}       		& 79.2 &	63.8 &	70.4   \\
        & $multi$  & \{str, mfcc, chroma\}    	& 82.7 &	70.3 &	75.3 \\
    \hline
    \multirow{3}{*}{$M^0$}
        & $text$ & \{str\}        					&  73.6 &	54.5 &	62.8\\
        & $audio$ & \{mfcc, chroma\}       		&  74.9 &	48.9 &	59.5 \\
        & $multi$ & \{str, mfcc, chroma\}   		&  75.8 &	59.4 &	66.5 \\
    \hline
    \multirow{3}{*}{$M^-$}
        & $text$ & \{str\}         					& 67.5 &	30.9 &	41.9 \\
        & $audio$ & \{mfcc, chroma\}        		& 66.1 &	24.7 &	36.1  \\
        & $multi$ & \{str, mfcc, chroma\}  		& 68.0 &	35.8 &	46.7 \\
    \hline
    \end{tabular}
    \caption[Lyrics segmentation results for the different versions of DALI]{Results with \gls{multimodal} lyrics lines for the alignment quality ablation test on the datasets $M^+$, $M^0$, $M^-$ in terms of Precision (\textit{P}), Recall (\textit{R}) and F-measure$(F_1)$ in \%.}
    \label{tab:results_multimodal_scaling}
\end{table}

In Table~\ref{tab:results_multimodal_scaling}, we give the performances on the three subsets of \gls{dali}, $M^+$, $M^0$ and $M^-$ using the feature that performed best on the $M^+$ i.e. $text_\mathit{str}$, $audio_\mathit{mfcc, chroma}$, and $multi_\mathit{str, mfcc, chroma}$.
Each subset follows its own 5-fold cross-validations procedure.
We find that independent of the modality (text, audio, mult.), all models perform significantly better when the merge value is higher.
Moreover, even if the results decrease, the $multi$ version always performs better than the one domain versions.

\section{Conclusions}

In this chapter, we used \gls{dali} for the first time to address the task of lyrics segmentation on synchronized text-audio representations.
Since each song contains the lyrics aligned with the audio, we can compute several \glspl{ssm} for the two domains.
We showed that exploiting both textual and audio-based features outperforms the systems that rely on purely text-based features proving that each domain captures complementary structures that benefit to the overall performance.
This chapter is the result of a collaboration with Michael Fell and Elena Cabrio who trained and tested the models.

\chapter{Conditioned U-Net}
\label{sec:cunet}

\graphicspath{{figs/}{cunet/figs/}}

Data-driven models for audio \gls{source_separation} such as \gls{unet} (see Chapter~\ref{sec:unet}) or Wave-U-Net are usually models separated to and specifically trained for a single task, e.g. a particular source separation.
Training them for various tasks at once commonly results in worse performances than training them for a single specialized task.
In this chapter, we introduce the \gls{cunet} which adds a control mechanism to the standard \gls{unet}.
The control mechanism allows us to train a unique and generic \gls{unet} to perform the separation of various sources.
The \gls{cunet} decides the source to isolate according to a one-hot-encoding input vector.
The input vector is embedded to obtain the parameters that control \gls{film} layers.
\gls{film} layers modify the \gls{unet} feature maps in order to separate the desired source via affine transformations.
The \gls{cunet} performs different source separations, all with a single model achieving the same performances as the separated ones at a lower cost.
The multitask formalization also serves as an exploration of how \gls{film} can effectively control the behavior of a generic \gls{unet}.
This idea will be extended in the next chapter for improving vocal \gls{source_separation} using the \gls{dali}.

\section{Introduction}
\label{sec:cunet_intro}

Generally, in \gls{mir} we develop separated systems for specific tasks. Facing new (but similar) tasks require the development of new (but similar) specific systems.
This is the case of data-driven music \gls{source_separation} systems.

Music is usually distributed as mono channel or 2-channel stereo (left and right) where all the instruments are mixed.
However, there are many cases in which we are only interested in listening to or working with a single instrument or a particular combination of them.
It serves also as an intermediary step for other \gls{mir} tasks such as fundamental frequency estimation or score transcription.
The music \gls{source_separation} task aims to isolate the different instruments that appear in an audio mixture (a mixed music track), i.e. reversing the mixing process~\citep{Cano_2018}.

\Gls{source_separation} is one of the main \gls{mir} tasks. For decades, it has awakened the interest of researchers who have developed many approaches. Source position (to exploit the spatial position of the sources), kernel additive (to find local features presented in the spectrogram), non-negative matrix factorization (to factor the mix into a combination of basic sources activated at different time positions) or sinusoidal models (to approximate the mixture by several sinusoids) have been successfully applied~\citep{Rafii_2018, Cano_2018, Pardo_2018}.
Currently and thanks to the increase of annotated datasets, data-driven methods have taken the lead.
These methods use supervised learning where the mixture signals and the isolated instruments are available for training.
The usual approach is to build separated models for each source to be isolated~\citep{Jansson_2017, Stoller_2018}.
This has been proved to show great results.
However, since isolating an instrument requires a specific system, we can easily run into problems such as scaling issues (100 instruments = 100 systems).
Besides, these models do not use the commonalities between instruments.
If we modify them to do various tasks at once i.e., adding fix numbers of output masks in the last layer, they reduce their performance~\citep{Stoller_2018}.

\Gls{conditioning} has appeared as a solution to problems that need the integration of multiple resources of information.
Concretely, when we want to process one in the context of another i.e., modulating a system computation by the presence of external data.
\Gls{conditioning} divides problems into two elements: a \textbf{generic system} and a \textbf{control mechanism} that governs it according to external data.
Although there is a large diversity of domains that use it, it has been developed mainly in the image processing field for tasks such as visual reasoning or style transfer.
There, it has been proved very \-effective, improving the state of the art results~\citep{Perez_2017, Vries_2017, Strub_2018}.
This paradigm can be integrated into \gls{source_separation}, creating a generic model that adapts itself to isolate a particular instrument via a control mechanism.
We believe that this paradigm can benefit to a great variety of MIR tasks such as multi-pitch estimation, music transcription or music generation.

\begin{figure}[t!]
  \centerline{
    \includegraphics[width=.8\textwidth]{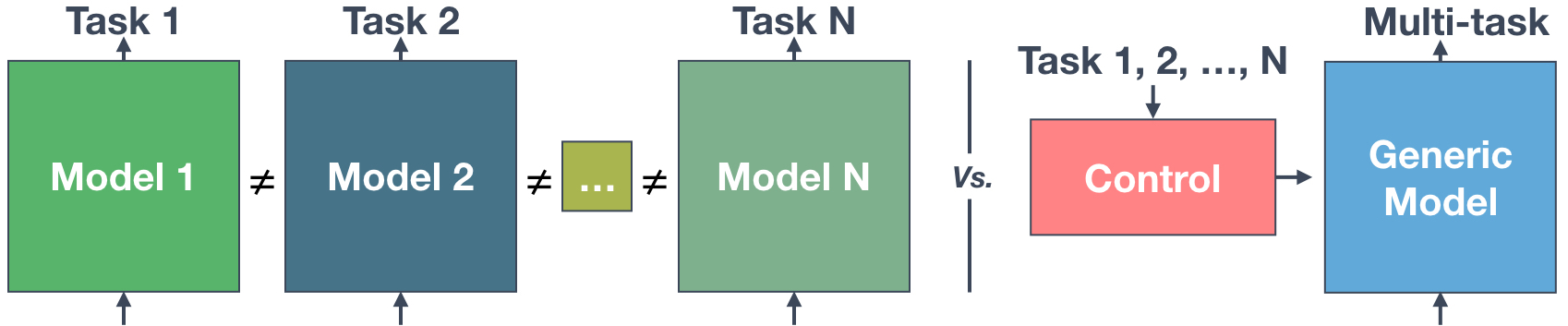}}
  \caption[Source separation based on DNN comparison]{[Left part] Traditional approach: a separated \gls{unet} is trained to separate a specific source.
  [Right part] Our proposition based on \gls{conditioning}. The problem is divided in two: a standard \gls{unet} (which provides generic \gls{source_separation} filters) and a control mechanism.
  This division allows the same model to deal with different tasks using the commonalities between them.
  %
  }
  \label{fig:general-schema}
\end{figure}


\section{Formalization}\label{sec:cunet_formalization}

In this chapter, we study the application of \gls{conditioning} for music \gls{source_separation} where an external context vector decides which instrument is to be separated.
As described in Chapter~\ref{sec:c_multimodal}, we can define our problem satisfying certain properties summarized as~\citep{Bengio_2013}:

\begin{description}[.15cm]
  \item[How is the multimodal model constructed?]
  Our system relies on
  a \textbf{standard \gls{unet} system} not specialized in a specific task but rather in finding a set of generic \gls{source_separation} filters, that we
  \textbf{control} differently for isolating a particular instrument,
  as illustrated in Figure \ref{fig:general-schema}. The \textit{conditioning} itself is performed using Feature-wise Linear Modulation (FiLM) layers~\citep{Perez_2017}, controlling the generic model model to perform different instrument source separations. Hence, this is a \textbf{model-based} multimodal learning approach.
  \item[What context information?] Our system takes as input the spectrogram of the mixed audio signal and the control vector as context information for guiding the separation.
  It gives as output (only one) the separated instrument defined by the control vector.
  The main advantages of our approach are
  - direct use of commonalities between different instruments,
  - a constant number of parameters no matter how many instruments the system is dealing with
  - and scalable architecture, in the sense that new instruments can be potentially added without training from scratch a new system.
  \item[When is the context information used?] We insert these layers in the encoder of the \textit{generic} network for creating different latent spaces concerning the context information that defines the instrument to separate.
\end{description}

\section{Related work}
\label{sec:souce_literature}
In this thesis, we focus only on the data-driven \gls{source_separation} approaches.
We review only works related to this approach as well as conditioning in audio and multitask \gls{source_separation}.
Note that many of the reviewed papers have been published after carrying out the work presented in this chapter.

\subsubsection{Source separation based on supervised learning}
Over the past decade, researchers have developed  a wide variety of \gls{source_separation} methods for many different scenarios.
We refer the reader to~\cite{Rafii_2018, Cano_2018, Pardo_2018} for an extensive and comprehensive overview.
In data-driven approaches, we have access to the mixture and the isolated sources that compose it.
Thereby, models learn in a \gls{sl} way to either compute a mask to isolate the source from the background (for instance Ideal Binary/Ratio Mask or Spectral Magnitude Mask) or to obtain directly the clean spectral representations (e.g. magnitude/power spectrum or mel spectrum)~\citep{Wang_2018}.
The appearance of large and representative datasets (The Music Audio Signal Separation (MASS) dataset~\citep{MTGMASSdb}, the Musdb dataset~\citep{musdb_2018} for the Signal Separation Evaluation Campaigns (SiSEC)~\citep{Vincent_2009} or \gls{medley}~\citep{Bittner_2014} among others) has accelerated the progress and boosted the separation performance.
Nowadays, neural networks have taken the lead.

The first approaches rely on global features across the entire frequency spectrum, over a longer period.
\cite{Wang_2014} proposes to separate speech signals from a noisy mixture, estimating the Ideal Ratio Mask (IRM) and the short-time Fourier transform Spectral Mask (FFT-MASK) with a \gls{fully} network.
\cite{Huang_2014, Huang_2015} implement a method for monaural source separation for speech separation, singing voice separation, and speech denoising.
They use a Recurrent Neural Network (RNN) to model the temporal evolution of timbre features for each source, taking as inputs single frames of the magnitude spectrogram of a mixture.
Timbre features are used to compute a soft masking function that isolates the desired source.
\cite{Nugraha_2016, Nugraha2016m} work with multichannel source separation, using both phase and magnitude information.
They use a \gls{fully} architecture to estimate the magnitude spectra of the sources combined with spatial covariance matrices to encode the spatial characteristics and a final multichannel Wiener filter.
\cite{Uhlich_2015} also implement a \gls{fully} architecture where the input concatenates multiple frames of the magnitude spectrogram of a mixture, modeling timbre features across multiple time frames.
While these approaches work well, they do not completely  exploit local time-frequency features.

The encoder-decoder architecture together with \gls{cnn} has become one of the most popular choices.
\cite{Chandna_2017} estimates time-frequency soft masks that are applied to the magnitude spectrogram.
The encoder consist of a `vertical' \gls{cnn} layer to capture local timbre information followed by a `horizontal' \gls{cnn} layer to capture temporal evolution.
The output is connected to a \gls{fully} layer to perform a dimensional reduction.
The decoder has a \gls{fully} layer with as many neurons as the size of the `horizontal' \gls{cnn} and a sucession of deconvolution layers to inverse the convolution that compute the final soft masks.
\cite{Jansson_2017} adapts the \gls{unet} architecture to separate the vocal and accompaniment components, training a specific model for each task.
It consists of an encoder-decoder with \gls{skip} between blocks at similar depths.
The architecture is described in detail in the following sections.
The two previous approaches use the phase of the original mixture to recompute the waveform from the magnitude spectrogram.
\cite{uhlich_2017} computes also masking on the magnitude spectrogram, using a bi-directional \textit{LSTM} model after compressing the frequency and channel information with a \gls{fully} block.
The core of the model are three \textit{LSTM} layers with a \gls{skip} between the input to the first \textit{LSTM} and the output of the last one. The output is decoded back to its original input dimensionality with two \gls{fully} blocks, computing the final mask.
A post-processing step with a multichannel Wiener filter obtains the final waveform.
This work has been open-source implemented in Open-Unmix~\citep{stoter_19}.
\cite{Wang_2018p} propose an iterative phase reconstruction procedure where the STFT and its inverse are replaced by trainable linear convolutional layers.

Since the output of these works is the magnitude spectrogram, they need to reconstruct the audio signal using a phase that is not the original one.
This potentially leads to artifacts.
Wave-U-Net proposes to apply an architecture similar to the U-Net one but on the audio-waveform~\citep{Stoller_2018}.
They also adapt their model for isolating different sources at once by adding as many outputs as sources to separate.
However, this multi-instruments version performs worse than the separated one (for vocal isolation) and has to be retrained to different source combinations.
\cite{Defossez_2019} add to the Wave-U-Net architecture a two layers bi-directional \textit{LSTM} in-between the encoder and the decoder and introduce a $1x1$ convolution after each convolution block to halve the number of channels. They also propose a semi-supervised approach to leverage unlabeled multitrack songs.
Conv-TasNet~\citep{Luo_2019} is also a time-domain approach that adds a separator to the encoder-decoder architecture.
Once the encoder computes a latent representation, the separator learns a mask in this latent space.
The separator is based on a temporal convolutional network (TCN)~\citep{lea_2016} with 8 stacked dilated convolutional blocks with exponentially increasing dilation factors.
\citep{chandna_2019} propose a different paradigm. Instead of extracting the vocals from the mixture, they use a convolutional network with \gls{skip} and dilated convolutions to estimate the parameters of a synthesizer which synthesizes the vocal track, without any interference from the backing track.
Finally, the separation can be improved using a generative adversarial training process~\citep{Stoller_2018a}.

\subsection{Conditioning in Audio}

Conditioning has been mainly explored in \textbf{speech generation}.
In the WaveNet approach~\citep{Oord_2016, Oord_2017} the speaker identity is fed to a conditional distribution adding a learnable bias to the gated activation units.
A WaveNet modified version is presented in~\citep{Shen_2017}. The time-domain waveform generation is conditioned by a sequence of Mel spectrogram computed from an input character sequence (using a recurrent sequence-to-sequence network with attention).
In \textbf{speech recognition} conditions are used in~\citep{Kim_2017}, applying conditional normalisation to a deep bidirectional LSTM (Long Short Term Memory) for dynamically generating the parameters in the normalisation layer.
This model adapts itself to different acoustic scenarios.
In~\citep{Kim_2017}, the conditions do not come from any external source but rather from utterance information of the model itself.
They have been also used in \textbf{music generation} for accompaniments conditioned on melodies~\citep{Cheng-Zhi_2018} or incorporating history information (melody and chords) from previous measures in a generative adversarial network (GAN)~\citep{Yang_2017}.
It has been also proved to be very efficient for \textbf{piano transcription}~\citep{Hawthorne_2018}: the pitch onset detection is internally concatenated to the frame-wise pitch prediction controlling if a new pitch starts or not. Both, onset detection and frame-wise prediction are trained together.

\subsubsection{Multitask and conditioning source separation}

Early work study informed source separation in an encoding/decoding framework~\citep{Liutkus_2012, Parvaix_2010, Parvaix_2009}.
In the encoding stage, it is assumed to have access to the isolated sources to extract characteristic descriptors (parameters that describe the structure of the source signals, or their contribution to the mixture) for each source.
This extra information is imperceptibly embedded within the mixture, using a watermarking technique.
The decoding stage has as input the new `watermarked' mixture (with the descriptors of each source) and extracts the characteristic descriptors, using them to perform the final source separation.
Note how this framework has two distinct and independent processes that differ from the encoder/decoder architecture used in deep neural network architectures.
This approach has been studied together with molecular grouping (gather of close Modified Discrete Cosine Transform coefficients)~\citep{Parvaix_2009, Parvaix_2010}
or modeling the sources as independent and locally stationary gaussian process, and the mixture as linear filter~\citep{Liutkus_2012}.

\cite{Slizovskaia_2019} extend the Wave-U-Net architecture to perform a non-fixed number of separations. Unlike previous works that isolate popular music instruments (vocals, drums or bass), they concentrate on classical instruments (violin or viola).
They add an external model that computes the necessary parameter to perform a multiplicative condition in the bottleneck (latent space) of the model.
This block receives an additional condition vector that specifies the instrument to be separated.
\cite{Kadandale_2020} explore different loss strategies to allow the model to have a fixed number of multiple outputs.
\cite{kavalerov_2019, Tzinis_2019} aim to develop a universal sound separation, i.e. separating acoustic sources from an open domain, regardless of their class.
They adopt a Conv-TasNet architecture and add a sound classifier that conditions the separation based on the semantic information extracted from the mixture.
Assuming that it is known that the mixture has $m$ sources (from a list of 527 categories) the classifiers predict the classes over time to condition the generic model.
\cite{Lee_2019} add a query-net to control the separation using a \gls{unet} architecture. Given an audio query that defines the source to be isolated, query-net computes a latent representation that is both added as an input to the actual model that performs the separation and to condition the decoder part.
\cite{Seetharaman_2019} embed the mixture in gaussian spaces using \textit{BLSTM} stack.
A conditioning mask applied on the gaussian spaces is generated from a one-hot vector indicating the class as input.
%
%
Meta-TasNet adds a parameter generator to the Conv-TasNet~\citep{Samuel_2020}. As before, a one-hot encoder codifies the desired instrument. The conditioning is performed in the latent space learning how to mask the important features. They apply a multi-stage architecture to iteratively upsampl low-resolution audio to high resolution.


Finally, there is a group of papers in which researchers, instead of finding generic source separation models independent of sources to isolate, investigate multitasking \gls{source_separation} by having the model jointly solve some additional (but related) task. They hypothesize that the insights learned for solving this extra task help in isolating a particular instrument.
Most of the works focus on vocal separation. Early works such as \citep{Durrieu_2008} propose a two-step process where the vocal is isolated from the mixture before extracting the fundamental \gls{f0}.
\cite{Nakano_2019} invert the order and perform first \gls{f0} extraction to enhance the vocal \gls{source_separation}.
Possible errors in the \gls{f0} extraction can be manually solved~\citep{Nakano_2020}.
\cite{Jansson_2019} explore many different configurations such as shared-encoder, cross-stitch, or stack operations.
\cite{Stoller_2018j} aims to extract vocal activity detection to stabilize and improve the performance of vocal separation.
All these previous works used the \gls{unet} as a base network.
Finally, \gls{source_separation} is also quite related to music transcription (notating in a musical score).
\cite{Manilow_2020} simultaneously transcribe and separate multiple instruments, learning a shared musical representation for both tasks.

\section{Conditioning learning methodology}
\label{sec:methodology}

\begin{figure}[t!]
  \centerline{
    \includegraphics[width=.7\textwidth]{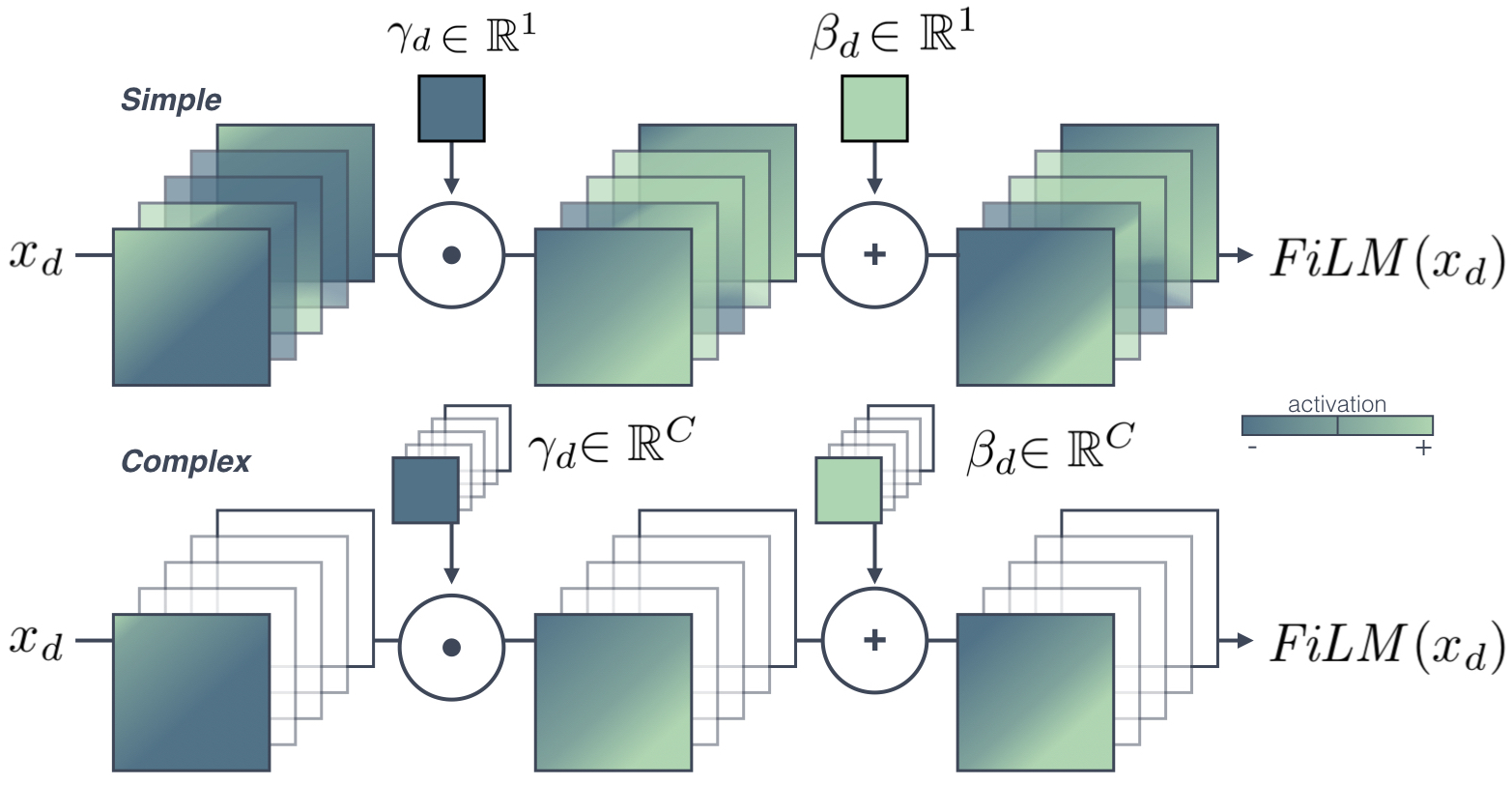}}
  \caption[Different FiLM layers]{[Top] \textit{\gls{film} simple} layer applies the same affine transformation to all the input feature maps $x_d$.
  [Bottom] In the \textit{\gls{film} complex} layer, independent affine transformations are applied to each feature map $C$ and a given depth of the net $d$.}
  \label{fig:film}
\end{figure}

\subsection{Conditioning mechanism.}
There are many ways to condition a network (see~\cite{Dumoulin_2018} for a wide overview) but most of them can be formalized as affine transformations denoted by the acronym \gls{film} (Feature-wise Linear Modulation)~\citep{Perez_2017}.
\gls{film} permits to modulate any neural network architecture inserting one or several \gls{film} layers at any depth of the original model.
A \gls{film} layer conditions the network computation by applying an affine transformation to intermediate features:

\begin{equation}
    \mathit{FiLM}(x_d) = \gamma_d(z) \cdot x_d + \beta_d(z)
\end{equation}

\noindent where $x_d \in \mathbb{R}^{W \times H \times C}$ is the input of the \gls{film} layer, the intermediate features of the \textbf{conditioned network}, at a particular depth $d$ in the architecture,
$W$ and $H$ represent the `time' and `frequency' dimension and $C$ the number of feature channels (or feature maps) and $(\gamma_d, \beta_d)$ and are parameters to be learned.
They scale and shift $x_d$ based on the external information, $z$.
The output of a \gls{film} layer has the same dimension as the intermediate feature input $x_d$.
\gls{film} layers can be inserted at any depth $d$ in the controlled network.

As described in Figure \ref{fig:film},
the original \gls{film} layer applies an independent affine transformation to each feature map (dimension $C$)\footnote{Or element-wise.}:  $(\gamma_d, \beta_d) \in \mathbb{R}^{C}$~\citep{Perez_2017}. $\gamma_d$ and $\beta_d$ are scalars applied to the feature map $C$ of the input $x$ at a given depth of the network $d$, needing as many parameters as input channels of features has.
We call this a \textit{\gls{film} complex} layer (\textbf{Co}).
We propose a simpler version that applies the same $(\gamma_d, \beta_d) \in \mathbb{R}^{1}$ to all the feature maps (therefore $\gamma_d$ and $\beta_d$ do not depend on $C$).
We call it a \textit{\gls{film} simple} layer (\textbf{Si}).
The \textit{\gls{film} simple} layer decreases the degrees of freedom of the transformations to be carried out forcing them to be generic and less specialized. It also reduces drastically the number of parameters to be trained.
As \gls{film} layers do not change the shape of $x_d$, \gls{film} is transparent and can be used in any particular architecture providing flexibility to the network by adding a control mechanism.

Even if this transformation (an affine operation with two parameters) is identical to \gls{batch_norm} (see Chapter~\ref{sec:batch_norm}), it differs in some essential points.
\Gls{batch_norm} is fundamentally a feature standardization (zero mean and standard deviation of one) so that each feature has the same contribution.
It additionally applies an affine transformation to modify its mean and variance in order to find the best values to take advantage of the non-linearity function.
Therefore, the $\gamma_d$ and $\beta_d$ obtained do not depend on any external factor, but they serve to an optimization purpose.
On the other hand, the $\gamma_d$ and $\beta_d$ computed for \gls{film} depend on the task at hand and they have different values for each task.
Its main goal is to describe how to modulate the generic architecture, finding different specializations to carry on several tasks.
Although they are inserted after \gls{batch_norm}, this is not mandatory and they also apply to any type of feature (not necessarily standardized), showing similar results~\citep{Perez_2017}.

\subsection{Conditioning architecture}
A conditioning architecture has two components:

\subsubsection{The conditioned network}
It is the network that carries out the core computation and obtains the final output.
It is usually a generic network that we want to behave differently according to external data.
Its behavior is altered by the condition parameters, $\gamma_d$ and $\beta_d$ via \gls{film} layers.

\subsubsection{The control mechanism - condition generator}
It is the system that produces the parameters ($\gamma_d$'s and $\beta_d$'s) for the \gls{film} layers with respect to the external information $z$: the input conditions.
It codifies the task at hand and provides the instructions to control the conditioned network.
The condition generator can be trained jointly~\citep{Perez_2017, Strub_2018} or separately with the conditioned network~\citep{Vries_2017}.

This paradigm clearly separates, from the main core computation, the task description and the control instructions.

\section{Conditioned-U-Net for multitask source separation}

\newcommand{\CMCG}[0]{control mechanism/condition generator }

We formalize \gls{source_separation} as a multi-tasks problem where one task corresponds to the isolation of one instrument.
We assume that while the tasks are different they share many similarities, hence they will benefit from a conditioned architecture.
We name our approach the \textbf{Conditioned-U-Net} (\gls{cunet}).
It differs from the previous works where a separated model is trained for a single task~\citep{Jansson_2017} or where it has a fixed number of outputs~\citep{Stoller_2018}.

\begin{figure}[t!]
  \centerline{
    \includegraphics[width=.5\textwidth]{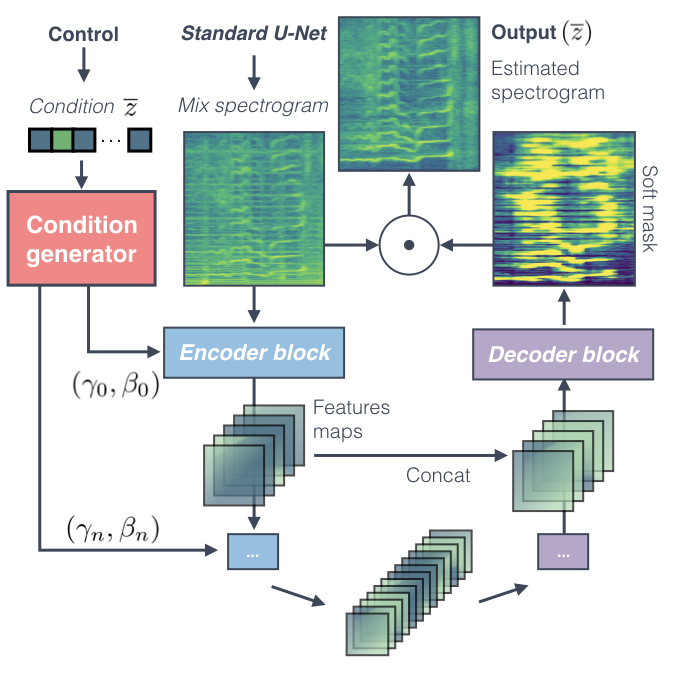}}
    \caption[C-U-Net architecture]{The \gls{cunet} has two distinct parts: the condition generator and a standard \gls{unet}. The former codifies $\overline{z}$ (with the instrument to isolate) for getting the needed $\gamma_d$ and $\beta_d$. The generic \gls{unet} has as input the magnitude spectrum. It adapts its behaviour via \gls{film} layers inserted in the encoder part. The system outputs the desired instrument defined by $\overline{z}$.}
  \label{fig:C-U-Net}
\end{figure}

As in~\citep{Jansson_2017, Stoller_2018}, our \textbf{conditioned network} is a standard \gls{unet} (see Chapter~\ref{sec:unet}) that computes a set of generic \gls{source_separation} filters that we use to separate the various instruments.
It adapts itself through the control mechanism (the condition generator) with \gls{film} layers inserted at different depths.
Our external data is a conditional vector $\overline{z}$ (a one-hot-encoding) which specifies the instrument to be separated.
For example, $\overline{z}=[0,1,0,0]$ corresponds to the drums.
The vector $\overline{z}$ is the input to the \CMCG that has to learn the best  $\gamma_d$ and $\beta_d$ values such that, when they modify the feature maps (in the \gls{film} layers) the \gls{cunet} separates the indicated instrument i.e., it decides which features maps information is useful to get each instrument.
The \CMCG is itself a neural network that embeds $\overline{z}$ into the best $\gamma_d$ and $\beta_d$.
The conditioned network and the condition generator are trained jointly.
A diagram is shown in Figure\ref{fig:C-U-Net}.

Our \gls{cunet} can perform different instrument \glspl{source_separation} as it alters its behavior depending on the value of the external condition vector $\overline{z}$.
The inputs of our system are the mixture and the vector $\overline{z}$.
There is only one output, which corresponds to the isolated instrument defined by $\overline{z}$.
While training, the output corresponds to the desired isolated instrument that matches the $\overline{z}$ activation.

\subsection{Conditioned network: U-Net architecture}

We used the \gls{unet} architecture proposed for vocal separation~\citep{Jansson_2017}, which is an adaptation of the \gls{unet} for microscopic images~\citep{Ronneberger_2015} (see Chapter~\ref{sec:unet}). 
The \gls{unet} follows an encoder-decoder architecture and adds \glspl{skip} to it (see Chapter~\ref{sec:skip}).
The input and output are the normalized magnitude spectrograms of the \gls{dft} of the monophonic mixture and the instrument to isolate.
\gls{dft} represents a signal as a sum of complex-valued Fourier coefficients (magnitude and phase) for sinusoids of varying frequency (see Figure~\ref{fig:freq_domains}).

\subsubsection{Encoder} It creates a compressed and deep representation of the input by reducing its dimensionality while preserving the relevant information for the separation.
It consists of a stack of convolutional layers, where each layer halves the size of the input but doubles the number of channels.

\subsubsection{Decoder} It reconstructs and interprets the deep features and transforms it into the final spectrogram.
It consists of a stack of deconvolutional layers.

\subsubsection{Residual/skip-connections}
As the encoder and decoder are symmetric i.e., feature maps at the same depth have the same shape, the \gls{unet} adds \glspl{skip} between layers of the encoder and decoder of the same depth.
This refines the reconstruction by progressively providing finer-grained information from the encoder to the decoder.
Namely, feature maps of a layer in the encoder are concatenated to the equivalent ones in the decoder.

The final layer is \textbf{a soft mask} (sigmoid function $\in [0,1]$) $f(X, \theta)$ which is applied to the input $X$ to get the isolated source $Y$. The loss of the \gls{unet} is defined as:
  \begin{equation}
    \mathcal{L}(X, Y; \theta) = \|f (X, \theta) \odot X - Y \|_{1,1}
  \end{equation}
 where $\theta$ are the parameters of the system.

\subsubsection{Architecture details}
Our implementation mimics the original one~\citep{Jansson_2017}.
The \textbf{encoder} consists in $6$ encoder blocks.
Each one is made of a $2D$ convolution with filters of size $(5,5)$, stride $2$, \gls{batch_norm}, and leaky rectified linear units (ReLu) with leakiness $0.2$.
The first layer has $16$ filters and we double them for each new block up to a total of $512$. This defines features maps with dimensions: $(256,64,16)$, $(128,32,32)$, $(64,16,64)$, $(32,8,128)$, $(16,4,256)$ and $(8,2,512)$.
The \textbf{decoder} maps the encoder, with $6$ decoders blocks with stride deconvolution, stride $2$ and filters of size $(5,5)$, \gls{batch_norm}, plain ReLu, and a 50\% \gls{dropout} in the first three blocks.
The final block (the soft mask) uses a sigmoid activation. The feature maps of each decoder block have a mirror dimension regarding the \textbf{encoder} $(16,4,256)$, $(32,8,128)$, $(64,16,64)$, $(128,32,32)$, $(256,64,16)$, $(512,128,1)$. Except for the first \textbf{decoder} block, the input from the rest is the concatenation via \glspl{skip} of the feature maps generated by its predecessor and the feature map of the corresponding \textbf{encoder} block.
The model is trained using the ADAM optimiser~\citep{Diederik_2014} and a $0.001$ learning rate.
As in~\citep{Jansson_2017}, we downsample to 8192 Hz, compute the Short Time Fourier Transform with a window size of 1024 and hop length of 768 frames.
The input is a patch of 128 frames (roughly 11 seconds) from the normalised magnitude spectrogram for both the mixture spectrogram and the isolated instrument.
The normalization is performed for the whole song not for the individual patch. We also keep the ratio between sources meaning that, if the maximum of a source before normalization is $0.86\%$ the maximum in the mixture, it continues to be the same after the normalization.

\begin{figure}[t!]
  \centerline{
    \includegraphics[width=.7\textwidth]{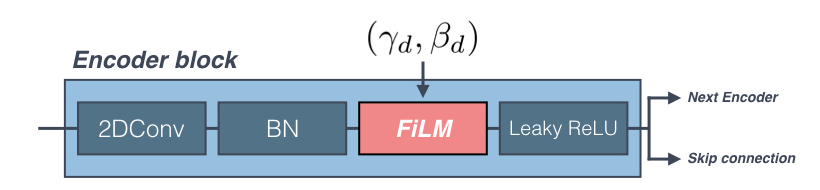}}
    \caption[FiLM layer possition]{\gls{film} layers are placed after the batch normalisation. The output of a encoding block is connected to the next encoding block and the equivalent layer in the decoder via the \glspl{skip}.}
  \label{fig:encoding-block}
\end{figure}

\subsubsection{Inserting FiLM} The \gls{unet} has two well differentiated stages: the \textbf{encoder} and \textbf{decoder}.
The \textbf{enconder} is the part that transforms the mixture magnitude input into a deep representation capturing the key elements to isolate an instrument.
The \textbf{decoder} interprets this representation for reconstructing the final audio.
We hypothesise that, if we can have a different way of encoding each instrument i.e., obtaining different deep representations, we can use a common `universal' decoder to interpret all of them.
Following this reasoning, we decided to condition only the \gls{unet} encoder part.
In the \gls{cunet}, a \gls{film} layer is inserted inside each encoding block after the batch normalisation and before the Leaky ReLu, as described in Figure \ref{fig:encoding-block}.
This decision relies on previous works where features are modified after the normalisation~\citep{Perez_2017, Vries_2017, Kim_2017}.
Batch normalisation normalises each feature map so that it has zero mean and unit variance~\citep{Ioffe_2015}.  Applying \gls{film} after batch normalisation re-scale and re-shift feature maps after the activations.
This allows the net to specialise itself to different tasks.
As the output of our encoding blocks is transformed by the \gls{film} layer the data that flows through the \glspl{skip} carries on also the transformations.
If we use the \textit{\gls{film} complex} layer, the \CMCG needs to generate $2016$ parameters ($1008$ $\gamma_d$ and $1008$ $\beta_d$). On the other hand, \textit{\gls{film} simple} layers imply $12$ parameters: one $\gamma_{d}$ and one $\beta_{d}$ for each of the 6 different encoding blocks, which means 2002 parameters less than for \textit{\gls{film} complex} layers.

\subsection{Condition generator: Embedding nets}

The \CMCG computes the $\gamma_d(\overline{z})$ and $\beta_d(\overline{z})$ that modify our standard \gls{unet} behavior.
Its architecture has to be flexible and robust to generate the best possible parameters.
It has also to be able to find relationships between instruments.
That is to say, we want it to produce similar $\gamma_d$ and $\beta_d$ for instruments that have similar spectrogram characteristics.
Hence, we explore two different embeddings: a fully connected version and a convolutional one (CNN).
Each one is adapted for the \textit{\gls{film} complex} layer as well as for the \textit{\gls{film} simple} layer.
In every \CMCG configuration, the last layer is always made of two concatenated fully connected layers. Each one has as many parameters ($\gamma_d$'s or $\beta_d$'s) as needed. With this distinction we can control $\gamma_d$ and $\beta_d$ individually (different activations).

\subsubsection{Fully-connected embedding (F)}

\Gls{fully} embedding (F) is formed of a first dense layer of 16 neurons and two fully connected blocks (dense layer, 50\% \gls{dropout} and \gls{batch_norm}) with 64 and 256 neurons for \textit{\gls{film} simple} and 256 and 1024 for \textit{\gls{film} complex}. All the neurons have ReLu activations. The last fully connected block is connected with the final \CMCG layer i.e., the two fully connected ones ($\gamma_d$ and $\beta_d$).
We call \textit{C-U-Net-SiF} and \textit{C-U-Net-CoF}, respectively, the \gls{cunet} that uses each one of these architectures.

\subsubsection{CNN embedding (C)}

Similarly to the previous one and inspired by~\citep{Shen_2017}, \gls{cnn} embedding (C) consists in a 1D convolution with $lenght(\overline{z})$ filters followed by two convolution blocks (1D convolution with also $lenght(\overline{z})$ filters, 50\% \gls{dropout} and \gls{batch_norm}). The first two convolutions have `same' padding and the last one, `valid'. Activations are also ReLu. The number of filters are 16, 32 and 64 for the \textit{\gls{film} simple} version and 32, 64, 256 for the \textit{\gls{film} complex} one.
Again, the last CNN block is connected with the two fully connected ones.
The \gls{cunet} that uses these architectures are called \textit{C-U-Net-SiC} and \textit{C-U-Net-CoC}.
This embedding is specially designed for dealing with several instruments because it seems more appropriated to find common $\gamma_d$ and $\beta_d$ values for similar instruments.

\begin{table}[t]
  \centering
  \caption[Comparsion of the number of parameters]{Parameters number in millions. With separated \glspl{unet}, each task needs a model with 10M params. \glspl{cunet} are multi-task and the number of params remains constant. SiF = Simple fully connected embedding,  CoF =  Complex fully connected embedding, SiC= Simple CNN embedding, CoC= Complex CNN embedding}
  \label{table:cunet-overview}
  \begin{tabular}{ c | c | c | c | c | c  }
    \hline
        MODEL      &  Non-conditioned      &  SiF     & CoF    & SiC    & CoC    \\
    \hline
        PARAM      &   39,30 M   (4 tasks x 9,825 M)   &   9,85 M     & 12 M   & 9,84 M    & 10,42 M         \\

    \hline
  \end{tabular}
\end{table}

The various control mechanisms only introduce a reduced number of parameters to the standard \gls{unet} architecture remaining constant regardless of the instruments to separate (see Table \ref{table:cunet-overview}).
Additionally, they make direct use of the commonalities between instruments.

\section{Evaluation}

\subsection{Evaluation protocol}
\label{sec:miscellaneous}

Our objective is to prove that conditioned learning via \gls{film} (generic model+control) allows us to transform the \gls{unet} into a multi-task system without losing performances with respect to each separated model.
In Section \ref{sec:miscellaneous} we review our experiment design aspects and we detail the experiment to validate the multi-task capability of the \gls{unet} in Section \ref{sec:exp-multitask}.

\subsubsection{Dataset}
We use the Musdb18 dataset~\citep{musdb_2018}.
It consists of 150 tracks with a defined split of 100 tracks for training and 50 for testing.
In Musdb18, mixtures are divided into four different sources: \textbf{Vocals}, \textbf{Bass}, \textbf{Drums} and \textbf{Rest} of instruments. The 'Rest' task mixes every instrument that it is not vocal, bass or drums.
From the 100 tracks, we use 95 (randomly assigned) for training, and the remaining 5 for the validation set, which is used for early stopping.
The performance is evaluated on the 50 test tracks.
Consequently, the \gls{unet} is trained for four tasks (one task per instrument) and $\overline{z}$ has four elements.

\subsubsection{Audio Reconstruction method}
The system works exclusively on the magnitude of audio spectrograms.
The output magnitude is obtained by applying the mask to the mixture magnitude.
As in~\citep{Jansson_2017}, the final predicted source (the isolated audio signal) is reconstructed concatenating temporally (without overlap) the output magnitude spectra and using the original mix phase unaltered.
We compute the predicted accompaniment subtracting the predicted isolated signal to the original mixture.
Despite there are better phase reconstruction techniques such as~\citep{Mayer_2017}, errors due to this step are common to both methods (\gls{unet} and C-U-Net) and do not affect our main goal: to validate \gls{conditioning} for \gls{source_separation}.

\subsubsection{Evaluation metrics}
We evaluate the performances of the separation using the mir\_evaltoolbox~\citep{Raffel_2014}.
We compute three metrics: Source-to-Interference Ratios (SIR), Source-to-Artifact Ratios (SAR) and Source-to-Distortion Ratios (SDR)~\citep{Vincent_2006}.
These metrics compare each estimated source $\hat{s}_j$ to its given true source $s_j$. They first decompose $\hat{s}_j$ into four components: $\hat{s}_j = s_{target} + e_{interf} + e_{noise} + e_{artif}$ where $s_{target}$ is the part of estimated source that comes from the orginal target source and the interference (error coming from other unwanted sources), distortion (sensor noise, spatial or filtering distortion) and artifacts (other causes like forbidden distortions of the sources and/or `burbling' artifacts) error terms produced by the process~\citep{Vincent_2006}.
The metrics compute different energy ratios to evaluate the relative amount of each of these four terms.
In practice, SIR measures the interference from other sources, SAR the algorithmic artifacts introduce in the process and SDR resumes the overall performance:

\begin{align*}
  SIR &:= 10 \cdot log_{10} \frac{|| s_{target} ||^2}{|| e_{interf} || ^ 2}\\
  SAR &:= 10 \cdot log_{10} \frac{|| s_{target} + e_{noise} + e_{artif} ||^2}{|| e_{artif} || ^ 2} \\
  SDR &:= 10 \cdot log_{10} \frac{|| s_{target} ||^2}{|| e_{interf} + e_{noise} + e_{artif} || ^ 2} \\
\end{align*}

To compute the three measures we need the define the 'accompaniment' i.e. the mixture part that does not correspond to the target source. Each task has a different accompaniment e.g., for the drums the accompaniment is rest+vocals+bass. We create the accompaniments by adding the audio signal of the needed sources. These metrics are measured in $dB$.

\subsubsection{Activation function for $\gamma_d$ and $\beta_d$}
One of the most important design choices is the activation function for $\gamma_d$ and $\beta_d$.
We tested all the possible combinations of three activation functions (linear, sigmoid and tanh) in the \textit{C-U-Net-SiF} configuration.
As in~\citep{Perez_2017}, the \gls{unet} works better when $\gamma_d$ and $\beta_d$ are linear.
Hence, our $\gamma_d$'s and $\beta_d$'s have always linear activations.

\subsubsection{Training flexibility} The conditioning mechanism gives the flexibility to have continuous values in the input $\overline{z} \in [0,1]$, which weights the target output $Y$ by the same value.
We call this training method \textbf{progressive}.
In practice, while training, we randomly weight $\overline{z}$ and $Y$ by a value between 0 and 1 every 5 instances.
This is a way of dealing with ablations by making the control mechanism robust to noise.
Additionally, it is a natural way of doing \gls{data_aug}.
As shown in Table \ref{table:cunet-multitask}, this training procedure \textit{(p)} improves the models.
Thus, we adopt it in our training.
Moreover, preliminary results (not reported) show that the \gls{unet} can be trained for complex tasks like bass+drums or voice+drums.
These complex tasks could benefit from `in between-class learning' method~\citep{Tokozume_2017} where $\overline{z}$ will have different intermediate instrument combinations.

\subsection{Multitask experiment}
\label{sec:exp-multitask}

We want to prove that a given \gls{unet} can isolate the \textbf{Vocals}, \textbf{Drums}, \textbf{Bass}, and \textbf{Rest} as good as four separated \gls{unet} trained specifically for each task\footnote{with the same learning rate and optimizer as the \glspl{unet}.}
We call this set of separated \gls{unet}, \textit{Fix-U-Nets}.
Each \glspl{unet} version (one model) is compared with the Fix-U-Nets set (four models).
We review the results at Table \ref{table:cunet-multitask} and show a comparison per task in Table \ref{table:cunet-multitask-details}.

\begin{table}[t!]
  \centering
  \caption[Overall perfomances]{Overall performance (mean $\pm$ std) for the avegare value over the 4 tasks.
   \textit{Si}= simple FiLM, \textit{Co}= complex FiLM,  \textit{F}= Fully-embedded and \textit{C}= CNN-embedded,  \textit{p}= progressive training or \textit{np}= non-progressive training.}
   \label{table:cunet-multitask}
   \begin{tabular}{| c  | c|c|c |}
    \hline
     \multirow{2}{*}{MODEL}  & \multicolumn{3}{c|}{Total} \\ \cline{2-4}
      &  SIR   &  SAR  &  SDR     \\
    \hline

    \textit{Fix-U-Net(x4)}             & 7.31 $\pm$ 4.04 & 5.70 $\pm$ 3.10 & 2.36 $\pm$ 3.96 \\
    \hline
    \hline
    \textit{C-U-Net-SiC-np}            & 7.35 $\pm$ 4.13 & 5.74 $\pm$ 3.18 & 2.34 $\pm$ 3.69 \\
    \textit{C-U-Net-SiC-p}             & \textbf{8.00} $\pm$ 4.37 & \textbf{5.74} $\pm$ 3.63 & \textbf{2.54} $\pm$ 4.07  \\
    \hline
    \textit{C-U-Net-CoC-np}            & 7.27 $\pm$ 4.24 & 5.60 $\pm$ 2.88 & 2.36 $\pm$ 3.81  \\
    \textit{C-U-Net-CoC-p}             & 7.49 $\pm$ 4.54 & 5.67 $\pm$ 3.03 & 2.42 $\pm$ 4.21  \\
    \hline
    \textit{C-U-Net-SiF-np}            & 7.23 $\pm$ 3.97 & 5.59 $\pm$ 3.01 & 2.22 $\pm$ 3.67 \\
    \textit{C-U-Net-SiF-p}             & 7.64 $\pm$ 4.05 & 5.73 $\pm$ 2.88 & 2.46 $\pm$ 3.88 \\
    \hline
    \textit{C-U-Net-CoF-np}            & 7.42 $\pm$ 4.20 & 5.59 $\pm$ 3.07 & 2.32 $\pm$ 3.85 \\
    \textit{C-U-Net-CoF-p}             & 7.52 $\pm$ 4.04 & 5.71 $\pm$ 2.99 & 2.42 $\pm$ 3.97 \\

    \hline
  \end{tabular}
\end{table}

Results in Table \ref{table:cunet-multitask} for all 4 instruments highlight that \textit{\gls{film} simple} layers work as good as the complex ones.
This is quite interesting because it means that applying 6 affine transformations with just 12 scalars (6 $\gamma_{i}$ and 6 $\beta_{i}$) at a precise point allows the \gls{cunet} to do several \glspl{source_separation}.
With \textit{\gls{film} complex} layers it is intuitive to think that treating each feature map individually let the \gls{cunet} learn several deep representations in the encoder.
However, we have no intuitive explanation for \textit{\gls{film} simple} layers.
We did the Tukey test with no significant differences between the \textit{Fix-U-Nets} and the \textit{C-U-Nets} for any task and metric.
Another remark is that the four \glspl{cunet} benefit from the \textit{progressive} training.
Nevertheless, it impacts more the simple layers than in the complex ones.
We think that the restriction of the former (fewer parameters) helps them to find an optimal state.

\begin{figure*}[t!]
  \centerline{
    \includegraphics[width=\textwidth]{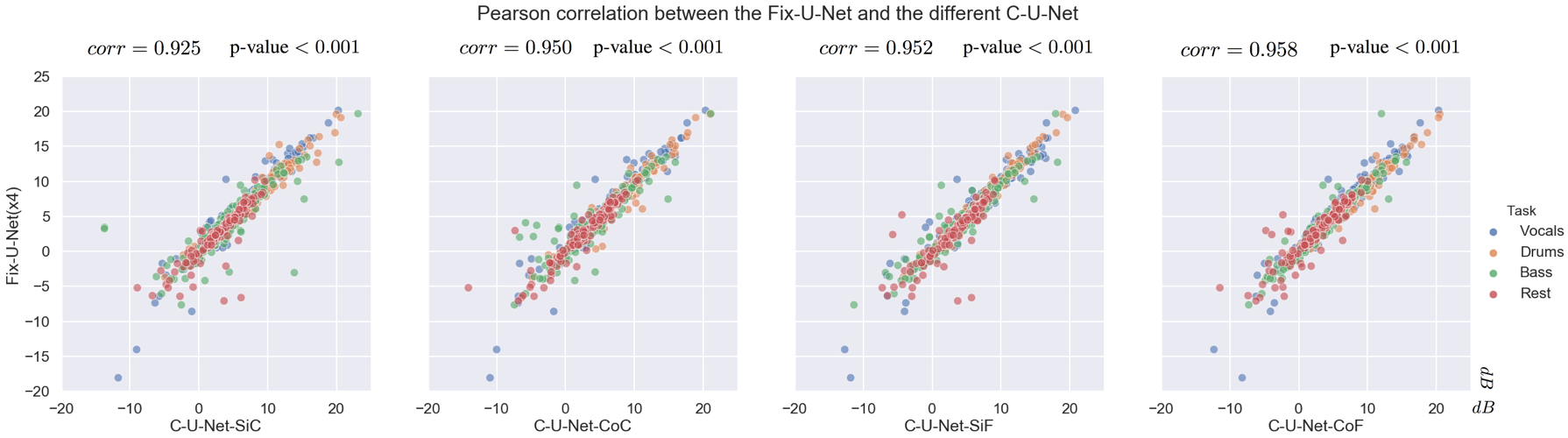}}
  \caption[Correlate performace of the different models]{
  Each graph correlates the performance of two models. On top of it, we show the correlation and p-value. The 'y' axis represents the fixed version (the four separated \glspl{unet}) and the 'x' one a different \gls{cunet} version (with progressive train).
  The coordinates of each dots correspond to the models' performance i.e., 'y' position for the Fix-U-Net performance and 'x' for \gls{cunet}.
  There are three dots per song, one per metric (SIR, SAR, and SDR) which does a total of 600 (50 songs x 3 metrics x 4 instruments).
  The dot alignment in the diagonal implies a strong correlation between models: if one works well, so do the others, and vice versa.
  Each color highlights the points of each \gls{source_separation} task.}
  \label{fig:corr}
\end{figure*}

However, these results do not prove nor discard the significant similarity between systems.
For demonstrating that, we have carried out a Pearson correlation experiment.
The results are detailed in Figure \ref{fig:corr}.
The Pearson coefficient measures the linear relationship between two sets of results (+1 implies an exact linear relationship).
It also computes the p-value that indicates the probability that uncorrelated systems have produced them.
Our distinct \gls{cunet} configurations have a global $corr > .9$ and $\textrm{p-value} < 0.001$.
Which means that there is always more than $90\%$ correlation between the performance of the four separated \textit{U-Nets} and the (various) conditional version(s).
Additionally, there is almost no probability that a \gls{cunet} version is not correlated with the separated models.
We have also computed the Pearson coefficient and p-value per task and per metric with the same results.
In Figure \ref{fig:corr} shows a strong correlation between the \textit{Fix-U-Net} results and the distinct \textit{\glspl{cunet}} (independently of the task or metric). Thus, if one works well, the others too and vice versa.

\begin{table}[t!]
  \centering
  \caption[Detailed perfomrances]{Task comparison between the \textit{separated U-Nets} and the \textit{C-U-Net-CoF}. Results indicate that they perform similarly for all the tasks. We also add the multi-instrument Wave-U-Net (M) results (median in parenthesis) and when possible the separated version (D). For vocals isolation the Wave-U-Net-M performs worse than the Wave-U-Net-D (from a mean 0.55 and median 4.58 to -2.10 and 3.0).}
  \label{table:cunet-multitask-details}
  \begin{tabular}{| c | c  || c|c|c |}
    \hline
     & Model                            &   SIR   &  SAR  &  SDR     \\
    \hline

    \hline
    \parbox[t]{2mm}{\multirow{4}{*}{\rotatebox[origin=c]{90}{Vocals}}}
    & \textit{Fix-U-Net(x4)}             & 10.70 $\pm$ 4.26  & 5.39 $\pm$ 3.58 & 3.52 $\pm$ 4.88  (4.72) \\
    & \textit{C-U-Net-CoF}               & 10.76 $\pm$ 4.39  & 5.32 $\pm$ 3.27 & 3.50 $\pm$ 4.37  (4.65) \\
    & \textit{Wave-U-Net-D}              & -                 &  -              & 0.55 $\pm$ 13.67 (4.58) \\
    & \textit{Wave-U-Net-M}              & -                 &  -              & -2.10 $\pm$ 15.41 (3.0) \\
    \hline
    \parbox[t]{2mm}{\multirow{3}{*}{\rotatebox[origin=c]{90}{Drums}}}
    & \textit{Fix-U-Net(x4)}             & 10.08 $\pm$ 4.28  & 6.42 $\pm$ 3.28 & 4.28 $\pm$ 3.65 (4.13) \\
    & \textit{C-U-Net-CoF}               & 10.03 $\pm$ 4.34  & 6.80 $\pm$ 3.25 & 4.30 $\pm$ 3.81 (4.38) \\
    & \textit{Wave-U-Net-M}              & -                 &  -              & 2.88 $\pm$ 7.68 (4.15) \\
    \hline
    \parbox[t]{2mm}{\multirow{3}{*}{\rotatebox[origin=c]{90}{Bass}}}
    & \textit{Fix-U-Net(x4)}             & 4.64 $\pm$ 4.76   & 6.51 $\pm$ 2.68 & 1.46 $\pm$ 4.31 (2.48)  \\
    & \textit{C-U-Net-CoF}               & 5.30 $\pm$ 4.73   & 6.29 $\pm$ 2.39 & 1.65 $\pm$ 4.07 (2.60)  \\
    & \textit{Wave-U-Net-M}              & -                 &  -              & -0.30 $\pm$ 13.50 (2.91) \\
    \hline
    \parbox[t]{2mm}{\multirow{3}{*}{\rotatebox[origin=c]{90}{Rest}}}
    & \textit{Fix-U-Net(x4) }            & 3.83 $\pm$ 2.84   & 4.47 $\pm$ 2.85 & 0.19 $\pm$ 3.00 (0.97) \\
    & \textit{C-U-Net-CoF}               & 4.00 $\pm$ 2.70   & 4.37 $\pm$ 3.06 & 0.24 $\pm$ 3.64 (1.71) \\
    & \textit{Wave-U-Net-M}              & -                 &  -              & 1.68 $\pm$ 6.14 (2.03) \\
    \hline
  \end{tabular}
\end{table}

In Table \ref{table:cunet-multitask-details} we detail the results per task and metric for the \textit{Fix-U-Net} and the \textit{C-U-Net-CoF} which is not the best \gls{cunet} but the one with the highest correlation with the separated ones.
There we can see how their performances are almost identical.
Nevertheless, our vocal isolation (in any case) is not as good as the one reported in~\citep{Jansson_2017}, we believe that this is mainly due to the lack of data.
These results can only be compared with the Wave-U-Net~\citep{Stoller_2018}.
Although they report the results (only the SDR) for the four tasks in the multi-instrument version (multiple outputs layers) they only have a separated version for vocals.
For vocal separation, the performance of the multi-instrument version decreases more than 2.5 dB in mean, 1.5 dB in the median and the std increase in almost 2 dB.
Furthermore, the \gls{cunet} performs better than the multi-instrument in three out of four tasks (vocals, bass, and drums)\footnote{Our experiment conditions are different in training data size (95 Vs 75) and in sampling rate (8192 Hz Vs 22050 Hz) than Wave-U-Net.}.
For the 'Rest' task, the multi-instrument Wave-U-Net outperforms our \glspl{cunet}.
This is normal because the separated \gls{unet} has already problems with this class and the \glspl{cunet} inherits the same issues.
We believe that they come from the vague definition of this class with many different instruments combinations at once.

This proves that the various \glspl{cunet} behave in the same way as the separated \glspl{unet} for each task and metric.
It also demonstrates that conditioned learning via \gls{film} is robust to diverse control mechanisms/condition generators and \gls{film} layers.
Moreover, it does not introduce any limitations which are due to other factors.

\section{Conclusions and future work}
In this chapter, we have applied \gls{conditioning} to the problem of instrument \glspl{source_separation} by adding a control mechanism to the \gls{unet} architecture. The \glspl{cunet} can do several \gls{source_separation} tasks without losing performance as it does not introduce any limitation and makes use of the commonalities of the distinct instruments. It has a fixed number of parameters (much lower than the separated model approach) independently of the number of instruments to separate.
We believe that \gls{conditioning} via \gls{film} will benefit many \gls{mir} problems because it defines a transparent and direct way of inserting external data to modify the behavior of a network.
Our key contributions are:
\begin{enumerate}
  \item the \gls{cunet}, a joint model that changes its behavior depending on external data and performs for any task as good as a separated model trained for it. \gls{cunet} has a fixed number of parameters no matter the number of output sources.
  \item The \gls{cunet} proves that \gls{conditioning} via \gls{film} layers is an efficient way of inserting external information to control deep neural network architectures.
  \item The \textit{\gls{film} simple}, a new conditioning layer that works as good as the original one but requires less $\gamma_d$'s and $\beta_d$'s.
\end{enumerate}


\Gls{conditioning} faces problems providing a generic model and a control mechanism.
This gives flexibility to the systems but introduces new challenges.
We plan to extend the \gls{cunet} to more instruments to find its limitations and to explore the performance for complex tasks i.e., separating several instruments combinations (e.g., vocals+drums).
We also intend to integrate it in other architectures such as Wave-U-Net and \gls{data_aug} techniques~\citep{Cohen_2019}.
Currently, the multitrack version of \gls{dali} divides the mixture into `vocals' + `accompaniment'. However, the rest of the sources are also presented but they not prepared yet to use (see Chapter~\ref{sec:multitracks}). We plan to add these sources to work with more instruments.
Likewise, we are exploring ways of adding new conditions (namely new instrument isolation) to a trained \gls{cunet} and how to separate the joint training, with the goal of creating a generic model than can be easily adapted to several control mechanisms.


\graphicspath{{figs/}{vunet/figs/}}

\chapter{Vocal Source Separation}
\label{sec:vunet}


Informed source separation has recently gained renewed interest with the introduction of neural networks and the availability of large multitrack datasets containing both the mixture and the separated sources.
These approaches use prior information about the target source to improve separation.
Historically, \gls{mir} researchers have focused primarily on score-informed source separation, but more recent approaches explore lyrics-informed source separation.
However, because of the lack of multitrack datasets with time-aligned lyrics, models use weak conditioning with the non-aligned lyrics.
In this chapter, we present a \gls{multimodal} multitrack dataset with lyrics aligned in time at the phoneme level as well as explore strong conditioning using the aligned phonemes.
Our model explores the \gls{cunet} architecture and takes as input both the magnitude spectrogram of a musical mixture and a matrix with aligned phoneme information.
The phoneme matrix is embedded to obtain the parameters that control \gls{film} layers.
These layers condition the \gls{cunet} feature maps to adapt the separation process to the presence of different phones via affine transformations
We show that phoneme conditioning can be successfully applied to improve singing voice source separation.

\section{Introduction}

Music \gls{source_separation} aims to isolate the different instruments that appear in an audio mixture (a mixed music track), reversing the mixing process.
Informed-\gls{source_separation} uses prior information about the target source to improve separation.
Researchers have shown that deep neural architectures can be effectively adapted to this paradigm~\citep{Kinoshita_2015, miron_2017}.
Music \gls{source_separation} is a particularly challenging task.
Instruments are usually correlated in time and frequency with many different harmonic instruments overlapping at several possitions and with dynamics variations.
Without additional knowledge about the sources the separation is often infeasible.
To address this issue, \gls{mir} researchers have integrated into the \gls{source_separation} process prior knowledge about the different instruments presented in a mixture, or musical scores that indicate where sounds appear.
This prior knowledge improves the performances~\citep{Slizovskaia_2020, Ewert_2014, miron_2017}. Recently, conditioning learning has shown that neural networks architectures can be effectively controlled for performing different music source isolation tasks~\citep{meseguerbrocal_2019, Tzinis_2019,Slizovskaia_2019, Seetharaman_2019, Samuel_2020, Schulze_2019}

Various \gls{multimodal} context information can be used.
Although \gls{mir} researchers have historically focused on score-informed \gls{source_separation} to guide the separation process, lyrics-informed \gls{source_separation} has become an increasing research area~\citep{Chandna_2020, Schulze_2019}.
Singing voice is one of the most important elements in a musical piece~\citep{demetriou_2018}.
Singing voice tasks (e.g. lyric or note transcription) are particularly challenging given its variety of timbre and expressive versatility.
Fortunately, recent data-driven machine learning techniques have boosted the quality and inspired many recent discovers~\citep{gomez_2018, humphrey_2018}.
Singing voice works as a musical instrument and at the same time conveys a semantic meaning through the use of language~\citep{humphrey_2018}.
The relationship between sound and meaning is defined by a finite phonetic and semantic representations~\citep{goldsmith_1976, ladd_2008}.
Singing in popular music usually has a specific sound based on phonemes, which distinguishes it from the other musical instruments.
This motivates researchers to use prior knowledge such as a text transcript of the utterance or linguistic features to improve the singing voice \gls{source_separation}~\citep{Chandna_2020, Schulze_2019}.
However, the lack of multitrack datasets with time-aligned lyrics has limited them to develop their ideas and only weak conditioning scenarios have been studied i.e. using the context information without explicitly informing where it occurs in the signal.
Time-aligned lyrics provide abstract and high-level information about the phonetic characteristics of the singing signal.
This prior knowledge can facilitate the separation and be beneficial to the final isolation.


Looking for combining the power of data-driven models with the adaptability of informed approaches, we propose a multitrack dataset with time-aligned lyrics.
Then, we explore how we can use strong conditioning where the content information about the lyrics is available frame-wise to improve vocal sources separation (see part~\ref{sec:multitracks}).
We investigate strong and weak conditioning using the aligned phonemes via \gls{film} layer in \gls{unet} based architecture (see Chatper~\ref{sec:cunet}).
We show that phoneme conditioning can be successfully applied to improve standard singing voice \gls{source_separation} and that the simplest strong conditioning outperforms any other scenario.

\section{Formalization}\label{sec:vunet_formalization}
We use the multimodal information as context to guide and improve the separation.
We formalize our problem satisfying certain properties summarized as~\citep{Bengio_2013} (see part~\ref{sec:c_multimodal}):

\begin{description}[.15cm]
  \item[How is the multimodal model constructed?] We divide the model into two distinct parts~\citep{Dumoulin_2018}: a \textit{generic} network that carries on the main computation and a \textit{control mechanism} that conditions the computation regarding context information and adds additional flexibility. The \textit{conditioning} itself is performed using Feature-wise Linear Modulation (FiLM) layers~\citep{Perez_2017}. FiLM can effectively modulate a generic source separation model by some external information, controlling a single model to perform different instrument source separations~\citep{meseguerbrocal_2019, Slizovskaia_2020}.
  With this strategy, we can explore the \textit{control} and \textit{conditioning} parts regardless of the \textit{generic} network used.
  \item[When is the context information used?] We can see this as at which depth in the \textit{generic} network we insert the context information, and when it affects the computation i.e. weak (or strong) conditioning without (or with) explicitly informing where it occurs in the signal.
  \item[What context information?] We explore here prior information about the phonetic evolution of the singing voice, aligned in time with the audio. To this end, we introduce a novel multitrack dataset with lyrics aligned in time.
\end{description}

\section{Related work}

Informed source separation uses context information about the sources to improve the separation quality, introducing in models additional flexibility to adapt to observed signals.
Researchers have explored different approaches for integrating different prior knowledge in the separation~\citep{liutkus_2013}.
In this section we review previous works related to informed source separation in speech and then in music, where we review both score-informed and text-informed.

\subsection{In speech}

\cite{LeMagoarou_2015} present one of the first text-informed source separation approaches. They propose a speech example-based paradigm where the text information generates (via synthesizer or human) a speech example aligned with the original mixture using \gls{dtw}. The example guides the separation exploiting linguistic similarities between the target speech and the example speech signal.
\cite{Kinoshita_2015} use text features derived from forced-aligned phonemes with noisy speech together with the audio features to train a deep neural network that predicts enhanced spectrum parameters. The authors show that distortion in the separation is smaller when using text.
Automatic Speech Recognition (ASR) can directly identify phonemes at each frame from the mixture without using text-transcript~\citep{chazan_2016, wang_2016}. Then, pre-trained phoneme-specific networks perform the separation.
\cite{biadsy_2019} uses an end-to-end sequence-to-sequence encoder/decoder architecture with an additional ASR decoder to predict the (grapheme or phoneme) transcript, which conditions the encoder latent representation.
Although its primary application is voice conversion, it is useful in a source extraction scenario. However, it requires a dataset of parallel paired input-output speech utterances.
\cite{Schulze_2020} describe a multitask model that jointly perform text-informed speech separation and phoneme alignment.
Their model uses a bidirectional recurrent neural network (BRNN) where the context information is extracted from the text via an attention mechanism.
The context information is refined with an additional loss for solving phoneme alignment task.
They show that jointly solving both tasks leads to mutual benefits.
\cite{Takahashi_2020} explicitly incorporates the phonetics using transfer learning. They first train an ASR encoder/decoder model on a large clean speech corpus. They adapt the intermediate features obtained from the encoder into a suitable representation for voice separation using a domain translation network.
These features condition the separation of a net based on Tasnet~\citep{Luo_2019}.

\subsection{In music}
Most of the recent data-driven music source separation methods use weak conditioning with prior knowledge about the different instruments presented in a mixture~\citep{Slizovskaia_2020, meseguerbrocal_2019, Slizovskaia_2019, Seetharaman_2019, Samuel_2020}.
Strong conditioning has been primarily used in score-informed source separation. In this section, we review works related to this topic as well as novel approaches that explore lyrics-informed source separation.

\subsection{Score-informed music source separation}

Scores provide prior knowledge for source separation in various ways. For each instrument (source), it defines which notes are played at which time, which can be linked to audio frames.
This information can be used to guide the estimation of the harmonics of the sound source at each frame~\citep{Ewert_2014, miron_2017}.
Pioneer approaches rely on non-negative matrix factorization (NMF).
Note activities in the score can constrain the NMF-based model by setting to zero the harmonic values outside the frequency range of each nominal musical note~\citep{ewert_2012}.
Authors introduce a multi-excitation per instrument (MEI) source-filter NMF model that uses pre-learned timbre models for each instrument~\citep{duan_2011, rodriguez_2015}.
They also learn the NMF components used for the isolation on synthetic signals with temporal and harmonic constraints generated from the score~\citep{fritsch_2013}.
These methods assume that the audio is synchronized with the score and use different alignment techniques to achieve this.
Nevertheless, alignment methods introduce errors.
Local misalignments influence the quality of the separation~\citep{duan_2011, miron_2015}.
This is compensated by allowing a tolerance window around note onsets and offsets~\citep{ewert_2012, fritsch_2013} or with context-specific methods to refine the alignment~\citep{miron_2016}.
Current approaches use deep neural network architectures and use the scores to filter the spectrograms, generating masks for each source~\citep{miron_2017}. The score-filtered spectrum is used as input to an encoder-decoder Convolutional  Neural  Network (CNN) architecture similar to \citep{Chandna_2017}.
\cite{Ewert_2017} propose an unsupervised method where scores guide the representation learning to induce structure in the separation that adds class activity penalties and structured dropout extensions to the encoder-decoder architecture.
Class activity penalties capture the uncertainty about the target label value and structured dropout uses labels to enforce a specific structure, canceling activity related to unwanted note.

\subsection{Text-informed music source separation}

Due to the importance of singing voice in a musical piece~\citep{demetriou_2018}, it is one of the most useful source to separate in a music track.
Researchers have integrated the vocal activity information to constrain a robust principal component analysis (RPCA) method, applying a vocal/non-vocal mask or ideal time-frequency binary mask~\citep{Chan_2015}.
\cite{Schulze_2019}~propose a bidirectional recurrent neural networks (BRNN) method that includes context information extracted from the text via attention mechanism.
The method takes as input a whole audio track and its associated text information and learn alignment between mixture and context information that enhance the separation.
Recently, \citep{Chandna_2020} extract a representation of the linguistic content related to cognitively relevant features such as phonemes (but they do not explicitly predict the phonemes) in the mixture. The linguistic content guides the synthesis of the vocals.

\section{Methodology}\label{sec:vunet_method}

\begin{figure}[t]
  \centerline{
    \includegraphics[width=.7\columnwidth]{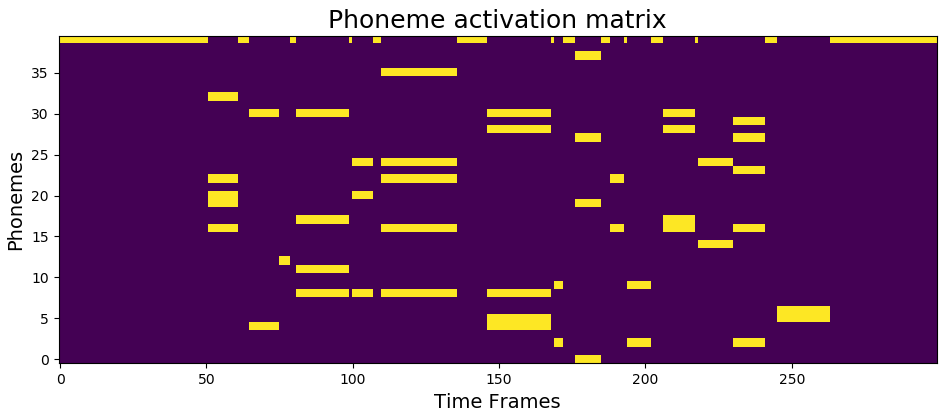}}
  \caption[Binary phoneme activation matrix]{Binary phoneme activation matrix. Note how words are represented as a bag of simultaneous phonemes.}
  \label{fig:phoneme}
\end{figure}

Our method adapts the \gls{cunet} architecture~\citep{meseguerbrocal_2019} to the singing voice separation task, exploring how to use the prior knowledge defined by the phonemes to improve the vocal separation.
Let $X \in \mathbb{R}^{T \times M}$  be the magnitude of the Short-Time Fourier Transform (STFT) with $M=512$ frequency bands and $T$ time frames. We compute the STFT on an audio signal down-sampled at 8192~Hz using a window size of 1024 samples and a hop size of 768 samples. Let $Z \in \mathbb{R}^{T \times P}$ be the aligned phoneme activation matrix with $P=40$ phoneme types as defined in the Carnegie MellonPronouncing Dictionary (CMUdict)\footnote{url{https://github.com/cmusphinx/cmudict}} and $T$ the same time frames as in $X$.
For computing this matrix we use the phoneme information $A_{\mathit{phoneme}} = (a_{k, \mathit{phoneme}})_{k=1}^{K_\mathit{phoneme}}$ (see Chapter~\ref{sec:dali_creation}) for the multitrack version of \gls{dali} (see Chapter~\ref{sec:multitracks}). After selecting the desired time resolution, we can derive a time frame based phoneme context activation matrix $Z$, which is a binary matrix that indicates the phoneme activation over time.
Note that the corresponding phonemes are active during all segments defined by $(t_\mathit{min}, t_\mathit{max})_k$
We add an extra row with the 'non-phoneme' activation with $1$ at time frames with no phoneme activation and $0$ otherwise. Figure~\ref{fig:phoneme} illustrates the final activation matrix.

Our model consider the music track as a sucession of patch segments of duration $N$.
It takes as inputs two submatrix $x \in \mathbb{R}^{N \times M}$ and $z \in \mathbb{R}^{N \times P}$ of $N=128$ frames (11 seconds) derived from $X$ and $Z$.
The \gls{cunet} model has two components: a \textbf{conditioned network} that processes $x$ and a \textbf{control mechanism} that conditions the computation with respect to $z$.
We denote by $x_d \in \mathbb{R}^{W \times H \times C}$ the intermediate features of the \textbf{conditioned network}, at a particular depth $d$ in the architecture.
$W$ and $H$ represent the `time' and `frequency' dimension and $C$ the number of feature channels (or feature maps).
A \textit{FiLM} layer conditions the network computation by applying an affine transformation to $x_d$:

\begin{equation}
    \label{eq_film}
    \mathit{FiLM}(x_d) = \gamma_{d}(z) \cdot x_d + \beta_{d}(z)
\end{equation}

\noindent where $\odot$ denotes the element-wise multiplication and $\gamma_{d}(z)$ and $\beta_{d}(z)$ are learnable parameters with respect to the input context $z$.
A \textit{FiLM} layer can be inserted at any depth of the original model and its output has the same dimension as the  $x_d$ input, i.e. $\in \mathbb{R}^{W \times H \times C}$.
To perform this, $\gamma_{d}(z)$ and $\beta_{d}(z)$ must have the same dimensionaly as $x_d$, i.e. $\in \mathbb{R}^{W \times H \times C}$.
However, we can define them omitting some dimensions. This results in a non-matching dimensionality with $x_d$, solved by broadcasting (repeating) the existing information to the missing dimensions.

As in~\citep{Jansson_2017, Stoller_2018, meseguerbrocal_2019}, we use a standard \gls{unet} as \textbf{conditioned network}.
This model follows an encoder-decoder mirror architecture based on CNN blocks with skip connections between layers at the same hierarchical level in the encoder and decoder.
Each convolutional block in the encoder halves the size of the input and doubles the number of channels.
The decoder is made of a stack of transposed convolutional operation, its output has the same size as the input of the encoder.
Following the original \gls{cunet} architecture, we insert the \textit{FiLM} layers at each encoding block after the batch normalization and before the Leaky ReLU~\citep{meseguerbrocal_2019}.

\newpage

\subsection{Control mechanism for weak conditioning}
Weak conditioning refers to the cases where
\begin{itemize}
  \item $\gamma_{d}(z)$ and $\beta_{d}(z)$  $\in \mathbb{R}^1$: they are scalar parameters applied independently of the times $W$, the frequencies $H$ and the channel $C$ dimensions. They depend only on the depth $d$ of the layer within the network~\citep{meseguerbrocal_2019}.
  \item $\gamma_{d}(z)$ and $\beta_{d}(z)$  $\in \mathbb{R}^C$: this is the orginal configuration proposed by~\citep{Perez_2017} with different parameters for each channel $c \in {1, ..., C}$.
\end{itemize}

We call them \textit{FiLM simple} ($\mathit{W}_\mathit{si}$) and \textit{FiLM complex} ($\mathit{W}_\mathit{co}$) respectively. Note how they apply the same transformation  without explicitly informing where it occurs in the signal (same value over the dimension $W$ and $H$)~\citep{meseguer2020cunet}.




Starting from the context matrix $z \in \mathbb{R}^{N \times P}$, we first apply  the autopool layer proposed by~\citep{mcfee_2018}\footnote{The auto-pool layer is a tuned soft-max pooling that automatically adapts the pooling behavior to interpolate between mean and max-pooling for each dimension} to reduce the input matrix to a time-less vector.
We then fed this vector into a dense layer and two dense blocks each composed by a dense layer, 50\% dropout and batch normalization.
For \textit{FiLM simple}, the number of units of the dense layers are 32, 64 and 128.
For \textit{FiLM simple}, they are 64, 256 and 1024.
All neurons have ReLU activations.
The output of the last block is then used to feed two parallel and independent dense layers with linear activation which outputs all the needed $\gamma_{d}(z)$ and $\beta_{d}(z)$.
While for the \textit{FiLM simple} configuration we only need $12$ $\gamma_{d}$ and $\beta_{d}$ (one $\gamma_{d}$ and $\beta_{d}$ for each of the 6 different encoding blocks) for the \textit{FiLM complex} we need $2016$ (the encoding blocks feature channel dimensions are $16$, $32$, $64$, $128$, $256$ and $512$, which adds up to $1008$).

\begin{table*}
\centering
\footnotesize
\resizebox{\textwidth}{!}{%
\begin{tabular}{r|c|c|c|c|c|c|c|c|c|c|c}
Model      & U-Net & $\mathit{W}_\mathit{si}$ & $\mathit{W}_\mathit{co}$ & $\mathit{S}_\mathit{fv}$ & $\mathit{S}_\mathit{fv*}$ & $\mathit{S}_\mathit{cs}$ & $\mathit{S}_\mathit{cs*}$ & $\mathit{S}_\mathit{fs}$ & $\mathit{S}_\mathit{fs*}$  & $\mathit{S}_\mathit{rs}$ & $\mathit{S}_\mathit{rs*}$    \\
\hline
$\theta$ &    $9.83 \cdot 10^6$     &   $+14,060$     &  $+2.35 \cdot 10^6$    &  $+1.97 \cdot 10^6$ & $+327,680$     &   $+80,640$   & $+40,960$   &  $+40,320$  & $+640$   & $+480$ & $+80$
\end{tabular}%
}
\caption[Number of parameters for the different vocal conditioning models]{Number of parameters ($\theta$) for the different configurations. We indicate the increase in the number of parameters w.r.t. the baseline \gls{unet} architecture.}
\label{table:param}
\end{table*}

\subsection{Control mechanism for strong conditioning}

\begin{figure}[t]
  \centerline{
    \includegraphics[width=.7\textwidth]{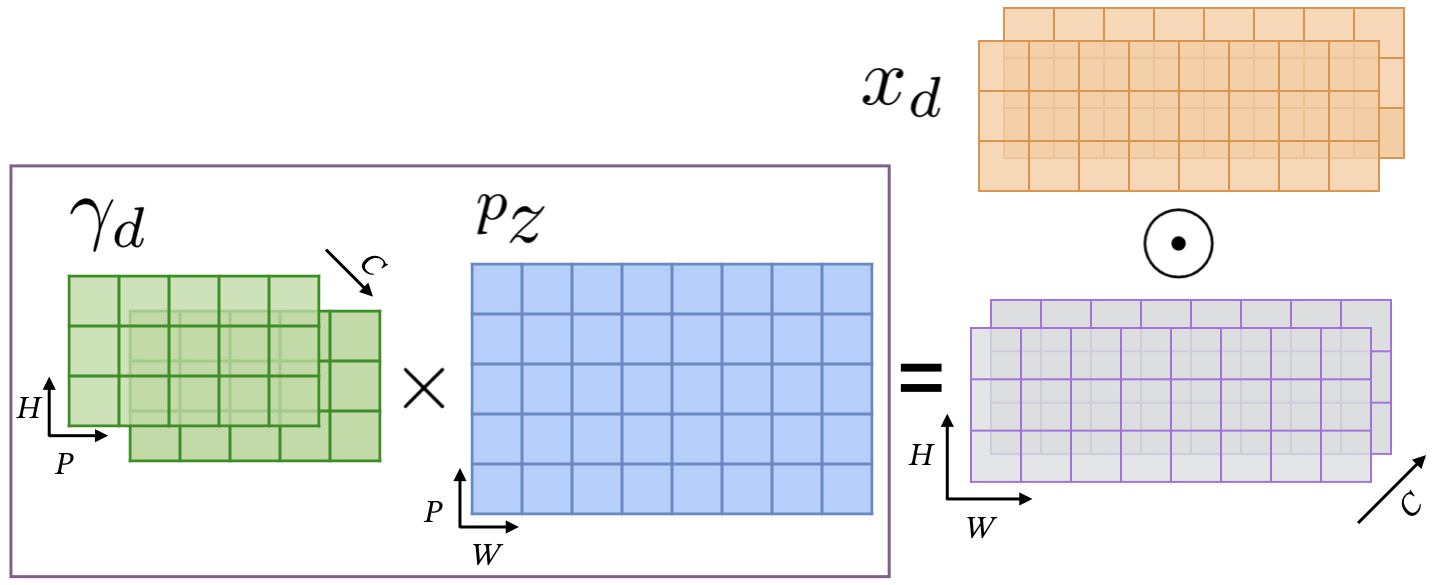}}
  \caption[Strong conditioning example]{Strong conditioning example with $(\gamma_{d} \times{} z_d)\odot x_d$. The phoneme activation $z_d$ defines how the basis tensors ($\gamma_d$) are employed for performing the conditioning on $x_d$.}
  \label{fig:strong}
\end{figure}

In this section, we extend the original \textit{FiLM} layer mechanism to adapt it to the strong conditioning scenario.
The context information represented in the input matrix $z$ describes the presence of the phonemes $p \in \{1,\ldots,P\}$ over time $n \in \{1, \ldots N\}$.
As in the popular Non-Negative Matrix factorization~\citep{lee_2001} (but without the non-negativity constraint), our idea is to represent this information as the product of tensors: an activation and two basis tensors~\citep{meseguer2020cunet}.

The \textbf{activation tensor $z_d$} indicates which phoneme occurs at which time: $z_{d} \in \mathbb{R}^{W \times P}$  where $W$ is the dimension which represents the time at the current layer $d$ (we therefore need to map the time range of $z$ to the one of the layer $d$) and $P$ the number of phonemes.

The \textbf{two basis tensors $\gamma_d$ and $\beta_d$} $\in \mathbb{R}^{H \times C \times P}$  where $H$ is the dimension which represents the frequencies at the current layer $d$, $C$ the number of input channels and $P$ the number of phonemes.
In other words, each phoneme $p$ is represented by a matrix in $\mathbb{R}^{H \times C}$.
This matrix represents the specific conditioning to apply to $x_d$ if the phoneme exists (see Figure~\ref{fig:strong}).
These matrices are learnable parameters (neurons with linear activations) but they do not depend on any particular input information (at a depth $d$ they do not depend on $x$ nor $z$), they are rather ``activated'' by $z_d$ at specific times.
As for the `weak' conditionning, we can define different versions of the tensors
\begin{itemize}
  \item the \textbf{full-version} ($\mathit{S}_\mathit{fv}$) (described so far) which has three dimensions: $\gamma_d , \beta_d \in \mathbb{R}^{H \times C \times P}$
  \item the \textbf{channel simple version} ($\mathit{S}_\mathit{cs}$): each phoneme is represented by a vector over input channels (therefore constant over frequencies): $\gamma_d , \beta_d \in \mathbb{R}^{C \times P}$
  \item the \textbf{frequency simple version} ($\mathit{S}_\mathit{fs}$): each phoneme is represented by a vector over input frequencies (therefore constant over channels): $\gamma_d , \beta_d \in \mathbb{R}^{H \times P}$
  \item the \textbf{really-simple version} ($\mathit{S}_\mathit{rs}$): each phoneme is represented as a scalar (therefore constant over frequencies and channels): $\gamma_d , \beta_d \in \mathbb{R}^{P}$
\end{itemize}

\noindent The global conditioning mechanism can then be written as
\begin{equation} \label{eq:act_mul}
  \mathit{FiLM}(x_d, z_d) = (\gamma_{d} \times{} z_d) \odot x_d + (\beta_{d} \times{} z_d)
\end{equation}

\noindent where $\odot$ is the element-wise multiplication and $\times$ the matrix multiplication.
We broadcast $\gamma_d$ and $\beta_d$ for missing dimensions and transpose them properly to perform the matrix multiplication.
We test two different configurations: inserting FiLM at each encoder block as suggested in~\citep{meseguerbrocal_2019} and inserting FiLM only at the last encoder block as proposed at~\citep{Slizovskaia_2020}. We call the former `complete' and the latter `bottleneck' and
denote it with $*$ after the model acronym.
We resume the different configurations at Table~\ref{table:param}.

\section{Experiments}

\begin{table}
\centering
\small
\begin{tabular}{c|c|c|c}
          & Train & Val & Test  \\
\hline
Threshold   &  $ .88  > \mathit{NCC} >= .7$  &  $.89 > \mathit{NCC} >= .88$     &   $.89  > \mathit{NCC} $      \\
\hline
Songs       & 357   &   30   & 101
\end{tabular}
\caption[DALI multitracks split]{\gls{dali} split according to agreement score $\mathit{NCC}$.}
\label{table:multi_split}
\end{table}

\subsection{Training}

\textbf{DATA.} As described in Table~\ref{table:multi_split}, we split the multitrack version of \gls{dali} into three sets according to the normalized agreement score $\mathit{NCC}$ (see chapter~\ref{sec:svd}).
This score provides a global indication of the global alignment correlation between the annotations and the vocal activity.
\\~\\
\noindent \textbf{DETAILS.}
We train the model using batches of $128$ spectrograms randomly drawn from the training set with $1024$ batches per epoch.
The loss function is the mean absolute error between the predicted vocals (masked input mixture) and the original vocals. We use a learning rate of $0.001$ and the reduction on plateau and early stopping callbacks evaluated on the validation set. We set the `patience' parameter to 15 and 30 respectively and a min delta variation for early stopping to $1e-5$.
Our output is a Time/Frequency mask to be applied to the magnitude of the input STFT mixture. We use the phase of the input STFT mixture to reconstruct the waveform with the inverse STFT algorithm.

For the strong conditioning, we apply a {\texttt{softmax}} on the input phoneme matrix $z$ over the phoneme dimension $P$ to constrain the outputs to sum to $1$, meaning it lies on a hyperplane, which helps in the optimization.

\subsection{Evaluation metrics}
We evaluate the performances of the separation using the mir\_evaltoolbox~\citep{Raffel_2014}.
As in chapter~\ref{sec:miscellaneous}, we compute three metrics: Source-to-Interference Ratios (SIR), Source-to-Artifact Ratios (SAR), and Source-to-Distortion Ratios (SDR)~\citep{Vincent_2006}.
In practice, SIR measures the interference from other sources, SAR the algorithmic artifacts introduce in the process and SDR resumes the overall performance.
We obtain them globally for the whole track.
However, these metrics are ill-defined for silent sources and targets.
Hence, we compute also the Predicted Energy at Silence (PES) and Energy at Predicted Silence (EPS) scores~\citep{Schulze_2019}. PES is the mean of the energy in the predictions at those frames with silent target and EPS is the opposite, the mean of the target energy of all frames with silent prediction and non-silent target.
For numerical stability, in our implementation, we add a small constant $\epsilon = 10^{-9}$ which results in a lower boundary of the metrics to be $-80$ dB~\citep{Slizovskaia_2020}.
We consider as silent segments those that have a total sum of less than $-25$ dB of the maximum absolute in the audio.
We report in Table~\ref{table:data_aug} the \textbf{median} values of these metrics over the all tracks in the \gls{dali} test set.
For SIR, SAR, and SDR larger values indicate better performance, for PES and EPS smaller values, mean better performance.

\begin{table}
\centering
\begin{tabular}{c|c|c|c|c|c}
Training & Test  & Aug & SDR & SIR & SAR \\

\hline\hline
Musdb18     & Musdb18   & False        & 4.27        & 13.17    & 5.17    \\
(90)        & (50)      & True         & 4.46        & 12.62    & 5.29    \\
\hline\hline
\multirow{4}{*}{\begin{tabular}[c]{@{}l@{}} DALI \\ (357) \end{tabular}}     & Musdb18     & False        & 4.60        & 14.03    & 5.39    \\
                          &           (50)                  & True         & 4.96        & 13.50    & 5.92    \\
\cline{2-6}
                          & DALI       & False        & 3.98        & 12.05    & 4.91    \\
                          & (101)      & True         & 4.05        & 11.40    & 5.32    \\
\end{tabular}
\caption[Data augmentation experiment]{Data augmentation experiment for the \gls{unet} architecture.}
\label{table:data_aug}
\end{table}

\subsection{Data augmentation}
To augment the data, we randomly created `fake' input mixtures every $4$ real mixtures.
In normal training, we employ the mixture as input and the vocals as a target.
However, we do not make use of the accompaniment (which is only employed during evaluation).
We can integrate it creating `fake' inputs by automatically mixing (mixing meaning simply adding) the target vocals to a random sample accompaniment from our training set.

We test this data augmentation process using the standard \gls{unet} architecture and checked that it improves the performance (see Table~\ref{table:data_aug}).
We train two models on \gls{dali} and Musdb18 dataset~\citep{musdb_2018}\footnote{We use 10 songs of the training set for the early stopping and reduction on plateau callbacks}.
This data augmentation enables models to achieve better SDR and SAR but lower SIR.

This technique does not reflect a large improvement when the model trained on \gls{dali} is tested on \gls{dali}. However, when this model is tested on Musdb18, it shows a better generalization (we have not seen any song of Musidb18 during training) than the model without data augmentation (we gain $0.36$ dB).
One possible explanation for not having a large improvement on \gls{dali} is the larger size of the test set. It also can be due to the fact that vocal targets in \gls{dali} still contain leaks such as low volume music accompaniment that come from the singer headphones.
We adopt this technique for training all the following models.

Finally, we confirmed a common belief that training with a large dataset and clean separated sources improves the separation over a small dataset~\citep{Pretet_2019}.
Both models trained on \gls{dali} (with and without augmentation) improve the results obtained with the models trained on Musdb18.

\section{Results}

\begin{table}
\centering
\begin{tabular}{c|c|c|c|c|c}
Model & SDR & SIR & SAR & PES & EPS  \\
\hline\hline
U-Net                     & 4.05  & 11.40    &  5.32   &  -42.44   &   -64.84   \\
\hline
\hline
$\mathit{W}_\mathit{si}$  & \textbf{4.24}  & \textbf{11.78}    & 5.38   &  \textbf{-49.44}    &  -65.47    \\
\hline
$\mathit{W}_\mathit{co}$  & \textbf{4.24}  & \textbf{12.72}    & 5.15   & \textbf{-59.53}    & -63.46     \\
\hline
\hline
$\mathit{S}_\mathit{fv}$  & 4.04    & \textbf{12.14}  & 5.13   & \circletext{\textbf{-59.68}}   & -61.73    \\
$\mathit{S}_\mathit{fv*}$ & \textbf{4.27}    & \textbf{12.42}   & 5.26    & \textbf{-54.16}   & -64.56     \\
\hline
$\mathit{S}_\mathit{cs}$  & \textbf{4.36}    & \textbf{12.47}   & 5.34    & \textbf{-57.11}   & -65.48   \\
$\mathit{S}_\mathit{cs*}$ & \textbf{4.32}  & \textbf{12.86}    &  5.15    &  \textbf{-54.27}   &  \textbf{-66.35}    \\
\hline
$\mathit{S}_\mathit{fs}$  & 4.10    & 11.40   & 5.24    & \-47.75   & -62.76    \\
$\mathit{S}_\mathit{fs*}$ & 4.21       & \circletext{\textbf{13.13}}    & 5.05    & \textbf{-48.75}    &  \circletext{\textbf{-72.40}}     \\
\hline
$\mathit{S}_\mathit{rs}$  & \textbf{\circletext{4.45}}    & \textbf{11.87}  &  \circletext{\textbf{5.52}}   & \textbf{-51.76}   & -63.44 \\
$\mathit{S}_\mathit{rs*}$ & \textbf{4.26}      & \textbf{12.80}    &  5.25   & \textbf{-57.37}    &   -65.62  \\
\end{tabular}
\caption[Conditioning vocal source separaiton results]{Median performance in dB of the different models on the \gls{dali} test set.
In bold are the results that significantly improve over the \gls{unet} ($p < 0.001$) and inside the circles the best results for each metric.
}
\label{table:results}
\end{table}

\begin{figure*}[t]
  \centerline{
    \includegraphics[width=\textwidth]{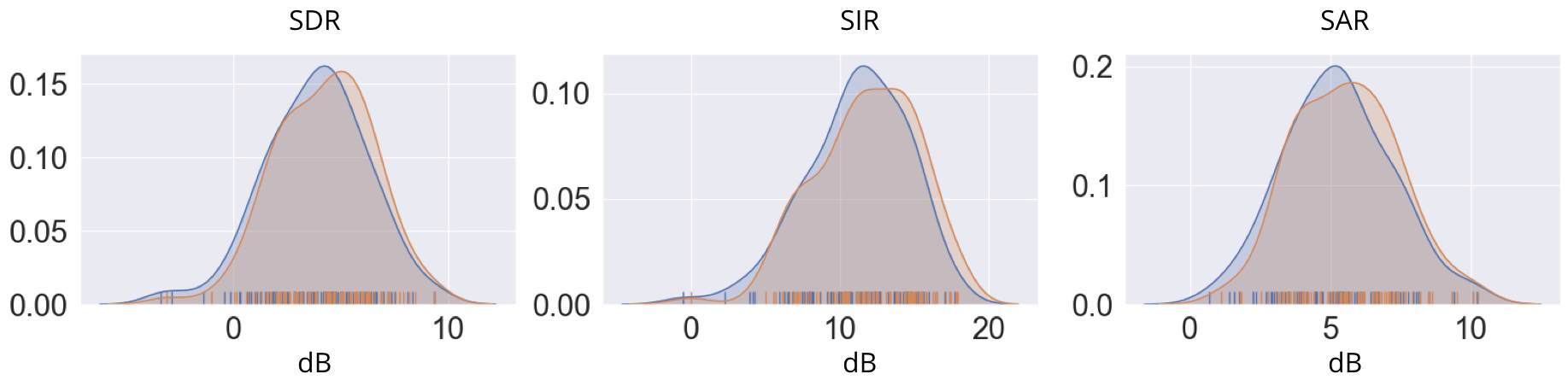}}
  \caption[Distribution of scores for the the standard U-Net and $\mathit{S}_\mathit{rs}$]{Distribution of scores for the the standard \gls{unet} (Blue) and $\mathit{S}_\mathit{rs}$ (Orange).}
  \label{fig:comparing}
\end{figure*}

We report the median source separation metrics (SDR, SAR, SIR, PES, ESP) in Table~\ref{table:results}. To measure the significance of the improvement differences between the results, we performed a paired t-test between each conditioning model and the standard \gls{unet} architecture, the baseline. This test measures ($p$-value) if the differences could have happened by chance. A low $p$-value indicates that data did not occur by chance.
As expected, the improvement is consistent over most of the proposed methods.
The statistical significance ($p<0.001$) for the SDR, SIR, and PES is generalized except for the versions where the basis tensors have a `frequency' $H$ dimension.
This is an expected result since when singing, the same phoneme can be sung at different frequencies (appearing at many frequency positions in the feature maps). Hence, these systems have difficulties to find generic basis tensors.
This also explains why the `bottleneck' versions (for both $\mathit{S}_\mathit{fs*}$ and $\mathit{S}_\mathit{fv*}$) outperforms the `complete' while this is not the case for the other versions.
Most versions also improve the performance on silent vocal frames with a much lower PES.
However, there is no difference in predicting silence at the right time (same EPS).
The only metric that does not consistently improve is SAR, which measures the algorithmic artifacts introduced in the process.
Our conditioning mechanisms can not reduce the artifacts that seem more dependent on the quality of the training examples (it is the metric which has the highest improvement in the data augmentation experiment Table~\ref{table:data_aug}).
Figure \ref{fig:comparing} shows a comparison with the distribution of SDR, SIR, and SAR for the best model $\mathit{S}_\mathit{rs}$ and the \gls{unet}. We can see how the distributions move toward higher values.

One relevant remark is the fact that we can effectively control the network with just a few parameters.
$\mathit{S}_\mathit{rs}$ just adds $480$ (or just $80$! for $\mathit{S}_\mathit{rs*}$) new learnable parameters and have significantly better performance than $\mathit{S}_\mathit{fv}$ that adds $1.97 \cdot 10^6$ parameters.
We believe that the more complex control mechanisms tend to find complex basis tensors that do not generalize well. In our case, it is more effective to perform a simple global transformation.
In the case of weak conditioning, both models behave similarly although $\mathit{W}_\mathit{si}$ has $1.955 \cdot 10^6$ fewer parameters than $\mathit{W}_\mathit{co}$.
This seems to indicate that controlling channels is not particularly relevant.

Regarding the different types of conditioning, when repeating the paired t-test between weak and strong models only $\mathit{S}_\mathit{rs}$ outperforms the weak systems.
We believe that strong conditioning can lead to higher improvements but several issues need to be addressed. First of all, there are missalignments in the annotations that force the system to perform an unnecessary operation which damages the computation. This is one of the possible explanations of why models with fewer parameters perform better. They are forced to find more generic conditions. The weak conditioning models are robust to these problems since they process $z$ and compute an optimal modification for a whole input patch (11 seconds). We also need to disambiguate the phonemes inside words. Currently, a word is described as a bag of phonemes that occur at the same time. This prevents strong conditioning models to learn properly the phonemes in isolation, instead, they consider them jointly with the other phonemes.

\section{Conclusions}

The goal of this chapter was to improve singing voice separation using the prior knowledge defined by the phonetic characteristics.
We use the phoneme activation as side information and show that it helps in the separation.

In future works, we intend to use other prior aligned knowledge such as vocal notes or characters also defined in \gls{dali}.
Regarding the conditioning approach and since it is transparent to the conditioned network, we are determined to explore recent state-of-the-art source separation methods such as Conv-Tasnet\citep{Luo_2019}.
The current formalization of the two basis tensors $\gamma_d$ and $\beta_d$ does not depend on any external factor.
A way to exploit the complex control mechanisms is to make these basis tensors dependent on the input mixture $x$ which may add additional flexibility.
Finally, we plan to jointly learn how to infer the alignment and perform the separation~\citep{Schulze_2020, Takahashi_2020}.

The general idea of lyrics-informed source separation leaves room for many possible extensions.
The formalization we presented relies on time-aligned lyrics which is not the real-world scenario.
Features similar to the phoneme activation~\citep{vaglio_2020, Stoller_2019} can replace them or be used to align the lyrics as a pre-processing step. These two options would allow adapting the current system to the real-world scenario.
These feature can also help in properly placing and desambiaguating the phonemes of a word to improve the current annotations.

%

\chapter{Conclusions and Future Work}
\label{sec:con_fut}

\section{Summary of Contributions}

\Gls{multimodal} learning is a growing and exciting field in the \gls{mir} community that aims to understand music through its various facets.
Since the earlier works, researchers have investigated and developed \gls{multimodal} learning systems.
Almost every \gls{mir} task can be formalized in a \gls{multimodal} set up, showing that integrating the natural multidimensionality of music can improve the results~\citep{essid_2012, simonetta_2019}.
Nevertheless, it still grows at a much slower rate than other topics in the community.
We hypothesize that the lack of development stems from (1) the lack of good quality \gls{multimodal} datasets and (2) the costly process of integrating data from different domains (see chapter~\ref{sec:c_intro}).

Apart from the obvious audio signal, music can be expressed in many different dimensions such as scores, videos, or motion.
Among them, text is one of the most used information.
It provides many different and complementary sources of information. From editorial reviews that describe music, social web content that provides user interactions to metadata-tags that group music in categories, or ratings that classify it from a subjective point of view, whether someone likes it or not.
All these sources add many useful insights not only about music but also about how we interact to it.
In this thesis, we focused on lyrics because they have a direct connection to the audio signal via the singing voice.
The singing voice acts as a musical instrument and at the same time conveys semantic meaning through the lyrics, adding a linguistic dimension that complements the abstraction of the musical instruments.
Additionally, singing voice tasks (e.g. lyrics or note transcription) are particularly challenging given its variety of timbre and expressive versatility. Getting access to complementary information helps in developing better systems.

During the development of our work we have investigated different \gls{multimodal} strategies and problem formalizations.
To summarize them, we outlined the four questions that guide this dissertation in Chapter\ref{sec:c_intro}. In this section, we will outline our answers to each of those questions.

\subsubsection{How can we obtain large amounts of labeled data where lyrics and its melodic representation aligned in time with the audio to train data-driven methods?}
In Chapter~\ref{sec:dali_description}, we highlighted the need for more \gls{multimodal} datasets with lyrics and notes aligned in time, reviewing the available datasets.  We saw that prior datasets were very small, inconsistent regarding the levels of alignment and without note annotations. To address this, we released the \gls{dali} dataset that contains lyrics and vocal notes aligned in time at different levels of granularity and includes a multitrack subset with separated vocal and accompaniment separated. \gls{dali} remains the largest, most complete dataset with lyrics and vocal notes aligned in time annotations to date.
We outlined the different versions and the characteristics of the data, formally defined the dataset and introduced the developed tools that help us to work with this complex data.

In Chapter~\ref{sec:dali_creation}, we detailed our methodology to create \gls{dali} automatically. We leverage data from the Internet and integrate active learning and weakly-supervised learning techniques in the process.
Non-expert karaoke time-aligned lyrics and notes describe the lyrics as a sequence of time-aligned notes with their associated textual information.
We then linked each annotation to the correct audio and globally aligned the annotations to it using the normalized cross-correlation on the voice annotation sequence and the singing voice probability vector.
We used the \gls{teacher-student} to create an interaction between the dataset creation and model learning that benefits each other.
This helped us to improve our \gls{svd} system which allows a better selection and global alignment.

In Chapters~\ref{sec:structures} and \ref{sec:vunet} we used this data to train and evaluate a variety of data-driven methods for lyrics segmentation and source separation. The resulting approaches sparked several new research directions.

In summary, we can obtain large amounts of aligned lyrics and notes by collecting data from the Internet and developing methods that can filter and adapt the annotations to the audio. However, this process comes with a price and it is error prone.

\subsubsection{How can we automatically identify and solve errors in these labels?}
In Chapter~\ref{sec:dali_errors}, we deepened into the different types of labeling errors presented in \gls{dali}.
We proposed automatic solutions to global alignment issues with simple correlations. These are effective but limited.
We explored alignment techniques to solve different local alignment issues. We treated the annotations as `scores' employing the \emph{musical note events} labels and aligned them with the audio. However, there is not an efficient way of measuring if the new labels are better or worse than the original ones and we cannot properly measure the different proposed methods.

To address this, we proposed in Chapter~\ref{sec:dali_correction} a novel \gls{cleansing} technique that considers the time-varying structure of labels.
We train our model in a self-supervised way to automatically tell if a note label is correct for a given audio signal.
Our method exploits the local structure of the labels to find possible errors in vocal note event annotations.
We evaluated this at the frame level and obtained the \gls{epf} vector that measures the `quality' of the annotations.
We showed that this vector can be successfully applied to improve the performance of estimating the \gls{f0}, a downstream inference task.
We improved the Raw Pitch Accuracy by over $10$ percentage points simply by filtering the training dataset using our data cleansing model.
Our approach is particularly useful when training on very noisy datasets such as those collected from the Internet and automatically aligned.
We also used our proposed error detection model to estimate the error rate, knowing the current status of the dataset.

\subsubsection{How can we exploit the inherent relationships between lyrics and audio to improve the performance for lyrics segmentation?}

In Chapter~\ref{sec:structures}, we used \gls{dali} for improving lyrics segmentation using a model-agnostic and early fusion approach that integrates the audio and text-domain.
As pre-processing step, we created a coordinate representation that results in \glspl{ssm} of the same dimensions for both domains. This allowed us an easy adaptation of an existing model to capture the complementary structure of both domains.
Through experiments, we showed that the \gls{multimodal} system outperforms the previous existing model based only on text.

\subsubsection{How can we effectively control data-driven models by context information? Can we then use the prior knowledge about the audio signal defined by the lyrics to improve the isolation of the singing voice from the mixture?}

We centered our effort in the task of music \gls{source_separation} that aims to isolate the different instruments that appear in an audio mixture.
Concretely, we adapted data-driven approaches to be controlled by some external information.
In this paradigm, we have access to the mixture and the isolated sources that compose it. Thus, models learn in a \gls{sl} way to either compute a mask to isolate the source from the background or to obtain directly the clean spectral representations.

Our approach consisted of a model-based system that uses external information to condition the behavior of a generic model.
We presented a novel approach where the conditioning is implemented using \gls{film}, a well-known conditioning approach that comes from the image processing domain.
Chapter~\ref{sec:cunet} provides a first approximation to \gls{conditioning} for music source separation.
We explored the multitask source separation and used a weak conditioning system where we used the prior knowledge about the different instruments to perform several instrument separations with a single model without losing performance and adding just a small number of parameters.
This showed that we can effectively control a generic neural network by some external information. In this case for performing several instrument isolations.

In Chapter~\ref{sec:vunet}, we extended this approach to singing vocal \gls{source_separation} in an informed-source separation scenario, where we aimed to use prior information about the target source to improve separation.
Although previous approaches employed the prior information provided by the scores, we explored the phonetic characteristics.
Looking for combining the power of data-driven models with the adaptability of informed approaches, we used the multitrack version of \gls{dali} and the phonetic information defined in the annotations to introduce additional flexibility in our model and adapt it to the observed signals.
We used this information to improve vocal separations and proposed a weak and strong conditioning strategy that outperforms the standard architecture without external information.

Overall, we found that \gls{conditioning} was an effective method for controlling the behavior of a generic model. We showed that it is not necessary to introduce complex architectures and just with a few but precise parameters we can improve the performances of the current architectures.

\section{Future work}

There is a large number of interesting new research directions hinted at in this work. We group them according to the different topics treated in this dissertation.

\subsection{Dataset}
In this work, we created two different versions of the \gls{dali} dataset.
We plan also to extend the data contained in the annotations. Currently \gls{dali} has \emph{musical note event} labels where a note event consists of a start time, end time and pitch. This is quantized at the note level but it does not contain any information about the \gls{f0} i.e. the frequency evolution of notes in the audio signal. This can be added by separating the vocals and tracking the \gls{f0} in a similar way to what was done in \gls{medley}.
Another area of improvement is the phoneme information, which was extracted automatically from the word level. There are two directions. First of all, this process is error prone in both word annotations during the labelling and the automatic phonetization. Further metrics to estimate the quality of these annotations are needed. Secondly, the phonemes of a word are represented as a bag of phonemes that occurr at the same time without explict onsets or offsets rather than as a succession of phonemes. Providing detailed alignment for individual phonemes will benefit to the strong conditioning source separation and lead to better results. This will also be helpful for the work detailed in Section~\ref{sec:approaching}.

Concerning the \gls{epf}, we believe the error detection model could be applied to scenarios other than training. A natural future work is to integrate it as a guide or objective measure to evaluate the performance of the different proposed methods for solving the local errors (see Chapter~\ref{sec:dali_correction}). This will allow the evaluation of many different alignment configurations as well as exploring ways of combining the different information (text and notes) to create a more robust error solving system.

Regarding the proposed \gls{cleansing} strategy itself, we plan to directly apply it to any kind of note event annotation (not only frame level), as well as extend it for other types of time-varying annotations such as chords or beats.
We can also use it to streamline manual annotation efforts by using the model to select time regions that are likely wrong and send them to an expert for correction.
We would also like to explore how this idea can be generalized to other domains beyond music and to test the contribution of different factors including the amount of noise in a dataset and the nature of the noise.

Finally, the multitrack subset contains the instruments that define the accompaniment section. Cleaning and preparing these instruments will help in future work related to multitask source separation. Additionally, the \emph{musical note event} labels are only employed for creating the \gls{epf}. Scores are also one of the most important representation of music. There is a wide area of research integrating this information in the current setup.

\subsection{Structures}
We only explored a scenario in which the audio information helped to improve the structure segmentation of text information.
The first and most clear extension of this work would be to use the text information to help in detecting musical structures.
The hierarchical information defined by the different levels of granularity would (1) help in disambiguating many of the inconsistencies (meaning having difficulties in selecting when to segment) presented in music structures detection systems, (2) define hierarchical relationships between musical motives (lyrics lines define melodic lines) and (3) add a semantic dimension to each structure i.e. being able to label each section and analyze the topic finding correlations between the evolution of the lyrics and the music, various lyric sections define musical structures: verses reveal stories and choruses sum up the emotional message.
This \gls{multimodal} formalization of the structure analysis will help in improving the understanding that methods have about what a song is.

\subsection{Source separation}

The \gls{cunet} model was tested only for four instruments. We plan to extend it to more instruments to explore its limitations. We can achieve this after cleaning and preparing the other tracks presented in the multitrack version of \gls{dali}.
We would like also to explore the performance for complex tasks i.e. separating several instruments combinations (e.g. vocals+drums).
Regarding \gls{film} layers, we plan to explore how they affect the computation to deeply understand their behavior to use it more effectively. For instance, we would like to visualize the latent spaces and the encoder blocks to better understand where to apply the conditioning.

As we mentioned when exploring vocal isolation, the strong conditioning worked better when the basis have limited flexibility, especially when the same value was applied to all the frequency dimension. To overcome this limitation, we would like to make the bases dependent on other data such as the notes annotations defined in \gls{dali}.
Additionally, we are not considering possible errors in the alignment that may lead to errors in the separation. We would like to integrate attention mechanisms to add further flexibility to the model and be robust to these issues.
As mentioned in the previous sections, once we have access to the explicit alignment per phoneme rather than per word, the system will be able to see each phoneme in isolation which will help in the performance.
An interesting line of work is to adapt the model to directly obtain the phoneme activations matrix from the mixture so that it will not need it for facing the real-world scenario.

We also would like to integrate our conditioning mechanism to other architectures such as Wave-U-Net or TASNet.
Likewise, we aim to develop ways of adding new conditions (namely new instrument isolation) to a trained model and find ways of separating the joint training, to create a generic model that can be easily adapted to new control mechanisms.

\subsection{Approaching the multimodal scenario}
\label{sec:approaching}
A line of research not explored in this work is how to use \gls{dali} to create systems that automatically generate \gls{multimodal} data i.e. lyrics and vocal notes aligned in time. Tasks such as automatic lyric alignment, singing to text/notes transcription can greatly benefit from this data. Recently, there have been many advances in these fields such as the use of alignment losses like Connectionist temporal classification (CTC), the improvement of \gls{f0} methods and speech transcription methods that make us think that soon we will be able to automatically obtain aligned lyrics or notes to formulate \gls{multimodal} models that do not depend on having this data.

Additionally, we can formulate tasks such as \gls{source_separation} to perform both tasks, improving the separation as well as learning how to create features that capture the phonetic information over time.

\subsection{Other multimodal scenarios}
Finally, there are many other \gls{mir} tasks that can benefit from the insights we can derive from such datasets.
Problems like cover detection, genre classification, or mood estimation are directly connected with the lyrics.
In covers, the vocal melody and the lyrics are usually some of the main elements that remain from the original song.
The topics of lyrics are highly related to music genres even defining concept albums where tracks are part of a single central narrative that holds a collective meaning.
Additionally, in popular music, lyrics and music work together for conveying defined emotions which make mood estimation another natural \gls{multimodal} scenario to explore.

Until now we have talked only about discriminative approaches where we aim to label frames or the song as a whole. However, there is a vast range of applications that can be investigated in the generative field. For example, exploring ways of automatically generating lyrics given a particular audio melody or the other way round, to provide vocal melodies given lyrics.

\section{Conclusions}

In this dissertation, we faced music \gls{multimodal} learning as a whole. We developed novel methods for automatically creating a large dataset that fitted our needs and evaluating the quality of the annotations. We presented several strategies that combined different \gls{multimodal} formalizations as well as made use of the different dimensions and levels of hierarchy presented in our dataset.
The body of work presented in this dissertation can be summarized as follows.
We showed that data creation and model learning can work together and benefit each other. We further developed ways of dealing with nosiy data.
We then saw how \gls{multimodal} data-driven methods can exploit inherent relationships between domains and were able to outperform existing models based on a single modality. We have also integrated conditioning mechanisms for effectively controlling standard architectures with respect to some external information, verifying that we can control them with just a few but effective parameters.

We hope the results and methods proposed in this thesis prompt novel research into \gls{multimodal} learning methods for efficiently and effectively developing new work and encourage researchers to continue in this direction.

 \appendix
 \printglossary[type=\acronymtype]
 \printglossary[type=main]
 \chapter{Publications and code}

\section{Publications}

\begin{enumerate}
  \item Gabriel Meseguer-Brocal, Alice Cohen-Hadria and Geoffroy Peeters. \textbf{\textit{DALI: a large Dataset of synchronized Audio, LyrIcs and notes, automatically created using teacher-student machine learning paradigm}}. In 19th International Society for Music Information Retrieval (ISMIR) Conference, 2018

  \item Gabriel Meseguer-Brocal and Geoffroy Peeters. \textbf{\textit{Conditioned-U-Net: Introducing a Control Mechanism in the U-Net for Multiple Source Separations}}. In 20th International Society for Music Information Retrieval (ISMIR) Conference, 2019

  \item Gabriel Meseguer-Brocal, Alice Cohen-Hadria, Geoffroy Peeters. \textbf{\textit{Creating DALI, a large dataset of synchronized audio, lyrics, and notes}}. Transactions of the International Society for Music Information Retrieval Journal (TISMIR).

  \item Michael Fell E, Yaroslav Nechaev, Gabriel Meseguer-Brocal, Elena Cabrio, Fabien Gandon and Geoffroy Peeters. \textbf{\textit{Lyrics Segmentation via Bimodal Text-audio Representation}}.  Natural Language Engineering.

  \item Gabriel Meseguer-Brocal and Geoffroy Peeters. \textbf{\textit{Content based singing voice source separation via strong conditioning using aligned phonemes}}. In 21th International Society for Music Information Retrieval (ISMIR) Conference, 2020.

  \item Gabriel Meseguer-Brocal, Rachel Bittner, Simon Duran, Brian Brost. \textbf{\textit{Data Cleansing with Contrastive Learning for Vocal Note Event Annotations}}. In 21th International Society for Music Information Retrieval (ISMIR) Conference, 2020.
\end{enumerate}

\section{Code}
During the course of this thesis, a great among of tools have been developed.
Most of them are publicly available at my GitHub: \url{https://github.com/gabolsgabs}

This includes the package for working with \gls{dali} as well as all the necessary functions to reproduce the methods from scratch.
This covers preprocessing the data, the creation of the pipelines embedded in the graph for efficiently handling the data, the creation of the different architectures, the training process as well as the evaluation.
The code has been developed using {\tt Python}. The main toolbox employed is \textbf{librosa}~\citep{mcfee_2015} and \textbf{mir\_eval}~\citep{Raffel_2014}.
Deep neural networks have been exploited using \textbf{TensorFlow}~\citep{abadi_2016}.
Each project is self-contained and can be installed used {\tt pip}.


 \bibliographystyle{apalike} 
 \bibliography{main}

\end{document}